%% file: Manuskript.tex
\documentclass[a4paper,twoside,11pt]{scrbook}
\usepackage{a4}
\usepackage[ngerman,english]{babel}
\usepackage{multirow}
\usepackage{graphicx}
\usepackage{amsmath}
\usepackage{exscale}
\usepackage{float}
\usepackage{here}
\usepackage{textcomp}
\usepackage{epsfig}

\begin{document}

\include{Title}

\clearpage
\thispagestyle{empty}
\setlength{\parindent}{0.0cm}

\selectlanguage{english}
\tableofcontents{}

\pagestyle{headings}

\setlength{\parskip}{6pt} 

\include{Chapter_1}

\include{Chapter_2}

\include{Chapter_3}

\include{Chapter_4} 

\include{Chapter_5}

\include{Chapter_6}

\include{Chapter_7}

\include{Bibliography}

\end{document}

%% file: Title.tex
\begin{titlepage}
\setlength{\topskip}{1cm}
\begin{center}
\begin{huge}
\vspace{1cm}
\begin{bf} 

Relativistic Nucleus-Nucleus

Collisions and the QCD Matter 

Phase Diagram

\end{bf} 
\end{huge}
\end{center}
\vspace{1.5cm}
\begin{center}
\begin{large}
Reinhard Stock, Physics Department, University of Frankfurt

\vspace{15cm}

Dedicated to Andr\'{e}s Sandoval at the occasion of his 60th birthday
\end{large}
\end{center}
\end{titlepage}

%% file: Chapter_1.tex
\chapter{Introduction}
\label{chap:Introduction} 

\section{Overview}
\label{sec:Overview}
This review will be concerned with our knowledge of extended matter
under the governance of strong interaction, in short: QCD matter.
Strictly speaking, the hadrons are representing the first layer of
extended QCD architecture. In fact we encounter the characteristic
phenomena of confinement as distances grow to the scale of 1 $fm$
(i.e. hadron size): loss of the chiral symmetry property of the
elementary QCD Lagrangian via non-perturbative generation of
''massive'' quark and gluon condensates, that replace the bare QCD
vacuum \cite{1}. However, given such first experiences of transition
from short range perturbative QCD phenomena (jet physics etc.),
toward extended, non perturbative QCD hadron structure, we shall
proceed here to systems with dimensions far exceeding the force
range: matter in the interior of heavy nuclei, or in neutron stars,
and primordial matter in the cosmological era from electro-weak
decoupling ($10^{-12}$ s) to hadron formation (0.5 $\cdot
10^{-5}s$). This primordial matter, prior to hadronization, should
 be deconfined in its QCD sector, forming a plasma (i.e. color
conducting) state of quarks and gluons \cite{2}: the Quark Gluon
Plasma (QGP). 

In order to recreate matter at the corresponding high
energy density in the terrestial laboratory one collides heavy
nuclei (also called ``heavy ions") at ultrarelativistic energies. Quantum Chromodynamics
predicts \cite{2,3,4} a phase transformation to occur between
deconfined quarks and confined hadrons. At near-zero net baryon
density (corresponding to big bang conditions) non-perturbative
Lattice-QCD places this transition at an energy density of about $1
\: GeV/fm^3$, and at a critical temperature, $T_{crit}$ $\approx$ $170 \: MeV$
\cite{4,5,6,7,8} (see the article on Lattice QCD in this Volume). The ultimate goal of the physics with
ultrarelativistic heavy ions is to locate this transition, elaborate
its properties, and gain insight into the detailed nature of the
deconfined QGP phase that should exist above. What is meant by the
term ''ultrarelativistic'' is defined by the requirement that the
reaction dynamics reaches or exceeds the critical density $\epsilon
\approx 1 \: GeV/fm^3$. Required beam energies turn out \cite{8} to
be $\sqrt{s} \ge 10\: GeV$, and various experimental programs have
been carried out or are being prepared at the CERN SPS (up to about
$20 \: GeV$), at the BNL RHIC collider (up to $200 \:GeV)$ and
finally reaching up to $5.5 \:TeV$ at the LHC of CERN.

QCD confinement-deconfinement is of course not limited to the domain
that is relevant to cosmological expansion dynamics, at very
small excess of baryon over anti-baryon number density and, thus,
near zero baryo-chemical potential $\mu_B$. In fact, modern QCD
suggests \cite{9,10,11} a detailed phase diagram of QCD matter and
its states, in the plane of $T$ and baryo-chemical potential $\mu_B$. 
For a map of the QCD matter phase diagram we are thus employing the terminology of the grand canonical Gibbs ensemble that describes an extended volume $V$ of partonic or hadronic matter at temperature $T$. 
In it, total particle number is not conserved at relativistic energy, due to particle production-annihilation processes occurring at the microscopic level. However, the probability distributions (partition functions)
describing the relative particle species abundances have to respect the presence of certain, to be conserved net quantum numbers ($i$), notably non-zero net baryon number and zero net strangeness and charm.
Their global conservation is achieved by a thermodynamic trick, adding to the system Lagrangian a so-called Lagrange multiplier term, for each of such quantum number conservation tasks. This procedure enters a "chemical potential" $\mu_i$ that modifies the partition function via an extra term $\exp{\left(-\mu_i/T\right)}$ occuring in the phase space integral (see chapter~\ref{chap:hadronization} for detail). It modifies the canonical "punishment factor" ($\exp{\left(-E/T\right)}$), where
E is the total particle energy in vacuum, to arrive at an analogous grand canonical factor for the extended medium,of $\exp{\left(-E/T - \mu_i/T\right)}$. 
This concept is of prime importance for a description of the state of matter created in heavy ion collisions, where net-baryon number (valence quarks) carrying objects are considered --- extended "fireballs" of QCD matter.
The same applies to the matter in the interior of neutron stars. The corresponding conservation of net baryon number is introduced into the grand canonical statistical model of QCD matter via the "baryo-chemical potential" $\mu_B$.

\begin{figure}[h]   
\begin{center}
\includegraphics[scale=0.5]{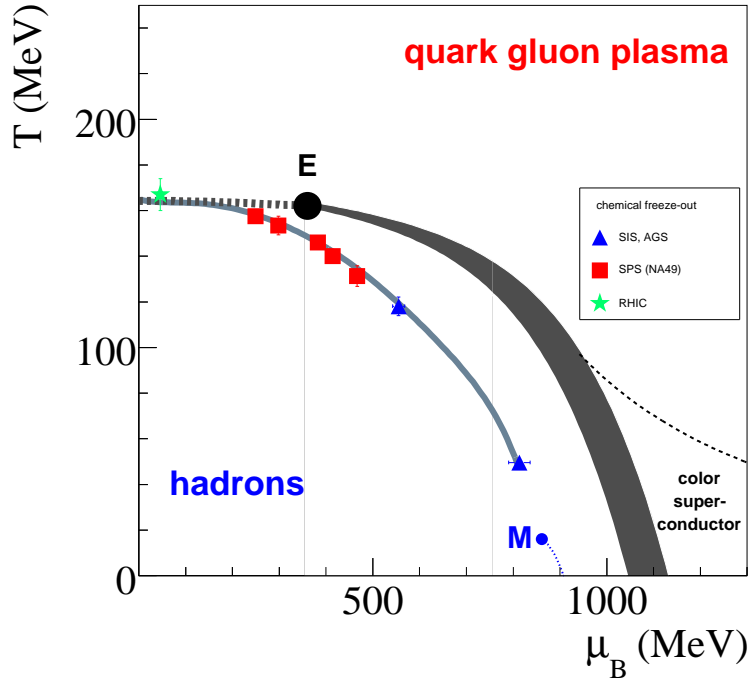}
\caption{Sketch of the QCD matter phase diagram in the plane of temperature
$T$ and baryo-chemical potential $\mu_B$. The parton-hadron phase
transition line from lattice QCD  \cite{8,9,10,11} ends in a
critical point $E$. A cross-over transition occurs at smaller $\mu_B$. Also shown are the points of hadro-chemical
freeze-out from the grand canonical statistical model.}
\label{fig:Figure1}
\end{center}
\end{figure}

We employ this terminology to draw a phase diagram of QCD matter in figure~\ref{fig:Figure1}, in the variables $T$ and $\mu_B$. Note that $\mu_B$ is high at low energies of collisions creating a matter fireball. In a head-on collision
of two mass 200 nuclei at $\sqrt{s}=15 GeV$ the fireball contains about
equal numbers of newly created quark-antiquark pairs (of zero net baryon number), and of initial valence quarks. The accomodation of the latter,
into created hadronic species, thus requires a formidable redistribution task of net baryon number, reflecting in a high value of $\mu_B$. Conversely, at LHC energy (5.5TeV for Pb+Pb collisions), the initial valence quarks constitute a mere 5\% fraction of the total quark density, correspondingly requiring a small value of $\mu_B$. In the extreme, big bang matter evolves toward hadronization (at $T$=170 MeV) featuring a quark over antiquark density excess of $10^{-9}$ only, resulting in $\mu_B \approx 0$.

Note that the limits of
existence of the hadronic phase are not only reached by temperature
increase, to the so-called Hagedorn value $T_H$ (which coincides
with $T_{crit}$ at $\mu_B \rightarrow 0$), but also by density
increase to $\varrho > (5-10)\: \varrho_0$: ''cold compression''
beyond the nuclear matter ground state baryon density $\varrho_0$ of
about 0.16 $B/fm^3$. We are talking about the deep interior sections
of neutron stars or about neutron star mergers \cite{12,13,14}. A
sketch of the present view of the QCD phase diagram \cite{9,10,11}
is given in Fig.~\ref{fig:Figure1}. It is dominated by the parton-hadron phase
transition line that interpolates smoothly between the extremes of
predominant matter heating (high $T$, low $\mu_B$) and predominant
matter compression ($T \rightarrow 0, \: \mu_B > 1 \: GeV$). Onward
from the latter conditions, the transition is expected to be of
first order \cite{15} until the critical point of QCD matter is
reached at $200 \le \mu_B \: (E)\: \le 500 \: MeV$. The relatively
large position uncertainty reflects the preliminary character of
Lattice QCD calculations at finite $\mu_B$ \cite{9,10,11}. Onward from the critical point, E, the phase transformation at lower $\mu_B$ is a cross-over\cite{11}.\\
 \\
We note, however, that these estimates represent a major recent
advance of lattice theory which was, for two decades, believed to be
restricted to the $\mu_B=0$ situation. Onward from the critical
point, toward lower $\mu_B$, the phase transformation should acquire
the properties of a rapid cross-over \cite{16}, thus also including
the case of primordial cosmological expansion. This would finally
rule out former ideas, based on the picture of a violent first order
''explosive'' cosmological hadronization phase transition, that
might have caused non-homogeneous conditions, prevailing during
early nucleo-synthesis \cite{17}, and fluctuations of global matter
distribution density that could have served as seedlings of galactic
cluster formation \cite{18}. However, it needs to be stressed that
the conjectured order of phase transformation, occuring along the
parton - hadron phase boundary line, has not been unambiguously
confirmed by experiment, as of now.

On the other hand, the {\it position} of the QCD phase boundary at
low $\mu_B$ has, in fact, been located by the hadronization points
in the $T, \: \mu_B$ plane that are also illustrated in Fig.~\ref{fig:Figure1}. They
are obtained from statistical model analysis \cite{19} of the
various hadron multiplicities created in nucleus-nucleus collisions, which results in a [$T, \: \mu_B$]
determination at each incident energy, which ranges from SIS
via AGS and SPS to RHIC energies, i.e. $3\le \sqrt{s} \le 200 \:
GeV$. Toward low $\mu_B$ these hadronic freeze-out points merge with
the lattice QCD parton-hadron coexistence line: hadron formation
coincides with hadronic species freeze-out. These points also
indicate the $\mu_B$ domain of the phase diagram which is accessible
to relativistic nuclear collisions. The domain at $\mu_B \ge 1.5 \:
GeV$ which is predicted to be in a further new phase of QCD
featuring color-flavor locking and color superconductivity \cite{20}
will probably be accessible only to astrophysical observation.

One may wonder how states and phases of matter in
thermodynamical equilibrium - as implied by a description in grand
canonical variables - can be sampled via the dynamical evolution of
relativistic nuclear collisions. Employing heavy nuclei, $A  \approx
200$, as projectiles/targets or in colliding beams (RHIC, LHC),
transverse dimensions of the primordial interaction volume do not
exceed about $8 \: fm$, and strong interaction ceases after about
$20 \: fm/c$. We shall devote an entire later section to the aspects
of equilibrium (\ref{subsec:Transvers_phase_space}) but note, for now, that the time and
dimension scale of primordial perturbative QCD interaction at the
microscopic partonic level amounts to subfractions of $1 \: fm/c$,
the latter scale, however, being representative of non perturbative
processes (confinement, ''string'' formation etc.). The A+A
 fireball size thus exceeds, by far, the elementary non perturbative scale. 
An equilibrium quark gluon plasma represents an extended non-perturbative
QCD object, and the question whether its relaxation time scale can be
provided by the expansion time scale of an A+A collision, needs
careful examination. Reassuringly, however, the hadrons that are
supposedly created from such a preceding non-perturbative QGP phase
at top SPS and RHIC energy, do in fact exhibit perfect hadrochemical
equilibrium, the derived [$T, \: \mu_B$] values \cite{19} thus
legitimately appearing in the phase diagram, Fig.~\ref{fig:Figure1}.

In the present review we will order the physics observables to be
treated, in sequence of their origin from successive stages that
characterize the overall dynamical evolution of a relativistic
nucleus-nucleus collision. In rough outline this evolution can be
seen to proceed in three major steps. An initial period of matter
compression and heating occurs in the course of interpenetration of
the projectile and target baryon density distributions. Inelastic
processes occuring at the microscopic level convert initial beam
longitudinal energy to new internal and transverse degrees of
freedom, by breaking up the initial baryon structure functions.
Their partons thus acquire virtual mass, populating transverse phase
space in the course of inelastic perturbative QCD shower
multiplication. This stage should be far from thermal equilibrium,
initially. However, in step two, inelastic interaction between the
two arising parton fields (opposing each other in longitudinal phase
space) should lead to a pile-up of partonic energy density centered
at mid-rapidity (the longitudinal coordinate of the overall center
of mass). Due to this mutual stopping down of the initial target and
projectile parton fragmentation showers, and from the concurrent
decrease of parton virtuality (with decreasing average square
momentum transfer $Q^2$) there results a slowdown of the time scales
governing the dynamical evolution. Equilibrium could be approached
here, the system ''lands'' on the $T, \: \mu$ plane of Fig.~\ref{fig:Figure1}, at
temperatures of about 300 and $200 \: MeV$ at top RHIC and top SPS
energy, respectively. The third step, system expansion and decay,
thus occurs from well above the QCD parton-hadron boundary line.
Hadrons and hadronic resonances then form, which decouple swiftly
from further inelastic transmutation so that their yield ratios
become stationary (''frozen-out''). A final expansion period dilutes
the system to a degree such that strong interaction ceases all
together.

In order to verify in detail this qualitative overall model, and to
ascertain the existence (and to study the properties) of the
different states of QCD that are populated in sequence, one seeks
observable physics quantities that convey information imprinted
during distinct stages of the dynamical evolution, and
''freezing-out'' without significant obliteration by subsequent
stages. Ordered in sequence of their formation in the course of the
dynamics, the most relevant such observables are briefly
characterized below:
\begin{enumerate}
\item Suppression of $J/\Psi$ and $Y$ production by Debye-screening
in the QGP. These vector mesons result from primordial, pQCD
production of $c\overline{c}$ and $b\overline{b}$ pairs that would
hadronize unimpeded in elementary collisions but are broken up if
immersed into a npQCD deconfined QGP, at certain characteristic
temperature thresholds.
\item Suppression of dijets which arise from primordial
$q\overline{q}$ pair production fragmenting into partonic showers
(jets) in vacuum but being attenuated by QGP-medium induced gluonic
bremsstrahlung: Jet quenching in A+A collisions.
\begin{enumerate}
\item A variant of this: {\it any} primordial hard parton suffers a
high, specific loss of energy when traversing a deconfined medium:
High $p_T$ suppression in A+A collisions.
\end{enumerate}
\item Hydrodynamic collective motion develops with the onset of
(local) thermal equilibrium. It is created by partonic pressure
gradients that reflect the initial collisional impact geometry via
non-isotropies in particle emission called ''directed'' and
''elliptic'' flow. The latter reveals properties of the QGP, seen
here as an ideal partonic fluid.
\begin{enumerate}
\item Radial hydrodynamical expansion flow (''Hubble expansion'')
is a variant of the above that occurs in central, head on collisions
with cylinder symmetry, as a consequence of an isentropic expansion.
It should be sensitive to the mixed phase conditions characteristic
of a first order parton-hadron phase transition.
\end{enumerate}
\item Hadronic ''chemical'' freeze-out fixes the abundance ratios of
the hadronic species into an equilibrium distribution. Occuring very
close to, or at hadronization, it reveals the dynamical evolution
path in the [$T, \:\mu_B$] plane and determines the critical
temperature and density of QCD. The yield distributions in A+A
collisions show a dramatic strangeness enhancement effect,
characteristic of an extended QCD medium.
\item Fluctuations, from one collision event to another (and even
within a single given event) can be quantified in A+A collisions due
to the high charged hadron multiplicity density (of up to 600 per
rapidity unit at top RHIC energy). Such event-by-event (ebye)
fluctuations of pion rapidity density and mean transverse momentum
(event ''temperature''), as well as event-wise fluctuations of the
strange to non-strange hadron abundance ratio (may) reflect the
existence and position of the conjectured critical point of QCD
(Fig.~\ref{fig:Figure1}).
\item Two particle Bose-Einstein-Correlations are the analog of the
Hanbury-Brown, Twiss (HBT) effect of quantum optics. They result
from the last interaction experienced by hadrons, i.e. from the
global decoupling stage. Owing to a near isentropic hadronic
expansion they reveal information on the overall
space-time-development  of the ''fireball'' evolution.
\end{enumerate}

In an overall view the first group of observables (1 to 2a) is
anchored in established pQCD physics that is well known from
theoretical and experimental analysis of elementary collisions
($e^+e^-$ annihilation, $pp$ and $p\overline{p}$ data). In fact, the
first generation of high $Q^2$ baryon collisions, occuring at the
microscopic level in A+A collisions, should closely resemble such
processes. However, their primary partonic products do not escape
into pQCD vacuum but get attenuated by interaction with the
concurrently developing extended high density medium , thus serving
as diagnostic tracer probes of that state. The remaining observables
capture snapshots of the bulk matter medium itself. After initial
equilibration we may confront elliptic flow data with QCD during the
corresponding partonic phase of the dynamical evolution employing 
thermodynamic \cite{21} and hydrodynamic \cite{22}
models of a high temperature parton plasma. The hydro-model stays
applicable well into the hadronic phase. Hadron formation
(confinement) occurs in between these phases (at about 5
microseconds time in the cosmological evolution). In fact
relativistic nuclear collision data may help to finally pin down the
mechanism(s) of this fascinating QCD process \cite{23,24,25} as we
can vary the conditions of its occurence, along the parton-hadron
phase separation line of Fig.~\ref{fig:Figure1}, by proper choice of collisional
energy $\sqrt{s}$, and system size A, while maintaining the overall
conditions of an extended imbedding medium of high energy density
within which various patterns \cite{9,10,11,15,16} of the
hadronization phase transition may establish. The remaining physics
observables (3a, 5 and 6 above) essentially provide for auxiliary
information about the bulk matter system as it traverses (and
emerges from) the hadronization stage, with special emphasis placed
on manifestations of the conjectured critical point.

The present review will briefly cover each of the above physics
observables in a separate chapter, beginning with the phenomena of
confinement and hadronization (chapter~\ref{chap:hadronization}), then to turn to the preceding
primordial dynamics, e.g. to elliptical flow (chapter~\ref{chap:Elliptic_flow}), high $p_T$ and
jet quenching (chapter~\ref{chap:In_medium_high_pt}) and quarkonium suppression (chapter~\ref{chap:Vector_Mesons}) as well as
in-medium $\varrho$-meson ''melting''. We then turn to the late
period, with correlation and fluctuation studies (chapter~\ref{chap:Fluctuations}). We
conclude (chapter~\ref{chap:Conclusion}) with a summary, including an outlook to the future of the research field.

However, before turning to such specific observables we shall
continue this introductory chapter, with a look at the
origin, and earlier development of the ideas that have shaped this
field of research (section~\ref{sec:History}). Then we turn to a detailed
description of the overall dynamical evolution of relativistic
nucleus-nucleus collisions, and to the typical overall patterns
governing the final distributions in transverse and longitudinal
(rapidity) phase space (chapter~\ref{sec:Bulk_Hadron_Prod}). The aspects of an approach toward
equilibrium, at various stages of the dynamical evolution (which are
of key importance toward the intended elucidation of the QCD matter
phase diagram), will be considered, in particular.

\section {History}
\label{sec:History}
The search for the phase diagram of strongly interacting matter
arose in the 1960s, from a coincidence of ideas developing - at
first fairly independently - in nuclear and astrophysics. In fact,
the nuclear proton-neutron matter, a quantum liquid at $T=0$ and
energy density $\epsilon = 0.15 \: GeV/fm^3$, represents the ground
state of extended QCD matter. Of course, QCD was unknown during the
development of traditional nuclear physics, and the extended matter
aspects of nuclei - such as compressibility or the equation of
state, in general - did not receive much attention until the advent,
in the 1960s, of relativistic nuclear mean field theory, notably
s-matrix theory by Brueckner \cite{26} and the $\sigma$-model of
Walecka \cite{27}. These theories developed the novel view of ''{\it
infinite} nuclear matter'' structure, based on in-medium properties
of the constituent baryons that share parts of their vacuum mass and
surface structure with the surrounding, continuous field of
relativistic scalar and vector mesons. Most importantly, in the
light of subsequent development, these theories allowed for a
generalization away from ground state density and zero temperature.
Such developments turned out to be of key relevance for acute
nuclear astrophysics problems: the dynamics of type II super-novae
and the stability of neutron stars, which both required the relation
of pressure to density and temperature of hadronic matter, i.e. the
hadronic matter equation of state (EOS). H. A. Bethe, G. Brown and
their collaborators \cite{28} postulated that the final stages of
supernova collaps should evolve through the density interval $0.1
\le \varrho/\varrho_0 \le 5$ where $\varrho_0=0.16$ (baryons per
$fm^3$) is the nuclear matter ground state density, and a similar
domain was expected for the neutron star density variation from
surface to interior \cite{29}. It was clear that, at the highest
thus considered densities the EOS might soften due to strange hadron
production caused by increasing Fermi energy. However the field
theoretical models permitted no reliable extrapolation to such high
densities (which, in retrospect, are perhaps not reached in supernova
dynamics \cite{30}), and the experimental information concerning the
EOS from study of the giant monopole resonance - a collective
density oscillation also called ''breathing mode'' - covered only
the parabolic minimum occuring in the related function of energy vs.
density at $T=0$, $\varrho = \varrho_0$.

The situation changed around 1970 due to the prediction made by W.
Greiner and his collaborators \cite{31} that nucleus-nucleus
collisions, at relatively modest relativistic energies, would result
in shock compression. This mechanism promised to reach matter
densities far beyond those of a mere superposition (i.e.
$\varrho/\varrho_0 \le 2 \gamma$) of initial target and projectile
densities. Coinciding in time, the newly developed Bevalac
accelerator at LBL Berkeley offered projectiles up to $^{38}Ar$, at
just the required energies, $100 \: MeV \le E_{Lab}/nucleon \le 2 \:
GeV$. The field of ''relativistic heavy ion physics'' was born. The
topic was confronted, at first, with experimental methods available
both from nuclear and particle physics. It was shown that particle
production (here still restricted to pions and kaons) could indeed
be linked to the equation of state \cite{32} and that, even more
spectacularly, the entire ''fireball'' of mutually stopped hadrons
developed decay modes very closely resembling the initial
predictions of hydrodynamical shock flow modes \cite{33} which
directly link primordial pressure gradients with collective velocity
fields of matter streaming out, again governed by the
nuclear/hadronic matter EOS. Actually, both these statements do, in
fact, apply (mutatis mutandis) up to the present ultra-relativistic
energies (see chapters~\ref{sec:Bulk_Hadron_Prod},~\ref{chap:hadronization},~\ref{chap:Elliptic_flow}). However it turned out soon that the equation of state at low or even zero
temperature (as required in supernova and neutron star studies)
could only be obtained in a semi-empirical manner \cite{34}. The
reason: compression can, in such collisions, be only accomplished
along with temperature and entropy increase. In an ideal baryon gas
$exp(s/A) \propto T^{3/2}/\varrho$, i.e. $T^{3/2}$ will grow faster
than $\varrho$ in a non-isentropic compression. Thus the reaction
dynamics will be sensitive to various {\it isothermes} of the ground
state EOS  $P=f(\varrho, \: T=0)$ , staying at $T \gg 0$,
throughout, and, moreover, not at constant $T$. Thus a relativistic
dynamical mean field model is required in order to interactively
deduce the $T=0$ EOS from data \cite{34}. The EOS result thus
remains model dependent.

The ideas concerning creation of a quark gluon plasma arose almost
concurrent with the heavy ion shock compression proposal. In 1974 T.
D. Lee formulated the idea that the non-perturbative vacuum condensates could be
''melted down . . by distributing high energy or high nucleon
density over a relatively large volume'' \cite{35}. Collins and
Perry \cite{36} realized that the asymptotic freedom property of QCD
implies the existence of an ultra-hot form of matter with deconfined
quarks and gluons, an idea that gained wide recognition when S.
Weinberg \cite{37} proposed an asymptotic freedom phase at the
beginning of ''The first Three minutes''. In fact, this idea of
deconfinement by asymptotic freedom (with implied temperature of several GeV) was correct, but somewhat
besides the point, as everybody expected, likewise, that
deconfinement sets in right above the limiting hadron temperature of
R. Hagedorn \cite{38}, $T_H \approx 160 \: MeV$. A medium existing
down to that temperature would, however, feature an average momentum
square transfer $Q^2 < 1 \: GeV^2$, i.e. be far into the non
perturbative  domain, and very far from asymptotic freedom. Right
above the hadron to parton transition the ''quark gluon plasma'' (as
it was named by E. Shuryak \cite {39}) is not a weakly coupled ideal
pQCD gas as soon became obvious by Lattice QCD calculations for extended
matter \cite{40}. Seen in retrospect one obviously can not defend a
picture of point like quarks (with ''current'' masses) at $Q^2 \le
0.2 \: GeV^2$ where size scales of 0.5 to 1 $fm$ must play a
dominating role.

An analytic QCD description of deconfinement does not exist. For
heavy quarkonia, $c \overline{c} \: (J/\Psi)$ and $b \overline{b} \:
(Y)$ deconfinement in partonic matter, Matsui and Satz proposed
\cite{41} a Debye screening mechanism, caused by the high spatial
density of free color carriers, that removes the confining long
range potential as $T$ increases toward about $2 \: T_c$, an effect
reproduced by modern lattice QCD  \cite{42}. However, light hadron
deconfinement can not be understood with a non-relativistic
potential model. Such critical remarks not withstanding, we shall
demonstrate in chapters~\ref{chap:hadronization} to~\ref{chap:Vector_Mesons} that the very existence, 
and also crucial properties of the QGP can in fact be inferred from
experiment, and be confronted with corresponding predictions of
recent lattice QCD theory.

Our present level of an initial understanding of the phase diagram
of QCD matter (Fig.~\ref{fig:Figure1}), is the result of a steady development of both
experiment and theory, that began about three decades ago, deriving
initial momentum from the Bevalac physics at LBL which motivated -
along with the developing formulation of the quark gluon plasma
research goals - a succession of experimental facilities progressing
toward higher $\sqrt{s}$. Beginning with the AGS at BNL ($^{28}Si$
and $^{197}Au$ beams with $\sqrt{s} \le 5 \: GeV$), the next steps
were taken at the CERN SPS ($\sqrt{s}$ from 6 to $20 \: GeV; \:
^{16}O, \: ^{32}S, \: ^{208}Pb$ beams), and at the Relativistic
Heavy Ion Collider RHIC (the first facility constructed explicitly
for nuclear collisions) which offers beams of $^{64}Cu$ and
$^{197}Au$ at $20 \le \sqrt{s} \le 200 \: GeV$. A final, gigantic
step in energy will be taken 2008 with the CERN Large Hadron
Collider: $^{208}Pb$ beams at $\sqrt{s}= 5.5 \: TeV$.

\chapter {Bulk Hadron Production in A+A Collisions}
\label{sec:Bulk_Hadron_Prod}
\begin{figure}[htbp!]   
\begin{center}
\includegraphics[scale=0.375]{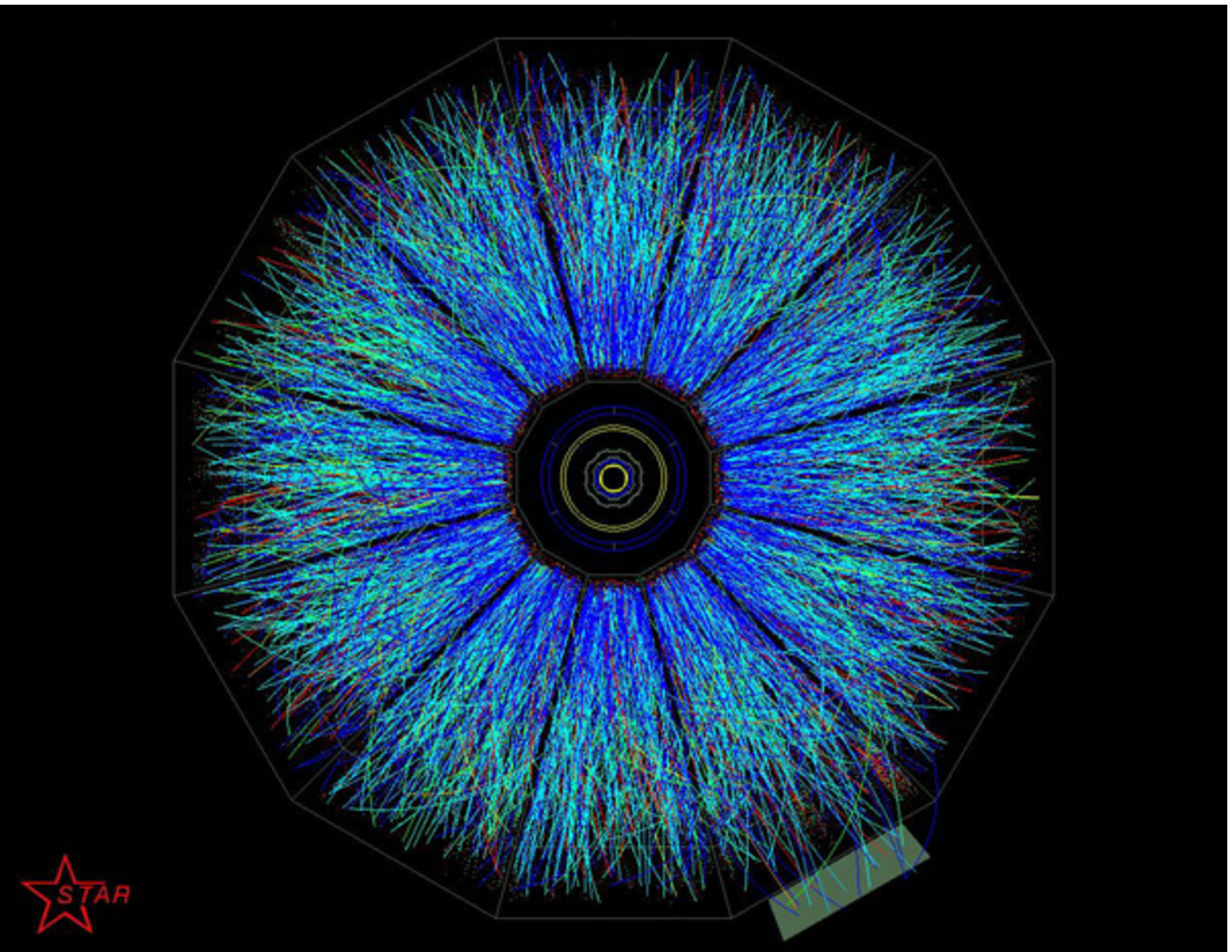}
\includegraphics[scale=0.77]{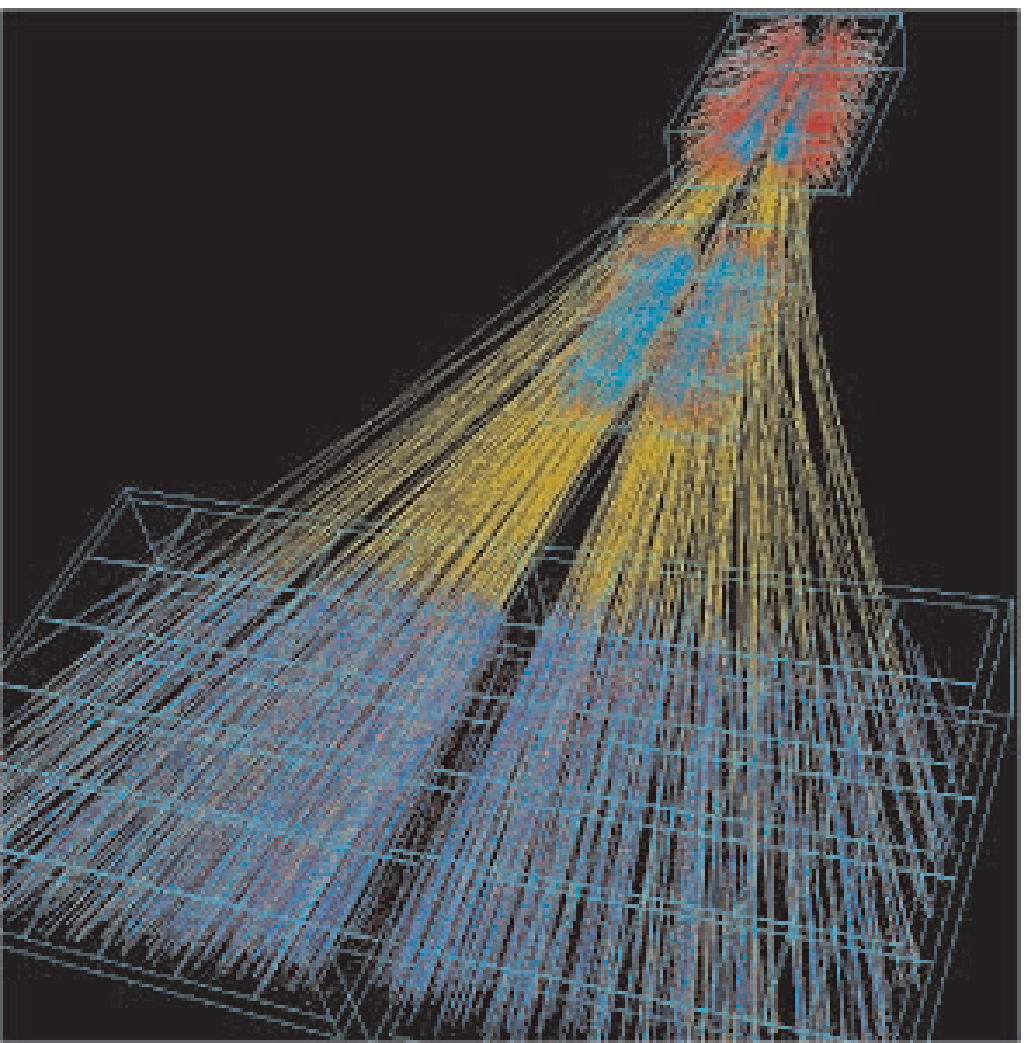}
\caption{Charged particle tracks in central Au+Au and Pb+Pb collision events,
in collider geometry (top) from RHIC STAR TPC tracking at
$\sqrt{s}=200 \: GeV$, and in fixed target geometry (bottom) from
NA49 at the SPS, $\sqrt{s}=17.3 \: GeV$.}
\label{fig:Figure2}
\end{center}
\end{figure}

In this section we take an overall look at bulk hadron production in
nucleus-nucleus collisions. In view of the high total c.m. energies
involved at e.g. top SPS $(E ^{tot}_{cm} \approx  3.3 \:TeV$) and
top RHIC (38 $TeV$) energies, in central Pb+Pb (SPS) and Au+Au
(RHIC) collisions, one can expect an extraordinarily high spatial
density of produced particles. Thus, as an overall idea of analysis,
one will try to relate the observed flow of energy into transverse
and longitudinal phase space and particle species to the high energy
density contained in the primordial interaction volume, thus to
infer about its contained matter. The typical experimental patterns
of such collisions, both in collider mode at RHIC and in a fixed
target configuration at the SPS, are illustrated in Fig.~\ref{fig:Figure2} which
shows a fractional view of the total distribution of charged
particles (about 4000 and 1600, respectively) within the tracking
volume of the STAR and NA49 experiments. \\
 \\
Most of these tracks correspond to ''thermal'' pions ($p_T$ up to $2 \:GeV$) and, in
general, such thermal hadrons make up for about 95\% of the observed
multiplicity: the bulk of hadron production. Their distributions in
phase space will be illustrated in the subsections below. This will
lead to a first insight into the overall reaction dynamics, and also
set the stage for consideration of the rare signals, imbedded in
this thermal bulk production: correlations,jets, heavy flavors,
fluctuations, which are the subject of later chapters.

\section {Particle Multiplicity and Transverse Energy Density}
\label{subsec:Particle_Multiplicity}
\begin{figure}[h]   
\begin{center}
\includegraphics[scale=1.4]{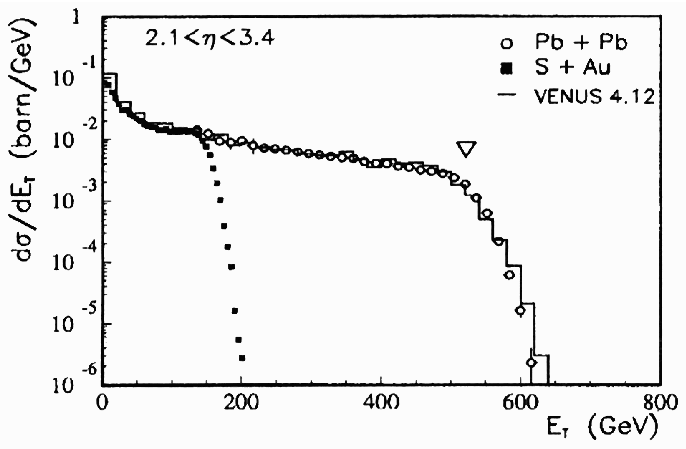}
\caption{Minimum bias distribution of total transverse energy in Pb+Pb
collisions at $\sqrt{s}=17.3 \: GeV$, and S+Au collisions at
$\sqrt{s}=20 \: GeV$, in the rapidity interval $2.1 < y < 3.4$, from
\cite{43}.}
\label{fig:Figure3}
\end{center}
\end{figure}

Particle production can be assessed globally by the total created
transverse energy, the overall result of the collisional creation of
{\it transverse} momentum $p_T$ or transverse mass ($m_T=\sqrt{
p^2_T + m^2_0}$), at the microscopic level. Fig.~\ref{fig:Figure3} shows the
distribution of total transverse energy $E_T=\sum\limits_i \:
E(\Theta_i) \cdot sin \Theta$ resulting from a calorimetric
measurement of energy flow into calorimeter cells centered at angle
$\Theta_i$ relative to the beam \cite{43}, for $^{32}S + ^{197}$Au
collisions at $\sqrt{s} = 20 \: GeV$, and for $^{208}Pb+^{208}$Pb
collisions at $\sqrt{s}=17.3\: GeV$. \\
 \\
The shape is characteristic of the impact parameter probability distribution (for equal size
spheres in the Pb+Pb case). The turnoff at $E_T=520 \: GeV$
indicates the point where geometry runs out of steam, i.e. where $b
\rightarrow 0$, a configuration generally referred to as a ''central
collision''. The adjacent shoulder results from genuine event by
event fluctuations of the actual number of participant nucleons from
target and projectile (recall the diffuse Woods-Saxon nuclear
density profiles), and from experimental factors like calorimeter
resolution and limited acceptance. The latter covers 1.3 units of
pseudo-rapidity and contains mid-rapidity $\eta_{mid}=2.9$.
Re-normalizing \cite{43} to $\Delta \: \eta=1$ leads to $dE_T/d\eta
(mid)= 400 \:GeV$, in agreement with the corresponding WA80 result
\cite{44}. Also, the total transverse energy of central Pb+Pb
collisions at $\sqrt{s}=17.3 \: GeV$ turns out to be about $1.2
\:TeV$. As the definition of a central collision, indicated in
Fig.~\ref{fig:Figure3}, can be shown \cite{42} to correspond to an average nucleon
participant number of $N_{part}=370$ one finds an average total
transverse energy per nucleon pair, of $E_T/\left<0.5 \: N_{part}\right>=6.5 \:
GeV$. After proper consideration of the baryon pair rest mass (not
contained in the calorimetric $E_T$ response but in the
corresponding $\sqrt{s}$) one concludes \cite {43} that the observed
total $E_T$ corresponds to about 0.6 $E_T^{max}$, the maximal $E_T$
derived from a situation of ''complete stopping'' in which the
incident $\sqrt{s}$ gets fully transformed into internal excitation
of a single, ideal isotropic fireball located at mid-rapidity. The
remaining fraction of $E_T^{max}$ thus stays in longitudinal motion,
reflecting the onset, at SPS energy, of a transition from a central
fireball to a longitudinally extended ''fire-tube'', i.e. a
cylindrical volume of high primordial energy density. In the limit
of much higher $\sqrt{s}$ one may extrapolate to the idealization of
a boost invariant primordial interaction volume, introduced by
Bjorken \cite{45}.

\begin{figure}[h]   
\begin{center}
\includegraphics[scale=0.6]{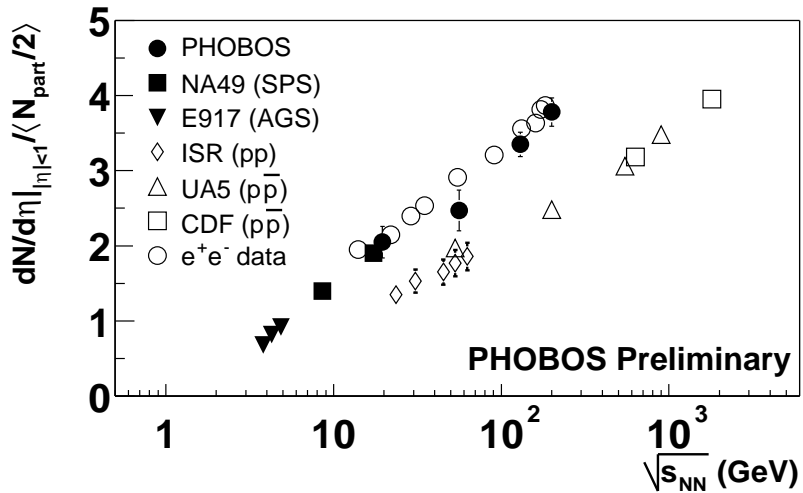}
\caption{Charged hadron rapidity density at mid-rapidity vs. $\sqrt{s}$,
compiled from $e^+e^-, \: pp, \: p \overline{p}$ and A+A collisions
\cite{53}.}
\label{fig:Figure4}
\end{center}
\end{figure}

\begin{figure}[h]   
\begin{center}
\includegraphics[scale=0.55]{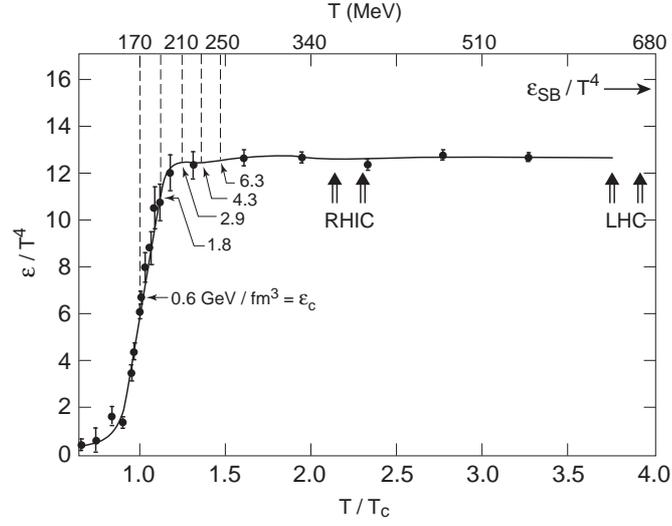}
\caption{Lattice QCD results at zero baryon potential for energy density
$\epsilon/T^4$ versus $T/T_c$ with three light quark flavors,
compared to the Stefan-Boltzmann-limit $\epsilon_{SB}$ of an ideal
quark-gluon gas \cite{48}.}
\label{fig:Figure5}
\end{center}
\end{figure}

We shall show below (section~\ref{subsec:Rap_Distribution}) that the charged particle rapidity
distributions, from top SPS to top RHIC energies, do in fact
substantiate a development toward a boost-invariant situation. One
may thus employ the Bjorken model for an estimate of the primordial
spatial energy density $\epsilon$, related to the energy density in
rapidity space via the relation \cite{45}
\begin{equation}
\epsilon(\tau_0) = \frac{1}{\pi R^2} \: \frac{1}{\tau_0} \:
\frac{dE_T}{dy}
\label{eq:equation1}
\end{equation}
where the initially produced collision volume is considered as a
cylinder of length $dz=\tau_0 dy$ and transverse radius $R \propto
A^{1/3}$. Inserting for $\pi R^2$ the longitudinally projected
overlap area of Pb nuclei colliding near head-on (''centrally''),
and assuming that the evolution of primordial pQCD shower
multiplication (i.e. the energy transformation into internal degrees
of freedom) proceeds at a time scale $\tau_0 \le 1 fm/c$, the above
average transverse energy density, of $dE_T/dy=400 \: GeV$ at top
SPS energy \cite{43,44} leads to the estimate
\begin{equation}
\epsilon (\tau_0=1fm)=3.0 \pm 0.6 \:GeV/fm^3,
\label{eq:equation2}
\end{equation}
thus exceeding, by far, the estimate of the critical energy density
$\epsilon_0$ obtained from lattice  QCD (see below), of about 1.0
$GeV/fm^3$. Increasing the collision energy to $\sqrt{s}=200 \: GeV$
for Au+Au at RHIC, and keeping the same formation time, $\tau_0=1 \:
fm/c$ (a conservative estimate as we shall show in section~\ref{subsec:Gluon_Satu_in_AA_Coll}), the
Bjorken estimate grows to $\epsilon \approx 6.0 \pm 1 \: GeV/fm^3$.
This statement is based on the increase of charged particle
multiplicity density at mid-rapidity with $\sqrt{s}$, as illustrated
in Fig.~\ref{fig:Figure4}. 
From top SPS to top RHIC energy \cite{46} the density per
participant nucleon pair almost doubles. However, at $\sqrt{s}=200
\: GeV$ the formation or thermalization time $\tau_0$, employed in
the Bjorken model \cite{45}, was argued \cite{47} to be shorter by a
factor of about 4. We will return to such estimates of $\tau_0$ in
section~\ref{subsec:Transvers_phase_space} but note, for now, that the above choice of $\tau_0=1 \:
fm/c$ represents a conservative upper limit at RHIC energy.\\
 \\
These Bjorken-estimates of spatial transverse energy density are
confronted in Fig.~\ref{fig:Figure5} with lattice QCD results obtained for three
dynamical light quark flavors \cite{48}, and for zero baryo-chemical
potential (as is realistic for RHIC energy and beyond but still
remains a fair approximation at top SPS energy where $\mu_B \approx
250 \: MeV)$. The energy density of an ideal, relativistic parton
gas scales with the fourth power of the temperature,
\begin{equation}
\epsilon = gT^4
\label{eq:equation3}
\end{equation}
where $g$ is related to the number of degrees of freedom. For an
ideal gluon gas, $g=16 \: \pi^2/30$; in an interacting system the
effective $g$ is smaller. The results of Fig.~\ref{fig:Figure5} show, in fact, that
the Stefan-Boltzmann limit $\epsilon_{SB}$ is not reached, due to
non perturbative effects, even at four times the critical
temperature $T_c=170 \: MeV$. The density $\epsilon/T^4=g$ is seen
to ascend steeply, within the interval $T_c \pm 25 \: MeV$. At $T_c$
the critical QCD energy density $\epsilon=0.6-1.0 \: GeV/fm^3$.
Relating the thermal energy density with the Bjorken estimates
discussed above, one arrives at an estimate of the initial
temperatures reached in nucleus-nucleus collisions, thus implying
thermal partonic equilibrium to be accomplished at time scale
$\tau_0$ (see section~\ref{subsec:Transvers_phase_space}). For the SPS, RHIC and LHC energy domains
this gives an initial temperature in the range 190 $\le T^{SPS} \le
220 \: MeV, \: 220 \le T^{RHIC} \le 400 \: MeV$ (assuming \cite{47}
that $\tau_0$ decreases to about 0.3 $fm/c$ here) and $T^{LHC} \ge
600 \: MeV$, respectively. From such estimates one tends to conclude
that the immediate vicinity of the phase transformation is sampled
at SPS energy, whereas the dynamical evolution at RHIC and LHC
energies dives deeply into the ''quark-gluon-plasma'' domain of QCD.
We shall return to a more critical discussion of such ascertations
in section~\ref{subsec:Transvers_phase_space}. \\ \\     
One further aspect of the mid-rapidity charged particle densities
per participant pair requires attention: the comparison with data
from elementary collisions. Fig.~\ref{fig:Figure4} shows a compilation of $pp, \:
p\overline{p}$ and $e^+e^-$ data covering the range from ISR to LEP and 
Tevatron energies. \\
\begin{figure}[h!]   
\begin{center}
\includegraphics[scale=0.43]{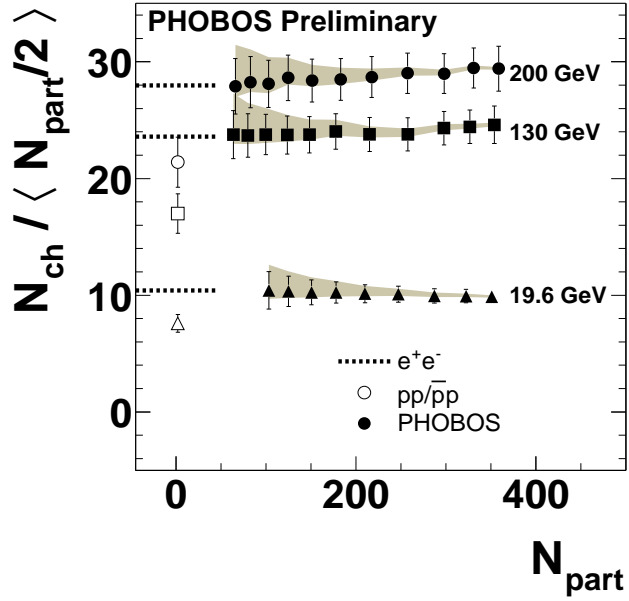}
\caption{The total number of charged hadrons per participant pair shown as a
function of $N_{part}$ in Au+Au collisions at three RHIC energies
\cite{53}.}
\label{fig:Figure6}
\end{center}
\end{figure}\\
The data from $e^+e^-$ represent $dN_{ch}/dy$,
the rapidity density along the event thrust axis, calculated
assuming the pion mass \cite{49} (the difference between $dN/dy$ and
$dN/d\eta$ can be ignored here). Remarkably, they superimpose with the central A+A collision data, 
whereas $pp$ and $p\overline{p}$
show similar slope but amount to only about 60\% of the AA and
$e^+e^-$ values. This difference between $e^+e^-$ annihilation to
hadrons, and $pp$ or $p\overline{p}$ hadro-production has been
ascribed \cite{50} to the characteristic leading particle effect of
minimum bias hadron-hadron collisions which is absent in $e^+e^-$.
It thus appears to be reduced in AA collisions due to subsequent
interaction of the leading parton with the oncoming thickness of the
remaining target/projectile density distribution. This naturally
leads to the scaling of total particle production with $N_{part}$
that is illustrated in Fig.~\ref{fig:Figure6}, for three RHIC energies and minimum
bias Au+Au collisions; the close agreement with $e^+e^-$
annihilation data is obvious again. One might conclude that,
analogously, the participating nucleons get ''annihilated'' at high
$\sqrt{s}$, their net quantum number content being spread out over
phase space (as we shall show in the next section).
\begin{figure}[h] 
\begin{center}
\includegraphics[scale=0.84]{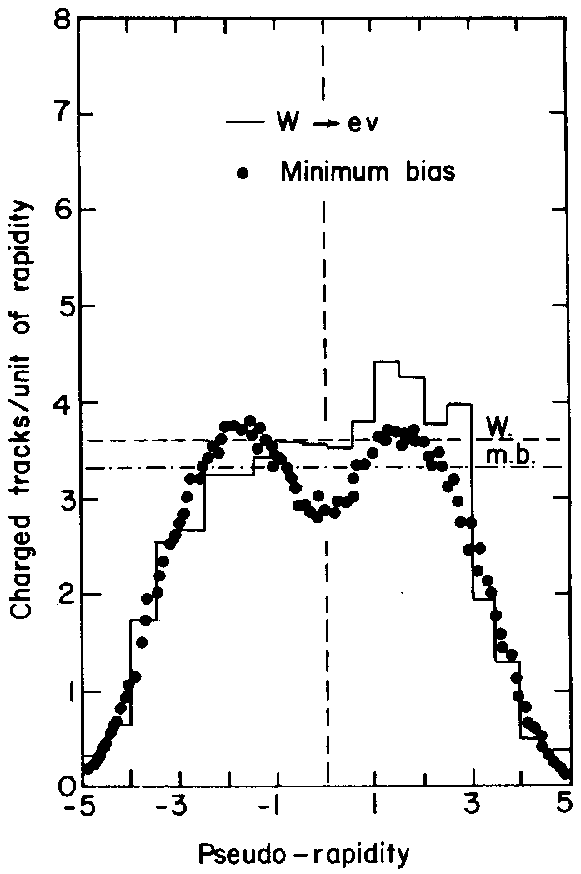}
\includegraphics[scale=0.4]{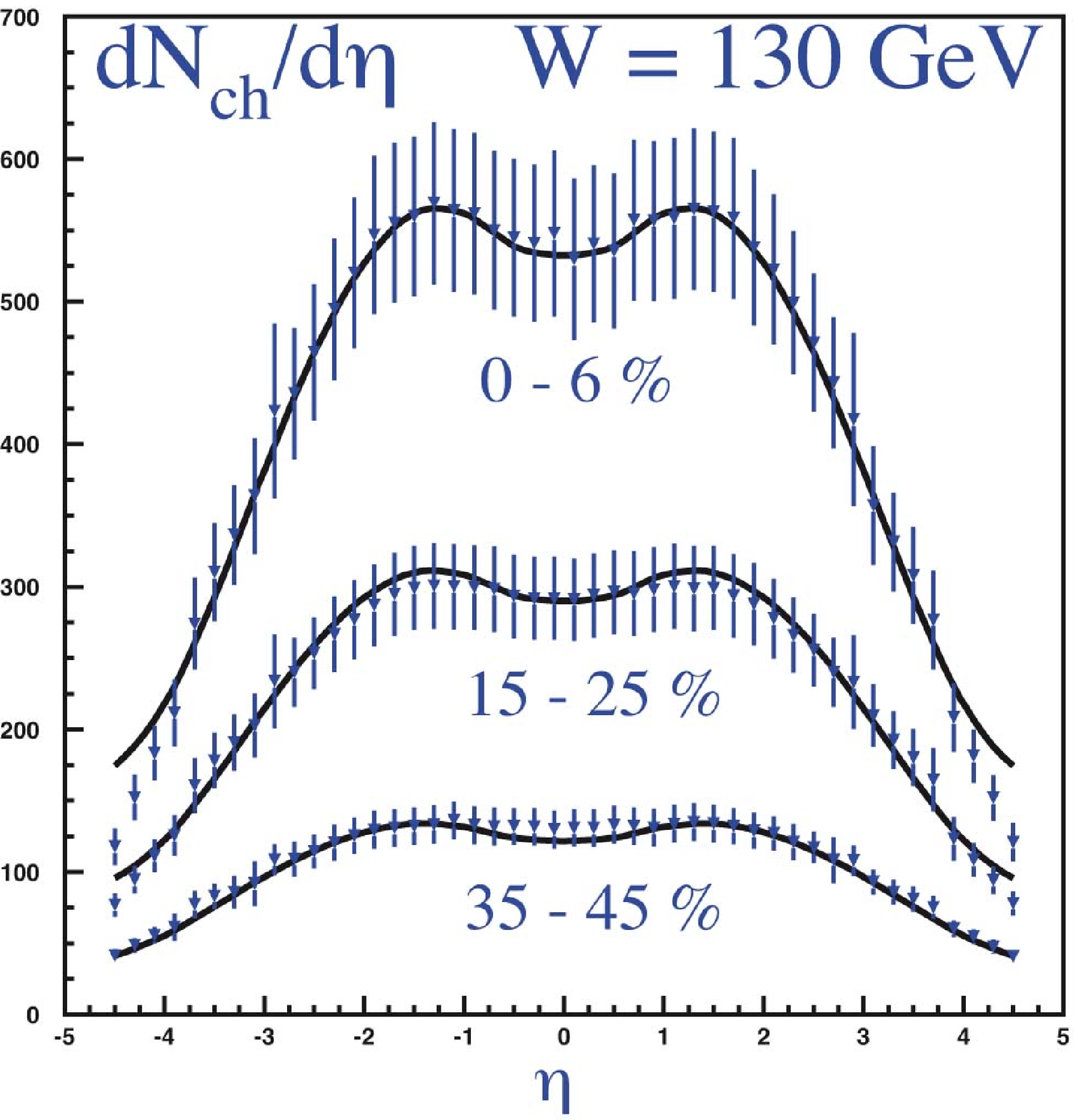}
\caption{Left panel: charged particle pseudo-rapidity distribution in $p
\overline{p}$ collisions at $\sqrt{s}=540 \: GeV$ \cite{51}. Right
panel: same in RHIC Au+Au collisions at $\sqrt{s}=130 \: GeV$ at
different centralities \cite{52}. Closed lines represent fits with
the color glass condensate model \cite{64}.}
\label{fig:Figure7}
\end{center}
\end{figure}

\section {Rapidity Distributions}
\label{subsec:Rap_Distribution}
\begin{figure}[h!]   
\begin{center}
\includegraphics[scale=0.4]{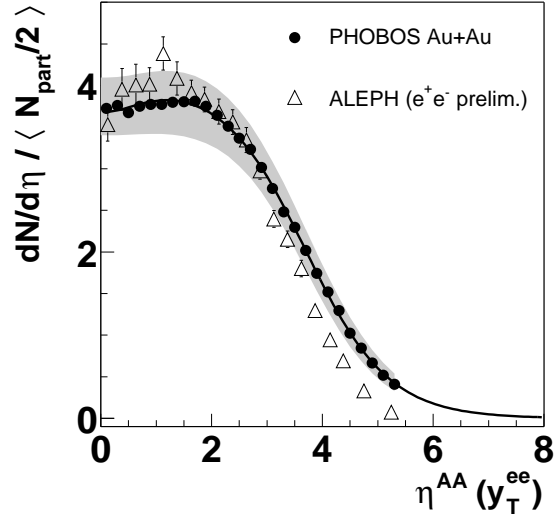}
\caption{Pseudo-rapidity distribution of charged hadrons produced in central
Au+Au collisions at $\sqrt{s}=200 \: GeV$ compared with $e^+e^-$
data at similar energy. The former data normalized by $N_{part}/2$.
From ref. \cite{53}.}
\label{fig:Figure8}
\end{center}
\end{figure}

\begin{figure}[h]   
\begin{center}
\includegraphics[scale=0.32]{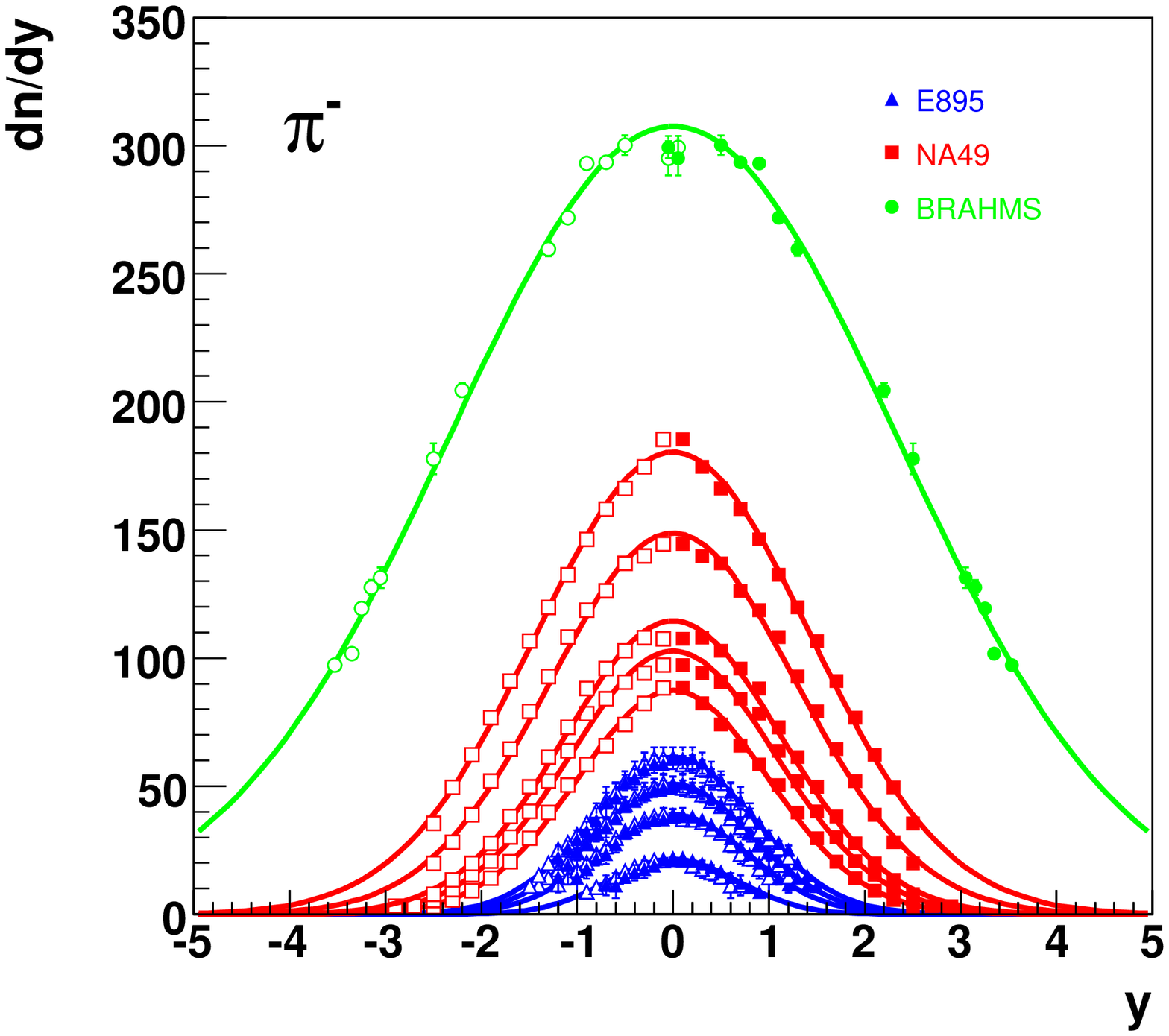}
\includegraphics[scale=0.32]{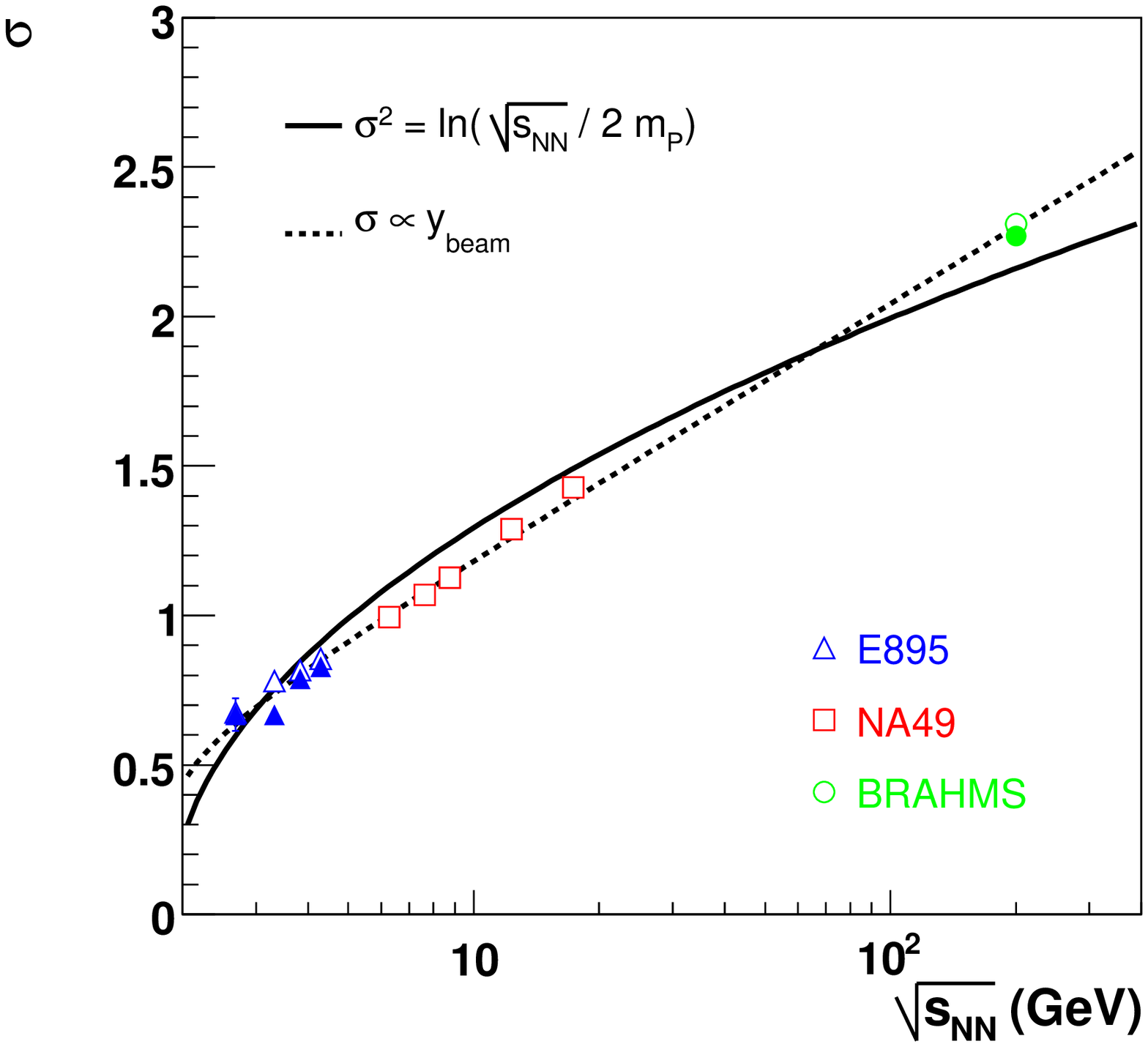}
\caption{Left panel: negative pion rapidity distributions in central Au+Au
and Pb+Pb collisions from AGS via SPS to RHIC energies \cite{54}.
Right panel: the Gaussian rapidity width of pions versus $\sqrt{s}$,
confronted by Landau model predictions (solid line) \cite{54}.}
\label{fig:Figure9}
\end{center}
\end{figure}

\begin{figure}[h]   
\begin{center}
\includegraphics[scale=0.4]{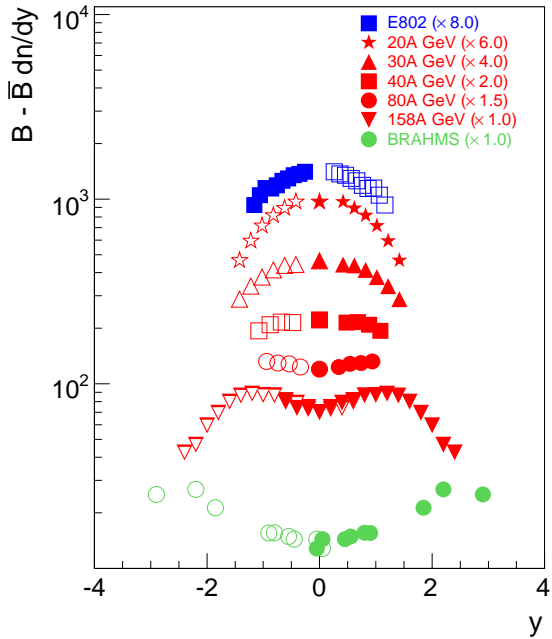}
\caption{Net proton rapidity distributions in central Au+Au/Pb+Pb collisions
at AGS, SPS and RHIC energies \cite{56, 57}.}
\label{fig:Figure10}
\end{center}
\end{figure}

Particle production number in A+A collisions depends globally on
$\sqrt{s}$ and collision centrality, and differentially on $p_T$ and
rapidity $y$, for each particle species $i$. Integrating over $p_T$
results in the rapidity distribution $dN_i/dy$. Particle rapidity\footnote{The rapidity variable represents a compact (logarithmic) description of longitudinal phase space. It is based on longitudinal particle velocity (derived from $p_{long}$ and $m$), $y=1/2 \ln ((1+\beta_L) / (1-\beta_L))$. The rapidity distribution $dN/dy$ is shape invariant under longitudinal Lorentz transformation, and centered at ``mid-rapidity" $y_{mid}=y_{CM}$, for all produced particle species; see figures~\ref{fig:Figure7} to~\ref{fig:Figure10}.},
$y=sin h^{-1} \: p_L/M_T$ (where $M_T=\sqrt{m^2 + p_T^2}$), requires mass identification. If that is
unknown one employs pseudo-rapidity ($\eta = - \mbox{ln} \:
[\mbox{tan}(\Theta/2)]$) instead. This is also chosen if the joint
rapidity distribution of several unresolved particle species is considered:
notably the charged hadron distribution. We show two 
examples in Fig.~\ref{fig:Figure7}. The left panel illustrates charged particle
production in $p\overline{p}$ collisions studied by UA1 at
$\sqrt{s}=540 \:GeV$ \cite{51}. Whereas the minimum bias
distribution (dots) exhibits the required symmetry about the center
of mass coordinate, $\eta=0$, the rapidity distribution
corresponding to events in which a $W$ boson was produced
(histogram) features, both, a higher average charged particle yield,
and an asymmetric shape. The former effect can be seen to reflect
the expectation that the $W$ production rate increases with the
''centrality'' of $p\overline{p}$ collisions, involving more
primordial partons as the collisional overlap of the partonic
density profiles gets larger, thus also increasing the overall,
softer hadro-production rate. The asymmetry should result from a
detector bias favoring $W$ identification at negative rapidity: the
transverse $W$ energy, of about $100 \: GeV$ would {\it locally}
deplete the energy store available for associated soft production.
If correct, this interpretation suggests that the wide rapidity gap
between target and projectile, arising at such high $\sqrt{s}$, of
width $\Delta y \approx 2 \: ln \:(2 \gamma_{CM})$, makes it
possible to define local sub-intervals of rapidity within which the
species composition of produced particles varies.

The right panel of Fig.~\ref{fig:Figure7} shows charged particle pseudo-rapidity
density distributions for Au+Au collisions at $\sqrt{s}=130 \: GeV$
measured by RHIC experiment PHOBOS \cite{52} at three different
collision centralities, from ''central'' (the 6\% highest charged
particle multiplicity events) to semi-peripheral (the corresponding
35-45\% cut). We will turn to centrality selection in more detail
below. Let us first remark that the slight dip at mid-rapidity and,
moreover, the distribution shape in general, are common to
$p\overline{p}$ and Au+Au. This is also the case for $e^+e^-$ annihilation as is shown in Fig.~\ref{fig:Figure8} which compares the ALEPH rapidity distribution along the mean $p_T$ (``thrust") axis
of jet production in $e^+e^-$ at $\sqrt{s}=200 \: GeV$ \cite{49}
with the scaled PHOBOS-RHIC distribution of central Au+Au at the
same $\sqrt{s}$ \cite{53}. Note that the mid-rapidity values contained in Figs.~\ref{fig:Figure7} and~\ref{fig:Figure8} have been employed already in Fig.~\ref{fig:Figure4}, which
showed the overall $\sqrt{s}$ dependence of mid-rapidity charged
particle production. What we concluded there was a perfect scaling
of A+A with $e^+e^-$ data at $\sqrt{s} \ge 20 \: GeV$ and a 40\%
suppression of the corresponding $pp, \: p\overline{p}$ yields. We
see here that this observation holds, semi-quantitatively, for the
entire rapidity distributions. These are not ideally boost invariant
at the energies considered here but one sees in $dN_{ch}/dy$ a
relatively smooth ''plateau'' region extending over $\mid y \mid \le
1.5 -2.5$. \\ \\
The production spectrum of charged hadrons is, by far, dominated by
soft pions ($p_T \le 1 \: GeV/c)$ which contribute about 85\% of the
total yield, both in elementary and nuclear collisions. The
evolution of the $\pi^-$ rapidity distribution with $\sqrt{s}$ is
illustrated in Fig.~\ref{fig:Figure9} for central Au+Au and Pb+Pb collisions from AGS
via SPS to RHIC energy, $2.7 \le \sqrt{s} \le 200 \: GeV$ \cite{54}. \\
\\
At lower $\sqrt{s}$ the distributions are well described by single
Gaussian fits \cite{54} with $\sigma (y)$ nearly linearly
proportional to the total rapidity gap $\Delta y \propto \: ln
\sqrt{s}$ as shown in the right hand panel of Fig.~\ref{fig:Figure9}. Also
illustrated is the prediction of the schematic hydrodynamical model
proposed by Landau \cite{55},
\begin{equation}
\sigma^2 \propto ln \: (\frac{\sqrt{s}}{2m_p})
\label{eq:equation4}
\end{equation}
which pictures hadron production in high $\sqrt{s}$ $pp$ collisions
to proceed via a dynamics of initial complete ''stopping down'' of
the reactants matter/energy content in a mid-rapidity fireball that
would then expand via 1-dimensional ideal hydrodynamics. Remarkably,
this model that has always been considered a wildly extremal
proposal falls rather close to the lower $\sqrt{s}$ data for central
A+A collisions but, as longitudinal phase space widens approaching
boost invariance we expect that the (non-Gaussian) width of the
rapidity distribution grows linearly with the rapidity gap $\Delta
y$. LHC data will finally confirm this expectation, but Figs.~\ref{fig:Figure7} to \ref{fig:Figure9}
clearly show the advent of boost invariance, already at $\sqrt{s} =
200 \: GeV$.

A short didactic aside: At low $\sqrt{s}$ the total rapidity gap
$\Delta y = 2-3$ does closely resemble the total rapidity width
obtained for a thermal pion velocity distribution at temperature
$T=120-150 \: MeV$, of a single mid-rapidity fireball, the
y-distribution of which represents the longitudinal component according to
the relation \cite{19}
\begin{equation}
\frac{dN}{dy} \propto (m^2T \:+ \frac{2mT^2}{coshy} \: +
\frac{2T^2}{cosh^2y}) \: exp \: [-mcoshy/T]
\label{eq:equation5}
\end{equation}
where $m$ is the pion mass. {\it Any} model of preferentially
longitudinal expansion of the pion emitting source, away from a
trivial single central ''completely stopped'' fireball, can be
significantly tested only once $\Delta y > 3$ which occurs upward from SPS
energy. The agreement of the Landau model prediction with the data
in Fig.~\ref{fig:Figure9} is thus fortuitous, below $\sqrt{s} \approx 10 \: GeV$, as \emph{any} created fireball occupies the entire rapidity gap with pions.\\
\\
The Landau model offers an extreme view of the mechanism of
''stopping'', by which the initial longitudinal energy of the
projectile partons or nucleons is inelastically transferred to
produced particles and redistributed in transverse and longitudinal
phase space, of which we saw the total transverse fraction in Fig.~\ref{fig:Figure3}.
Obviously $e^+e^-$ annihilation to hadrons represents the extreme
stopping situation. Hadronic and nuclear collisions offer the
possibility to analyze the final distribution in phase space of
their non-zero net quantum numbers, notably net baryon number. Figure~\ref{fig:Figure10} shows
the net-proton rapidity distribution (i.e. the proton rapidity
distribution subtracted by the antiproton distribution) for central
Pb+Pb/Au+Au collisions at AGS ($\sqrt{s}=5.5 \: GeV$), SPS
($\sqrt{s} \le 17.3 \: GeV$) and RHIC ($ \sqrt{s}=200 \: GeV$)
\cite{56}. With increasing energy we see a central (but
non-Gaussian) peak developing into a double-hump structure that
widens toward RHIC leaving a plateau about mid-rapidity. The RHIC-BRAHMS experiment
acceptance for $p, \: \overline{p}$ identification does
unfortunately not reach up to the beam fragmentation domain at
$y_p=5.4$ (nor does any other RHIC experiment) but only to $y
\approx 3.2$, with the consequence that the major fraction of
$p^{net}$ is not accounted for. However the mid-rapidity region is
by no means net baryon free. At SPS energy the NA49 acceptance
covers the major part of the total rapidity gap, and we observe in
detail a net $p$ distribution shifted down from $y_p=2.9$ by an
average rapidity shift \cite{56} of $\left<\delta y\right> = 1.7$. From Fig.~\ref{fig:Figure10}
we infer that $\left<\delta y\right>$ can not scale linearly with $y_p \approx
ln (2\gamma_{CM}) \approx ln \sqrt{s}$ for ever  - as it does up to
top SPS energy where $\left<\delta y\right>=0.58 \: y_p$ \cite{56}. Because
extrapolating this relation to $\sqrt{s}=200 \: GeV$ would result in
$\left<\delta y\right>=3.1$, and with $y_p \approx 5.4$ at this energy we would
expect to observe a major fraction of net proton yield in the
vicinity of $y=2.3$ which is not the case. A saturation must thus
occur in the $\left<\delta y\right>$ vs. $\sqrt{s}$ dependence.\\
\\
The re-distribution of net baryon density over longitudinal phase
space is, of course, only partially captured by the net proton yield
but a recent study \cite{57} has shown that proper inclusion of
neutron \footnote{Neutrons are not directly measured in the SPS and
RHIC experiments but their production rate, relative to protons,
reflects in the ratio of tritium to $^3He$ production measured by
NA49 \cite{57}, applying the isospin mirror symmetry of the
corresponding nuclear wave functions.} and hyperon production data
at SPS and RHIC energy scales up, of course, the $dN/dy$
distributions of Fig.~\ref{fig:Figure10} but leaves the peculiarities of their shapes
essentially unchanged. As the net baryon rapidity density
distribution should resemble the final valence quark distribution
the Landau model is ruled out as the valence quarks are seen to be
streaming from their initial position at beam rapidity toward
mid-rapidity (not vice versa). It is remarkable, however, to see
that some fraction gets transported very far, during the primordial
partonic non-equilibrium phase. We shall turn to its theoretical
description in section~\ref{subsec:Gluon_Satu_in_AA_Coll} but note, for now, that $pp$ collisions
studied at the CERN ISR \cite{58} lead to a qualitatively similar
net baryon rapidity distribution, albeit characterized by a smaller
$\left<\delta y\right>$.

The data described above suggest that the stopping mechanism universally resides in the primordial, first generation of collisions at the microscopic level. 
The rapidity distributions of charged particle
multiplicity, transverse energy and valence quark exhibit
qualitatively similar shapes (which also evolve similarly with
$\sqrt{s}$) in $pp, \: p\overline{p}, \: e^+e^-$ reactions, on the
one hand, and in central or semi-peripheral collisions of $A \approx
200$ nuclei, on the other. Comparing in detail we formulate a
nuclear modification factor for the bulk hadron rapidity distributions,
\begin{equation}
R^{AA}_y \:  \equiv \: \frac{dN^{ch}/dy \: (y) \:\: in \:\: A+A}{0.5
\:  N_{part} \:\: dN^{ch}/dy \:\: in \:\:  pp}
\label{eq:equation6}
\end{equation}
where $N_{part} < 2A$ is the mean number of ''participating
nucleons'' (which undergo at least one inelastic collision with another nucleon) which
increases with collision centrality. For identical nuclei colliding
$\left<N^{proj}_{part}\right> \simeq \left<N^{targ}_{part}\right>$ and thus $0.5 \: N_{part}$
gives the number of opposing nucleon pairs. $R^{AA}=1$ if each such
''opposing'' pair contributes the same fraction to the total A+A
yield as is produced in minimum bias $pp$ at similar $\sqrt{s}$.
From Figs.~\ref{fig:Figure4} and~\ref{fig:Figure6} we infer that for $\mid \eta \mid < 1, \: \: 
R^{AA}=1.5$ at top RHIC energy, and for the pseudo-rapidity
integrated total $N^{ch}$ we find $R^{AA} = 1.36$, in central Au+Au
collisions. AA collisions thus provide for a higher stopping power than $pp$
(which is also reflected in the higher rapidity shift $\left<\delta y\right>$
of Fig.~\ref{fig:Figure10}). The observation that their stopping  power resembles the
$e^+e^-$ inelasticity suggests a substantially reduced leading
particle effect in central collisions of heavy nuclei. This might
not be surprising. In a Glauber-view of
successive minimum bias nucleon collisions occuring during
interpenetration, each participating nucleon is struck $\nu > 3$
times on average, which might saturate the possible inelasticity,
removing the leading fragment.

This view naturally leads to the scaling of the total particle
production in nuclear collisions with $N_{part}$, as seen clearly in
Fig.~\ref{fig:Figure6}, reminiscent of the ''wounded nucleon model'' \cite{59} but
with the scaling factor determined by $e^+e^-$ rather than $pp$
\cite{60}. Overall we conclude from the still rather close
similarity between nuclear and elementary collisions that the
mechanisms of longitudinal phase space population occur
primordially, during interpenetration which is over after $0.15 \:
fm/c$ at RHIC, and after $1.5 \: fm/c$ at SPS energy. I.e. it is the
primordial non-equilibrium pQCD shower evolution that accounts for
stopping, and its time extent should be a lower limit to the
formation time $\tau_0$ employed in the Bjorken model \cite{45},
equation~\ref{eq:equation1}. Equilibration at the partonic level might begin at
$t>\tau_0$ only (the development toward a quark-gluon-plasma phase),
but the primordial parton redistribution processes set the stage for
this phase, and control the relaxation time scales involved in
equilibration \cite{61}. More about this in section~\ref{subsec:Transvers_phase_space}. We infer
the existence of a saturation scale \cite{62}
controlling the total inelasticity: with ever higher reactant
thickness, proportional to $A^{1/3}$, one does not get a total
rapidity or energy density proportional to $A^{4/3}$ (the number of
''successive binary collisions'') but to $A^{1.08}$ only \cite{63}.
Note that the lines shown in Fig.~\ref{fig:Figure7} (right panel) refer to such a
saturation theory: the color glass condensate (CGC) model \cite{64}
developed by McLerran and Venugopulan. The success of these models
demonstrates that ''successive binary baryon scattering'' is not an
appropriate picture at high $\sqrt{s}$. One can free the partons
from the nucleonic parton density distributions only {\it once},
and their corresponding transverse areal density sets the
stage for the ensuing QCD parton shower evolution \cite{62}.
Moreover, an additional saturation effect appears to modify this evolution at high transverse areal parton density (see chapter~\ref{subsec:Gluon_Satu_in_AA_Coll}).

\section {Dependence on system size}
\label{subsec:Dep_on_sys_size}
We have discussed above a first attempt toward a variable
($N_{part}$) that scales the system size dependence in A+A
collisions. Note that one can vary the size either by centrally
colliding a sequence of nuclei, $A_1 + A_1, \: A_2 + A_2$ etc., or
by selecting different windows in $N_{part}$ out of minimum bias
collision ensembles obtained for heavy nuclei for which BNL employs
$^{197}Au$ and CERN $^{208}Pb$. The third alternative, scattering a
relatively light projectile, such as $^{32}S$, from increasing $A$
nuclear targets, has been employed initially both at the AGS and SPS
but got disfavored in view of numerous disadvantages, of both
experimental (the need to measure the entire rapidity distribution,
i.e. lab momenta from about 0.3-100 $GeV$/c, with uniform efficiency)
and theoretical nature (different density distributions of projectile and target; 
occurence of an''effectiv'' center of mass, different for hard and soft collisions, 
and depending on impact parameter).

The determination of $N_{part}$ is of central interest, and thus we
need to look at technicalities, briefly. 
The approximate linear scaling with $N_{part}$ that we observed
in the total transverse energy and the total charged particle number
(Figs.~\ref{fig:Figure3},~\ref{fig:Figure6}) is a reflection of the primordial redistribution of
partons and energy. Whereas all observable properties that refer to
the system evolution at later times, which are of interest as
potential signals from the equilibrium, QCD plasma ''matter'' phase,
have different specific dependences on $N_{part}$, be it
suppressions (high $p_T$ signals, jets, quarkonia production) or
enhancements (collective hydrodynamic flow, strangeness production).
$N_{part}$ thus emerges as a suitable common reference scale.

$N_{part}$ captures the number of potentially directly
hit nucleons. It is estimated from an eikonal straight trajectory Glauber
model as applied to the overlap region arising, in dependence of
impact parameter $b$, from the superposition along beam direction of
the two initial Woods-Saxon density distributions of the interacting
nuclei. To account for the dilute surfaces of these distributions
(within which the intersecting nucleons might not find an
interaction partner) each incident nucleon trajectory gets equipped
with a transverse radius  that represents the total inelastic NN
cross section at the corresponding $\sqrt{s}$. The formalism is
imbedded into a Monte Carlo simulation (for detail see \cite{66})
starting from random microscopic nucleon positions within the
transversely projected initial Woods-Saxon density profiles. 
Overlapping cross sectional tubes of target and projectile nucleons 
are counted as a participant nucleon pair. Owing to the statistics of nucleon initial position
sampling each considered impact parameter geometry thus results in a
probability distribution of derived $N_{part}$. Its width $\sigma$
defines the resolution $\Delta (b)$ of impact parameter $b$
determination within this scheme via the relation
\begin{equation}
\frac{1}{\Delta (b)} \: \sigma (b) \approx \frac{d\left<N_{part}
(b)\right>}{db}
\label{eq:equation7}
\end{equation}
which, at A=200, leads to the expectation to determine $b$ with
about 1.5 $fm$ resolution \cite{66}, by measuring $N_{part}$.

\begin{figure}[h]   
\begin{center}
\includegraphics[scale=0.85]{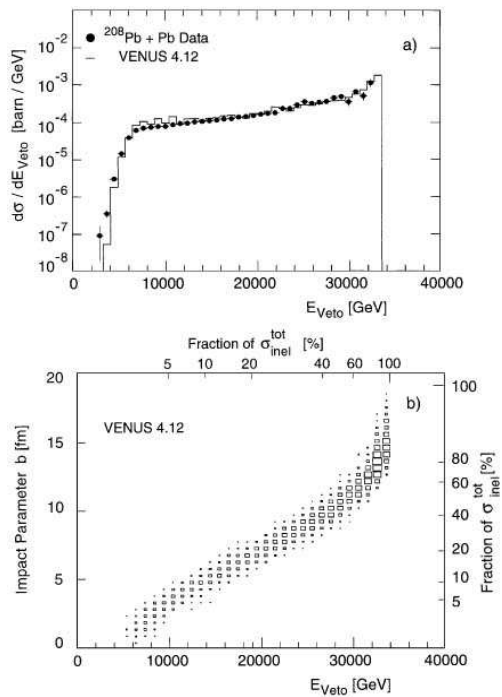}
\caption{(a) Energy spectrum of the forward calorimeter in Pb+Pb collisions
at 158$A$ GeV; (b) impact parameter and fraction of
total inelastic cross section related to forward energy from the
VENUS model \cite{67,68}.}
\label{fig:Figure11}
\end{center}
\end{figure}

How to measure $N_{part}$? In fixed target experiments one can
calorimetrically count all particles with beam momentum per nucleon
and superimposed Fermi momentum distributions of nucleons, i.e. one
looks for particles in the beam fragmentation domain $y_{beam} \pm
0.5, \:\: p_T\le 0.25 \: GeV/c$. These are identified as spectator
nucleons, and $N^{proj}_{part}=A-N^{proj}_{spec}$. For identical
nuclear collision systems $\left<N_{part}^{proj}\right> = \left<N_{part}^{targ}\right>$,
and thus $N_{part}$ gets approximated by 2 $N_{part}^{proj}$. This
scheme was employed in the CERN experiments NA49 and WA80, and
generalized \cite{67} in a way that is illustrated in Fig.~\ref{fig:Figure11}. \\
\\
The top panel shows the minimum bias distribution of total energy
registered in a forward calorimeter that covers the beam fragment domain in 
Pb+Pb collisions at lab. energy of 158 $GeV$ per projectile nucleon, $\sqrt{s}=17.3 \:
GeV$. The energy spectrum extends from about $3 \: TeV$ which
corresponds to about 20 projectile spectators (indicating a
''central'' collision), to about 32 $TeV$ which is close to the
total beam energy and thus corresponds to extremely peripheral
collisions. Note that the shape of this forward energy spectrum is
the mirror image of the minimum bias transverse energy distribution
of Fig.~\ref{fig:Figure3}, both recorded by NA49. From both figures we see that the
{\it ideal} head-on, $b \rightarrow 0$ collision can not be selected
from these (or any other) data, owing to the facts that $b=0$
carries zero geometrical weight, and that the diffuse Woods-Saxon
nuclear density profiles lead to a fluctuation of participant
nucleon number at given finite $b$. Thus the $N_{part}$ fluctuation
at finite weight impact parameters overshadows the genuinely small
contribution of near zero impact parameters. Selecting ''central''
collisions, either by an on-line trigger cut on minimal forward
energy or maximal total transverse energy or charged particle
rapidity density, or by corresponding off-line selection, one thus faces a compromise between event
statistics and selectivity for impact parameters near zero. In the
example of Fig.~\ref{fig:Figure11} these considerations suggest a cut at about $8 \:
TeV$ which selects the 5\% most inelastic events, from among the
overall minimum bias distribution, then to be labeled as ''central''
collisions. This selection corresponds to a soft cutoff at $b \le 3
\: fm$.

\begin{figure}[h]
   \begin{minipage}[t]{70mm}
      \includegraphics*[width=6.8cm, height=6.0cm]{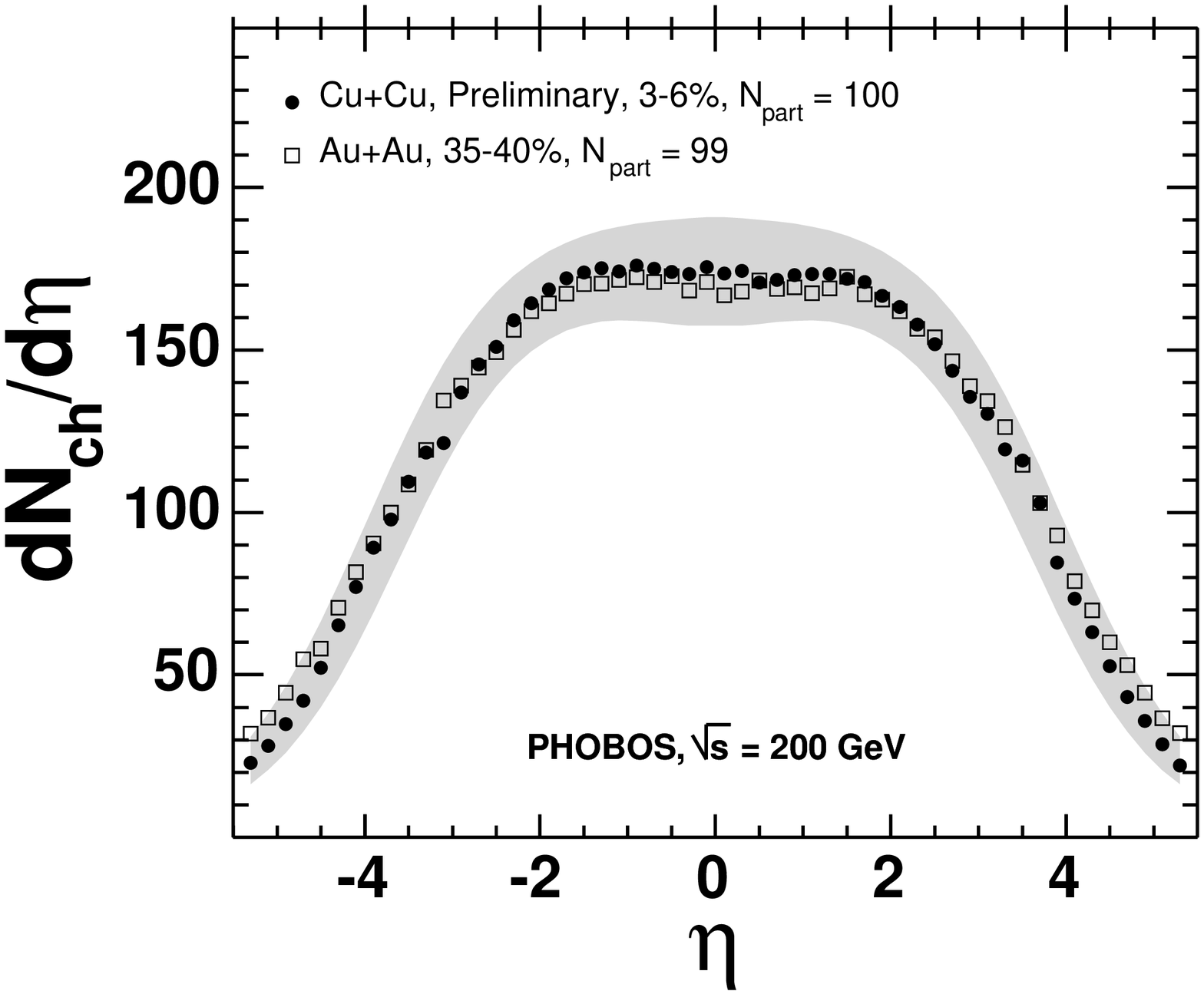}    
      \caption{Charged hadron pseudo-rapidity distributions in Cu+Cu and Au+Au
collisions at $\sqrt{s}=200 \: GeV$, with similar $N_{part} \approx
100$ \cite{52}.}
      \label{fig:Figure12}
   \end{minipage}
      \hspace{\fill}
  \begin{minipage}[t]{70mm}
      \includegraphics[width=6.6cm]{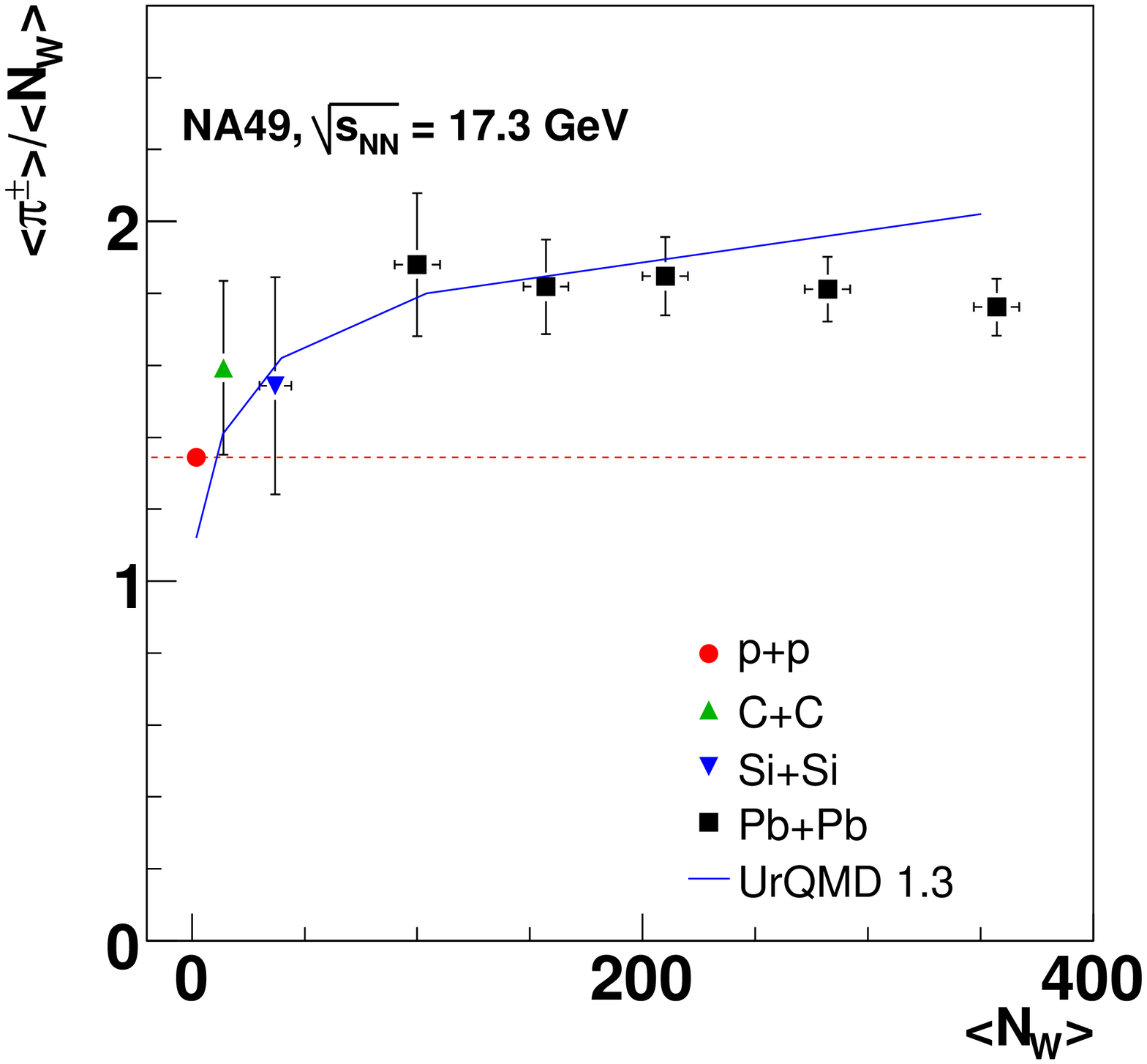}
      \caption{Charged pion multiplicity normalized by $N_W$ vs. centrality in p+p,
C+C, Si+Si and Pb+Pb collisions at $\sqrt{s}=17.3 \: GeV$
\cite{67,69}.}
      \label{fig:Figure13}
   \end{minipage}
\end{figure}

The selectivity of this, or of other less stringent cuts on
collision centrality is then established by comparison to a Glauber
or cascade model. The bottom panel of Fig.~\ref{fig:Figure11} employs the VENUS
hadron/string cascade model \cite{68} which starts from a Monte
Carlo position sampling of the nucleons imbedded in Woods-Saxon
nuclear density profiles but (unlike in a Glauber scheme with
straight trajectory overlap projection) following the cascade of
inelastic hadron/string multiplication, again by Monte Carlo
sampling. It reproduces the forward energy data reasonably well and
one can thus read off the average impact parameter and participant
nucleon number corresponding to any desired cut on the percent
fraction of the total minimum bias cross section. Moreover, it is
clear that this procedure can also be based on the total minimum
bias transverse energy distribution, Fig.~\ref{fig:Figure3}, which is the mirror image
of the forward energy distribution in Fig.~\ref{fig:Figure11}, or on the total, and
even the mid-rapidity charged particle density (Fig.~\ref{fig:Figure6}). The latter
method is employed by the RHIC experiments STAR and PHENIX. \\ \\
How well this machinery works is illustrated in Fig.~\ref{fig:Figure12} by RHIC-PHOBOS
results at $\sqrt{s}=200 \: GeV$ \cite{52}. The charged particle
pseudo-rapidity density distributions are shown for central (3-6\%
highest $N_{ch}$ cut) Cu+Cu collisions, with $\left<N_{part}\right>=100$, and
semi-peripheral Au+Au collisions selecting the cut window (35-40\%)
such that the same $\left<N_{part}\right>$ emerges. The distributions are
nearly identical. In extrapolation to $N_{part}=2$ one would expect
to find agreement between min. bias $p+p$, and ''super-peripheral''
A+A collisions, at least at high energy where the nuclear Fermi
momentum plays no large role. Fig.~\ref{fig:Figure13} shows that this expectation is
correct \cite{69}. As it is technically difficult to select
$N_{part}=2$ from A=200 nuclei colliding, NA49 fragmented the
incident SPS Pb beam to study $^{12}C+^{12}C$ and $^{28}Si+^{28}Si$
collisions \cite{67}. These systems are isospin symmetric, and
Fig.~\ref{fig:Figure13} thus plots $0.5(\left<\pi^+\right>+\left<\pi^-\right>)/\left<N_W\right>$ including
$p+p$ where $N_W=2$ by definition. We see that the pion multiplicity of A+A collisions
interpolates to the p+p data point. \\
\\
Note that NA49 employs the term ''wounded nucleon'' number ($N_{W}$) to count the 
nucleons that underwent at least one inelastic {\it nucleon-nucleon} collision. This is what
the RHIC experiments (that follow a Glauber model) call $N_{part}$
whereas NA49 reserves this term for nucleons that underwent {\it
any} inelastic collision.  Thus $N_W$ in
Fig.~\ref{fig:Figure13} has the same definition as $N_{part}$ in Figs.~\ref{fig:Figure4},~\ref{fig:Figure6},~\ref{fig:Figure8},~\ref{fig:Figure12}. We
see that a smooth increase joins the $p+p$ data, via the light A+A
central collisions, to a saturation setting in with semi-peripheral
Pb+Pb collisions, the overall, relative increase amounting to about
40\% (as we saw in Fig.~\ref{fig:Figure4}).

There is nothing like an $N_{part}^{1/3}$ increase (the
thickness of the reactants) observed here, pointing to the saturation
mechanism(s) mentioned in the previous section, which are seen from
Fig.~\ref{fig:Figure13} to dampen the initial, fast increase once the primordial
interaction volume contains about 80 nucleons. In the Glauber model
view of successive collisions (to which we attach only symbolical
significance at high $\sqrt{s}$) this volume corresponds to $\left<\nu\right>
\approx 3$, and within the terminology of such models we might thus
argue, intuitively, that the initial {\it geometrical} cross
section, attached to the nucleon structure function as a whole, has
disappeared at $\left<\nu\right> \approx 3$, all constituent partons being
freed.

\section {Gluon Saturation in A+A Collisions}
\label{subsec:Gluon_Satu_in_AA_Coll}
We will now take a closer look at the saturation phenomena of high
energy QCD scattering, and apply results obtained for deep inelastic
electron-proton reactions to nuclear collisions, a procedure that
relies on a universality of high energy hadron scattering. This
arises at high $\sqrt{s}$, and at relatively low momentum
transfer squared $Q^2$ (the condition governing bulk charged particle
production near mid-rapidity at RHIC, where Feynman $x \approx 0.01$
and $Q^2 \le 5 \: GeV^2$). Universality comes
about as the transverse resolution becomes higher and higher, with
$Q^2$, so that within the small area tested by the collision there
is no difference whether the partons sampled there belong to the
transverse gluon and quark density projection of any hadron species,
or even of a nucleus. And saturation arises once the areal
transverse parton density exceeds the resolution, leading to
interfering QCD sub-amplitudes that do not reflect in the total
cross section in a manner similar to the mere summation of separate,
resolved color charges~\cite{61,62,63,64,65,70,71}.

\begin{figure}[h]   
\begin{center}
\includegraphics[scale=0.35]{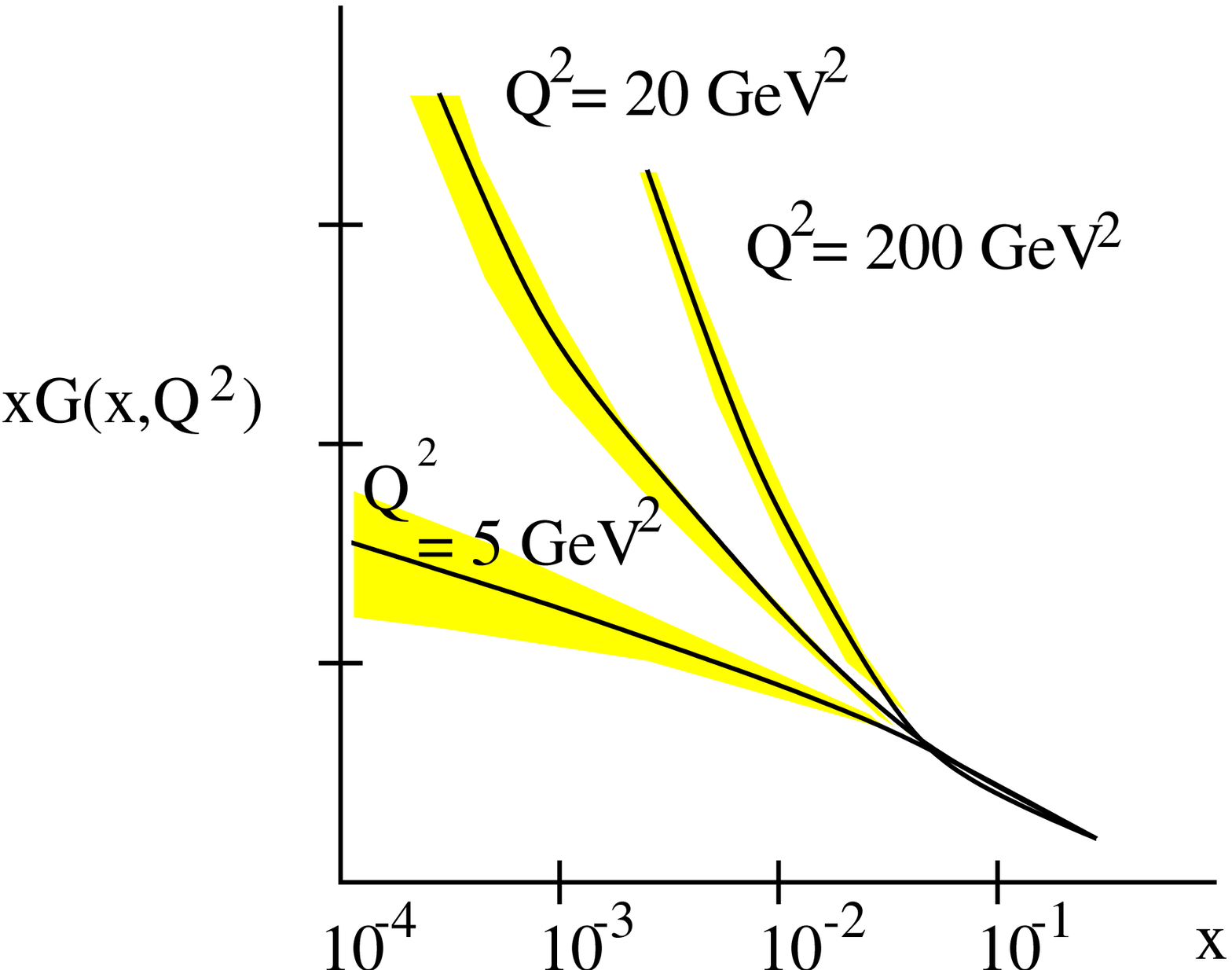}
\includegraphics[scale=0.3]{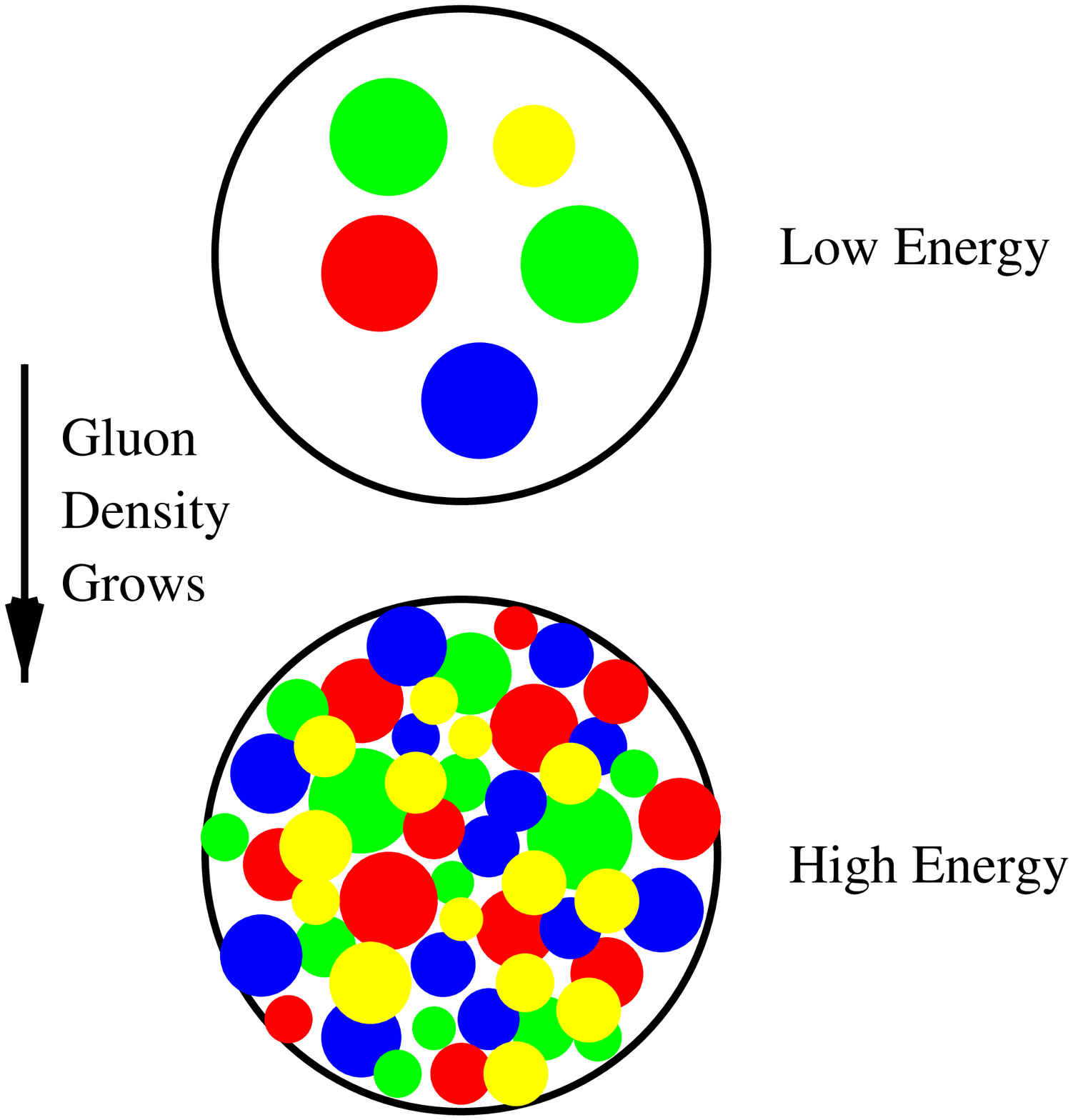}
\caption{(left) The HERA data for the gluon distribution function as a function of
fractional momentum $x$ and square momentum transfer $Q^2$ \cite{72}. (right)
Saturation of gluons in a hadron; a head on view as $x$ decreases
\cite{75}.}
\label{fig:Figure14}
\end{center}
\end{figure}

The ideas of saturation and universality are motivated by HERA deep
inelastic scattering (DIS) data \cite{72} on the gluon distribution
function shown in Fig.~\ref{fig:Figure14} (left side). The gluon rapidity density,
$xG(x,Q^2)=\frac{dN^{gluon}}{dy}$ rises rapidly as a function of
decreasing fractional momentum, $x$, or increasing resolution,
$Q^2$. The origin of this rise in the gluon density is, ultimately,
the non-abelian nature of QCD. Due to the intrinsic non-linearity of
QCD \cite{70,71}, gluon showers generate more gluon showers,
producing an avalanche toward small $x$. As a consequence of this
exponential growth the spatial density of gluons (per unit
transverse area per unit rapidity) of any hadron or nucleus must
increase as $x$ decreases \cite{65}. This follows because
the transverse size, as seen via the total cross section, rises more
slowly toward higher energy than the number of gluons. This is
illustrated in Fig.~\ref{fig:Figure14} (right side). In a head-on view of a hadronic
projectile more and more partons (mostly gluons) appear as $x$
decreases. This picture reflects a representation of the hadron in
the ''infinite momentum frame'' where it has a large light-cone
longitudinal momentum $P^+ \gg M$. In this frame one can describe
the hadron wave function as a collection of constituents carrying a
fraction $p^+=xP^+, \: 0 \le x<1$, of the total longitudinal
momentum \cite{73} (''light cone quantization'' method~\cite{74}). In 
DIS at large $sqrt{s}$ and $Q^2$ one measures the quark distributions $dN_q/dx$ at small
$x$, deriving from this the gluon distributions $xG(x, Q^2)$ of
Fig.~\ref{fig:Figure14}.

It is useful \cite{75} to consider the rapidity distribution implied
by the parton distributions, in this picture. Defining $y=y_{hadron}
- ln(1/x)$ as the rapidity of the potentially struck parton, the
invariant rapidity distribution results as
\begin{equation}
dN/dy = x \:dN/dx = xG(x,Q^2).
\label{eq:equation8}
\end{equation}
At high $Q^2$ the measured quark and gluon structure functions are
thus simply related to the number of partons per unit rapidity,
resolved in the hadronic wave function.\\
\\
The above textbook level \cite{74,75} recapitulation leads, however,
to an important application: the $dN/dy$ distribution of constituent
partons of a hadron (or nucleus), determined by the DIS experiments,
is similar to the rapidity distribution of produced particles in
hadron-hadron or A+A collisions as we expect the initial gluon
rapidity density to be represented in the finally observed, produced
hadrons, at high $\sqrt{s}$. Due to the longitudinal boost
invariance of the rapidity distribution, we can apply the above
conclusions to hadron-hadron or A+A collisions at high $\sqrt{s}$,
by replacing the infinite momentum frame hadron rapidity by the center of mass
frame projectile rapidity, $y_{proj}$, while retaining the result
that the rapidity density of potentially interacting partons grows
with increasing distance from $y_{proj}$ like
\begin{equation}
\Delta y \equiv y_{proj} - y = ln(1/x).
\label{eq:equation9}
\end{equation}
At RHIC energy, $\sqrt{s}= 200 \: GeV$, $\Delta y$ at mid-rapidity
thus corresponds to $x<10^{-2}$ (well into the domain of growing
structure function gluon density, Fig.~\ref{fig:Figure14}), and the two intersecting
partonic transverse density distributions thus attempt to
resolve each other given the densely packed situation that is
depicted in the lower circle of Fig.~\ref{fig:Figure14} (right panel). At given $Q^2$
(which is modest, $Q^2 \le 5 \: GeV^2$, for bulk hadron production
at mid-rapidity) the packing density at mid-rapidity will increase
toward higher $\sqrt{s}$ as
\begin{equation}
\Delta y^{midrap} \approx ln(\sqrt{s}/M), \: i.e. \: 1/x \approx
\sqrt{s}/M
\label{eq:equation10}
\end{equation}
thus sampling smaller $x$ domains in Fig.~\ref{fig:Figure14} according to equation~\ref{eq:equation9}. It
will further increase in proceeding from hadronic to nuclear
reaction partners A+A. Will it be in proportion to $A^{4/3}$?
We know from the previous sections (\ref{subsec:Rap_Distribution} and \ref{subsec:Dep_on_sys_size}) that this is not the case, the data indicating an increase with $A^{1.08}$. This observation is, in fact caused by the parton saturation effect, to which we turn now.

For given transverse resolution $Q^2$ and increasing $1/x$ the
parton density of Fig.~\ref{fig:Figure14} becomes so large that one can not neglect
their mutual interactions any longer. One expects such interactions
to produce ''{\it shadowing}'', a decrease of the scattering cross
section relative to incoherent independent scattering \cite{70,71}.
As an effect of such shadowed interactions there occurs \cite{75} a
{\it saturation} \cite{61,62,63,64,65,70,71,75} of the cross section
at each given $Q^2$, slowing the increase with $1/x$ to become
logarithmic once $1/x$ exceeds a certain critical value $x_s(Q^2)$.
Conversely, for fixed $x$, saturation occurs for transverse momenta
below some critical $Q^2(x)$,
\begin{equation}
Q_s^2(x) = \alpha_s N_c \: \frac{1}{\pi R^2} \: \frac{dN}{dy}
\label{eq:equation11}
\end{equation}
where $dN/dy$ is the $x$-dependent gluon density (at $y=y_{proj} -
ln(1/x))$. $Q^2_s$ is called the {\it saturation scale}. In equation~\ref{eq:equation11}
$\pi R^2$ is the hadron area (in transverse projection), and $\alpha
_s N_c$ is the color charge squared of a single gluon. More
intuitively, $Q^2_s (x)$ defines an inversely proportional
resolution area $F_s(x)$ and at each $x$ we have to choose $F_s(x)$
such that the ratio of total area $\pi R^2$ to $F_s(x)$ (the number
of resolved areal pixels) equals the number of single gluon charge
sources featured by the total hadron area. As a consequence the
saturation scale $Q^2_s(x)$ defines a critical areal resolution,
with two different types of QCD scattering theory defined, at each
$x$, for $Q^2 > Q^2_s$ and $Q^2 < Q^2_s$, respectively
\cite{62,65,75}.

As one expects a soft transition between such theories, to occur
along the transition line implied by $Q^2_s(x)$, the two types of
QCD scattering are best studied with processes featuring typical
$Q^2$ well above, or below $Q^2_s(x)$. Jet production at $\sqrt{s}
\ge 200 \: GeV$ in $p\overline{p}$ or AA collisions with typical
$Q^2$ above about $10^3 \: GeV^2$, clearly falls into the former
class, to be described e.g. by QCD DGLAP evolution of partonic
showers \cite{76}. The accronym DGLAP refers to the inventors of the perturbative QCD evolution
of parton scattering with the ''runing'' strong coupling constant $\alpha_{s}(Q^{2})$, Dokshitzer,
Gribov, Levine, Altarelli and Parisi. On the other hand, mid-rapidity bulk hadron
production at the upcoming CERN LHC facility ($\sqrt{s}=14 \: TeV$
for $pp$, and $5.5 \: TeV$ for A+A), with typical $Q^2 \le 5 \:
GeV^2$ at $x \le 10^{-3}$, will present a clear case for QCD
saturation physics, as formulated e.g. in the ''Color Glass
Condensate (CGC)'' formalism developed by McLerran, Venugopalan and
collaborators \cite{64,65,75,77}. This model develops a classical
gluon field theory for the limiting case of a high areal occupation
number density, i.e for the conceivable limit of the situation
depicted in Fig.~\ref{fig:Figure14} (right hand panel) where the amalgamating small
$x$ gluons would overlap completely, within any finite resolution
area at modest $Q^2$. Classical field theory captures, by
construction, the effects of color charge coherence, absent in DGLAP
parton cascade evolution theories \cite{75}. This model appears to
work well already at $\sqrt{s}$ as ''low'' as at RHIC, as far as
small $Q^2$ bulk charged particle production is concerned. We have
illustrated this by the CGC model fits \cite{64} to the PHOBOS
charged particle rapidity distributions, shown in Fig.~\ref{fig:Figure7}.

Conversely, QCD processes falling in the transition region between
such limiting conditions, such that typical $Q^2 \approx Q^2_s(x)$, should present observables that are functions of the ratio between
the transferred momentum $Q^2$ and the appropriate saturation scale,
expressed by $Q^2_s(x)$. As $Q^2$ defines the effective transverse
sampling area, and $Q^2_s(x)$ the characteristic areal size at which
saturation is expected to set in, a characteristic behaviour of
cross sections, namely that they are universal functions of
$Q^2/Q^2_s$, is called ''{\it geometric scaling}''. The HERA ep
scattering data obey this scaling law closely \cite{78}, and the
idea arises to apply the universality principle that we mentioned
above: at small enough $x$, all hadrons or nuclei are similar, their
specific properties only coming in via the appropriate saturation
scales $Q^2_s(x,h)$ or $Q^2_s(x,A)$. Knowing the latter for RHIC
conditions we will understand the systematics of charged particle
production illustrated in the previous chapter, and thus also be
able to extrapolate toward LHC conditions in $pp$ and AA collisions.\\
\begin{figure}[h]   
\begin{center}
\includegraphics[scale=0.6]{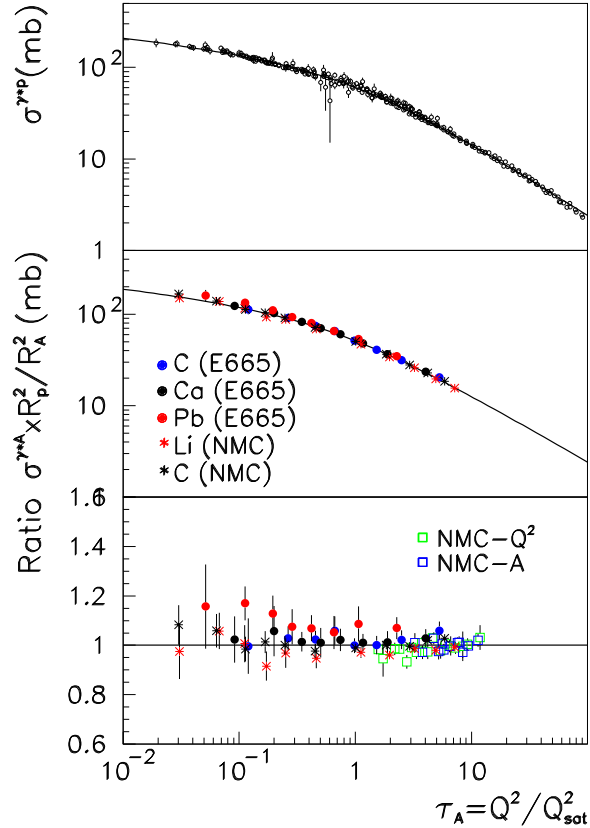}
\caption{(top) Geometric scaling of the virtual photo-absorption cross section
$\sigma ^{\gamma p}$ on protons; (middle) cross sections for nuclei
normalized according to equation~\ref{eq:equation13}; (bottom) the ratio of
$\sigma^{\gamma A}$ to a fit of $\sigma^{\gamma p}$ (see \cite{63}
for data reference).}
\label{fig:Figure15}
\end{center}
\end{figure}\\
All data for the virtual photo-absorption cross section $\sigma^{\gamma p}
(x,Q^2)$ in deep inelastic ep scattering with $x \le 0.01$
(which is also the RHIC mid-rapidity $x$-domain) have been found
\cite{78} to lie on a single curve when plotted against $Q^2/Q^2_s$,
with
\begin{equation}
Q_s^2(x) \sim  (\frac{x_0}{x})^\lambda \: 1 GeV^2
\label{eq:equation12}
\end{equation}
with $\lambda \simeq 0.3$ and $x_0 \simeq 10^{-4}$. This scaling
\cite{79} with $\tau=Q^2/Q^2_s$ is shown in Fig.~\ref{fig:Figure15} (top panel) to
interpolate all data. A chain of arguments, proposed by Armesto,
Salgado and Wiedemann \cite{63} connects a fit to these data with
photo-absorption data for (virtual) photon-A interactions \cite{80}
via the geometrical scaling ansatz
\begin{equation}
\frac{\sigma^{\gamma A} (\tau_A)}{\pi R^2_A} = \frac{\sigma^{\gamma
p} (\tau_p=\tau_A)}{\pi R^2_p}
\label{eq:equation13}
\end{equation}
assuming that the scale in the nucleus grows with the ratio of the
transverse parton densities, raised to the power $1/\delta$ (a free
parameter),
\begin{equation}
Q^2_{s,A}=Q^2_{s,p}\left(\frac{A \pi R^2_p}{\pi
R^2_A}\right)^{1/\delta}, \: \: \tau_A = \tau_h \left(\frac{\pi
R^2_A}{A \pi R^2_h}\right)^{1/\delta}.
\label{eq:equation14}
\end{equation}
Fig.~\ref{fig:Figure15} (middle and bottom panels) shows their fit to the nuclear
photo-absorbtion data which fixes $\delta=0.79$ and $\pi R^2_p=1.57
\: fm^2$ (see ref. \cite{63} for detail). The essential step in
transforming these findings to the case of A+A collisions is then
taken by the empirical ansatz
\begin{equation}
\frac{dN^{AA}}{dy} \: (at \: y \simeq 0) \: \propto \: Q^2_{s,A}(x)
\label{eq:equation15}
\pi R^2_A
\end{equation}
by which the mid-rapidity parton (gluon) density $dN/dy$ in equation~\ref{eq:equation11}
gets related to the charged particle mid-rapidity density at $y
\approx 0$ \cite{70,81}, measured in nucleus-nucleus collisions.
Replacing, further, the total nucleon number 2A in a collision of
identical nuclei of mass A by the number $N_{part}$ of participating
nucleons, the final result is \cite{63}
\begin{equation}
\frac{1}{N_{part}} \: \:  \frac{dN^{AA}}{dy} \: (at \: y \approx 0)
= N_0 (\sqrt{s})^{\lambda} \: N^{\alpha}_{part}
\label{eq:equation16}
\end{equation}
where the exponent $\alpha \equiv (1 - \delta)/3 \delta=0.089$, and
$N_0=0.47$. The exponent $\alpha$ is {\it far} smaller than 1/3, a
value that represents the thickness of the reactants, and would be
our naive guess in a picture of ''successive'' independent nucleon
participant collisions, whose average number $\left<\nu \right> \: \propto \:
(N_{part}/2)^{1/3}$. The observational fact (see Fig.\ 1.13) that $\alpha < 1/3$ for
mid-rapidity low $Q^2$ bulk hadron production in A+A collisions
illustrates the importance of the QCD saturation effect. This is
shown \cite{63} in Fig.~\ref{fig:Figure16} where equation~\ref{eq:equation16} is applied to the RHIC
PHOBOS data for mid-rapidity charged particle rapidity density per
participant pair, in Au+Au collisions at $\sqrt{s}=19.6$, 130 and
200 $GeV$ \cite{82}, also including a {\it prediction} for LHC
energy. Note that the {\it factorization of energy and centrality
dependence}, implied by the RHIC data \cite{52}, is well captured by equation~\ref{eq:equation11}
 and the resulting fits in Fig.~\ref{fig:Figure16}. Furthermore, the steeper
slope, predicted for $N_{part} \le 60$ (not covered by the employed
data set), interpolates to the corresponding $pp$ and
$p\overline{p}$ data, at $N_{part}=2$. It resembles the pattern
observed in the NA49 data (Fig.~\ref{fig:Figure13}) for small $N_{part}$ collisions
of light A+A systems, at $\sqrt{s}=17-20 \: GeV$, and may be seen,
to reflect the onset of QCD saturation. Finally we note that the
conclusions of the above, partially heuristic approach~\cite{63},
represented by equations~\ref{eq:equation13} to~\ref{eq:equation16}, have been backed up by the CGC
theory of McLerran and Venugopulan \cite{64,65,75}, predictions of
which we have illustrated in Fig.~\ref{fig:Figure7}.\\
\begin{figure}[h!]   
\begin{center}
\includegraphics[scale=0.6]{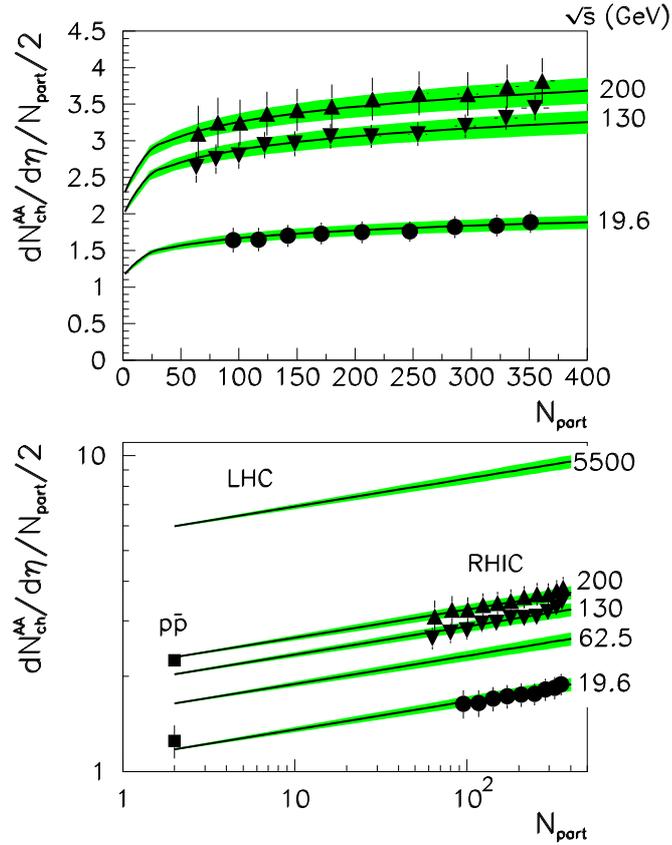}
\caption{Saturation model fit \cite{63} applied to RHIC charged hadron
multiplicity data at mid-rapidity normalized by number of
participant pairs, at various energies \cite{82}. Also shown is an
extrapolation to $p \overline{p}$ data and a prediction for minimum
bias Pb+Pb collisions at LHC energy, $\sqrt{s}=5500 \: GeV$.}
\label{fig:Figure16}
\end{center}
\end{figure}\\
Bulk hadron production in AA
collisions at high $\sqrt{s}$ can be related, via the assumption of
universality of high energy QCD scattering, to the phenomenon of
geometric scaling first observed in HERA deep inelastic ep cross
sections. The underlying feature is a QCD saturation effect arising
from the diverging areal parton density, as confronted with the
limited areal resolution $Q^2$, inherent in the considered
scattering process. The ''saturation scale'' $Q^2_s(x,A)$ captures
the condition that a single partonic charge source within the
transverse partonic density profile can just be resolved by a
sufficiently high $Q^2$. Bulk hadron production in A+A collisions falls below this scale.

\section {Transverse phase space: equilibrium and the QGP state}
\label{subsec:Transvers_phase_space}
At RHIC energy, $\sqrt{s}=200 \: GeV$, the Au+Au collision reactants
are longitudinally contracted discs. At a nuclear radius $R \approx
A^{1/3} \: fm$ and Lorentz $\gamma \approx 100$ their primordial
interpenetration phase ends at time $\tau_0 \le 0.15 \: fm/c$.
This time scale is absent in $e^+e^-$ annihilation at similar
$\sqrt{s}$ where $\tau_0 \approx 0.1 \: fm/c$ marks the end of the
primordial pQCD partonic shower evolution \cite{83} during which the
initially created $q \overline{q}$ pair, of ''virtually''
$Q=\sqrt{s}/2$ each, multiplies in the course of the QCD DGLAP
evolution in perturbative vacuum, giving rise to daughter partons of
far lower virtuality, of a few $GeV$. In A+A collisions this shower era
should last longer, due to the interpenetrational spread of primordial collision time.
It should be over by about 0.25 $fm/c$. The shower partons in $e^{+}e^{-}$ annihilation 
are localized within back to back cone geometry reflecting the directions of the
primordial quark pair. The eventually observed ''jet'' signal,
created by an initial $Q^2$ of $10^4 \: GeV^{2}$, is established by
then. Upon a slow-down  of the dynamical evolution time scale to
$\tau \approx 1 \: fm/c$ the shower partons fragment further,
acquiring transverse momentum and yet lower virtuality, then to
enter a non perturbative QCD phase of color neutralization during
which hadron-like singlet parton clusters are formed. Their net
initial pQCD virtuality, in pQCD vacuum, is recast in terms of
non-perturbative vacuum hadron mass. The evolution ends with
on-shell, observed jet-hadrons after about $3 \:fm/c$ of overall
reaction time.

\begin{figure}[h]   
\begin{center}
\includegraphics[scale=0.6]{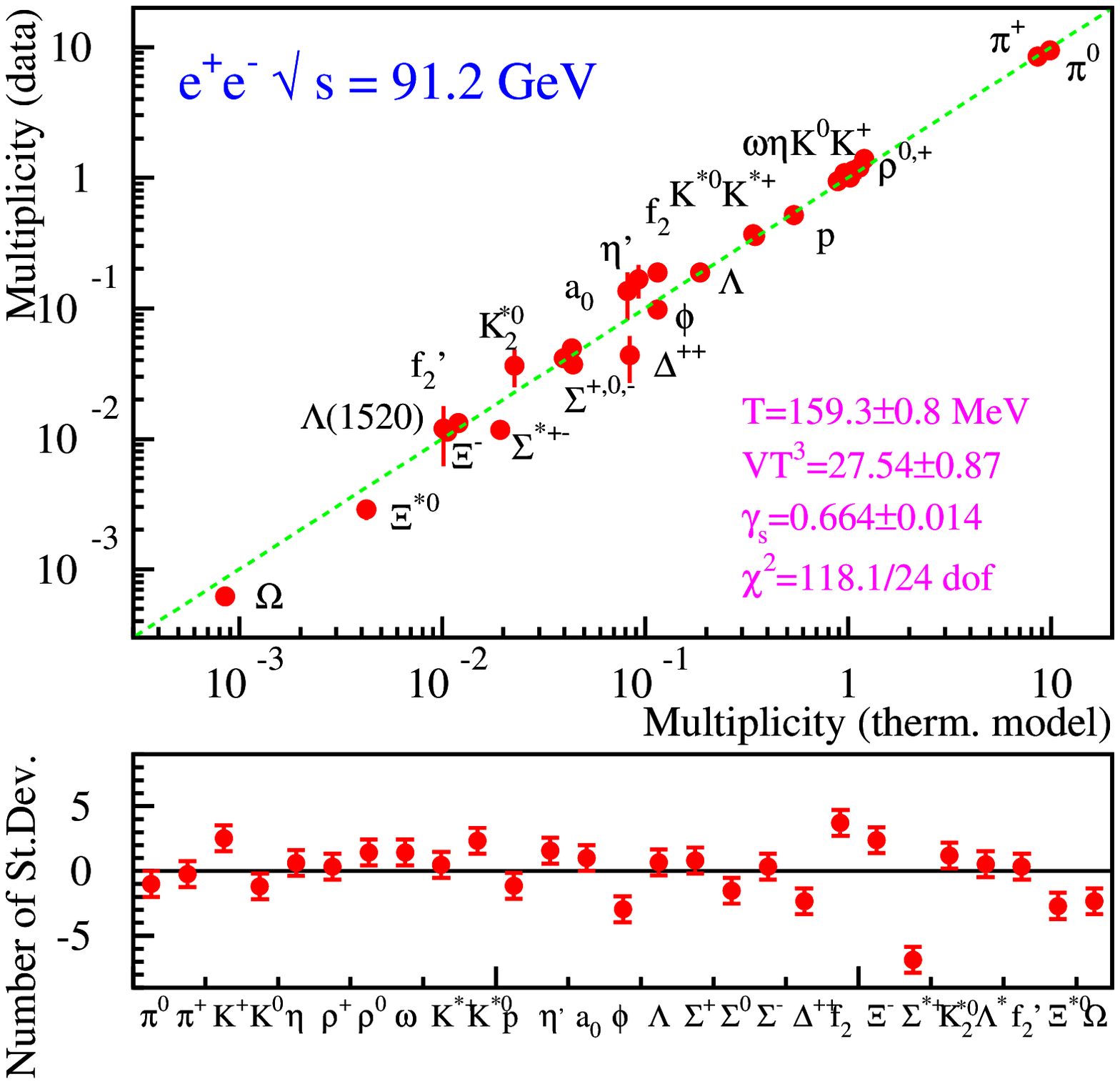}
\caption{Hadron multiplicities in LEP $e^+e^-$ annihilation at $\sqrt{s}=91.2
\: GeV$ confronted with the predictions of the canonical statistical
hadronization model \cite{84}.}
\label{fig:Figure17}
\end{center}
\end{figure}

\begin{figure}[h]   
\begin{center}
\includegraphics[scale=0.63]{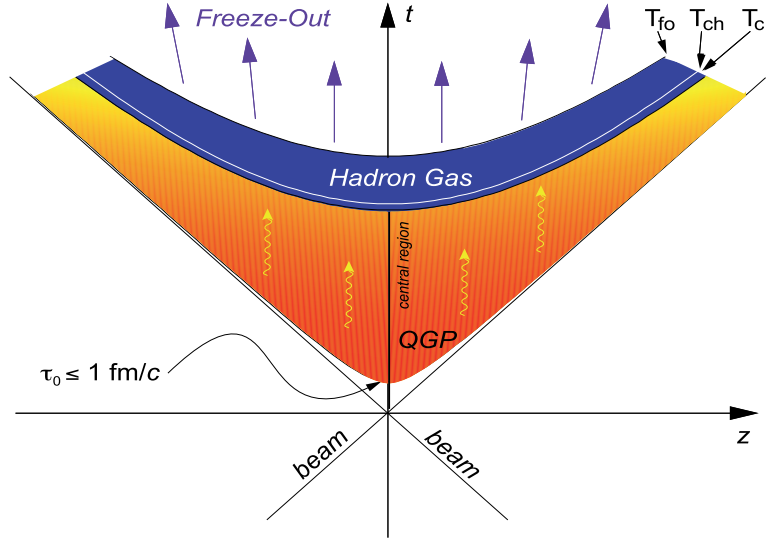}
\caption{Schematic light cone diagram of the evolution of a high energy heavy
ion collision, indicating a formation phase $\tau_0$ (see text).}
\label{fig:Figure18}
\end{center}
\end{figure}

Remarkably, even in this, somehow most elementary process of QCD
evolution, an aspect of equilibrium formation is observed, not in
the narrowly focussed final dijet momentum topology but in the
relative production rates of the various created hadronic species.
This so-called ''hadrochemical'' equilibrium among the hadronic
species is documented in Fig.~\ref{fig:Figure17}. The hadron multiplicities per
$e^+e^-$ annihilation event at $\sqrt{s}=91.2 \: GeV$  \cite{38} are
confronted with a Hagedorn \cite{38} canonical statistical Gibbs
ensemble prediction \cite {84} which reveals that the apparent
species equilibrium was fixed at a temperature of $T=165 \: MeV$,
which turns out to be the universal hadronization temperature of all
elementary and nuclear collisions at high $\sqrt{s}$ (Hagedorns
limiting temperature of the hadronic phase of matter). We shall
return to this topic in chapter~\ref{chap:hadronization} but note, for now, that reactions
with as few as 20 charged particles exhibit such statistical
equilibrium properties, a pre-requisite for application of
thermodynamic or hydrodynamic concepts.\\
\\
What happens with parton (and hadron) dynamics in A+A collisions after
$\tau_0$? There will not be a QCD evolution in vacuum (which would be over
after $3 \:fm/c$) as the transverse radius of the interacting system is large. 
It may grow to about twice the nuclear radius, i.e. to about $15 \:
fm$ before interactions cease; i.e. the system needs about $15 \:
fm/c$ to decouple. This simple fact is the key to our expectation
that the expansive evolution of the initial high energy density
deposited in a cylinder of considerable diameter (about $10 \: fm$),
may create certain equilibrium properties that allow us to treat the
contained particles and energy in terms of thermodynamic phases of
matter, such as a partonic QGP liquid,  or a hadronic liquid or gas,
etc.. Such that the expansion dynamics makes contact to the phase
diagram illustrated in Fig.~\ref{fig:Figure1}. This expectation turns out to be
justified as we shall describe in chapters~\ref{chap:hadronization} and~\ref{chap:Elliptic_flow}. 
What results for the evolution after $\tau_0$ in a central A+A collision is sketched in
Fig.~\ref{fig:Figure18} by means of a schematic 2-dimensional light cone diagram,
which is entered by the two reactant nuclei along $z=t$ trajectories
where $z$ is the beam direction and Lorentz contraction has been
taken to an extreme, such that there occurs an idealized $t=z=0$
interaction ''point''. Toward positive $t$ the light cone proper
time profiles of progressing parton-hadron matter evolution are
illustrated. The first profile illustrated here corresponds to the
end of formation time $\tau_0$. From our above discussion of the
$e^+e^-$ annihilation process one obtains a first estimate, $\tau_0
\ge 0.25 \: fm/c$ (including interpenetration time of 0.15 $fm/c$ at RHIC)
which refers to processes of very high $Q^2 \ge 10^3 \: GeV^{2}$, 
far above the saturation scale $Q^2_s$ discussed in the previous section. 
The latter scale has to be taken into account for low $p_{T}$ hadron production.\\
\\
It is the specific resolution scale $Q^2$ of a QCD
sub-process, as enveloped in the overall collision dynamics of two
slabs of given transverse partonic structure function density, that
determines which fraction of the constituent partons enters
interaction. In the simple case of extremely high $Q^2$ processes
the answer is that all constituents are resolved. However, at modest
$Q^2$ (dominating bulk hadron production) the characteristic QCD
saturation scale $Q^2_s (x)$ gains prominence, defined such that
processes with $Q^2 < Q^2_s$ do not exploit the initial transverse
parton densities at the level of independent single constituent
color field sources (see equation~\ref{eq:equation11}). For such processes the proper
formation time scale, $\tau_0$, is of order of the inverse
saturation momentum \cite{61}, $1/Q_s \sim 0.2 \: fm/c$ at
$\sqrt{s}=200 \: GeV$. The first profile of the time evolution, sketched in
Fig.~\ref{fig:Figure18}, should correspond to proper time $t=\tau_0=0.25 \: fm/c$ at
RHIC energy. At top SPS energy, $\sqrt{s}=17.3 \: GeV$, we can not
refer to such detailed QCD considerations. A pragmatic approach
suggests to take the interpenetration time, at $\gamma \approx 8.5$,
for guidance concerning the formation time, which thus results as
$\tau_0 \approx 1.5 \: fm/c$.

In summary of the above considerations we assume that the initial
partonic color sources, as contained in the structure functions
(Fig.~\ref{fig:Figure14}), are spread out in longitudinal phase space after light
cone proper time $t=\tau_0 \approx 0.2 \: fm/c$, at top RHIC energy,
and after $\tau_0 \approx 1.5 \: fm/c$ at top SPS energy. No
significant transverse expansion has occured at this early stage, in
a central collision of $A \approx 200$ nuclei with transverse
diameter of about 12 $fm$. The Bjorken estimate \cite{45} of initial
energy density $\epsilon$ (equation~\ref{eq:equation1}) refers to exactly this
condition, after formation time $\tau_0$. In order to account for
the finite longitudinal source size and interpenetration time, at RHIC, we finally 
put the average $\tau_0\approx 0.3 \: fm$, at $\sqrt{s}=200 \: GeV$, indicating the
''initialization time'' after which all partons that have been
resolved from the structure functions are engaged in shower
multiplication. As is apparent from Fig.~\ref{fig:Figure18}, this time scale is
Lorentz dilated for partons with a large longitudinal momentum, or
rapidity. This means that the slow particles are produced first
toward the center of the collision region, and the fast (large
rapidity) particles are produced later, away from the
collision region. This Bjorken ''inside-out'' correlation \cite{45}
between coordinate- and momentum-space is similar to the Hubble
expansion pattern in cosmology: more distant galaxies have higher
outward velocities. This means that the matter created in A+A
collisions at high $\sqrt{s}$ is also born expanding, however with
the difference that the Hubble flow is initially one dimensional
along the collision axis. This pattern will continue, at
$\sqrt{s}=200 \: GeV$, until the system begins to feel the effects
of finite size in the transverse direction which will occur at some
time $t_0$ in the vicinity of $1 \: fm/c$. However, the tight
correlation between position and momentum initially imprinted on the
system will survive all further expansive evolution of the initial
''firetube'', and is well recovered in the expansion pattern of the
finally released hadrons of modest $p_T$ as we shall show when
discussing radial flow and pion pair Bose-Einstein momentum
correlation (see section~\ref{subsec:Bulk_hadron_transverse_spectra} and chapter~\ref{chap:Fluctuations}).

In order to proceed to a more quantitative description of the
primordial dynamics (that occurs onward from $\tau_0$ for as long the time period
of predominantly longitudinal expansion might extend) we return to
the Bjorken estimate of energy density, corresponding to this
picture~\cite{45}, as implied by equation~\ref{eq:equation1}, which we now recast as
\begin{equation}
\epsilon = \left(\frac{dN_h}{dy} \right) \left<E^T_h\right> (\pi \: R^2_A \:
t_0)^{-1}
\label{eq:equation18}
\end{equation}
where the first term is the (average) total hadron multiplicity per
unit rapidity which , multiplied with the average hadron transverse
energy, equals the total transverse energy recorded in the
calorimetric study shown in Fig.~\ref{fig:Figure3}, as employed in~equation~\ref{eq:equation1}. 
The quantity $R_A$ is, strictly speaking, {\it not} the radius parameter of the
spherical Woods-Saxon nuclear density profile but the $rms$ of the
reactant overlap profiles as projected onto the transverse plane
(and thus slightly smaller than $R_A \approx A^{1/3} \:fm$).
Employing $A^{1/3}$ here (as is done throughout) leads to a
conservative estimate of $\epsilon$, a minor concern. However, the
basic assumption in~equation~\ref{eq:equation18} is to identify the primordial transverse
energy ''radiation'', of an interactional cylindric source of radius
$R_A$ and length $t_0$ (where $\tau_0 \le
t_0 \le 1 \: fm/c$, not Lorentz dilated at midrapidity), with the finally emerging bulk hadronic
transverse energy. We justify this assumption by the two
observations, made above, that
\begin{enumerate}
\item the bulk hadron multiplicity density per unit rapidity
$\frac{dN_h}{dy}$ resembles the parton density, primordially
released at saturation scale $\tau_0$ (Figs.~\ref{fig:Figure7},~\ref{fig:Figure16}) at $\sqrt{s}=200
\: GeV$, and that
\item the global emission pattern of bulk hadrons (in rapidity and
$p_T$) closely reflects the initial correlation between coordinate
and momentum space, characteristic of a primordial period of a
predominantly longitudinal expansion, as implied in the Bjorken
model.
\end{enumerate}
Both these observations are surprising, at first sight. The Bjorken
model was conceived for elementary hadron collisions where the
expansion proceeds into vacuum, i.e. directly toward observation.
Fig.~\ref{fig:Figure18} proposes that, to the contrary, primordially produced partons
have to transform through further, successive stages of partonic and
hadronic matter, at decreasing but still substantial energy density,
in central A+A collisions. The very fact of high energy density, with implied short mean free path
of the constituent particles, invites a hydrodynamic description
of the expansive evolution. With initial conditions fixed between
$\tau_0$ and $t_0$, an ensuing 3-dimensional hydrodynamic
expansion would preserve the primordial Bjorken-type correlation
between position and momentum space, up to lower density conditions
and, thus, close to emission of the eventually observed hadrons. We
thus feel justified to employ~equation~\ref{eq:equation1} or~\ref{eq:equation18} for the initial
conditions at RHIC, obtaining \cite{61,84}
\begin{equation}
6 \: GeV/fm^3 \le  \epsilon \le 20 \: GeV/fm^3
\label{eq:equation19}
\end{equation}
for the interval $0.3 \:fm/c \le t_0 \le 1 \: fm/c$, in central
Au+Au collisions at $y \approx 0$ and $\sqrt{s}=200 \:GeV$. The
energy density at top SPS energy, $\sqrt{s}=17.3 \: GeV$, can
similarly be estimated \cite{43,44} to amount to about $3 \:
GeV/fm^3$ at a $t_0$ of 1 $fm/c$ but we can not identify conditions
at $\tau_0<t_0$ in this case as the mere interpenetration of two Pb
nuclei takes $1.4 \: fm/c$. Thus the commonly accepted $t_0=1 \:
fm/c$ may lead to a high estimate. An application of the
parton-hadron transport model of Ellis and Geiger \cite{85,86} to
this collision finds $\epsilon=3.3 \: GeV/fm^3$ at $t=1 \:fm/c$. A
primordial energy density of about $3 \: GeV/fm^3$ is twenty times
$\rho_0 \approx 0.15 \: GeV/fm^3$, the average energy density of
ground state nuclear matter, and it also exceeds, by far, the
critical QCD energy density, of $0.6 \le \epsilon_c \le 1 \:
GeV/fm^3$ according to lattice QCD \cite{48}. The initial dynamics
thus clearly proceeds in a deconfined  QCD system also at top SPS energy,
and similarly so with strikingly higher energy density, at RHIC,
where time scales below $1 \: fm/c$ can be resolved.

However, in order now to clarify the key question as to whether, and
when conditions of partonic dynamical equilibrium may arise under
such initial conditions, we need estimates both of the proper
relaxation time scale (which will, obviously, depend on energy
density and related collision frequency), and of the expansion time
scale as governed by the overall evolution of the collision volume.
Only if $\tau (relax.) < \tau (expans.)$ one may conclude that the
''deconfined partonic system'' can be identified with a ''deconfined
QGP {\it state} of QCD matter'' as described e.g. by lattice QCD,
and implied in the phase diagram of QCD matter suggested in Fig.~\ref{fig:Figure1}.

For guidance concerning the overall time-order of the system
evolution we consider information \cite{87} obtained from
Bose-Einstein correlation analysis of pion pair emission in momentum
space (for detail see chapter~\ref{chap:Fluctuations}). Note that pions should be emitted at
{\it any} stage of the evolution, after formation time, from the
surface regions of the evolving ''fire-tube''. Bulk emission of
pions occurs, of course, after hadronization (the latest stages
illustrated in the evolution sketch given in Fig.~\ref{fig:Figure18}). The dynamical
pion source expansion models by Heinz \cite{88} and Sinyukov
\cite{89} elaborate a Gaussian emission time profile, with mean
$\tau_f$ (the decoupling time) and width $\Delta \tau$ (the duration
of emission). \\
\begin{figure}[h]   
\begin{center}
\includegraphics[scale=0.8]{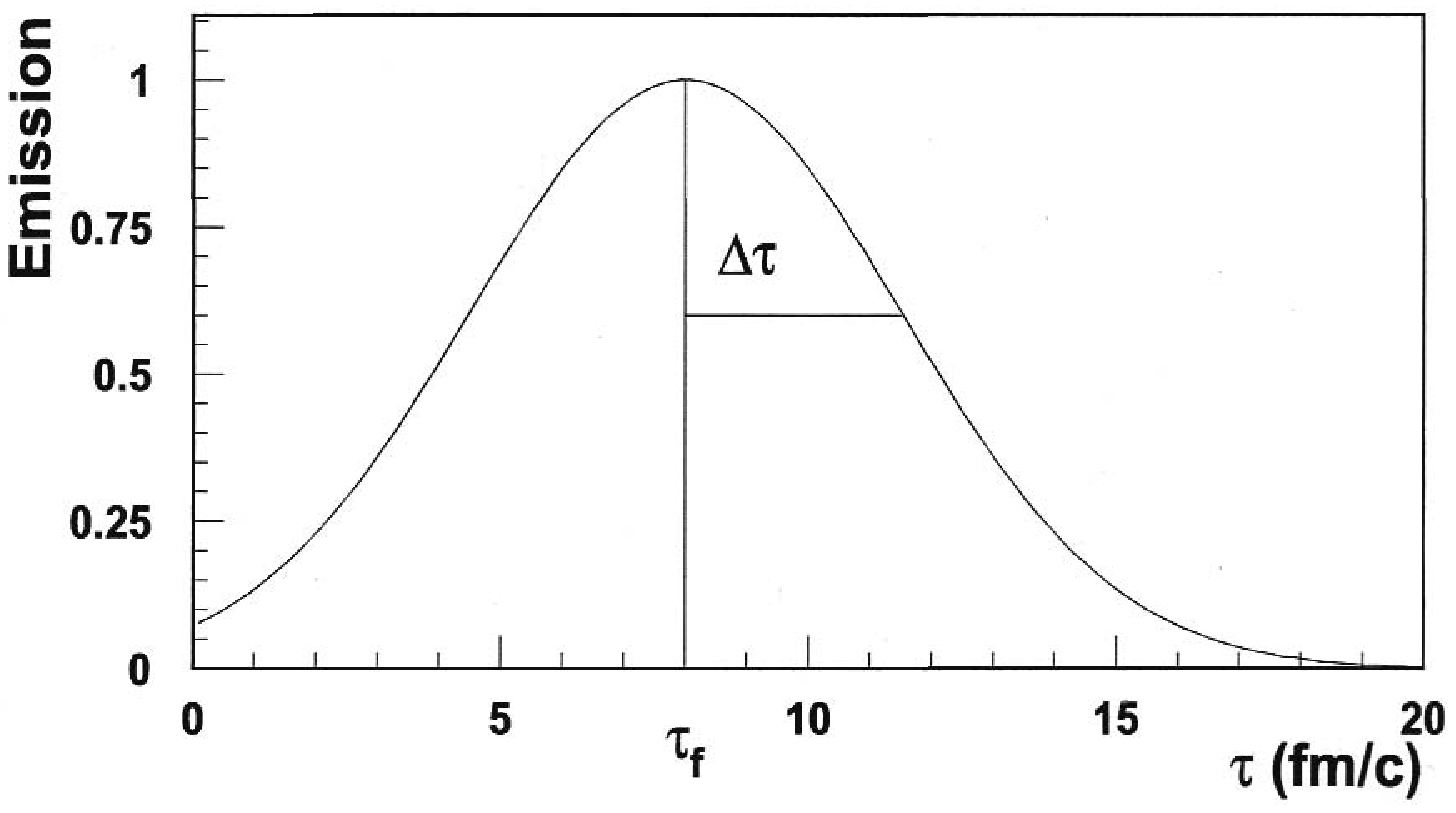}
\caption{Time profile of pion decoupling rate from the fireball in a central
Pb+Pb collision, with $\tau=0$ the end of the formation phase.
Bose-Einstein correlation of $\pi^- \pi^-$pairs yields an average
Gaussian decoupling profile with $\tau_f=8 \: fm/c$ and duration of
emission parameter $\Delta \tau=4 \: fm/c$ \cite{87,88}.}
\label{fig:Figure19}
\end{center}
\end{figure}\\
Fig.~\ref{fig:Figure19} shows an application of this analysis to
central Pb+Pb collision negative pion pair correlation data obtained
by NA49 at top SPS energy, $\sqrt{s}=17.3 \: GeV$ \cite{90}, where
$\tau_f \approx 8 \: fm/c$ and $\Delta \tau \approx 4\: fm/c$ (note that $\tau=0$
in Fig.~\ref{fig:Figure19} corresponds, not to interaction time $t=0$ but to $t
\approx 1.4 \: fm/c$, the end of the interpenetration phase). We
see, first of all, that the overall dynamical evolution of a central
Pb+Pb collision at $\sqrt{s}=17.3 \: GeV$ is ending at about $15
\:fm/c$; the proper time defines the position of the last,
decoupling profile illustrated in Fig.~\ref{fig:Figure18}, for the SPS collisions
considered here. While the details of Fig.~\ref{fig:Figure19} will turn out to be
relevant to our later discussion of hadronization (chapter~\ref{chap:hadronization}) and
hadronic expansion (chapter~\ref{chap:Elliptic_flow}), we are concerned here with the
average proper time at which the partonic phase ends. After
consideration of the duration widths of these latter expansion
phases \cite{86,87} one arrives at an estimate for the average
time, spent before hadronization, of $\Delta t=3-4 \:fm/c$, again in
agreement with the parton cascade model mentioned above \cite{86}.
This model also leads to the conclusion that parton thermal
equilibrium is, at least, closely approached locally in these central Pb+Pb
collisions as far as mid-rapidity hadron production is concerned (at
forward-backward rapidity the cascade re-scattering processes do not
suffice, however).

This finding agrees with earlier predictions of $\tau_{relax}=1-2 \:
fm/c$ at top SPS energy \cite{91}. However we note that all such
calculations employ perturbative QCD methods, implying the
paradoxical consequence that equilibrium is closely approached only
{\it at the end} of the partonic phase, at such low $\sqrt{s}$, i.e.
in a QGP state at about $T=200 \: MeV$ which is, by definition, of
non-perturbative nature. We shall return to the question of partonic
equilibrium attainment at SPS energy in the discussion of the
hadronization process in nuclear collisions (chapter~\ref{chap:hadronization}).

Equilibrium conditions should set in earlier at top RHIC energy. As
transverse partonic expansion should set in after the proper time
interval $0.3 \:fm/c \le t_0 \le 1 \: fm/c$ (which is now resolved
by the early dynamics, unlike at top SPS energy), we take guidance
from the Bjorken estimate of primordial energy density which is
based on transverse energy production {\it data}. Conservatively
interpreting the result in equation~\ref{eq:equation19} we conclude that $\epsilon$ is
about four times higher than at $\sqrt{s}=17.3 \: GeV$ in the above proper
time interval. As the binary partonic
collision frequency scales with the square of with the square of the density $\rho$ (related to the 
energy density $\epsilon$ via the relation $\epsilon$ = $\left<E\right>\rho$ = $T\rho$), and is
inversely proportional to the relaxation time $\tau_{relax}$ we
expect
\begin{equation}
\tau_{relax}\propto (1/\rho)^2 \approx (T/\epsilon)^{2}
\label{eq:equation20}
\end{equation}
which implies that $\tau_{relax}(RHIC) \approx 0.25 \: \tau_{relax}(SPS)
\approx 0.5 \: fm/c$ if we employ the estimate $T(RHIC) \: = \: 2T(SPS)$. This crude estimate is, 
however, confirmed by the parton transport model of Molar and Gyulassy \cite{92}.

Partonic equilibration at $\sqrt{s}=200 \: GeV$ should thus set in
at a time scale commensurate to the (slightly smaller) formation
time scale, at which the to be participant partons are
resolved from the initial nucleon structure functions and enter shower multiplication. 
Extrapolating to the conditions expected at LHC energy ($\sqrt{s}=5.5 \: TeV$ for
A+A collisions), where the initial parton density of the structure
functions in Fig.~\ref{fig:Figure14} is even higher ($x \approx 10^{-3}$ at
mid-rapidity), and so is the initial energy density, we may expect
conditions at which the resolved partons are almost ''born into
equilibrium''.

Early dynamical local equilibrium at RHIC is required to understand the observations 
concerning elliptic flow, with which we shall deal, in detail, in chapter~\ref{chap:Elliptic_flow}. 
This term refers to a collective anisotropic azimuthal emission pattern of bulk hadrons in
semi-peripheral collisions, a hydrodynamical phenomenon that
originates from the initial geometrical non-isotropy of the
primordial interaction zone \cite{93,94}. 
A detailed hydrodynamic model analysis of the
corresponding elliptic flow signal at RHIC \cite{95} 
leads to the conclusion that local equilibrium (a prerequisite to 
the hydrodynamic description) sets in at $t_0 \approx 0.6 \: fm/c$. This conclusion agrees with the
estimate via equation~\ref{eq:equation20} above, based on Bjorken energy density and
corresponding parton collisions frequency.

We note that the concept of a hydrodynamic
evolution appears to be, almost necessarily ingrained in the physics
of a system born into (Hubble-type) expansion, with a primordial
correlation between coordinate and momentum space, and at extreme
initial parton density at which the partonic mean free path length
$\lambda$ is close to the overall spatial resolution resulting from
the saturation scale, i.e. $\lambda \approx 1/Q_s$.

The above considerations suggest that a quark-gluon plasma state should be created early
in the expansion dynamics at $\sqrt{s}=200 \: GeV$, at about $T= 300
\: MeV$, that expands hydrodynamically until hadronization is
reached, at $T \approx 165-170 \: MeV$. Its manifestations will be
considered in chapters~\ref{chap:hadronization} to \ref{chap:Vector_Mesons}. 
At the lower SPS energy, up to $17.3 \: GeV$, we can conclude, with some caution, that a deconfined hadronic
matter system should exist at $T \approx 200 \: MeV$, in the closer
vicinity of the hadronization transition. It may closely resemble
the QGP state of lattice QCD, near $T_c$.

\section {Bulk hadron transverse spectra and radial expansion flow} 
\label{subsec:Bulk_hadron_transverse_spectra}
In this chapter we analyze bulk hadron transverse momentum
spectra obtained at SPS and RHIC energy, confronting the data with
predictions of the hydrodynamical model of collective expansion
matter flow that we have suggested in the previous section, to arise, almost
necessarily, from the primordial Hubble-type coupling between
coordinate and momentum space that prevails at the onset of the
dynamical evolution in A+A collisions at high $\sqrt{s}$. As all
hadronic transverse momentum spectra initially follow an
approximately exponential fall-off (see below) the bulk hadronic
output is represented by thermal transverse spectra at $p_T \le 2 \: GeV/c$. We
shall turn to high $p_T$ information in later sections.

Furthermore
we shall focus here on mid-rapidity production in near central A+A
collisions, because hydrodynamic models refer to an
initialization period characterized by Bjorken-type longitudinal
boost invariance, which we have seen in Figs.~\ref{fig:Figure7} and~\ref{fig:Figure9} to be
restricted to a relatively narrow interval centered at mid-rapidity.
Central collisions are selected to exploit the azimuthal symmetry of
emission, in an ideal impact parameter $b \rightarrow 0$ geometry.
We thus select the predominant, relevant hydrodynamic ''radial
flow'' expansion mode, from among other, azimuthaly oriented
(directed) flow patterns that arise once this cylindrical symmetry
(with respect to the beam direction) is broken in finite impact
parameter geometries.

In order to define, quantitatively, the flow phenomena mentioned
above, we rewrite the invariant cross section for production of
hadron species $i$ in terms of transverse momentum, rapidity, impact
parameter $b$ and azimuthal emission angle $\varphi_p$ (relative to
the reaction plane),
\begin{equation}
\frac{dN_i(b)}{p_Tdp_T dy d\varphi_p} = \frac{1}{2 \: \pi}\:
\frac{dN_i(b)}{p_Tdp_Tdy} \: {\left[1 + 2v_1^i \: (p_T,b) cos \varphi_p +
2v_2^i \: (p_T,b) cos (2 \varphi_p)+ . . .\right]}
\label{eq:equation21}
\end{equation}
where we have expanded the dependence on $\varphi_p$ into a Fourier
series. Due to reflection symmetry with respect to the reaction
plane in collisions of identical nuclei, only cosine terms appear.
Restricting to mid-rapidity production all odd harmonics vanish, in
particular the ''directed flow'' coefficient $v^i_1$, and we have
dropped the  y-dependence in the flow coefficients $v^i_1$ and
$v^i_2$. The latter quantifies the amount of ''elliptic flow'', to
which we turn in chapter~\ref{chap:Elliptic_flow}. In the following, we will restrict to
central collisions which we shall idealize as near-zero impact
parameter processes governed by cylinder symmetry, whence all
azimuthal dependence (expressed by the $v^i_1, \: v^i_2, . . .$
terms) vanishes, and the invariant cross section reduces to the
first term in equation~\ref{eq:equation21}, which by definition also corresponds to all
measurements in which the orientation of the reaction plane is not
observed.

\begin{figure}[h]   
\begin{center}
\includegraphics[scale=0.53]{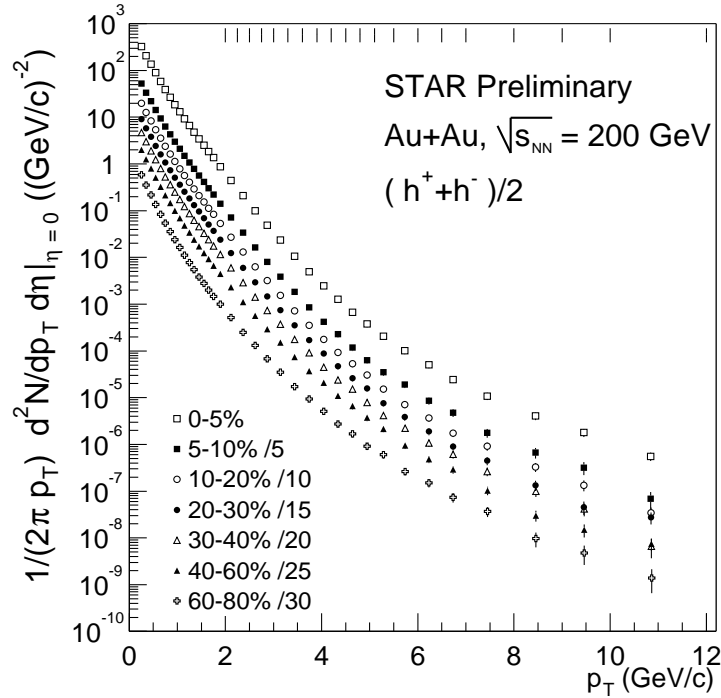}
\caption{Transverse momentum spectra of charged hadrons in Au+Au collisions
at $\sqrt{s}=200 \: GeV$, in dependence of collision centrality
\cite{97} (offset as indicated), featuring transition from
exponential to power law shape.}
\label{fig:Figure20}
\end{center}
\end{figure}

Typical transverse momentum spectra of the latter type are shown in
Fig.~\ref{fig:Figure20}, for charged hadron production in Au+Au collisions at
$\sqrt{s}=200 \: GeV$, exhibiting mid-rapidity data at various
collision centralities \cite{97}. We observe a clear-cut transition,
from bulk hadron emission at $p_T \le 2 \: GeV/c$ featuring a
near-exponential cross section (i.e. a thermal spectrum), to a high
$p_T$ power-law spectral pattern. Within the context of our previous
discussion (section~\ref{subsec:Gluon_Satu_in_AA_Coll}) we tentatively identify the low $p_T$ region
with the QCD physics near saturation scale. Hadron production at
$p_T \rightarrow 10 \: GeV/c$ should, on the other hand, be the
consequence of primordial leading parton fragmentation originating
from ''hard'', high $Q^2$ perturbative QCD processes.

We thus identify bulk hadron production at low $p_T$ as the
emergence of the initial parton saturation conditions that give rise
to high energy density and small equilibration time scale, leading
to a hydrodynamical bulk matter expansion evolution. Conversely, the
initially produced hard partons, from high $Q^2$ processes, are not
thermalized into the bulk but traverse it, as tracers, while being
attenuated by medium-induced rescattering and gluon radiation, the
combined effects being reflected in the high $p_T$ inclusive hadron
yield, and in jet correlations of hadron emission. We shall turn to
the latter physics observables in chapter~\ref{chap:In_medium_high_pt}, while staying here with
low $p_T$ physics, related to hydrodynamical expansion modes,
focusing on radially symmetric expansion.\\
\\
In order to infer from the spectral shapes of the hadronic species
about the expansion mechanism, we first transform to the transverse
mass variable, $m_T=(p^2_T + m^2)^{1/2}$, via
\begin{equation}
\frac{1}{2\pi} \: \frac{dN_i}{p_T dp_T dy} = \frac{1}{2 \pi} \:
\frac{dN_i}{m_T dm_T dy}
\label{eq:equation22}
\end{equation}
because it has been shown in p+p collisions \cite{98} near RHIC
energy that the $m_T$ distributions of various hadronic species
exhibit a universal pattern (''$m_T$ scaling'') at low $m_T$:
\begin{equation}
\frac{1}{2\pi} \: \frac{dN_i}{m_T dm_T dy} = A_i \: exp (-m^i_T / T)
\label{eq:equation23}
\end{equation}
with a universal inverse slope parameter $T$ and a species dependent
normalization factor $A$. Hagedorn showed \cite{99} that this
scaling is characteristic of an adiabatic expansion of a fireball at
temperature $T$. We recall that, on the other hand, an ideal
hydrodynamical expansion is isentropic.\\
\begin{figure}[h]
\begin{center}
\includegraphics[scale=0.65]{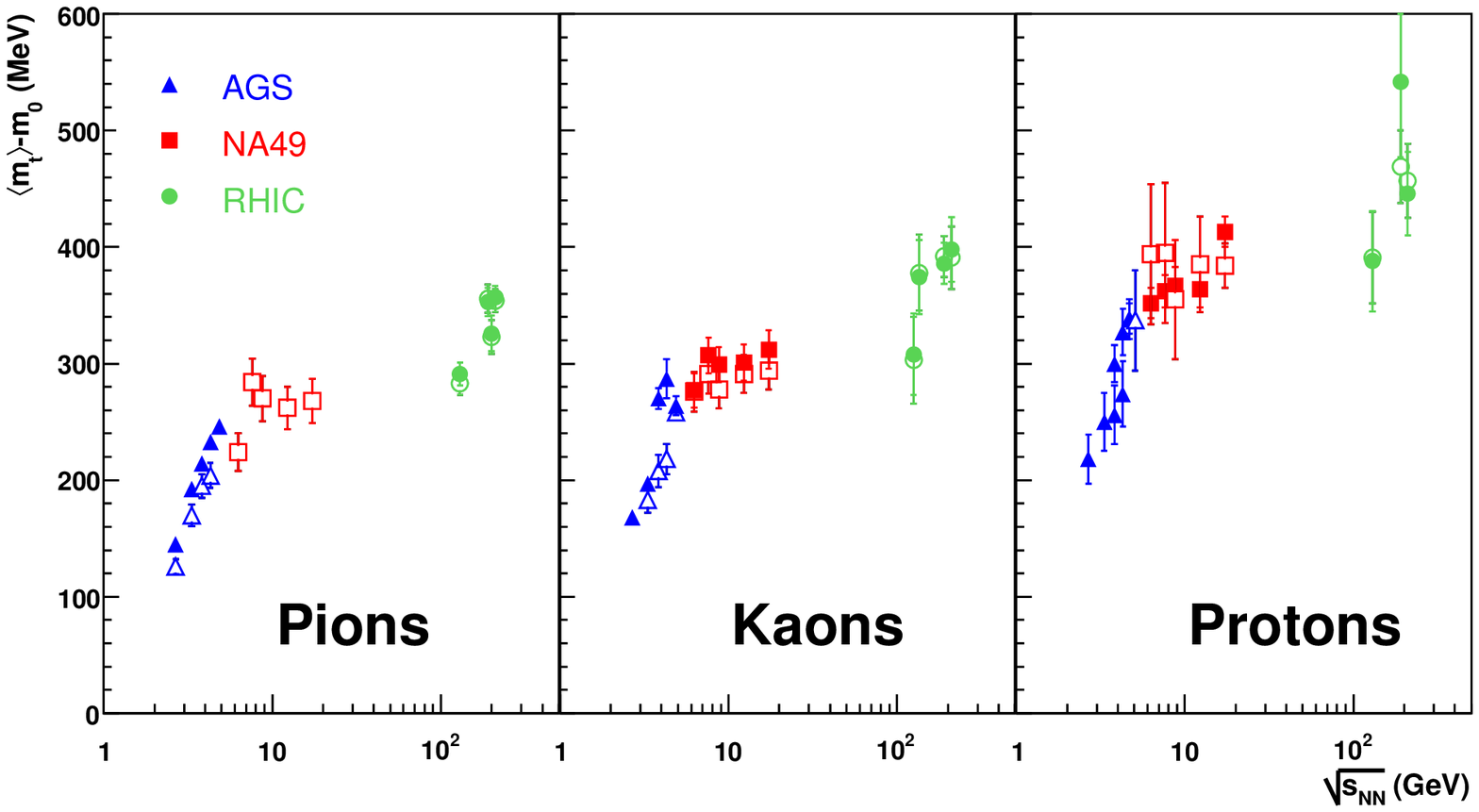}
\caption{The average transverse kinetic energy $\left<m_T\right> - m_0$ for pions, kaons
and protons vs. $\sqrt{s}$ in central Au+Au/Pb+Pb collisions
\cite{54}. Open symbols represent negative hadrons.}
\label{fig:Figure21}
\end{center}
\end{figure}\\
Fig.~\ref{fig:Figure21} shows the $\sqrt{s}$ dependence of the average transverse
kinetic energy $\left<m^i_T\right> -m^i$ for pions, kaons and protons observed
at mid-rapidity in central Au+Au/Pb+Pb collisions \cite{54}.
Similarly, the inverse slope parameter $T$ resulting from a fit of
equation~\ref{eq:equation23} to $K^+$ and $K^-$ transverse mass spectra  (at $p_T \le 2
\: GeV/c$) is shown in Fig.~\ref{fig:Figure22}, both for nuclear and p+p collisions
\cite {100}. We see, first of all, that $m_T$ scaling does not apply
in A+A collisions, and that the kaon inverse slope parameter, $T \approx
230 \: MeV$ over the SPS energy regime, can not be identified with
the fireball temperature at hadron formation which is $T_h \approx
165 \: MeV$ from Fig.~\ref{fig:Figure1}. The latter is seen, however, to be well represented
by the p+p spectral data exhibited in the left panel of Fig.~\ref{fig:Figure22}. There is, thus, not 
only thermal energy present in A+A transverse expansion, but also hydrodynamical radial
flow.

We note that the indications in Figs.~\ref{fig:Figure21} and~\ref{fig:Figure22}, of a plateau in both
$\left<m_T\right>$ and $T$, extending over the domain of SPS energies, $6 \le
\sqrt{s} \le 17 \: GeV$, have not yet been explained by any
fundamental expansive evolution model, including hydrodynamics.
Within the framework of the latter model, this is a consequence of
the {\it initialization problem} \cite{96} which requires a detailed
modeling, both of primordial energy density vs. equilibration time
scale, and of the appropriate partonic matter equation of state
(EOS) which relates expansion pressure to energy density. At top
RHIC energy, this initialization of hydro-flow occurs, both, at a
time scale $t_0 \approx 0.6 \: fm/c$ which is far smaller than the
time scale of eventual bulk hadronization ($t \approx 3 \: fm/c$),
and at a primordial energy density far in excess of the critical QCD
confinement density. After initialization, the partonic plasma phase
thus dominates the overall expansive evolution,over a time interval
far exceeding the formation and relaxation time scale. \\
\begin{figure}[h]
\begin{center}
\includegraphics[scale=0.32]{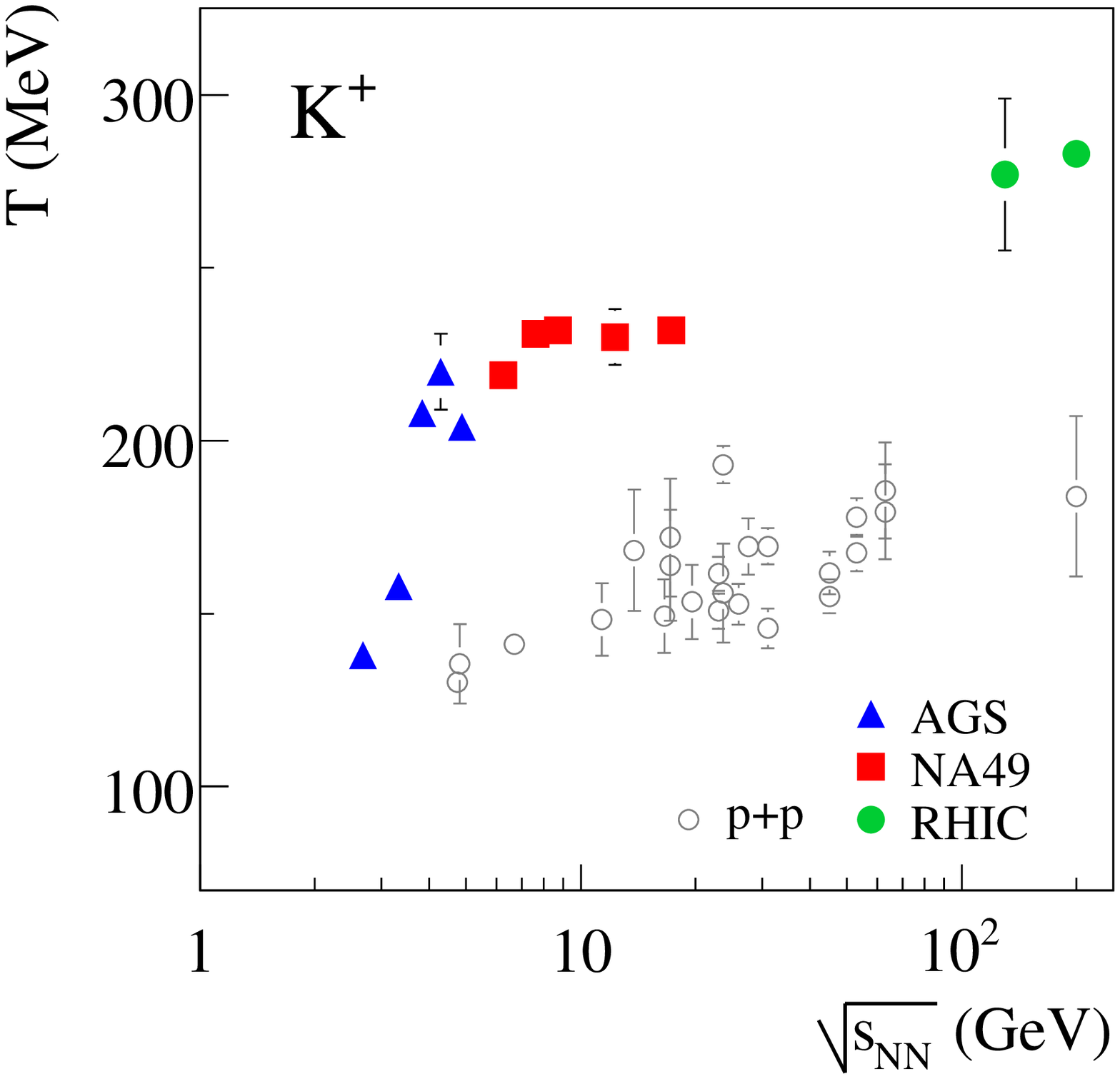}
\includegraphics[scale=0.32]{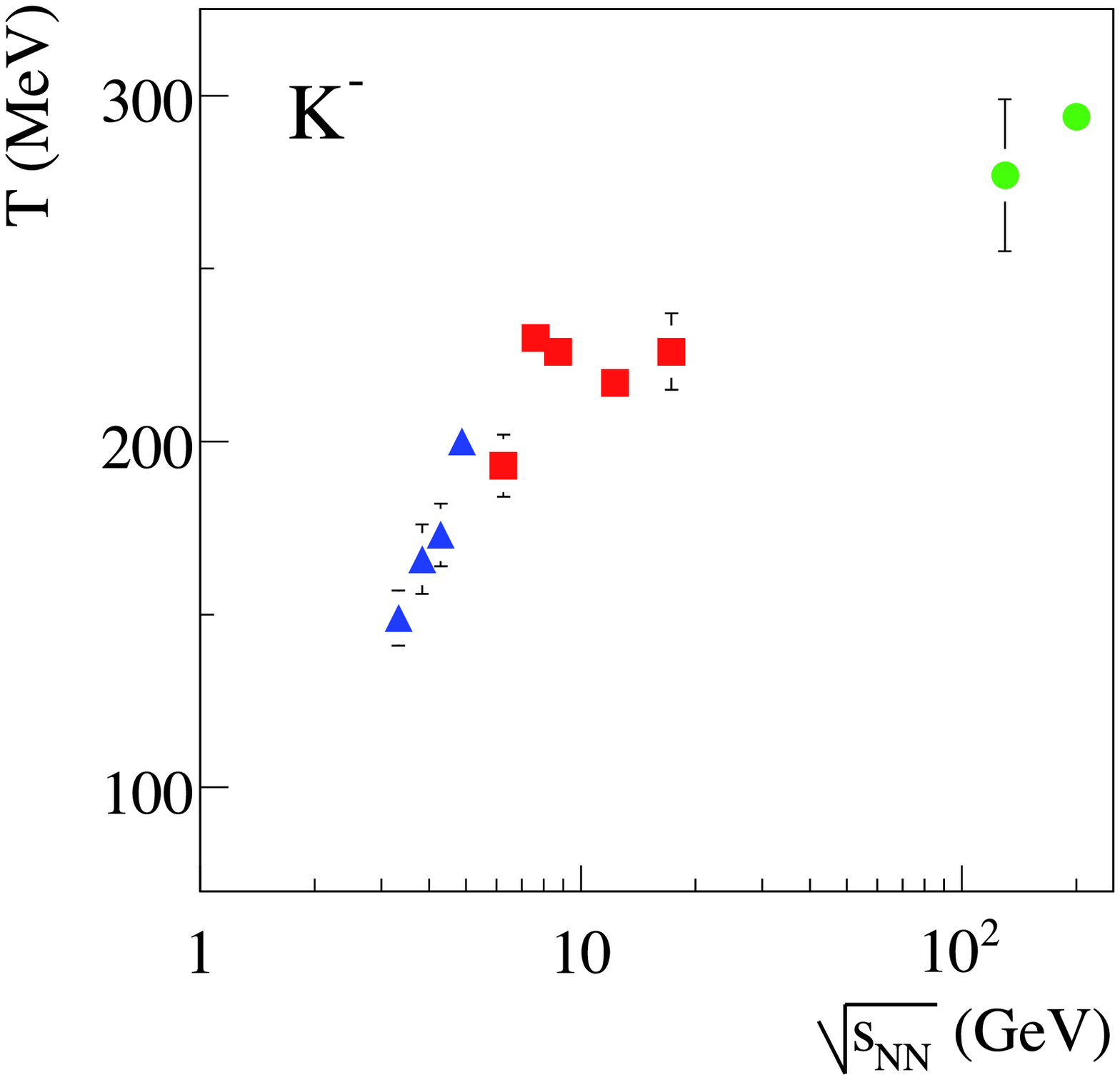}
\caption{The inverse slope parameter $T$ of equation~\ref{eq:equation23} for $K^+ \mbox{and} \:
K^-$ transverse mass spectra at $p_T < 2 \: GeV/c$ and mid-rapidity
in central A+A, and in minimum bias p+p collisions \cite{100}.}
\label{fig:Figure22}
\end{center}
\end{figure}
Thus, at RHIC energy, parton transport \cite{92} and relativistic hydrodynamic \cite{95, 96} 
models establish a well developed expansion mode that survives the
subsequent stages of hadronization and hadronic expansion. This is
reflected in their success in describing elliptic flow. On the other
hand, the hydrodynamical model far overestimates elliptic flow at
SPS energy \cite{96} at which, as we have shown in section\ref{subsec:Transvers_phase_space}, the
initialization period may be not well separated from the confinement
(hadronization) stage. Thus, whereas the expansion evolution at
$\sqrt{s}=200 \: GeV$ (occuring at near-zero baryo-chemical
potential in Fig.~\ref{fig:Figure1}) ''races'' across the parton-hadron phase
boundary with fully established flow patterns, near $\mu_B=0$ where
lattice QCD predicts the phase transformation to be merely a soft
cross-over \cite{16}, the dynamics at $\sqrt{s}=10-20 \: GeV$ may
originate from only slightly above, or even at the phase boundary,
thus sampling the domain $200 \le \mu_B \le 500 \: MeV$ where the
equation of state might exhibit a ''softest point'' \cite{96}. The hydrodynamic model thus faces 
formidable uncertainties regarding initialization at SPS energy. 

The plateau in Figs.~\ref{fig:Figure21},~\ref{fig:Figure22} may be the consequence of the fact
that not much flow is generated in, or transmitted from the partonic
phase, at SPS energies, because it is initialized close to the phase
boundary \cite{100} where the expected critical point \cite{9,10}
(Fig.~\ref{fig:Figure1}), and the corresponding adjacent first order phase transition
might focus \cite{101} or stall \cite{96} the expansion trajectory,
such that the observed radial flow stems almost exclusively from the
hadronic expansion phase. The SPS plateau, which we shall
subsequently encounter in other bulk hadron variables (elliptic
flow, HBT radii) might thus emerge as a consequence of the critical
point or, in general, of the flatness of the parton-hadron
coexistence line. RHIC dynamics, on the other hand, originates from
far above this line.

Hadronic expansion is known to proceed isentropically \cite{102}:
commensurate to expansive volume increase the momentum space volume
must decrease, from a random isotropic thermal distribution to a
restricted momentum orientation preferentially perpendicular to the fireball
surface, i.e. radial. The initial thermal energy, implied by the
hadron formation temperature $T_H=165 \: MeV$, will thus fall down
to a residual $T_F$ at hadronic decoupling from the flow field
(''thermal freeze-out'') plus a radial transverse kinetic energy
term $m_i \left< \beta _T\right>^2$ where $m_i$ is the mass of the considered
hadron species and $\left<\beta_T\right>$ the average radial velocity. We thus
expect \cite{103} for the slope of equation~\ref{eq:equation23}:
\begin{equation}
T=T_F+m_i\left<\beta _T\right>^2, \: p_T \le 2 \: GeV/c
\label{eq:equation24}
\end{equation}
and
\begin{equation}
T=T_F \left( \frac{1+\left<v_T\right>}{1-\left<v_T\right>} \right)^{1/2}, \: p_T \gg m_i
\label{eq:equation25}
\end{equation}
the latter expression valid at $p_T$ larger than hadron mass
scale ($T$ then is the ''blue-shifted temperature'' at decoupling
\cite{104} and $\left<v_T\right>$ the average transverse velocity). The
assumption that radial flow mostly originates from the hadronic
expansion phase is underlined by the proportionality of flow energy
to hadron mass (equation~\ref{eq:equation24}). \\
\begin{figure}[h] 
\begin{center}
\includegraphics[scale=1.05]{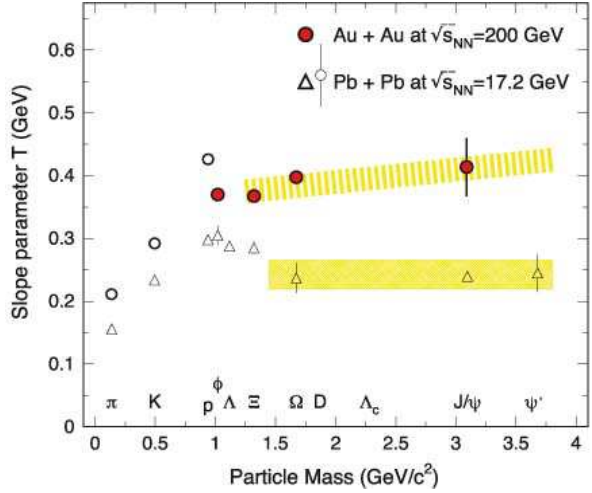}
\caption{Hadron slope parameters $T$ at mid-rapidity as a function of mass.
For Pb+Pb at $\sqrt{s}=17.3 \: GeV$ (triangles) and Au+Au at
$\sqrt{s}=200 \: GeV$ (circles); from \cite{103}.}
\label{fig:Figure23}
\end{center}\vspace{-0.6cm}
\end{figure}
Fig.~\ref{fig:Figure23} illustrates this proportionality, by a recent compilation \cite{103} of 
RHIC results for central Au+Au collisions at $\sqrt{s}= 200 \: GeV$, and SPS results for 
central Pb+Pb collisions at top SPS energy, $\sqrt{s}=17.3 \: GeV$. At the
latter energy the slope parameter of the $\Phi$ meson is seen to be
close to that of the similar mass baryons $p$ and $\Lambda$,
emphasizing the occurence of $m_i$  scaling as opposed to valence
quark number scaling that we will encounter in RHIC elliptic flow
data \cite{94}. As is obvious from Fig.~\ref{fig:Figure23} the multi-strange hyperons
and charmonia exhibit a slope saturation which is usually explained
\cite{103} as a consequence of their small total cross sections of
rescattering from other hadrons, leading to an early decoupling from
the bulk hadron radial flow field, such that $\left<\beta_T\right> _{\Omega} \:
< \: \left<\beta_T\right>_p$. \\
\begin{figure}[h]
\begin{center}
\includegraphics[scale=1.05]{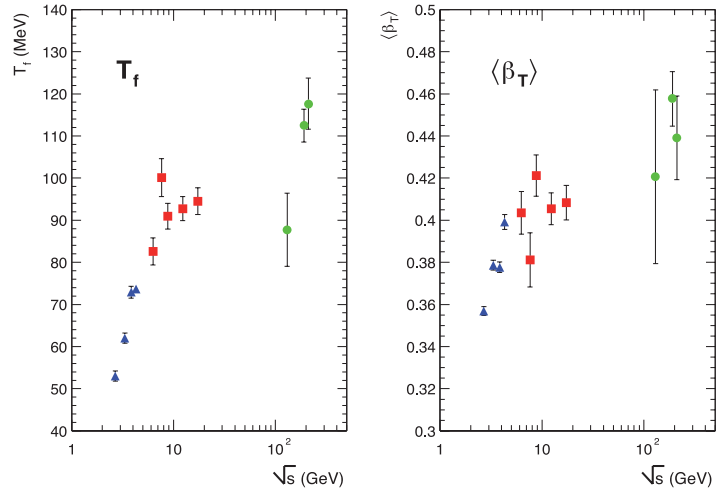}
\caption{Hadron decoupling temperature $T_f$, and average radial flow
velocity $\left<\beta_T\right>$ extracted from blast wave model (see equation~\ref{eq:equation26}) fits of
$m_T$ spectra vs. $\sqrt{s}$ \cite{54}.}
\label{fig:Figure24}
\end{center}
\end{figure}
According to our observations with equation~\ref{eq:equation24} a hydrodynamical ansatz
for the transverse mass spectrum of hadrons should thus contain the
variables ''true temperature'' $T_F$ at decoupling from the flow
field, and its average velocity $\left<\beta_T\right>$, common to all hadrons.
This is the case for the blast wave model \cite{104} developed as an
approximation to the full hydrodynamic formalism
\cite{96}, assuming a common decoupling or ''freeze-out'' from flow,
for all hadronic species, and a boost-invariant longitudinal
expansion:
\begin{equation}
\frac{dN_i}{m_T dm_T dy} = A_i \: m_T \: K_1 \: \left(\frac{m_T cosh
\rho}{T_F} \right) \: I_0 \left( \frac{p_T sinh \rho}{T_F} \right)
\label{eq:equation26}
\end{equation}
where $\rho=tanh ^{-1}\beta_T$. In an extended version of this model
a function is included that describes the radial profile of the
transverse velocity field, $\beta_T(r)=\beta_T^{max} \: r/R$,
instead of employing a fixed $\beta_T$ at decoupling \cite{106}.
Fig.~\ref{fig:Figure24} shows \cite{54} the resulting energy dependence of $T_F$ and
$\left<\beta_T\right>$, for the same set of data as implied already in Figs.~\ref{fig:Figure21}
and~\ref{fig:Figure22}. The ''true'' decoupling temperature rises steeply at the AGS and less so at
SPS energy (as does $\left<\beta_T\right>$), to a value of about $95 \:
MeV$ at top SPS energy, which is considerably lower than the
chemical freeze-out temperature, $T_H = 165 \: MeV$, at which the
hadronic species relative yield composition of the hadronic phase
becomes stationary (see chapter~\ref{chap:hadronization}, and Fig.~\ref{fig:Figure1}). Chemical decoupling
thus occurs early, near the parton-hadron phase boundary, whereas
hadronic radial flow ceases after significant further expansion and
cooling, whence the surface radial velocity (its average value given
by $\left<\beta_T\right>$ in Fig.~\ref{fig:Figure24}) approaches $\beta_T \approx 0.65$. Both
data sets again exhibit an indication of saturation, over the interval toward top SPS energy: the SPS plateau.
This supports our above conjecture that radial flow is,
predominantly, a consequence of isentropic bulk hadronic expansion
in this energy domain, which sets in at $T_H$. At RHIC energy, both
parameters exhibit a further rise, suggesting that primordial
partonic flow begins to contribute significantly to radial flow.

%% file: Chapter_2.tex
\chapter{Hadronization and hadronic freeze-out in A+A collisions}
\label{chap:hadronization}

Within the course of the global expansion of the primordial reaction volume 
the local flow ''cells'' will hit the parton-hadron phase boundary
as their energy density approaches $\epsilon_{crit} \approx 1 \:
GeV/fm^3$.  Hadronization will thus occur, not at an instant over
the entire interaction volume, but within a finite overall time
interval \cite{86} that results from the spread of proper time at
which individual cells, or coherent clusters of such cells (as
developed during expansion) arrive at the phase boundary. However,
irrespective of such a local-temporal occurence, the hadronization
process (which is governed by non perturbative QCD at
the low $Q^2$ corresponding to bulk hadronization) universally
results in a novel, {\it global} equilibrium property that concerns
the relative abundance of produced hadrons and resonances. This
so-called ''hadrochemical equilibrium state'' is directly
observable, in contrast to the stages of primordial parton
equilibration that are only indirectly assessed, via dynamical model
studies.

This equilibrium population of species occurs both in elementary and
nuclear collisions \cite{107}. We have seen in Fig.~\ref{fig:Figure17} a first
illustration, by $e^+e^-$ annihilation data at $\sqrt{s}=91.2 \:
GeV$ LEP energy, that are well reproduced by the partition functions
of the statistical hadronization model (SHM) in its canonical form
\cite{84}. The derived hadronization temperature, $T_H = 165 \:
MeV$, turns out to be universal to all elementary and nuclear
collision processes at $\sqrt{s} \ge 20 \: GeV$, and it agrees with
the limiting temperature predicted by Hagedorn \cite{38} to occur in
any multi-hadronic equilibrium system once the energy density
approaches about $0.6 \: GeV/fm^3$. Thus, the upper limit of
hadronic equilibrium density corresponds, closely, to the lower
limit, $\epsilon_{crit} = 0.6 - 1.0 \: GeV/fm^3$ of partonic
equilibrium matter, according to lattice QCD \cite{48}. In
elementary collisions only about 20 partons or hadrons participate:
there should be no chance to approach thermodynamic equilibrium of
species by rescattering cascades, neither in the partonic nor in the
hadronic phase. The fact that, nevertheless, the hadron formation
temperature $T_H$ coincides with the Hagedorn limiting temperature
and with the QCD confinement temperature, is a consequence of the
non-perturbative QCD hadronization process itself \cite{85}, which
''gives birth'' to hadrons/resonances in canonical equilibrium, at
high $\sqrt{s}$, as we shall see below. This process also governs
A+A collisions but, as it occurs here under conditions of high
energy density extended over considerable volume, the SHM
description now requires a {\it grand} canonical ensemble, with
important consequences for production of strange hadrons
(strangeness enhancement).

The grand canonical order of hadron/resonance production in central
A+A collisions, and its characteristic strangeness enhancement
shows that a state of extended matter that is quantum mechanically
coherent must exist at hadronization \cite{87, 88, 107}. Whether or not it
also reflects partonic equilibrium properties (including flavor
equilibrium), that would allow us to claim the direct observation of
a quark gluon plasma state near $T_c$, can not be decided on the
basis of this observation alone, as the hadronization
process somehow generates, by itself, the observed hadronic
equilibrium. This conclusion, however, is still the subject of
controversy \cite{107}.\\
\begin{figure}[h!]   
\begin{center}\vspace{-0.8cm}
\includegraphics[scale=0.5]{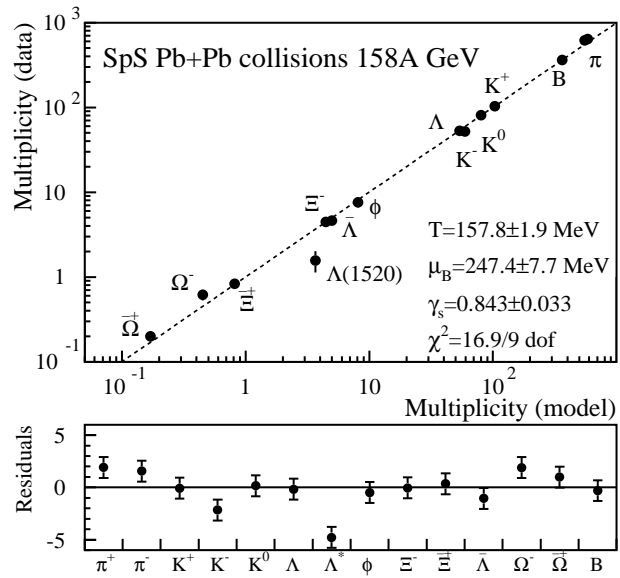}
\caption{Total hadron multiplicities in central Pb+Pb collisions at
$\sqrt{s}=17.3 \: GeV$ \cite{100} versus prediction of the grand
canonical statistical hadronization model \cite{19}.}
\label{fig:Figure25}
\end{center}\vspace{-0.6cm}
\end{figure} \\
Two typical examples of grand canonical SHM application are
illustrated in Figs.~\ref{fig:Figure25} and~\ref{fig:Figure26}, the first showing total hadron
multiplicities in central Pb+Pb collisions at $\sqrt{s}=17.3 \: GeV$
by NA49 \cite{100} confronted with SHM predictions by Becattini et
al.\ \cite{19}.
This plot is similar to Fig.~\ref{fig:Figure17} in which $e^+ e^-$ annihilation to hadrons is confronted with a SHM prediction derived from the \emph{canonical} ensemble \cite{84}.
Central Au+Au collision data at $\sqrt{s}=200 \: GeV$
from several RHIC experiments are compared to grand canonical model
predictions by Braun-Munzinger et al. \cite{108} in Fig.~\ref{fig:Figure26}. The key
model parameters, $T_H$ and the baryo-chemical potential $\mu_B$
result as $159 \: MeV$ ($160 \: MeV)$, and $247 \: MeV$ ($20 \:
MeV)$ at $\sqrt{s}=17.3 \: (200) \: GeV$, respectively. The
universality of the hadronization temperature is obvious from
comparison of the present values with the results of the canonical
procedure employed in $e^+e^-$ annihilation to hadrons at
$\sqrt{s}=91.2 \: GeV$ (Fig.~\ref{fig:Figure17}), and in canonical SHM fits
\cite{109} to p+p collision data at $\sqrt{s}=27.4 \: GeV$ where
$T_H=159$ and $169 \: MeV$, respectively.\\
\begin{figure}[h!]   
\begin{center}
\includegraphics[scale=0.7]{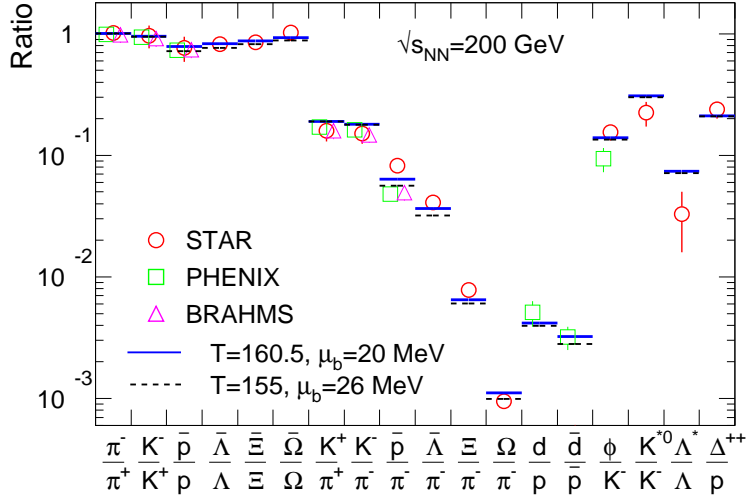}
\caption{Hadron multiplicity ratios at mid-rapidity in central Au+Au
collisions at $\sqrt{s}=200 \: GeV$ from RHIC experiments STAR,
PHENIX and BRAHMS, compared to predictions of the grand canonical
statistical model \cite{108}.}
\label{fig:Figure26}
\end{center}
\end{figure} \\
Figs.~\ref{fig:Figure25} and~\ref{fig:Figure26} illustrate two different approaches 
employed in grand canonical SHM application, the former addressing the values of the hadronic multiplicities as obtained in approximate
full $4 \pi$ acceptance (within limitations implied by detector
performance), the latter employing a set of multiplicity {\it
ratios} obtained in the vicinity of mid-rapidity as invited, at RHIC
energy, by the limited acceptance of the STAR and PHENIX
experiments. The latter approach is appropriate, clearly, in the
limit of boost-invariant rapidity distributions where hadron
production ratios would not depend on the choice of the observational
rapidity interval. We have shown in section~\ref{subsec:Rap_Distribution} that such conditions
do, in fact, set in at top RHIC energy, as referred to in Fig.~\ref{fig:Figure26}.
However, at low $\sqrt{s}$ the y-distributions are far from
boost-invariant, and the total rapidity gap $\Delta y$ may become
comparable, in the extreme case, to the natural rapidity widths of
hadrons emitted in the idealized situation of a single,
isotropically decaying fireball positioned at mid-rapidity. Its
rapidity spectra, equation~\ref{eq:equation5}, resemble Gaussians with widths $\Gamma_i
\approx 2.35 \: (T/m_i)^{1/2}$ for hadron masses $m_i$. Clearly, the
particle ratios $(dN_i/dy)/(dN_j/dy)$ then depend strongly on the
position of the rapidity interval $dy$: away from $y=0$ heavy
hadrons will be strongly suppressed, and particle yields in narrow
rapidity intervals are useless for a statistical model analysis
unless it is known a priori that the radiator is a single stationary
spherical fireball \cite{110}. This is not the case toward top SPS
energy (see Fig.~\ref{fig:Figure10}), due to significant primordial longitudinal
expansion of the hadron emitting source. Given such conditions, the
total multiplicity per collision event (the invariant yield divided
by the total overall inelastic cross section) should be employed in
the SHM analysis, as is exemplified in Fig.~\ref{fig:Figure25}.

\section{Hadronic freeze-out from expansion flow}
\label{sec:Hadronic_freeze_out}
The hadronic multiplicities result from integration of the
invariant triple differential cross section over $p_T$ and $y$. Instrumental, 
experiment-specific conditions tend to result in incomplete $p_{T}$ and/or $y$
acceptances. It is important to ascertain that the effects of hydrodynamic transverse
and longitudinal flow do not blast a significant part of the total hadron yield to outside the acceptance,
and that they, more generally, do not change the relative hadron yield composition, thus basically affecting 
the SHM analysis. To see that hadronization incorporates only the internal energy in the co-moving
frame \cite{110}, we first assume that hadrochemical freeze-out
occurs on a sharp hypersurface $\Sigma$, and write the total yield
of particle species $i$ as
\begin{equation}
N_i=\int \frac{d^3p}{E} \int_\Sigma \: p^{\mu}d^3 \sigma_{\mu} (x)\:
f_i (x,p)= \int_{\Sigma}d^3 \sigma_{\mu} (x) j_i^{\mu} (x)
\label{eq:equation27}
\end{equation}
where $d^3 \sigma$ is the outward normal vector on the surface, and
\begin{equation}
j_i^{\mu}(x) = g_i \int d^{4}p 2 \Theta (p^0) \delta(p^2-m^2_i) \:
p^{\mu}(exp \: [p \cdot u(x)-\mu_i]/T \pm 1)^{-1}
\label{eq:equation28}
\end{equation}
is the grand canonical number current density of species $i$, $\mu_i$
the chemical potential, $u(x)$ the local flow velocity, and $g_i$
the degeneracy factor. In thermal equilibrium it is given by
\begin{eqnarray}
j_i^{\mu}(x)& = & \rho_i(x) u^{\mu}(x) \: \: \mbox{with} \nonumber\\
 \rho_i(x) & = & u_{\mu}(x) j_i^{\mu}(x) = \int d^4 p 2 \Theta
(p^0)\delta (p^2-m_i^2) \: p \cdot u(x) \: f_i (p \cdot u (x); T;
\mu_i) \nonumber \\
 & = & \int d^3 p' \: f_i(E_{p'}; T, \mu_i)=\rho_i(T, \mu_i).
\label{eq:equation29}
\end{eqnarray}
Here $E_{p'}$ is the energy in the local rest frame at point $x$.
The total particle yield of species $i$ is therefore
\begin{equation}
N_i=\rho_i (T, \mu_i) \int_\Sigma d^3 \sigma_{\mu}(x) u^{\mu}(x) =
\rho_i (T,\mu_i)\: V_{\Sigma}(u^{\mu})
\label{eq:equation30}
\end{equation}
where only the total comoving volume $V_{\Sigma}$ of the freeze-out
hypersurface $\Sigma$ depends on the flow profile $u^{\mu}$. $V$ is
thus a common total volume factor at hadronization (to be determined separately), and the flow pattern
drops out from the yield distribution over species in $4 \pi$ acceptance \cite{110}.
For nuclear collisions at SPS energies and below one thus should
perform a SHM analysis of the total, $4 \pi$-integrated hadronic
multiplicities, as was done in Fig.~\ref{fig:Figure25}.

We note that the derivation above illustrates the termination
problem of the hydrodynamic description of A+A collisions, the
validity of which depends on conditions of a short mean free path,
$\lambda < 1 \: fm$. A precise argumentation suggests
that two different free paths are relevant here, concerning hadron
occupation number and hadron spectral freeze-out, respectively. As hadrochemical
freeze-out occurs in the immediate vicinity of $T_c$ (and $T_H
\approx 160-165\: MeV$ from Figs.~\ref{fig:Figure25},~\ref{fig:Figure26}), the hadron species
distribution stays constant throughout the ensuing hadronic phase,
i.e the ''chemical'' mean free path abruptly becomes infinite at
$T_H$, whereas elastic and resonant rescattering may well extend far
into the hadronic phase, and so does collective pressure and flow.
In fact we have seen in section~\ref{subsec:Bulk_hadron_transverse_spectra} that the decoupling from flow
occurs at $T_{F}$ as low as $90-100 \: MeV$ (Fig.~\ref{fig:Figure24}). Thus the hydrodynamic
evolution of high $\sqrt{s}$ collisions has to be, somehow
artificially, stopped at the parton-hadron boundary in order to get
the correct hadron multiplicities $N_i$, of equations~\ref{eq:equation27} to~\ref{eq:equation30}, which then stay frozen-out during the subsequent hadronic expansion.

The equations (\ref{eq:equation27}-\ref{eq:equation30}) demonstrate the application of the
Cooper-Frye prescription \cite{111} for termination of the
hydrodynamic evolution. The hyper-surface $\Sigma$ describes the
space-time location at which individual flow cells arrive at the
freeze-out conditions, $\epsilon = \epsilon_c$ and $T=T_c$, of hadronization. At this point, the resulting hadron/resonance spectra (for
species $i$) are then given by the Cooper-Frye formula
\begin{equation}
E \frac{dN_i}{d^3p} = \frac{dN_i}{dy p_Td p_T}  = \frac{g_i}{(2
\pi)^3} \: \int_{\Sigma} f_i (p \cdot u(x),x) p \cdot d^3 \sigma
(x),
\label{eq:equation31}
\end{equation}
where $p^{\mu} f_i d^3 \sigma_{\mu}$ is the local flux of particle
$i$ with momentum $p$ through the surface $\Sigma$. For the phase
space distribution $f$ in this formula one takes the local
equilibrium distribution at hadronic species freeze-out from the grand canonical SHM
\begin{equation}
f_i (E,x) = [exp \{(E_i - \mu_i(x))/T \} \pm 1]^{-1}
\label{eq:equation32}
\end{equation}
boosted with the local flow velocity $u^{\mu}(x)$ to the global
reference frame by the substitution $E \rightarrow p \cdot u(x)$.
Fixing $T=T_c$ (taken e.g. from lattice QCD) the hadron
multiplicities $N_i$ then follow from equation~\ref{eq:equation30}, and one compares to
experiment, as in Figs.~\ref{fig:Figure25},~\ref{fig:Figure26}. In order now to follow the further
evolution, throughout the hadronic rescattering phase, and to
finally compare predictions of equation~\ref{eq:equation31} to the observed flow data as represented by
the various Fourier-terms of equation~\ref{eq:equation21} one has to re-initialize (with
hadronic EOS) the expansion from $\Sigma$($T_{c}$ = 165 MeV) until final decoupling
\cite{96}, at $T \approx 100 \: MeV$, thus describing e.g. radial and elliptic flow.

\begin{figure}[h!]   
\begin{center}
\includegraphics[scale=0.56]{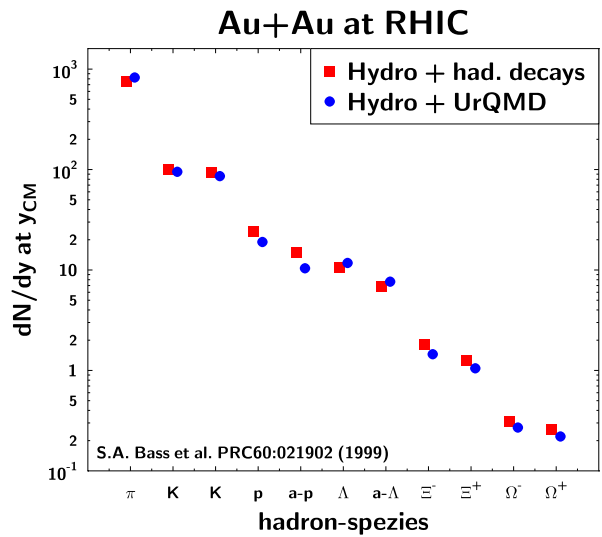}
\caption{Modification of mid-rapidity hadron multiplicities in central Au+Au
collisions at $\sqrt{s}=200 \: GeV$ after chemical freeze-out at
$T=T_c$. Squares show a hydrodynamic model prediction at $T=T_c$
(without further interaction); circles show the result of an
attached UrQMD hadronic cascade expansion calculation \cite{114}.}
\label{fig:Figure27}
\end{center}
\end{figure}

Alternatively, one might end the hydrodynamic description at
$T=T_c$ and match the thus obtained phase space distribution of
equation~\ref{eq:equation31} to a microscopic hadron transport model of the hadronic
expansion phase \cite{95,112}. This procedure is illustrated in
Fig.~\ref{fig:Figure27} by an UrQMD \cite{113} calculation of Bass and Dumitru
\cite{114} for central Au+Au collisions at top RHIC energy. We
select here the results concerning the survival of the hadronic
multiplicities $N_i$ throughout the dynamics of the hadronic
expansion phase, which we have postulated above, based on the
equality of the hadronization temperatures, $T_H \approx 160 \:
MeV$, observed in $e^+e^-$ annihilation (Fig.~\ref{fig:Figure17}), where no hadronic
expansion phase exists, and in central collisions of $A \approx 200$
nuclei (Figs.~\ref{fig:Figure25},~\ref{fig:Figure26}). In fact, Fig.~\ref{fig:Figure27} shows that the $\{N_i\}$
observed at the end of the hadronic cascade evolution agree,
closely, with the initial $\{N_i\}$ as derived from a Cooper-Frye
procedure (equation~\ref{eq:equation30}) directly at hadronization. On the other hand, $p_{T}$ spectra and 
radial flow observables change, drastically, during the hadronic cascade expansion phase. \\
 \\
The hadronic multiplicity distribution  $\{N_i\}$,
arising from the hadronization process at high $\sqrt{s}$,
freezes-out instantaneously also in A+A collisions, and is thus
preserved throughout the (isentropic) hadronic expansion phase. {\it
It is thus directly measurable} and, moreover, its hadrochemical
equilibrium features lend themselves to an analysis within the
framework of Hagedorn-type statistical, grand canonical models. As
we shall show below, the outcome of this analysis is contained in a
$[T_H, \mu_B]$ parameter pair that reflects the conditions of QCD
matter prevailing at hadronization, at each considered $\sqrt{s}$.
In fact, the $[T, \mu]$ points resulting from the SHM analysis
exhibited in Figs.~\ref{fig:Figure25},~\ref{fig:Figure26} (at $\sqrt{s} = 17.3$ and $200 \: GeV$,
respectively) have been shown in the QCD matter phase diagram of Fig.~\ref{fig:Figure1} to
approach, closely, the parton-hadron phase coexistence line
predicted by lattice QCD. Thus, $T_H \approx T_c$ at high
$\sqrt{s}$: hadrochemical freeze-out occurs in the immediate
vicinity of QCD hadronization, thus providing for a location of the QCD phase boundary.

\section{Grand canonical strangeness enhancement}
\label{sec:Grand_canonical_strange}
The statistical model analysis \cite{19,107,108} of the
hadronization species distribution ${N_i}$ in A+A collisions is
based on the grand canonical partition function for species $i$,
\begin{equation}
ln Z_i= \frac{g_i V}{6 \pi^2 T} \: \int^{\infty}_0 \:
\frac{k^4dk}{E_i (k) exp \left\{(E_i (k) - \mu_i)/T \right\} \pm 1}
\label{eq:equation33}
\end{equation}
where $E_i^2=k^2+m^2_i$, and $\mu_i \equiv \mu_B B_i + \mu_s S_i +
\mu_I I_3^i$ is the total chemical potential for baryon number $B$,
strangeness $S$ and isospin 3-component $I_3$. Its role in equation~\ref{eq:equation33} is
to enforce, {\it on average} over the entire hadron source volume,
the conservation of these quantum numbers. In fact, making use of
overall strangeness neutrality $(\sum_i N_i S_i = 0)$ as well as of
conserved baryon number (participant Z+N) and isospin (participant
(N-Z)/Z) one can reduce $\mu_i$ to a single effective potential
$\mu_b$. Hadronic freeze-out is thus captured in three parameters,
$T,V$ and $\mu_b$. The density of hadron/resonance species $i$ then
results as
\begin{equation}
n_i=\frac{T}{V} \frac {\delta}{\delta_{\mu}} \: ln Z_i
\label{eq:equation34}
\end{equation}
which gives
\begin{equation}
N_i=Vn_i=\frac{g_i V}{(2 \pi)^2} \: \int^{\infty}_0 \frac{k^2dk}{exp
\left\{(E_i(k)- \mu_i)/T \right\} \pm 1}.
\label{eq:equation35}
\end{equation}

We see that the common freeze-out volume parameter is canceled if one considers hadron
multiplicity ratios, $N_i/N_j$, as was done in Fig.~\ref{fig:Figure26}. Integration
over momentum yields the one-particle function
\begin{equation}
N_i= \frac{VTg_i}{2 \pi^2} \: m_i^2 \: \sum_{n=1}^{\infty} \:
\frac{(\pm 1)^{n+1}}{n} \: K_2 \left(\frac{nm_i}{T}\right) exp
\left( \frac {n \mu_i}{T} \right)
\label{eq:equation36}
\end{equation}
where $K_2$ is the modified Bessel function. At high $T$ the effects
of Bose or Fermi statistics (represented by the $\pm 1$ term in the
denominators of equation~\ref{eq:equation33} and equation~\ref{eq:equation35}) may be ignored
, finally leading to the Boltzmann approximation
\begin{equation}
N_i= \frac{VTgi}{2 \pi^2} \: m_i^2 \: K_2 \left(\frac{m_i}{T}\right)
exp \left( \frac {\mu_i}{T} \right)
\label{eq:equation37}
\end{equation}
which is the first term of equation~\ref{eq:equation36}. This approximation is employed
throughout the SHM analysis. It describes the {\it primary} yield of
hadron species $i$, directly at hadronization. The abundance of
hadronic resonance states is obtained convoluting equation~\ref{eq:equation35} with a
relativistic Breit-Wigner distribution \cite{19}. Finally, the
overall multiplicity, to be compared to the data, is determined as
the sum of the primary multiplicity equation~\ref{eq:equation37} and the contributions
arising from the unresolved decay of heavier hadrons  and resonances:
\begin{equation}
N_i^{observed}= N_i^{primary} + \sum_j Br (j \rightarrow i) \:N_j.
\label{eq:equation38}
\end{equation}

After having exposed the formal gear of grand canonical ensemble
analysis we note that equation~\ref{eq:equation37} permits a simple, first orientation
concerning the relation of $T$ to $\mu_B$ in A+A collisions by
considering, e.g., the antiproton to proton production ratio. From 
equation~\ref{eq:equation37} we infer the simple expression
\begin{equation}
N (\overline{p})/N(p) = exp (-2 \mu_B/T).
\label{eq:equation39}
\end{equation}
Taking the mid-rapidity value 0.8 for $\overline{p}/p$ (from Fig.~\ref{fig:Figure26})
at top RHIC energy, and assuming that hadronization occurs directly
at the QCD phase boundary, and hence $T \approx T_c \approx 165 \:
MeV$, we get $\mu_B \simeq 18 \: MeV$ from equation~\ref{eq:equation39}, in close agreement
with the result, $\mu_B = 20 \: MeV$, obtained \cite{108} from the
full SHM analysis.
Equation \ref{eq:equation39} illustrates the role played by $\mu_B$ in the grand canonical ensemble. It logarithmically depends on the ratio of newly created quark-antiquark pairs (the latter represented by the $\bar{p}$ yield), to the total number of quarks including the net baryon number-carrying valence quarks (represented by the $p$ yield).
\\
\begin{figure}[h!]   
\begin{center}
\includegraphics[scale=0.5]{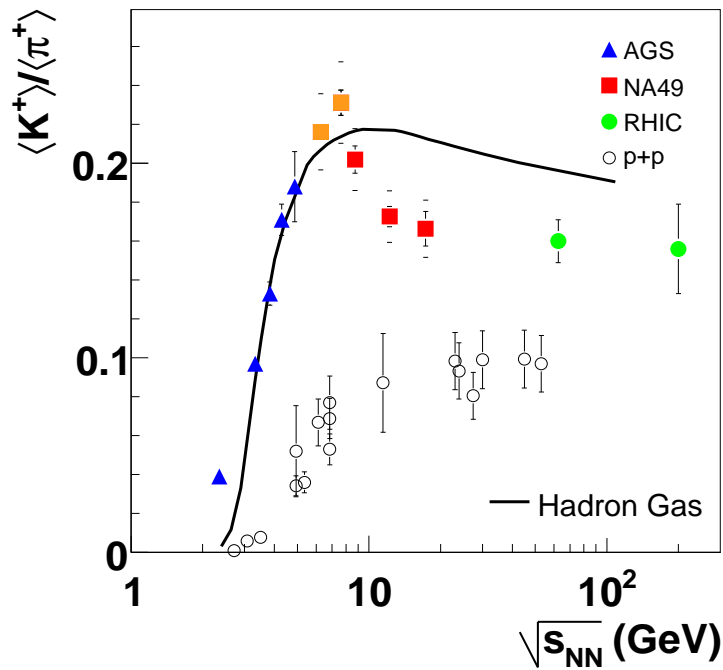}
\caption{The ratio of total $K^+$ to  total $\pi^+$ multiplicity as a
function of $\sqrt{s}$, in central Au+Au and Pb+Pb collisions and in
p+p minimum bias collisions \cite{100}.}
\label{fig:Figure28}
\end{center}
\end{figure} \\
The most outstanding property of the hadronic multiplicities
observed in central A+A collisions is the enhancement of all strange
hadron species, by factors ranging from about 2 to 20, as compared
to the corresponding production rates in elementary hadron-hadron
(and $e^+e^-$ annihilation) reactions at the same $\sqrt{s}$. I.e.
the nuclear collision modifies the relative strangeness output by a
''nuclear modification factor'', $R_s^{AA}=N_s^{AA}/0.5 \:
N_{part}\:  \cdot \: N_s^{pp}$, which depends on $\sqrt{s}$ and
$N_{part}$ and features a hierarchy with regard to the strangeness
number $s=1,2,3$ of the considered species, $R^{AA}_{s=1} <
R^{AA}_{s=2} < R^{AA}_{s=3}$. These properties are illustrated in
Figs.~\ref{fig:Figure28} and~\ref{fig:Figure29}. The former shows the ratio of total $K^+$ to
positive pion multiplicities in central Au+Au/Pb+Pb
collisions, from lower AGS to top RHIC energies, in comparison to
corresponding ratios from minimum bias p+p collisions \cite{100}. We
have chosen this ratio, instead of $\left<K^+\right>/N_{part}$, because it
reflects, rather directly, the ''Wroblewski ratio'' of produced
strange to non-strange quarks \cite{107}, contained in the produced hadrons, 
\begin{equation}
\lambda_s \equiv \frac
{2(\left<s\right>+\left<\overline{s}\right>)}{\left<u\right>+\left<d\right>+\left<\overline{u}\right>+\left<\overline{d}\right>}
\approx \begin{cases} 0.2\ \text{in } \text{pp}  \\ 0.45\ \text{in } \text{AA}. \end{cases}
\label{eq:equation40}
\end{equation}
The low value of $\lambda_s$ in $pp$ (and all other elementary)
collisions reflects a quark population far away from $u,d,s$ flavor
equilibrium, indicating {\it strangeness suppression} \cite{109}.\\
\begin{figure}[h!]   
\begin{center}
\includegraphics[scale=0.53]{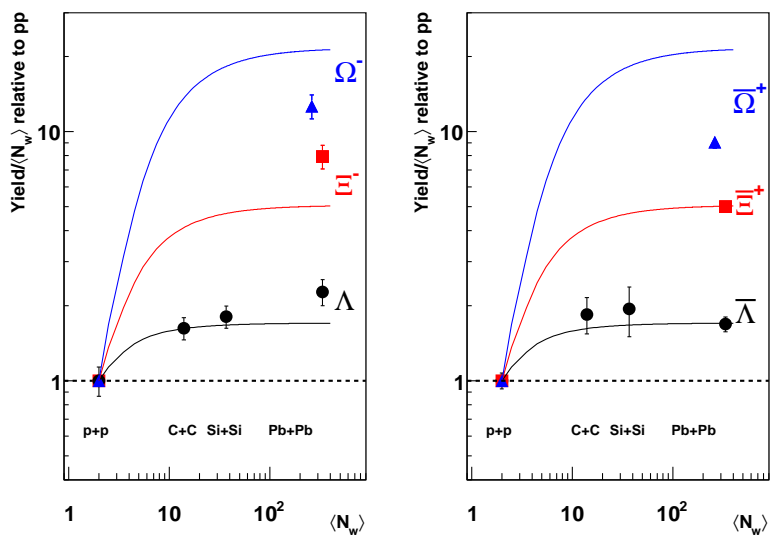}\vspace{-0.3cm}
\caption{The nuclear modification factors $R^{AA}_{s=1,2,3}$ for hyperon and
anti-hyperon production in nucleus-nucleus collisions at
$\sqrt{s}=17.3 \: GeV$, relative to the p+p reference at the same
energy scaled by $N_W(=N_{part})$. The NA49 data refer to total $4
\pi$ yields \cite{116}. Closed lines represent the inverse
strangeness suppression factors from ref. \cite{119}, at
this energy.}
\label{fig:Figure29}
\end{center}
\end{figure} \\
The so-called {\it strangeness enhancement} property of A+A
collisions (obvious from Figs.~\ref{fig:Figure28} and \ref{fig:Figure29}) is, thus, seen as the removal of
strangeness suppression; it is also referred to as a {\it
strangeness saturation}, in SHM analysis \cite{107,108}, for the
reason that $\lambda_s \approx 0.45$ corresponds to the grand
canonical limit of strangeness production, implicit in the analysis
illustrated in Figs.~\ref{fig:Figure25} and~\ref{fig:Figure26}. The average $R^{AA}_{s=1}$ at
$\sqrt{s} \ge 10 \: GeV$ thus is about 2.2, both in the data of
Fig.~\ref{fig:Figure28} and in the statistical model. It increases (Fig.~ref{fig:Figure29}) toward about 10 in
s = 3 production of $\Omega$ hyperons. \\ \\
In order to provide for a first guidance concerning the above facts
and terminology regarding strangeness production we propose an
extremely naive argument, based on the empirical fact of a universal
hadronization temperature (Figs.~\ref{fig:Figure17},~\ref{fig:Figure25},~\ref{fig:Figure26}) at high $\sqrt{s}$. Noting that $\left<s\right>=\left<\overline{s}\right>$ and $\left<u\right> \approx \left<\overline{u}\right> \approx\left<d\right>
\approx \left<\overline{d}\right>$ in a QGP system at $\mu_b$ near zero, and
$T=165 \: MeV$, just prior to hadronization, $\lambda_s \approx
\left<s\right>/\left<u\right> \approx exp \{(m_u - m_s)/T\} = 0.45$  at $p_T \rightarrow
0$ if we take current quark masses, $m_s - m_u \approx 130 \: MeV$.
I.e. the value of $\lambda_s$ in A+A collisions at high $\sqrt{s}$
resembles that of a grand canonical QGP at $\mu_b \rightarrow 0$, as
was indeed shown in a 3 flavor lattice QCD calculation \cite{115} at
$T \approx T_c$. On the contrary, a p+p collision features no QGP
but a small fireball volume, at $T \approx T_c$, within which local
strangeness neutrality, $\left<s\right>=\left<\overline{s}\right>$ has to be strictly
enforced, implying a canonical treatment \cite{109}. In our naive
model the exponential penalty factor thus contains twice the strangeness quark
mass in the exponent, $\lambda_s \:  \mbox {in pp collisions}
\approx exp \{2(m_u-m_s)/T \} \approx 0.2$, in agreement with the
observations concerning strangeness suppression, which are thus
referred to as canonical suppression. In a further extension of our
toy model, now ignoring the $u,d$ masses in comparison to $m_s
\approx 135 \: MeV$, we can estimate the hierarchy of hyperon
enhancement in A+A collisions,
\begin{equation}
R^{AA}_s \propto N_s^{AA}/N_s^{pp} \cdot 0.5 \ N_{part} \approx exp \{(-sm_s + 2sm_s)/T
\} = 2.2, \: 5.1, \:  11.6
\label{eq:equation41}
\end{equation}
for $s=1,2,3$, respectively. Fig.~\ref{fig:Figure29} shows that these estimates
correspond well with the data \cite{116} for $R^{AA}$ derived in $4
\pi$ acceptance for $\Lambda,\: \Xi \: \:  \mbox{and} \:  \: \Omega$
as well as for their antiparticles, from central Pb+Pb collisions at
$\sqrt{s}=17.3 \: GeV$. The p+p reference data, and C+C, Si+Si
central collisions (obtained by fragmentation of the SPS Pb beam)
refer to separate NA49 measurements at the same energy.

The above, qualitative considerations suggest that the relative
strangeness yields reflect a transition concerning the fireball
volume (that is formed in the course of a preceding dynamical
evolution) once it enters hadronization. Within the small volumes,
featured by elementary collisions (see section~\ref{sec:Origin_of_hadro}), phase space is
severely reduced by the requirement of {\it local} quantum number
conservation \cite{109,117} including, in particular, local
strangeness neutrality. These constraints are seen to be removed
in A+A collisions, in which extended volumes of high primordial
energy density are formed. Entering the hadronization stage, after
an evolution of expansive cooling, these extended volumes will decay
to hadrons under conditions of global quantum mechanical coherence,
resulting in quantum number conservation occuring, non-locally, and
{\it on average} over the entire decaying volume. This large
coherent volume decay mode removes the restrictions, implied by
local quantum number balancing. In the above, naive model we have
thus assumed that the hadronization of an Omega hyperon in A+A
collisions faces the phase space penalty factor of only three $s$
quarks to be gathered, the corresponding three $\overline{s}$ quarks
being taken care of elsewhere in the extended volume by global
strangeness conservation. In the framework of the SHM this situation
is represented by the grand canonical ensemble (equations~\ref{eq:equation35},~\ref{eq:equation37}); the
global chemical potential $\mu_b$ expresses quantum number
conservation {\it on average}. Strict, local conservation is
represented by the canonical ensemble.

The grand canonical (GC) situation can be shown to be the large
collision volume limit (with high multiplicities $\{N_i\}$) of the
canonical (C) formulation \cite{118,119}, with a continuous
transition concerning the degree of canonical strangeness
suppression \cite{119}. To see this one starts from a system that is
already in the GC limit with respect to baryon number and charge
conservation whereas strangeness is treated canonically. Restricting
to $s=1 \: \: \mbox {and} \: \: -1$ the GC strange particle
densities can be written (from equation~\ref{eq:equation37}) as
\begin{equation}
n_{s=\pm 1}^{GC} = \frac {Z_{s=\pm 1}}{V} \: \lambda ^{\pm 1}_s
\label{eq:equation42}
\end{equation}
with
\begin{equation}
Z_{s=\pm 1}= \frac {Vg_s}{2 \pi^2} \: m_s^2 \: K_2 \:
(\frac{m_s}{T}) \: exp \Big\{(B_s \mu_B + Q_s \mu_Q)/T \Big\}
\label{eq:equation43}
\end{equation}
and a ''fugacity factor'' $\lambda_s^{\pm 1}= exp \: (\mu_s/T)$. The
canonical strange particle density can be written as \cite{119}
\begin{equation}
n_s^C = n_s^{GC} \cdot (\tilde{\lambda_s})
\label{eq:equation44}
\end{equation}
with an effective fugacity factor
\begin{equation}
\tilde{\lambda_s} \: = \: \frac{S_{\pm 1}}{\sqrt{S_1 S_{-1}}} \:
\frac{I_1(x)}{I_0(x)}
\label{eq:equation45}
\end{equation}
where $S_{\pm 1}=\sum_{s=\pm 1} \: Z_{s=\pm1}$ is the sum over all
created hadrons and resonances with $s=\pm 1$, the $I_n(x)$ are
modified Bessel functions, and $x=2 \sqrt{S_1 S_{-1}}$ is
proportional to the total fireball volume $V$. In the limit $x
\approx V \rightarrow \infty$ the suppression factor $I_1(x)/I_0(x)
\rightarrow 1 $, and the ratio $S_{\pm1}/\sqrt{S_1 S_{-1}}$
corresponds exactly to the fugacity $\lambda_s$ in the GC
formulation (see equation~\ref{eq:equation42}). Thus the C and GC formulations are equivalent in
this limit, and the canonical strangeness suppression effect
disappears. Upon generalization to the complete strange hadron
spectrum, with $s=\pm 1, \pm 2, \pm 3,$ the strangeness suppression
factor results \cite{119} as
\begin{equation}
\eta(s) = I_s(x)/I_0(x).
\label{eq:equation46}
\end{equation}

In particular for small $x$ (volume), $\eta(s) \rightarrow (x/2)^s$,
and one expects that the larger the strangeness content of the
particle the smaller the suppression factor, and hence the larger
the enhancement in going from elementary to central A+A collisions.
This explains the hierarchy addressed in equation~\ref{eq:equation41}, and apparent from
the data shown in Fig.~\ref{fig:Figure29}. In fact, the curves shown in this figure
represent the results obtained from equation~\ref{eq:equation46}, for $s=1,2,3$ hyperon
production at $\sqrt{s}=17.3 \: GeV$ \cite{119}. They are seen to be
in qualitative agreement with the data. However the scarcity
of data, existing at top SPS energy for total hyperon yields,
obtained in $4 \pi$ acceptance (recall the arguments in section~\ref{sec:Hadronic_freeze_out})
both for A+A and p+p collisions does not yet permit to cover
the SHM strangeness saturation curves in detail, for $s>1$.

This saturation is seen in Fig.~\ref{fig:Figure29}, to set in already at modest
system sizes, but sequentially so, for ascending hyperon
strangeness. Note that SHM saturation is sequentially approached, from equation~\ref{eq:equation46},
with increasing fireball {\it volume} $V$. In order to make contact
to the experimental size scaling with centrality, e.g. $N_{part}$,
the model  of ref. \cite{119}, which is illustrated in Fig.~\ref{fig:Figure29}, has converted the genuine volume
scale to the $N_{part}$ scale by assuming a universal eigenvolume of
$7 \: fm^3$ per participant nucleon. I.e. $N_{part}=10$ really means
a coherent fireball volume of $70 \: fm^3$, in Fig.~\ref{fig:Figure29}. Within this
definition, saturation of $s=1,2,3$ sets in at fireball volumes at
hadronization of about 60, 240 and 600 $fm^3$, respectively: this is
the {\it real} message of the SHM curves in Fig.~\ref{fig:Figure29}.\\
\begin{figure}[h!]   
\begin{center}
\includegraphics[scale=1.4]{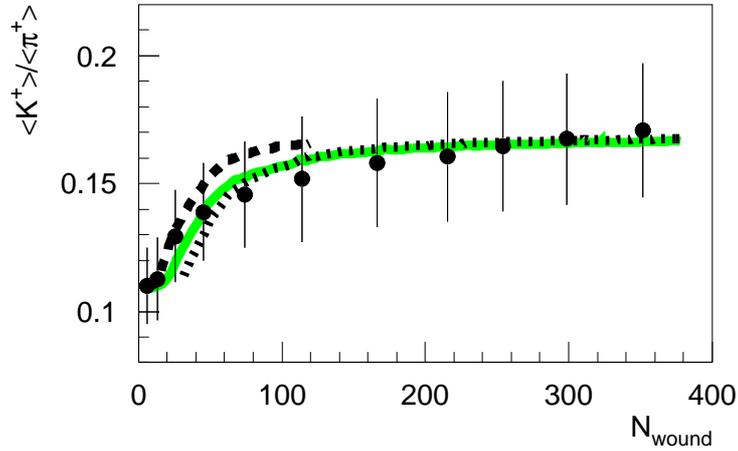}\vspace{-0.3cm}
\caption{The mid-rapidity $K^+$ to $\pi^+$ ratio vs. $N_{part}$ in minimum
bias Au+Au collisions at $\sqrt{s}=200 \: GeV$, compared to the
percolation model \cite{120} (solid line); a prediction of which for
Cu+Cu at similar energy is given by the long dashed line (see text
for detail).}
\label{fig:Figure30}
\end{center}
\end{figure} \\
The above direct translation of coherent fireball volume to
participant number is problematic \cite{120} as it assumes that all
participating nucleons enter into a single primordially coherent
fireball. This is, however, not the case \cite{120} particularly in
the relative small scattering systems that cover the initial, steep
increase of $\eta(s)$, where several local high density clusters are
formed, each containing a fraction of $N_{part}$. This is revealed
by a percolation model \cite{120} of cluster overlapp attached to a Glauber calculation
of the collision/energy density. At each $N_{part}$ an average
cluster volume distribution results which can be transformed by equation~\ref{eq:equation46}
to an average $\{\eta(s,V)\}$ distribution whose weighted mean is
the appropriate effective canonical suppression factor corresponding
to $N_{part}$. On the latter scale, the SHM suppression curve thus
shifts to higher $N_{part}$, as is shown in Fig.~\ref{fig:Figure30} for the $K^+/\pi^+$ ratio
vs. $N_{part}$, measured at mid-rapidity by  PHENIX in Au+Au
collisions at $\sqrt{s}=200 \: GeV$, which is reproduced by the
percolation model \cite{120}. Also included is a prediction for
Cu+Cu at this energy which rises more steeply on the common
$N_{part}$ scale because the collision and energy density reached in
central Cu+Cu collisions, at $N_{part} \approx 100$, exceeds that in
peripheral Au+Au collisions (at the same $N_{part}$) which share a
more prominent contribution from the dilute surface regions of the
nuclear density profile. We note, finally, that this deviation from
 universal $N_{part}$ scaling does not contradict the observations
of a perfect such scaling as far as overall charged particle
multiplicity densities are concerned (recall Fig.~\ref{fig:Figure12}) which are
dominated by pions, not subject to size dependent canonical
suppression.

\section{Origin of hadro-chemical equilibrium}
\label{sec:Origin_of_hadro}
The statistical hadronization model (SHM) is
{\it not} a model of the QCD confinement process leading to hadrons,
which occurs once the dynamical cooling evolution of the system
arrives at $T_c$. At this stage the partonic reaction volume, small
in elementary collisions but extended in A+A collisions, will decay
(by whatever elementary QCD process) to on-shell hadrons and
resonances. This coherent quantum mechanical decay results in a
de-coherent quasi-classical, primordial on-shell hadron-resonance
population which, at the instant of its formation, lends itself to a
quasi-classical Gibbs ensemble description. Its detailed modalities
(canonical for small decaying systems, grand canonical for extended
fireballs in A+A collisions), and its derived parameters [$T,
\mu_B$] merely recast the conditions, prevailing at hadronization.
The success of SHM analysis thus implies that the QCD hadronization
process ends in statistical equilibrium concerning the
hadron-resonance species population.

\begin{figure}[h!]   
\begin{center}
\includegraphics[scale=0.4]{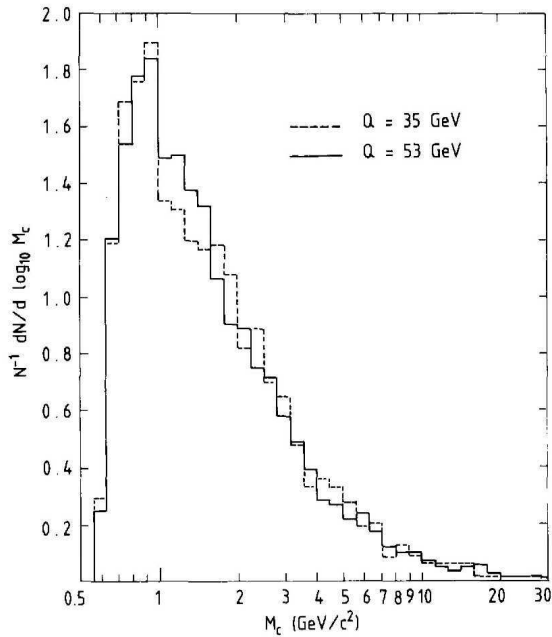}\vspace{-0.3cm}
\caption{Invariant mass spectrum of color neutralization clusters in the
Veneziano-Webber hadronization model \cite{83,121}.}
\label{fig:Figure31}
\end{center}
\end{figure}

In order to identify mechanisms in QCD hadronization that
introduce the hadro-chemical equilibrium we refer to jet
hadronization in $e^+e^-$ annihilation reactions, which we showed in
Fig.~\ref{fig:Figure17} to be well described by the canonical SHM. In di-jet
formation at LEP energy, $\sqrt{s}=92 \: GeV$, we find a charged
particle multiplicity of about 10 per jet, and we estimate that,
likewise, about 10 primordial partons participate on either side of
the back-to-back di-jet \cite{85}. There is thus no chance for
either a partonic or hadronic, extensive rescattering toward
chemical equilibrium. However, in the jet hadronization models
developed by Amati and Veneziano \cite{83}, Webber \cite{121} and
Ellis and Geiger \cite{85} the period of QCD DGLAP parton shower
evolution (and of perturbative QCD, in general) ends with local
color neutralization, by formation of spatial partonic singlet clusters.
This QCD ''color pre-confinement'' \cite{83} process reminds of a
coalescence mechanism, in which the momenta and the initial virtual
masses of the individual clustering partons get converted to
internal, invariant virtual mass of color neutral, {\it spatially
extended} objects. Their mass spectrum \cite{121} extends from about
0.5 to 10 $GeV$. This cluster mass distribution, shown in Fig.~\ref{fig:Figure31}, represents the first
stochastic element in this hadronization model.\\
 \\
The clusters are then re-interpreted within non-perturbative QCD:
their internal, initially perturbative QCD vacuum energy gets
replaced by non-perturbative quark and gluon condensates, making the
clusters appear like hadronic resonances. Their subsequent quantum
mechanical decay to on-shell hadrons is governed by the phase space
weights given by the hadron and resonance spectrum \cite{85,121}.
I.e. the clusters decay under ''phase space dominance'' \cite{85},
the outcome being a micro-canonical or a canonical hadron and
resonance ensemble \cite{84,107}. The apparent hadro-chemical
equilibrium thus is the consequence of QCD color neutralization to
clusters, and their quantum mechanical decay under local quantum
number conservation and phase space weights. We note that the
alternative description of hadronization, by string decay
\cite{122}, contains a quantum mechanical tunneling
mechanism, leading to a similar phase space dominance \cite{123}.

Hadronization in $e^+e^-$ annihilation thus occurs from local
clusters (or strings), isolated in vacuum, of different mass but
similar energy density corresponding to QCD confinement. These
clusters are boosted with respect to each other but it was shown
\cite{124} that for a Lorentz invariant scalar, such as
multiplicity, the contributions of each cluster (at similar $T$) can
be represented by a single canonical system with volume equal to the
sum of clusters. In the fit of Fig.~\ref{fig:Figure17} this volume sum amounts to
about 45 $fm^3$ \cite{84}; the individual cluster volumes are thus
quite small, of magnitude a few $fm^3$ \cite{85}. This implies
maximum canonical strangeness suppression but may, in fact, require
a micro-canonical treatment of strangeness \cite{109}, implying a
further suppression. These MC effects are oftentimes included
\cite{125} in the canonical partition functions by an extra
strangeness fugacity parameter $\gamma_s < 1$ which suppresses
$s=1,2,3$ in a hierarchical manner, $\left<N_i(s)\right> \approx
(\gamma_s)^{s_i}$. The fit of Fig.~\ref{fig:Figure17} requires $\gamma_s=0.66$, a
value typical of canonical multiplicity analysis in $p+p, \:
p+\overline{p}$ and $e^+e^-$ annihilation collisions \cite{109} at
$\sqrt{s} \ge 30 \: GeV$.

The above picture, of hadrochemical equilibrium resulting from the
combined stochastic features of QCD color neutralization by cluster
formation, and subsequent quantum mechanical decay to the on-shell
hadron and resonance spectrum (under phase space governance) lends
itself to a straight forward extension to A+A collisions. The
essential new features, of grand canonical hadronization including
strangeness enhancement, should result from the fact that extended
space-time volumes of $\epsilon > \epsilon_{crit}$ are formed in the
course of primordial partonic shower evolution, an overlap effect increasing
both with $\sqrt{s}$ and with the size of the primordial interaction
volume. As the volume of the elementary hadronization clusters
amounts to several $fm^3$ it is inevitable that the clusters
coalesce, to form extended ''super-cluster'' volumes prior to
hadronization \cite{120}. As these super-clusters develop toward
hadronization via non perturbative QCD dynamics, it is {\it
plausible} to assume an overall quantum mechanical coherence to
arise over the entire extended volume, which will thus decay to
hadrons under global quantum number conservation, the decay products
thus modeled by the GC ensemble.

Our expectation that space-time coalescence of individual hadronization clusters
will lead to a global, quantum mechanically coherent extended
super-cluster volume, that decays under phase space dominance, appears as an analogy to the dynamics and
quantum mechanics governing low energy nuclear fission from a preceding ``compound nucleus"~\cite{126}.
Note that the observation of a smooth transition from
canonical strangeness suppression to grand canonical saturation
(Figs.~\ref{fig:Figure29},~\ref{fig:Figure30}) lends further support to the above picture of a
percolative growth \cite{120} of the volume that is about to undergo
hadronization. 

An extended, coherent quark gluon plasma state would,
of course, represent an ideal example of such a volume \cite{127}
and, in fact, we could imagine that the spatial extension of
the plasma state results from a percolative overlap of primordial
zones of high energy density, which becomes more prominent with
increasing $\sqrt{s}$ and $N_{part}$. A QGP state preceding
hadronization will thus lead to all the observed features. However,
to be precise: the hadronizing QCD system of extended matter
decaying quantum coherently, could still be a non-equilibrium
precursor of the ideal equilibrium QGP, because we have seen above
that hadrochemical equilibrium also occurs in $e^+e^-$ annihilation,
where no partonic equilibrium exists. It gets established in the
course of hadronization, irrespective of the degree of equilibrium
prevailing in the preceding partonic phase.

\section{Hadronization vs. rapidity and $\sqrt{s}$}
\label{sec:Hadronization_vs_rap}
We have argued in section~\ref{sec:Hadronic_freeze_out} that, at relatively low $\sqrt{s}$, the
total rapidity gap $\Delta y$ does not significantly exceed the
natural thermal rapidity spreading width $\Gamma_i \approx 2.35 \:
(\Gamma/m_i)^{1/2}$ of a single, isotropically decaying fireball,
centered at mid-rapidity and emitting hadrons of mass $m_i$
\cite{110}.
However, this procedure involves an idealization
because in the real Pb+Pb collision the  intersecting dilute surface
sections of the nuclear density profiles will lead to a significant
contribution of single-scattering NN collisions, outside the central
high density fireball. The leading hadron properties of such
''corona collisions'' result in wider proper rapidity distributions,
quite different from those of the central fireball decay hadrons.
Their contribution will thus be prominent near target/projectile
rapidity, and will feature a canonically suppressed strangeness. The
one-fireball assumption, although inevitable at small $\Delta y$,
does not quite resemble the physical reality. This may explain the
need for an additional strangeness suppression factor in the GC
one-particle partition function (equation~\ref{eq:equation33}) that has, unfortunately,
also been labeled $\gamma_s$ but expresses physics reasons quite
different from the extra suppression factor that reflects
micro-canonical phase space constraints in elementary collisions. It turns out that all GC
analysis of central A+A collisions at low $\sqrt{s}$, and addressed
to total $4 \pi$ multiplicities, requires a $\gamma_s$ of 0.7 - 0.85
\cite{19}; in the fit of Fig.~\ref{fig:Figure25} $\gamma_s=0.84$.

\begin{figure}[h!]   
\begin{center}
\includegraphics[scale=0.7]{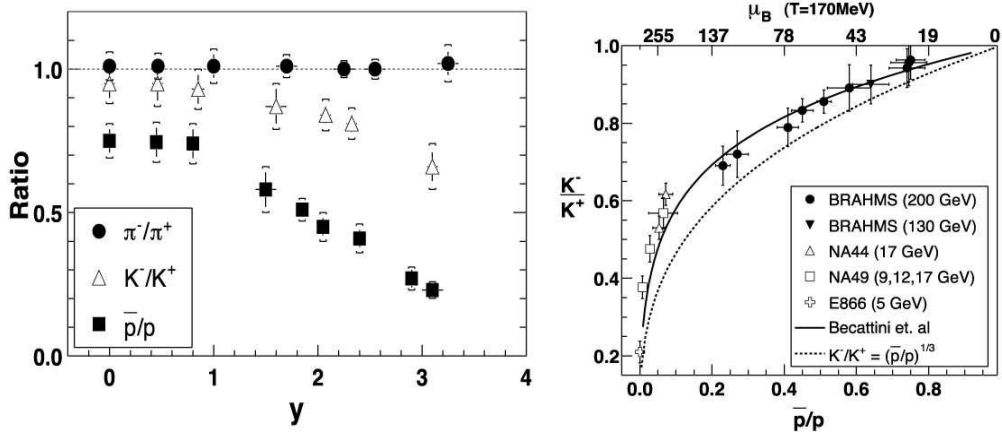}\vspace{-0.3cm}
\caption{(left) Anti-hadron to hadron ratios as a function of rapidity in
central Au+Au collisions at $\sqrt{s}=200 \: GeV$. The right panel
shows an interpretation of the correlation between $\overline{p}/p$
and $K^-/K^+$ in terms of baryo-chemical potential $\mu_B$ variation
in the grand canonical statistical model. From \cite{128}.}
\label{fig:Figure32}
\end{center}
\end{figure}

At RHIC, $\Delta y \approx 11 \gg \Gamma_i$, and such difficulties
disappear: $\gamma_s \approx 1$ at midrapidity and, moreover, the wide gap permits
a SHM analysis which is differential in $y$. Fig.~\ref{fig:Figure32} shows the
$y$-dependence of the ratios $\pi^-/\pi^+, \: K^-/K^+$ and
$\overline{p}/p$ as obtained by BRAHMS \cite{128} in central Au+Au
collisions at $\sqrt{s}=200 \: GeV$. The figure shows a dramatic
dependence of the $\overline{p}/p$ ratio, which reflects the local
baryochemical potential according to equation~\ref{eq:equation39}. At $y_{CM} > 1$ the
$\overline{p}/p$ ratio drops down steeply, to about 0.2 at $y
\approx 3.5$, thus making close contact to the top SPS energy value
obtained by NA49 \cite{129}. The $K^-/K^+$ ratio follows a similar
but weaker drop-off pattern, to about 0.65 again matching with the
top SPS energy value of about 0.6 \cite{130}. The deviation from
unity of these ratios reflects the rapidity densities of initial
valence $u,d$ quarks, relative to the densities of newly created
light and strange quark-antiquark pairs, i.e. the $y$ distribution
of the net baryon number density, and of the related baryo-chemical
potential of the GC ensemble. Thus, in analyzing successive bins of
the rapidity distributions in Fig.~\ref{fig:Figure32}, the major variation in the GC
fit concerns the baryo-chemical potential $\mu_B$(y) which increases
from about 20 $MeV$ (Fig.~\ref{fig:Figure26}) at mid-rapidity, to about 150 $MeV$ at
$y \ge 3$ while the hadronization temperature stays constant, at
$T=160 \: MeV$. This interplay between $K^-/K^+, \: \overline{p}/p$
and $\mu_B$ is illustrated \cite{128} in the right hand panel of
Fig.~\ref{fig:Figure32}, and shown to be well accounted for by the GC statistical
model \cite{131}.\\
 \\
These considerations imply that hadronization at RHIC (and LHC)
energy occurs {\it local} in $y$-space and {\it late} in time. The density distribution of net baryon
number results from the primordial pQCD shower evolution (c.f.
section~\ref{subsec:Gluon_Satu_in_AA_Coll}), and is thus fixed at formation time, $t_0 \le 0.6 \:
fm/c$ at RHIC. Hadronization of the bulk partonic matter occurs
later, at $t \ge 3 \: fm/c$ \cite{86,95}, and transmits the local
conditions in rapidity space by preserving the local net baryon quantum number
density. Most importantly we conclude that hadronization occurs, not
from a single longitudinally boosted fireball but from a succession
of ''super-clusters'', of different partonic composition depending
on $y$, and decaying at different time due to the Lorentz-boost that
increases with $y$, in an ''inside-outside'' pattern (c.f. Fig.~\ref{fig:Figure18}).
We are thus whitnessing at hadronization a Hubble expanding system
of local fireballs. The detailed implications of this picture have
not been analyzed yet. Note that a central RHIC collision
thus does not correspond to a single hadronization ''point'' in the
[$T, \: \mu$] plane of Fig.~\ref{fig:Figure1} but samples $\{T,\: \mu\}$ along the
QCD parton-hadron coexistence line \cite{132}.

\begin{figure}[h!]   
\begin{center}
\includegraphics[scale=0.4]{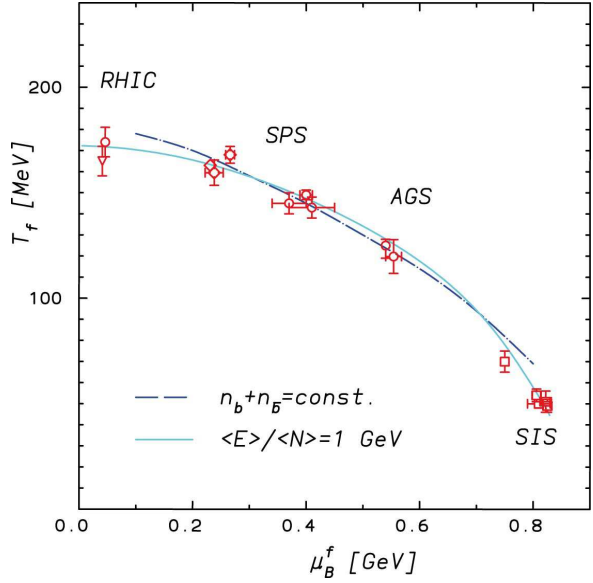}\vspace{-0.3cm}
\caption{Energy dependence of the hadro-chemical freeze-out points obtained
by grand canonical statistical model analysis in the plane [$T, \:
\mu_B$], with interpolating curve at fixed energy per particle of
about 1 $GeV$ \cite{107,139}.}
\label{fig:Figure33}
\end{center}
\end{figure}

Throughout this chapter we have discussed hadronic freeze-out at
high $\sqrt{s}$ only (top SPS to RHIC energy), because of the
proximity of the chemical freeze-out parameters [$T, \: \mu_b$] to
the QCD phase boundary from lattice QCD, which suggests an overall
picture of hadronization, to occur directly from a partonic cluster
or super-cluster. Our discussion of the GC statistical hadronization
model has been explicitly or implicitly based on the assumption that hadronic 
freeze-out coincides with hadronization. However, the GC model has also been applied successfully to
hadro-chemical freeze-out at $\sqrt{s}$ down to a few $GeV$
\cite{19,107,108} where it is not expected that the dynamical
evolution traverses the phase boundary at all, but grand canonical
multiplicity distributions, and their characteristic strangeness
enhancement pattern, are observed throughout. Toward lower
$\sqrt{s}, \: T$ decreases while $\mu_b$ increases, as is shown in
Fig.~\ref{fig:Figure33} which presents a compilation of all reported freeze-out
parameters \cite{108}.\\
 \\
These points have also been included in the phase diagram of Fig.~\ref{fig:Figure1}
which shows that they are gradually branching away from the phase
separation boundary line that could recently be predicted by lattice
QCD once new methods had been developed to extend the theory to
finite $\mu_B$ \cite{9,10}. At $\sqrt{s} \ge 20 \: GeV$ we see that
\begin{equation}
\epsilon_c (QCD) \approx \epsilon_H \approx \epsilon_{GC}
\label{eq:equation47}
\end{equation}
where $\epsilon_{GC}$ is the freeze-out density inferred from GC
analysis \cite{19,107,108}.

In turn, the GC hadronic freeze-out points drop below the lattice
QCD coexistence line at lower $\sqrt{s}$, implying that chemical
freeze-out now occurs within the hadronic expansion phase. This
requires a model of freeze-out, now governed by the properties of a
high density hadronic medium, upon expansive cooling and dilution.
Holding on to the model of a quantum mechanical de-coherence decay
to on-shell hadrons that we discussed in section~\ref{sec:Origin_of_hadro}, we argue that an
initial, extended high density hadronic fireball, given sufficient
life-time at $T$ smaller, but not far below $T_c$, could also be
seen as a quantum mechanically coherent super-cluster, as governed
by effective mean fields \cite{133}. In such a medium hadrons, at
$T$ near $T_c$, acquire effective masses and/or decay widths far off
their  corresponding properties in vacuum: they are off-shell,
approaching conditions of QCD chiral symmetry restoration as $T
\rightarrow T_c$ \cite{134}. This symmetry is inherent in the
elementary QCD Langrangian, and ''softly'' broken within the light
quark sector by the small non-zero current quark masses, but
severely broken at $T \rightarrow 0$ by the high effective
constituent quark masses that get dressed by non perturbative QCD
vacuum condensates. Pictorially speaking, hadrons gradually loose
this dressing as $T \rightarrow T_c$ \cite{135}, introducing a
change, away from in vacuum properties, in the hadronic mass and
width spectrum. Such in-medium chiral restoration effects have, in
fact, been observed in relativistic A+A collisions, by means of
reconstructing the in-medium decay of the $\rho$ vector meson to an
observed $e^+e^-$ pair \cite{136} (see chapter~\ref{sec:Low_mass_dilepton_spectra}).

A dense, high $T$ hadronic system, with mean-field
induced off-shell constituents is also, clearly, quantum
mechanically coherent. At a certain characteristic density,
$\epsilon < \epsilon_c$, and temperature $T<T_c$, as reached in the
course of overall hadronic expansion, this extended medium will
undergo a decoherence transition to classical on-shell hadrons. Its
frozen-out hadronic multiplicity distribution should be, again,
characterized by the phase space weights of a grand canonical
ensemble at $T<T_c$. Theoretical studies of such a mean field
hadronic expansion mode \cite{137} have also shown that such
mechanisms play essentially no role at $\sqrt{s} \ge 20 \: GeV$
because the expanding system is already in rapid flow once it
traverses the phase boundary, with an expansion time scale shorter
than the formation time scale of mean field phenomena. At lower
energies, on the other hand, the system might not even dive into the
deconfined phase but spend a comparatively long time in its direct
vicinity, at the turning point between compression and re-expansion
where all dynamical time constants are large, and the hadron density
is high, such that the inelastic hadronic transmutation rate becomes
high (particularly in collisions of more than two hadronic
reactants, with reaction rates \cite{138} proportional to
$\epsilon^n$), and sufficiently so for maintaining hadronic chemical
equilibrium after it is first established at maximum hadron density, in low $\sqrt{s}$
systems that do not cross the phase boundary at all.

The GC freeze-out parameters [$T, \mu$] at various $\sqrt{s}$ in
Fig.~\ref{fig:Figure33} permit a smooth interpolation in the $T, \mu$ plane
\cite{139}, which, in turn, allows for GC model predictions which
are continuous in $\sqrt{s}$. Such a curve is shown in Fig.~\ref{fig:Figure28}
compared to the $4 \pi$ data points for the $K^+/\pi^+$ multiplicity
ratio in central collisions Au+Au/Pb+Pb, at all $\sqrt{s}$
investigated thus far. It exhibits a smooth maximum, due to the
interplay of $T$ saturation and $\mu_B$ fall-off to zero, but does
not account for the sharp peak structure seen in the data at
$\sqrt{s} \approx 7 \: GeV$ and $\mu_B \approx 480 \: MeV$. This
behavior is not a pecularity of the $K^+$ channel only; it also is
reflected in an unusually high Wroblewski ratio (see equation~\ref{eq:equation40}) obtained
at $\sqrt{s} = 7.6 \: GeV$, of  $\lambda_s = 0.60$ \cite{19}. This
sharp strangeness maximum is unexplained as of yet. It implies
that hadron formation at this $\sqrt{s}$ reflects influences that
are less prominent above and below, and most attempts to understand
the effect \cite{141,142,143} are centered at the assumption that at
this particular $\sqrt{s}$ the overall bulk dynamics will settle
directly at the phase boundary where, moreover, finite $\mu_B$
lattice theory also expects a QCD critical point \cite{9,10,11}.
This would cause a softest point to occur in the equation of state,
i.e. a minimum in the relation of expansion pressure vs. energy
density, slowing down the dynamical evolution \cite{144,145}, and
thus increasing the sensitivity to expansion modes characteristic of
a first order phase transition \cite{143}, which occurs at $\mu_b \ge
\mu_b^{crit}$. Such conditions may modify the K/$\pi$ ratio (Fig.~\ref{fig:Figure28}) \cite{143}.

It thus appears that the interval from top AGS to lower SPS energy,
$5 \le \sqrt{s} \le 10 \: GeV$, promises highly interesting
information regarding the QCD phase diagram (Fig.~\ref{fig:Figure1}) in the direct
vicinity of the parton-hadron coexistence line. In particular, the
physics of a critical point of QCD matter deserves further study.
Observable consequences also comprise so-called ''critical
fluctuations'' \cite{146,147} of multiplicity density, mean
transverse momentum and hadron-chemical composition \cite{148}, the
latter in fact being observed near $\sqrt{s}=7 \: GeV$ in an event
by event study of the $K/\pi$ ratio in central Pb+Pb collisions
\cite{149}. We shall return to critical point physics in chapter~\ref{chap:Fluctuations}.

%% file: Chapter_3.tex
\chapter{Elliptic flow}
\label{chap:Elliptic_flow}

We have up to now mostly stressed the importance of central
collisions and mid-rapidity data because they provide for the
highest primordial energy density and avoid problems caused by
emission anisotropy and the presence of cold spectator sections of
target and projectile nuclei. On the other hand, a fundamentally new
window of observation is opened by non-central collisions as the
finite impact parameter breaks cylinder symmetry, defining emission
anisotropies with respect to the orientation of the impact vector
$\vec{b}$ as we have shown in equation~\ref{eq:equation21}. In a strongly interacting
fireball collision dynamics, the initial geometric anisotropy of the
reaction volume gets transferred to the final momentum spectra and
thus becomes experimentally accessible. Furthermore, the high
charged particle multiplicity allows for an event-by-event determination
of the reaction plane (direction of $\vec{b}$), enabling the study
of observables at azimuth $\varphi$, relative to the known reaction
plane. We shall show that this opens a window into the very early
stages of A+A collisions onward from the end of nuclear
interpenetration, at $\tau \approx 2 \: R (A)/\gamma_{CM}$. Our
observation thus begins at the extreme energy densities prevailing
right at formation time (section~\ref{subsec:Gluon_Satu_in_AA_Coll},~\ref{subsec:Transvers_phase_space}), 
i.e. concurrent with the initialization phase of relativistic hydrodynamic expansion.
We access the
phase diagram of Fig.~\ref{fig:Figure1} in regions {\it far above} the QCD phase
boundary.

\begin{figure}[h!]   
\begin{center}
\includegraphics[scale=0.6]{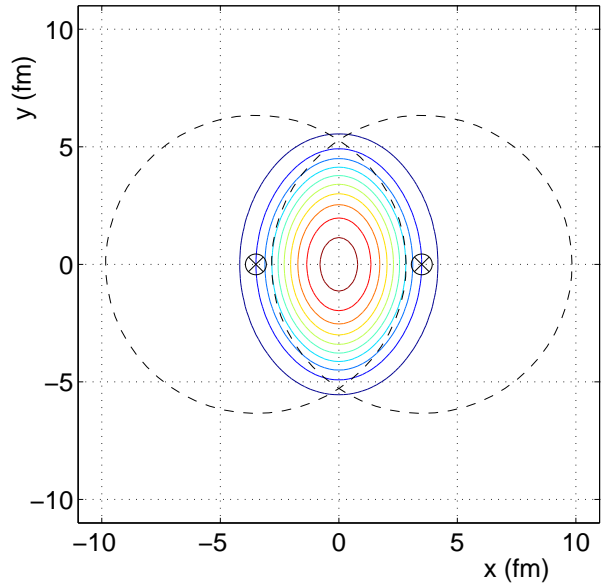}\vspace{-0.3cm}
\caption{Transverse projection of primordial binary collision density in an
Au+Au collision at impact parameter $7 \:fm$, exhibiting participant
parton spatial excentricity \cite{96}.}
\label{fig:Figure34}
\end{center}
\end{figure}

\begin{figure}[h!]   
\begin{center}
\includegraphics[scale=0.55]{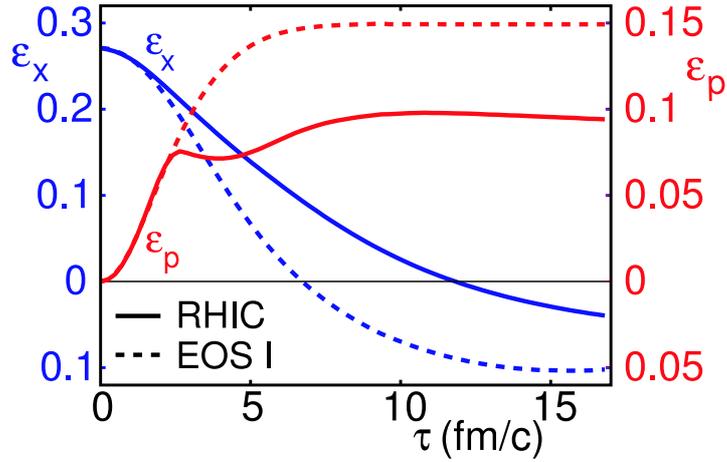}\vspace{-0.3cm}
\caption{Time evolution of the spatial excentricity $\epsilon_x$ and the
momentum space anisotropy $\epsilon_p$ (equations~\ref{eq:equation48} and \ref{eq:equation49}) in
the hydrodynamic model of an Au+Au collision at $b=7 \: fm$,
occuring at $\sqrt{s}=200 \: GeV$ \cite{96}. The dynamics is
illustrated with two equations of state.}
\label{fig:Figure35}
\end{center}
\end{figure}

Before turning to the details of elliptic flow data we wish to
illustrate \cite{96} the above statements. Fig.~\ref{fig:Figure34} exhibits the
transverse projection of primordial energy density, assumed to be
proportional to the number density of participant nucleons in the
overlap volume arising from a Au+Au collision at impact parameter
$b=7 \: fm$. The nuclear density profiles (assumed to be of
Woods-Saxon type) intersect in an ellipsoidal fireball, with minor
axis along the direction of $\vec{b}$ which is positioned at $y=0$.
The obvious geometrical deformation can be quantified by the spatial
excentricity (unfortunately also labeled $\epsilon$ in the
literature)
\begin{equation}
\epsilon_x(b)=\frac{\left<y^2-x^2\right>}{\left<y^2+x^2\right>}
\label{eq:equation48}
\end{equation}
where the averages are taken with respect to the transverse density
profiles of Fig.~\ref{fig:Figure34}. $\epsilon_x$ is zero for $b=0$, reaching a value
of about 0.3 in the case $b=7 \: fm$ illustrated in Fig.~\ref{fig:Figure34}.\\
 \\
Translated into the initialization of the hydrodynamic expansion
the density anisotropy implies a corresponding pressure
anisotropy. The pressure is higher in $x$ than in $y$ direction, and
thus is the initial acceleration, leading to an increasing momentum
anisotropy,
\begin{equation}
\epsilon_p(\tau)=\frac{\int dxdy \: (T^{xx}-T^{yy})}{\int dxdy \:
(T^{xx}+T^{yy})}
\label{eq:equation49}
\end{equation}
where $T^{\mu x}_{(x)}$ is the fluid's energy-momentum tensor.
Fig.~\ref{fig:Figure35} shows \cite{96,150} the time evolution of the spatial and
momentum anisotropies for the collision considered in Fig.~\ref{fig:Figure34},
implementing two different equations of state which are modeled with
(without) implication of a first order phase transition in ''RHIC''
(''EOS1''). A steep initial rise is observed for $\epsilon_p$, in
both cases: momentum anisotropy builds up during the early partonic
phase at RHIC, while the spatial deformation disappears. I.e. the
initial source geometry, which is washed out later on, imprints a
flow anisotropy which is preserved, and observable as ''elliptic
flow''. A first order phase transition essentially stalls the
buildup of $\epsilon_p$ at about $\tau=3 \: fm/c$ when the system
enters the mixed phase, such that the emerging signal is almost
entirely due to the partonic phase. We have to note here that the
''ideal fluid'' (zero viscosity) hydrodynamics \cite{150} employed
in Fig.~\ref{fig:Figure35} is, at first, a mere hypothesis, in view also of the fact
that microscopic transport models have predicted a significant
viscosity, both for a perturbative QCD parton gas \cite{92,151} and
for a hadron gas \cite{95,152}. Proven to be correct by the data, the ideal fluid description of 
elliptic flow tells us that the QGP is a non-perturbative liquid
\cite{153,154}.\\
\begin{figure}[h!]   
\begin{center}\vspace{-0.2cm}
\includegraphics[scale=0.45]{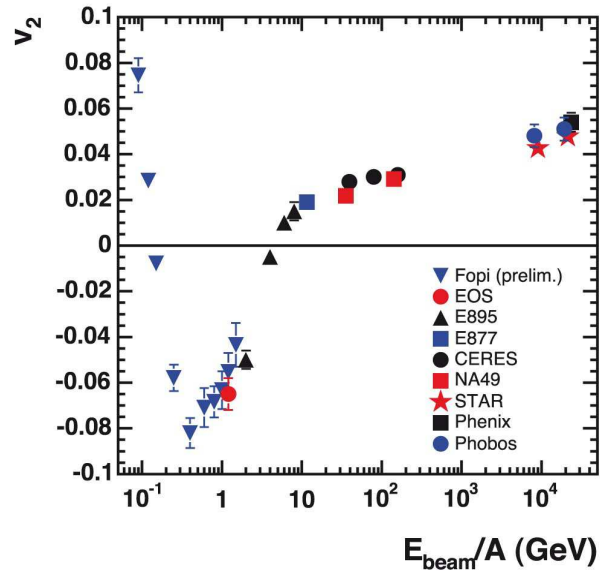}\vspace{-0.4cm}
\caption{Energy dependence of the elliptic flow parameter $v_2$ at
mid-rapidity and averaged over $p_T$, in Au+Au and Pb+Pb
semi-peripheral collisions \cite{155}.}
\label{fig:Figure36}
\end{center}
\end{figure} \\
Elliptic flow is quantified by the coefficient $v_2$ of the second
harmonic term in the Fourier expansion (see equation~\ref{eq:equation21}) of the invariant cross
section; it depends on $\sqrt{s}, b, y$ and $p_T$. Fig.~\ref{fig:Figure36} shows the
$\sqrt{s}$ dependence of $v_2$ at mid-rapidity and averaged over
$p_T$, in Au+Au/Pb+Pb semi-peripheral collisions \cite{93,155}. We 
see that the momentum space anisotropy is relatively small overall,
but exhibits a steep rise toward top RHIC energy. \\
\begin{figure}[h!]   
\begin{center}\vspace{-0.5cm}
\includegraphics[scale=0.45]{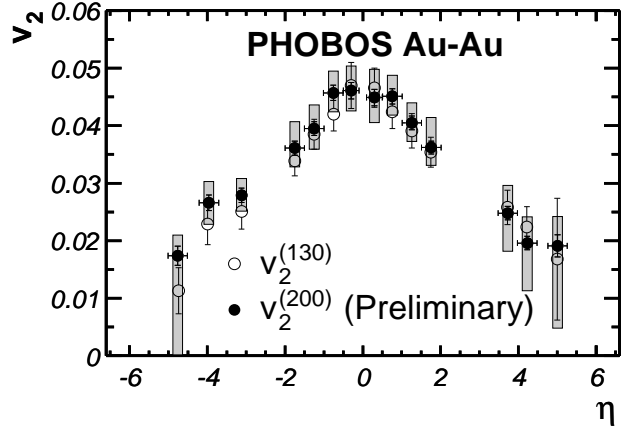}\vspace{-0.4cm}
\caption{Pseudo-rapidity dependence of the $p_T$-averaged elliptic flow
coefficient $v_2$ for charged hadrons at $\sqrt{s}=130$ and $200 \:
GeV$ \cite{156}.}
\label{fig:Figure37}
\end{center}\vspace{-0.2cm}
\end{figure} \\
Fig.~\ref{fig:Figure37} shows the (pseudo)-rapidity dependence of $v_2$ at $\sqrt{s}=130$ and 200
$GeV$ as obtained by PHOBOS \cite{156} for charged particles in
minimum bias Au+Au collisions. It resembles the corresponding
charged particle rapidity density distribution of Fig.~\ref{fig:Figure8}, suggesting
that prominent elliptic flow arises only at the highest attainable
primordial energy density. \\
\begin{figure}[h!]   
\begin{center}\vspace{-0.5cm}
\includegraphics[scale=0.45]{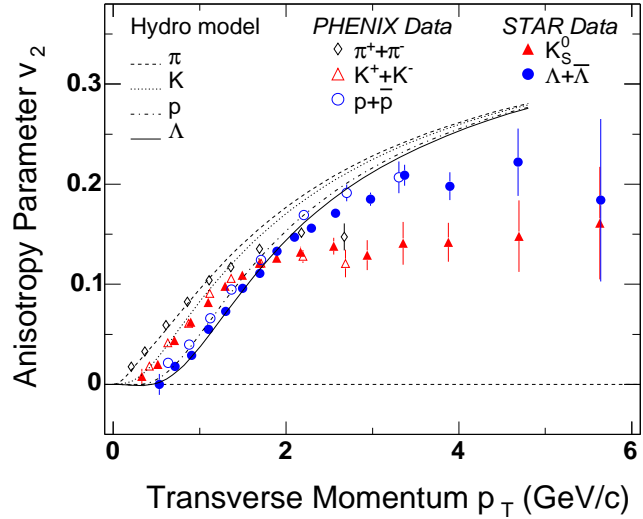}\vspace{-0.4cm}
\caption{Transverse momentum dependence of elliptic flow $v_2$ for mesons and
baryons in Au+Au collisions at $\sqrt{s}=200 \: GeV$. The
hydrodynamic model \cite{96,159} describes the mass dependence at
$p_T \le 2 \: GeV/c$.}
\label{fig:Figure38}
\end{center}
\end{figure} \\
That such conditions are just reached at top RHIC energy is shown in Figs.~\ref{fig:Figure38} and~\ref{fig:Figure39}. The former combines STAR \cite{157} and PHENIX \cite{158} data for the 
$p_T$ dependence of elliptic flow, observed for various identified hadron species
$\pi^±, K^±, p, K^0$ and $\Lambda, \overline{\Lambda}$ in Au+Au at
200 $GeV$. The predicted hydrodynamic flow pattern \cite{96,159}
agrees well with observations in the bulk $p_T  < 2 \: GeV/c$
domain. Fig.~\ref{fig:Figure39} (from \cite{155}) unifies average $v_2$ data from AGS
to top RHIC energies in a scaled representation \cite{93} where
$v_2$ divided by the initial spatial anisotropy $\epsilon_x$ is
plotted versus charged particle mid-rapidity density per unit
transverse area $S$, the latter giving the density weighted
transverse surface area of primordial overlap, Fig.~\ref{fig:Figure34}. Fig.~\ref{fig:Figure39}
includes the hydrodynamic predictions \cite{95,96,150,159,161} for
various primordial participant or energy densities as implied by the
quantity $(1/S)dn_{ch}/dy$ \cite{93}. Scaling $v_2$ by $\epsilon_x$
enhances the elliptic flow effect of near-central collisions where
$\epsilon_x$ is small, and we see that only such collisions at top
RHIC energy reach the hydrodynamical ideal flow limit in models that
include an EOS ansatz which incorporates \cite{96} the effect of a
first order phase transition, which reduces the primordial flow
signal as was shown in Fig.~\ref{fig:Figure35}. \\
\begin{figure}[h!]   
\begin{center}\vspace{-0.5cm}
\includegraphics[scale=0.45]{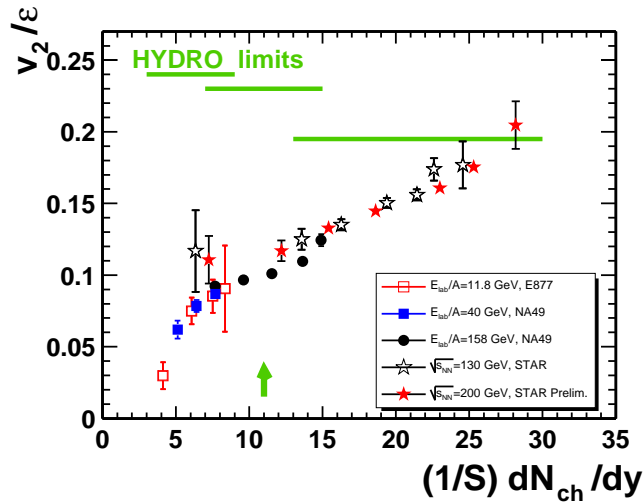}\vspace{-0.4cm}
\caption{Elliptic flow $v_2$ scaled by spatial excentricity $\epsilon$ as a
function of charged particle density per unit transverse area $S$,
from AGS to top RHIC energy. The hydrodynamic limit is only attained
at RHIC \cite{155}.}
\label{fig:Figure39}
\end{center}
\end{figure} \\
At top RHIC
energy, the interval between $t_0 \approx 0.6 \: fm/c$, and
hadronization time, $t_H \approx 3 \: fm/c$, is long enough to
establish dynamical consequences of an early approach toward local
equilibrium. The ''lucky coincidence'' of such a
primordial resolution of dynamical time scale, with the extreme
primordial density, offered by semi-central collisions of
heavy nuclei, results in an extremely short mean free path of the
primordial matter constituents, thus inviting a hydrodynamic
description of the expansive evolution. Consistent application of
this model reveals a low viscosity: the primordial matter resembles
an ideal fluid, quite different from earlier concepts, of a weakly
interacting partonic gas plasma state (QGP) governed by perturbative
QCD screening conditions \cite{36,41}.

\begin{figure}[h!]   
\begin{center}\vspace{-0.5cm}
\includegraphics[scale=1.9]{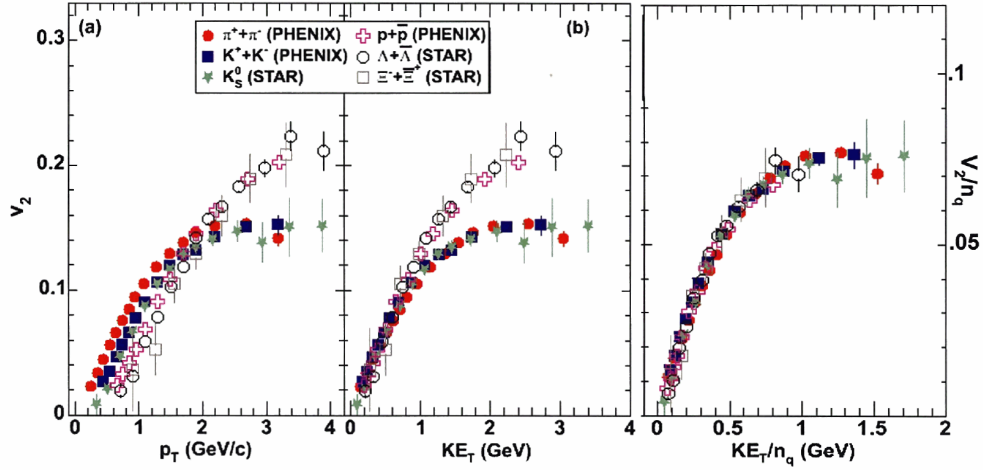}\vspace{-0.4cm}
\caption{$v_2$ vs. $p_T$ (left panel) and transverse kinetic energy
$KE_T=m_T-m_0$ (middle) for several hadronic species in min. bias
Au+Au collisions at $\sqrt{s}=200 \: GeV$, showing separate meson
and baryon branches. Scaling (right panel) is obtained by valence
quark number $n_q$, dividing $v_2$ and $KE_T$ \cite{94}.}
\label{fig:Figure40}
\end{center}
\end{figure}

A further, characteristic scaling property of elliptic
flow is derived from the $p_T$ dependence of $v_2$, observed
for the different hadronic species. In Fig.~\ref{fig:Figure38} one observes a hadron
mass dependence, the $v_2$ signal of pions and charged kaons rising
faster with $p_T$ than that of baryons. Clearly, within a
hydrodynamic flow {\it velocity} field entering hadronization,
heavier hadronic species will capture a higher $p_T$, at a given
flow velocity. However, unlike in hadronic radial expansion flow
phenomena (c.f.\ section~\ref{subsec:Bulk_hadron_transverse_spectra}) it is not the hadronic {\it mass} that
sets the scale for the total $p_T$ derived, per particle species,
from the elliptic flow field, but the hadronic valence quark
content. This conclusion is elaborated \cite{94} in Fig.~\ref{fig:Figure40}. \\
 \\
The left panel shows measurements of the $p_T$ dependence of $v_2$ for
several hadronic species, in minimum bias Au+Au collisions at
$\sqrt{s}=200 \: GeV$ \cite{161}. The middle panel bears out the
hydrodynamically expected \cite{162} particle mass scaling when
$v_2$ is plotted vs. the relativistic transverse kinetic energy
$KE_T \equiv m_T-m$ where $m_T=(p_T^2+m^2)^{1/2}$. For $KE_T \ge 1
\: GeV$, clear splitting into a meson branch (lower $v_2$) and a
baryon branch (higher $v_2$) occurs. However, both of these branches
show good scaling separately. The right panel shows the result
obtained after scaling both $v_2$ and $KE_T$ (i.e. the data in the
middle panel) by the constituent quark number, $n_q=2$ for mesons
and $n_q=3$ for baryons. The resulting perfect, universal scaling is
an indication of the inherent quark degrees of freedom in the
flowing matter as it approaches hadronization. We thus assert that
the bulk of the elliptic flow signal develops in the
pre-hadronization phase.

The above scaling analysis has been extended to $\Phi$ meson and
$\Omega$ hyperon production, and also to first PHENIX results
\cite{163} concerning elliptic flow of the charmed $D$ meson
\cite{94}, with perfect agreement to the observations made in
Fig.~\ref{fig:Figure40}, of a separation into meson/hadron branches on the $KE_T$
scale, which merge into a universal $v_2$ scaling once both $v_2$
and $KE_T$ per valence quark are considered. The observation that
the $D$ meson charm quark apparently shares in the universal flow
pattern is remarkable as its proper relaxation time is, in
principle, lengthened by a factor $M/T$ \cite{164}. A high partonic
rescattering cross section $\sigma$ is thus required in the
primordial QGP fireball, to reduce the partonic mean free path
$\lambda=1/n \sigma$ (where $n$ is the partonic density), such that
$\lambda \ll A^{1/3}$ (the overall system size) and, simultaneously,
$\lambda \: < \: 1 \: fm$ in order to conform with the near-zero
mean free path implication of the hydrodynamic description of the
elliptic flow, which reproduces the data gathered at RHIC energy.
The presence of a high partonic rescattering cross
section was born out in a parton transport model study \cite{92} of
the steep linear rise of the elliptic flow signal with $p_T$
(Fig.~\ref{fig:Figure38}). In such a classical Boltzmann equation approach the cross
sections required for the system to translate the initial spatial
into momentum space anisotropy, fast enough before the initial
deformation gets washed out during expansion (c.f.\ Fig.~\ref{fig:Figure35}), turn out
to exceed by almost an order of magnitude the values expected in a
perturbative QCD quark-gluon gas \cite{92}.

{\it The non-perturbative quark-gluon plasma is thus a strongly
coupled state} (which has been labeled sQGP \cite{165}). At RHIC
energy this reduces the partonic mean free path to a degree that makes
hydrodynamics applicable. A Navier-Stokes analysis \cite{166} of
RHIC flow data indicates that the viscosity of the QGP must be
about ten times smaller than expected if the QGP were a weakly
interacting pQCD Debye screened plasma. This justifies the use of
perfect fluid dynamics.

Considering first attempts to derive a quantitative estimate of the
dimensionless ratio of (shear) viscosity to entropy,we note that
$\eta/s$ is a good way to characterize the intrinsic ability of a
substance to relax toward equilibrium \cite{167}. It can be
estimated from the expression \cite{94}
\begin{equation}
\eta/s \approx T \lambda_f \: c_s
\label{eq:equation50}
\end{equation}
where $T$ is the temperature, $\lambda_f$ the mean free path, and
$c_s$ is the sound speed derived from the partonic matter EOS. A fit
by the perfect fluid ''Buda-Lund'' model \cite{168} to the scaled
$v_2$ data shown in Fig.~\ref{fig:Figure40} yields $T=165 \pm 3 \: MeV$; $c_s$ is
estimated as $0.35 \pm 0.05$ \cite{162,94}, and $\lambda_f \approx
0.30 \: fm$ taken from a parton cascade calculation including $2
\leftrightarrow 3$ scattering \cite{169}. The overall result is
\cite{94}
\begin{equation}
\eta/s = 0.09 \pm 0.02
\label{eq:equation51}
\end{equation}
in agreement with former estimates of Teaney and Gavin \cite{170}.
This value is very small and, in fact, close to the universal lower
bound of $\eta/s =1/4\pi$ recently derived field theoretically
\cite{171}.

Elliptic flow measurements thus confirm that
the quark-gluon matter produced as $\sqrt{s} \rightarrow 200 \: GeV$
is to a good approximation in local thermal equilibrium up to about
$3-4 \:fm/c$. In addition, the final hadron mass dependence of the
flow pattern is consistent with a universal scaling appropriate for
a nearly non-viscous hydrodynamic flow of partons, and the observed
$v_2$ signal reflects a primordial equation of state that is
consistent with first numerical QCD computations \cite{153,154,165}
of a strongly coupled quark-gluon plasma (sQGP) state. First
estimates of its proper shear viscosity to entropy ratio, $\eta/s$,
are emerging from systematic analysis of the elliptic flow signal.
At lower $\sqrt{s}$ precursor elliptic flow phenomena are observed,
as well, but are more difficult to analyze as the crucial, new
feature offered by top RHIC energies is missing here: a clear cut
separation in time, between primordial formation of local partonic
equilibrium conditions, and hadronization. At RHIC (and at future
LHC) energy elliptic flow systematics thus captures the emerging
quark-gluon state of QCD at energy densities in the vicinity of
$\epsilon = 6-15 \: GeV/fm^3$, at temperature $T \approx 300 \:
MeV$, and $\mu_B \rightarrow 0$, describing it as a strongly
coupled, low viscosity liquid medium.

%% file: Chapter_4.tex
\chapter{In-medium attenuation of high $p_T$ hadron and jet production}
\label{chap:In_medium_high_pt}

In the preceding sections we have followed the dynamical evolution
of bulk matter in A+A collisions (at $p_T \le 2 \: GeV$ which covers
about 95\% of the hadronic output), from initial formation time of
partonic matter which reflects in charged particle transverse energy
and multiplicity density, also giving birth to hadrons, and to the elliptic
expansion flow signal.

An alternative approach toward QCD plasma diagnostics exploits the
idea \cite{41,172} of implanting partonic products of primordial
high $Q^2$ processes into the evolving bulk medium, that could serve
as ''tracers'' of the surrounding, co-traveling matter. The ideal
situation, of being able to scatter well defined partons, or
electrons, from a plasma fireball, is approximated by employing
primordially formed charm-anticharm quark pairs \cite{41}, or
leading partons from primordial di-jet production \cite{172}. Both
processes are firmly anchored in perturbative QCD and well studied
in elementary collisions (where such partons are directly released
into vacuum), which thus serve as a reference in an analysis that
quantifies the in-medium modification of such tracer partons. Not a
surprise, in view of our above inferences, from elliptic flow, of a
high temperature, strongly coupled primordial medium: these
in-medium modifications are quite dramatic, leading to a suppression
of $J/\Psi$ production from primordial $c \overline{c}$ pairs
(chapter~\ref{chap:Vector_Mesons}), and to high $p_T$ hadron and jet quenching, the subject
of this chapter.

\section{High $p_T$ inclusive hadron production quenching}
\label{sec:High_pt_inclusive_hadron}
At top RHIC energy, $\sqrt{s}=200 \: GeV$, di-jet production from
primordial hard pQCD parton-parton collisions (of partons from
the initial baryonic structure functions) is the source of
''leading'' partons, with $E_T$ up to about $30 \: GeV$. They are
derived from the inclusive cross section arising if the A+A
collision is considered, first, as an {\it incoherent} superposition
of independent nucleon-nucleon collisions, as enveloped within the
target-projectile nucleon densities. In this framework, the pQCD
cross section for producing an $E_T$ parton in A+B takes the form of
''factorization'' \cite{173}
\begin{equation}
\frac{d \sigma}{dE_T dy}=\sum_{a,b}\: \int_{x_a} dx_a \int_{x_b}
dx_b \: f _{a/A} (x_a) f_{b/B} (x_b) \: \frac{d \sigma_{ab}}{dE_T
dy}
\label{eq:equation52}
\end{equation}
where the $f(x)$ are the parton distributions inside projectile A
and target B nuclei, and the last term is the pQCD hard scattering
cross section. This equation describes the primordial production
rate of hard partons, leading to the conclusion \cite{172} that, at
RHIC energy, all hadrons at $p_T \ge 6-10 \: GeV$ should arise from
initial pQCD parton production.

As partons are effectively frozen during the hard scattering, one
can treat each nucleus as a collection of free partons. Thus, with
regard to high $p_T$ production, the density of partons within the
parton distribution function of an atomic number A nucleus should be
equivalent to the superposition of $A$ independent nucleons $N$:
\begin{equation}
f_{a/A} \: (x, Q^2)= A \: f_{a/N} \: (x, Q^2).
\label{eq:equation53}
\end{equation}
From equations~\ref{eq:equation52} and~\ref{eq:equation53} it is clear that the primordial high $Q^2$
inclusive parton cross section in a minimum bias A+B reaction scales as 
A $\cdot$ B times the corresponding (N+N or) p+p cross
section. Furthermore, as each leading parton ends up in an observed
high $p_T$ hadron $h$ , we thus write the invariant hard hadron
cross section as
\begin{equation}
Ed \sigma_{AB \rightarrow h} / d^3p = A \cdot B \cdot E \: d \sigma_
{pp \rightarrow h}/d^3p.
\label{eq:equation54}
\end{equation}
Since nucleus-nucleus experiments usually measure invariant yields
$N_h$ for a given centrality bin, corresponding to an average impact
parameter $b$, one writes instead:
\begin{equation}
E d N_{AB \rightarrow h} (b) / d^3p = \left<T_{AB} (b)\right> E \: d \sigma_{pp
\rightarrow h}/d^3p,
\label{eq:equation55}
\end{equation}
where $T_{AB} (b)$ is the Glauber geometrical overlap function of
nuclei A, B at impact parameter $b$, which accounts for the average
number of participant parton collisions at given impact geometry
\cite{174}, $\left<N_{coll}(b)\right>$. One can thus quantify the attenuating
medium effects, as experienced by the primordially produced tracer
parton on its way toward hadronization, by the so-called {\it
nuclear modification factor} for hard collisions (analogous to equation~\ref{eq:equation6}, that refers to soft, bulk hadron production):
\begin{equation}
R_{AB} (p_T, y, b) =
\frac{d^2 N_{AB}/dy
dp_T}{\left<T_{AB}(b)\right> \: d^2 \sigma_{pp}/dydp_T}.
\label{eq:equation56}
\end{equation}
Obviously, this concept of assessing the in-medium modification of
hadron production at high $p_T$ requires corresponding p+p collision
data, as a reference basis. Such data have been, in fact, gathered
at top RHIC, and top SPS energies, $\sqrt{s}=200$ and $17.3 \: GeV$,
respectively. Alternatively, in situations where the relevant
reference data are not known, one considers the production ratio of
hadronic species $h$, observed in central relative to peripheral
collisions:
\begin{equation}
R_{CP}(p_T, y)=\frac{d^2 N_h(b_1)/dydp_t}{d^2 N_h (b_2)/dydp_t} \:
 \mbox{x} \: \frac{\left<T_{AB}(b_2)\right>}{\left<T_{AB}(b_1)\right>}
\label{eq:equation57}
\end{equation}
where $b_1 << b_2$ are the average impact parameters corresponding
to the employed trigger criteria for ''central'' and ''peripheral''
A+A collisions, respectively. This ratio recasts, to a certain
extent, the in-medium attenuation analysis, offered by $R_{AB}$,
insofar as peripheral collisions approach the limiting conditions,
of a few single nucleon-nucleon collisions occuring in the dilute
surface sections of the nuclear density profiles, i.e. essentially
in vacuum.\\
\begin{figure}[h!]   
\begin{center}\vspace{-0.5cm}
\includegraphics[scale=1.0]{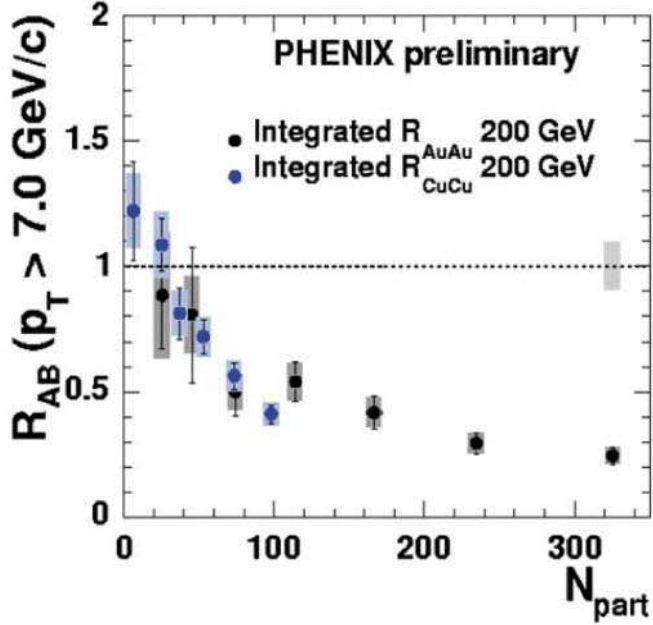}\vspace{-0.4cm}
\caption{The nuclear modification factor $R_{AA}$ for $\pi^0$ in min. bias
Cu+Cu and Au+Au collisions at $\sqrt{s}=200 \: GeV$, in the range
$p_T > 7 \: GeV/c$ \cite{175}, plotted vs. centrality as measured by participant nucleon number $N_{part}$.}
\label{fig:Figure41}
\end{center}
\end{figure} \\
Employing  the above analysis schemes, the RHIC experiments have, in
fact, demonstrated a dramatic in-medium suppression of the high
$p_T$ yield, for virtually all hadronic species. Fig.~\ref{fig:Figure41} shows
$R_{AA}$ for neutral pions produced in min. bias Cu+Cu and Au+Au
collisions at $\sqrt{s}=200 \: GeV$ where PHENIX extended the $p_T$
range up to $18 \: GeV/c$ \cite{175}; the nuclear modification factor
refers to the range $p_T > 7.0 \: GeV/c$ and is shown as a function
of centrality, reflected by $N_{part}$. We infer a drastic
suppression, by an $R_{AA} \approx 0.25$ in near central collisions.
Fig.~\ref{fig:Figure42} shows the $p_T$ dependence \cite{175} of neutral pion
$R_{AA}$ in central Au+Au collisions ($N_{part}=350$), with a
suppression to below 0.2 at $p_T \ge 4 \: GeV/c$.\\
\begin{figure}[h!]   
\begin{center}\vspace{-0.5cm}
\includegraphics[scale=0.4]{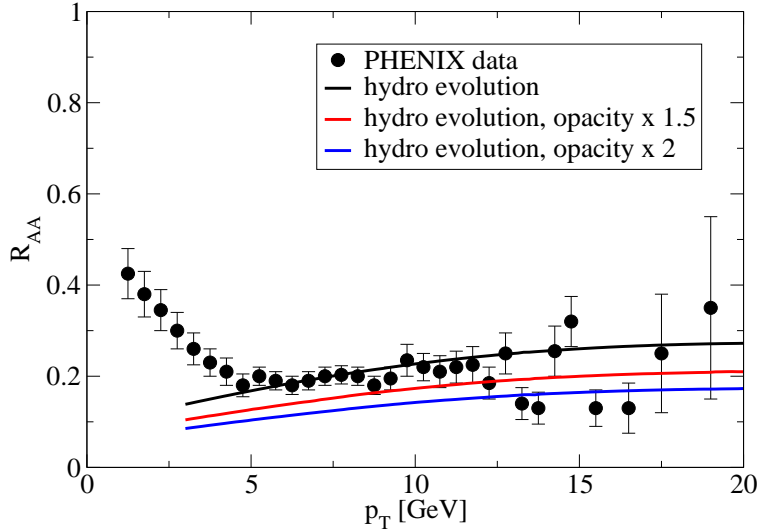}\vspace{-0.4cm}
\caption{$R_{AA}$ for $\pi^0$ production in central Au+Au collisions at
$\sqrt{s}=200 \: GeV$ \cite{175}, compared to a hydrodynamic
calculation for different opacities (transport coefficients
\cite{195}) of the plasma \cite{204}.}
\label{fig:Figure42}
\end{center}
\end{figure} \\
Note that $R_{AA}$ {\it can not reach arbitrarily small values}
because of the unattenuated contribution of quasi-in-vacuum surface
''corona'' single nucleon-nucleon collisions, closely resembling the
$p+p \rightarrow \pi^0 + X$ inclusive yield employed in the
denominator of $R_{AA}$, equation~\ref{eq:equation56}. Even in a central trigger
configuration this suggests a lower bound, $R_{AA} \approx 0.15$.
Nuclear attenuation of high $p_T$ pions thus appears to be almost
''maximal'', implying a situation in which the interior sections of
the high energy density fireball feature a very high opacity, i.e.
they are almost ''black'' to the high $p_T$ partons contained in
pions. 

We remark here, briefly, on a confusing feature that occurs in Fig.~\ref{fig:Figure42} (and in the
majority of other $R_{AA}$ vs. $p_T$ plots emerging from the RHIC
experiments): the pQCD number of collision scaling employed in
$R_{AA}$ does {\it not} describe pion production at low $p_T$ as we
have demonstrated in chapter~\ref{sec:Bulk_Hadron_Prod}. The entries at $p_T \le 3 \:
GeV/c$ are thus besides the point, and so are pQCD guided model
predictions, as shown here \cite{176}.

\begin{figure}[h!]   
\begin{center}\vspace{-0.5cm}
\includegraphics[scale=0.93]{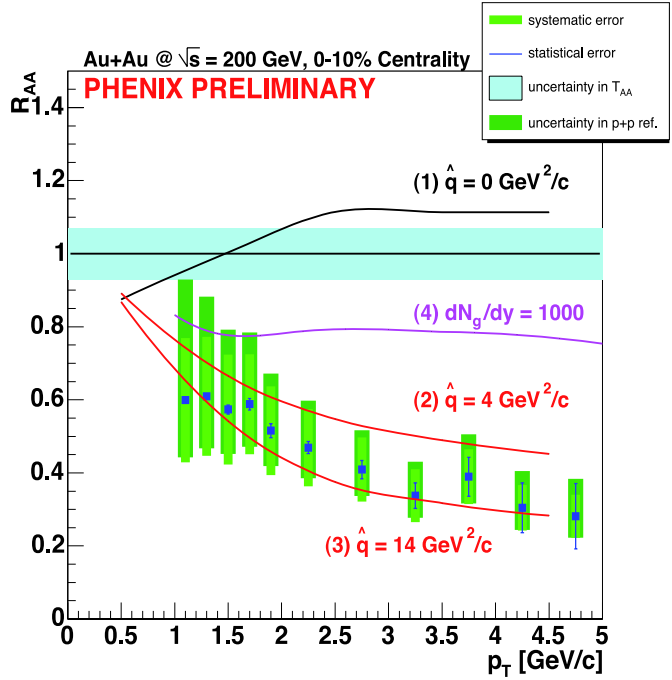}\vspace{-0.4cm}
\caption{The nuclear modification factor $R_{AA}$ vs. $p_T$ for electrons
from semi-leptonic decays of heavy flavor (mostly D) mesons in
central Au+Au collisions at $\sqrt{s}=200 \: GeV$ \cite{175}; with
calculations of in medium energy loss using different attenuation
models \cite{179,180}.}
\label{fig:Figure43}
\end{center}
\end{figure}

\begin{figure}[h!]   
\begin{center}\vspace{-0.1cm}
\includegraphics[scale=0.6]{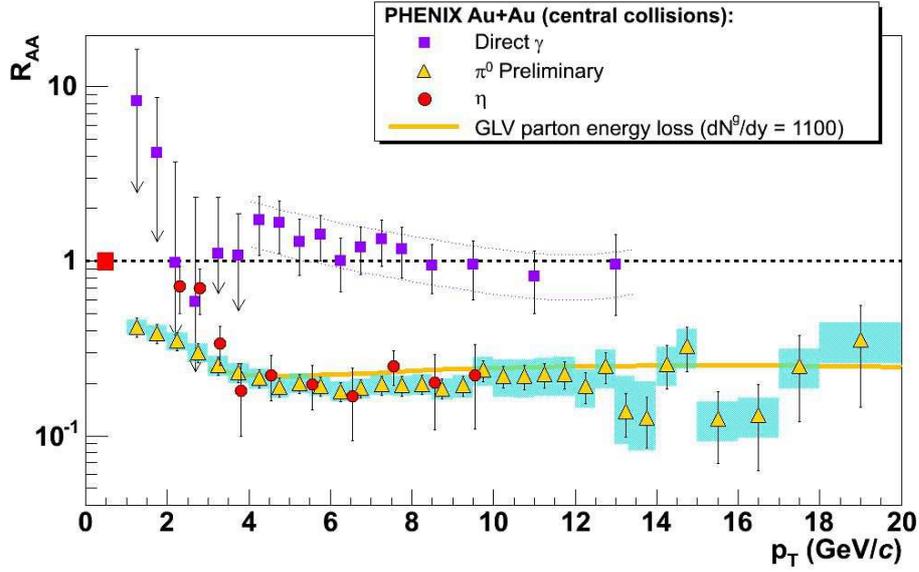}\vspace{-0.4cm}
\caption{Modification factor $R_{AA}$ vs. $p_T$ for direct photons in central
Au+Au at $200 \: GeV$ (squares). Also shown are $R_{AA}$ for $\pi^0$
(triangles) and $\eta$ (circles) \cite{181} fitted by the attenuation
model \cite{176,180}.}
\label{fig:Figure44}
\end{center}
\end{figure}

Nuclear modification analysis at RHIC covers, by now, the high $p_T$
production of a multitude of mesonic and baryonic species
\cite{177}, most remarkably even including the charmed D meson which
is measured via electrons from semi-leptonic heavy flavor decay by
PHENIX \cite{175} and STAR \cite{178}. Fig.~\ref{fig:Figure43} illustrates the first
PHENIX results from central Au+Au at $\sqrt{s}=200 \: GeV$, $R_{AA}$
falling to about 0.3 at 5 $GeV/c$. Heavy flavor attenuation thus
resembles that of light quarks, as is also attested by the
predictions of in-medium parton transport models \cite{179,180}
included in Fig.~\ref{fig:Figure43}, which cast the medium opacity into an effective
parton transport coefficient $\hat{q}$ which is seen here to
approach a value $14 \: GeV^2/c$ at high $p_T$, again corresponding
to a highly opaque medium. We shall describe this approach in more
detail below.\\
 \\
A most fundamental cross-check of the in-medium attenuation picture
of color charged partons consists in measuring $R_{AA}$ for
primordial, ''direct'' photons. Fig.~\ref{fig:Figure44} shows first PHENIX results
\cite{181} for direct photon $R_{AA}$ vs. $p_T$ in central Au+Au at
$\sqrt{s}=200 \: GeV$. {\it Photons obey pQCD number of collisions scaling},
$R_{AA} \approx 1$! Also included in Fig.~\ref{fig:Figure44} are the attenuation
ratios for neutral pions (already shown in Fig.~\ref{fig:Figure42}), and for $\eta$
mesons that follow the pattern of extreme suppression. In essence,
the PHENIX results in Figs.~\ref{fig:Figure43} and~\ref{fig:Figure44} wrap up all one needs to know
for a theoretical analysis of fireball medium opacity, for various
flavors, and indicate transparency for photons.

\begin{figure}[h!]   
\begin{center}\vspace{-0.1cm}
\includegraphics[scale=0.6]{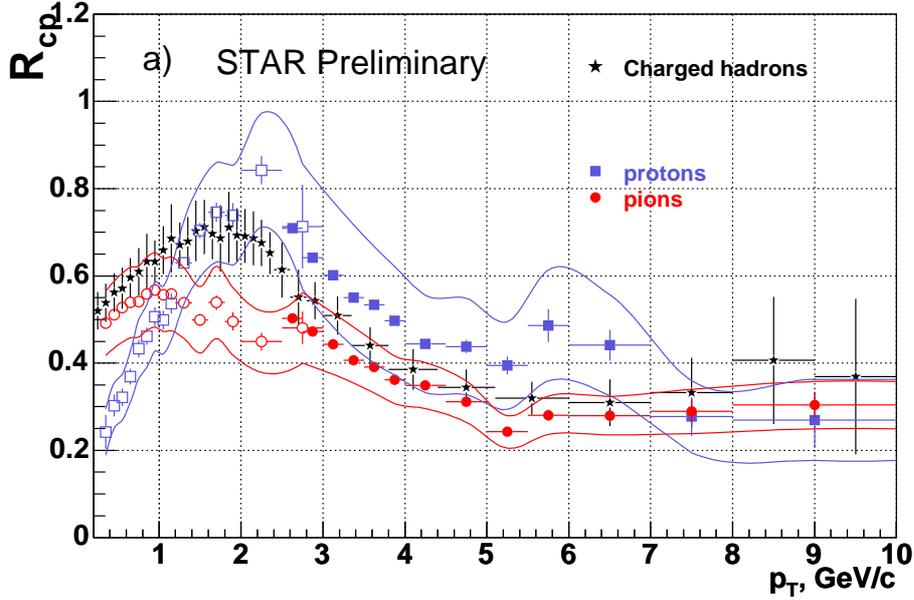}\vspace{-0.4cm}
\caption{$R_{CP}$, the ratio of scaled $\pi, \: p$ and charged hadron yields
vs. $p_T$ in central (5\%) and peripheral (60-80\%) Au+Au collisions
at $\sqrt{s}=200 \: GeV$ \cite{182}.}
\label{fig:Figure45}
\end{center}
\end{figure}

\begin{figure}[h!]   
\begin{center}\vspace{-0.1cm}
\includegraphics[scale=0.4]{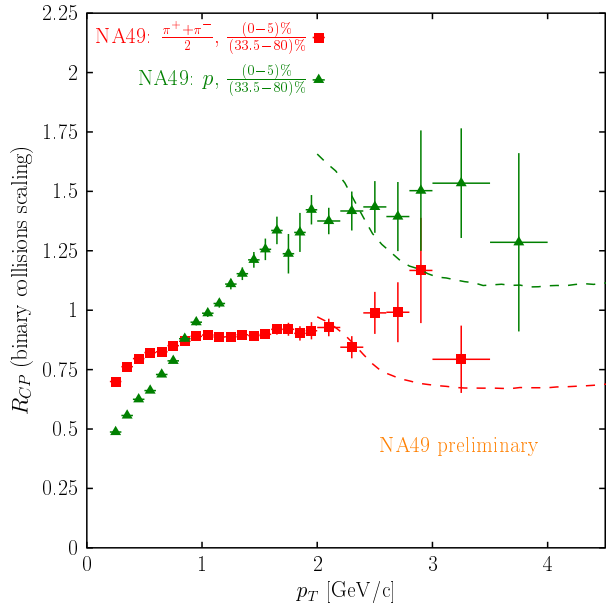}\vspace{-0.4cm}
\caption{$R_{CP}$ results from SPS Pb+Pb collisions at $\sqrt{s}=17.3 \:
GeV$, for pions and protons \cite{183}, with attenuation model fits
\cite{176}.}
\label{fig:Figure46}
\end{center}
\end{figure}

We turn, briefly, to $R_{CP}$ as opposed to $R_{AA}$ analysis, in
order to ascertain similar resulting conclusions at RHIC energy.
Fig.~\ref{fig:Figure45} illustrates an $R_{CP}$ analysis of $\pi, p$ and
charged hadron high $p_T$ production in Au+Au at $\sqrt{s}=200 \:
GeV$ by STAR \cite{182}, the central to peripheral yield ratio
referring to a 5\%, and a  60-80\% cut of minimum bias data. We
showed in equation~\ref{eq:equation57} that the $R_{CP}$ measure also refers to a
picture of pQCD number of binary collision scaling, inappropriate at
low $p_T$. Thus ignoring the features of Fig.~\ref{fig:Figure45} at $p_T \le 3 \:
GeV/c$ we conclude that the high $p_T$ data again suggest a
suppression by about 0.3, common to pions and protons, thus
approaching the ratio, of about 0.2, observed in Figs.~\ref{fig:Figure41} and~\ref{fig:Figure42}
which employ the ''ideal'' in-vacuum $p+p \rightarrow$ hadron + $X$
reference. \\
 \\
At top SPS energy, $\sqrt{s}=17.3 \: GeV$, the
experimentally covered $p_T$ range is fairly limited \cite{183},
$p_T < 4 \: GeV/c$. Fig.~\ref{fig:Figure46} shows NA49 results, $R_{CP}$ for p
and charged pions. Contrary to former expectations that such data
would be overwhelmed by Croonin-enhancement \cite{184} the same
systematic behaviour as at RHIC is observed, qualitatively: $R_{CP}$
(baryon)$
> R_{CP}$ (meson) at $p_T >3 \: GeV/c$. Note that, again, the data
do not approach unity at $p_T \rightarrow 0$ because of the employed
binary scaling, and that the strong rise of the proton signal at
$p_T < 2 \: GeV/c$ is largely the result of strong radial flow in
central Pb+Pb collisions, an effect much less prominent in pion
$p_T$ spectra. The high $p_T$ suppression is much weaker than at
RHIC but it is strong enough that the expected Croonin enhancement
of high $p_T$ mesons is not observed. These SPS data, as well as
first results obtained at the intermediate RHIC energy of
$\sqrt{s}=62.4 \: GeV$ \cite{185}  are reproduced by an attenuation
model based on the primordial gluon density $dN_g/dy$ that scales as
the charged particle midrapidity density $dN_{ch}/dy$ \cite{176},
and was also employed in Figs.~\ref{fig:Figure43} and~\ref{fig:Figure44}. \\

\section{Energy loss in a QCD medium}
\label{sec:Energy_loss_in_a_QCD}
The attenuation model that we have hinted at consists of a gluon radiative
energy loss theory of the primordially produced leading, high $p_T$
parton as it traverses a medium of color charges, by means of
emission of gluon bremsstrahlung. We expect that the resulting
partonic specific energy loss, per unit pathlength (i.e. its
$dE/dx$) should reflect characteristic properties of the traversed
medium, most prominently the spatial density of color charges
\cite{176} but also the average momentum transfer, per
in-medium collision of the considered parton or, more general, per unit pathlength at constant
density.  Most importantly, an aspect of non-abelian QCD leads to a
characteristic difference from the corresponding QED situation: the radiated gluon is itself
color-charged, and its emission probability is influenced, again, by
its subsequent interaction in the medium \cite{186} which, in turn,
is proportional to medium color charge density and traversed
pathlength $L$. Thus, the traversed path-length $L$ in-medium
occurs, both in the probability to emit a bremsstrahlung gluon, and
in its subsequent rescattering trajectory, also of length $L$, until
the gluon finally decoheres. Quantum mechanical coherence thus leads
to the conclusion that non-abelian $dE/dx$ is not proportional to
pathlength $L$ (as in QED) but to $L^2$ \cite{187}.

\begin{figure}[h!]   
\begin{center}\vspace{-0.1cm}
\includegraphics[scale=0.51]{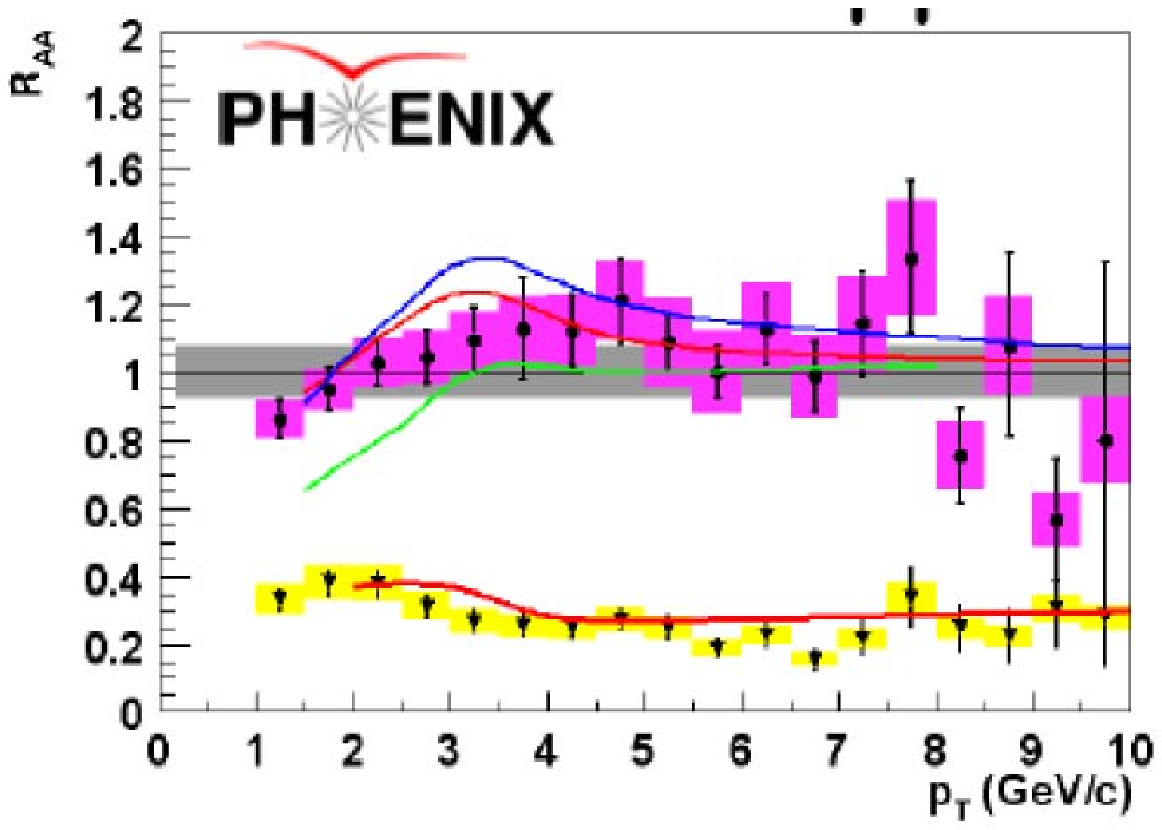}
\includegraphics[scale=0.33]{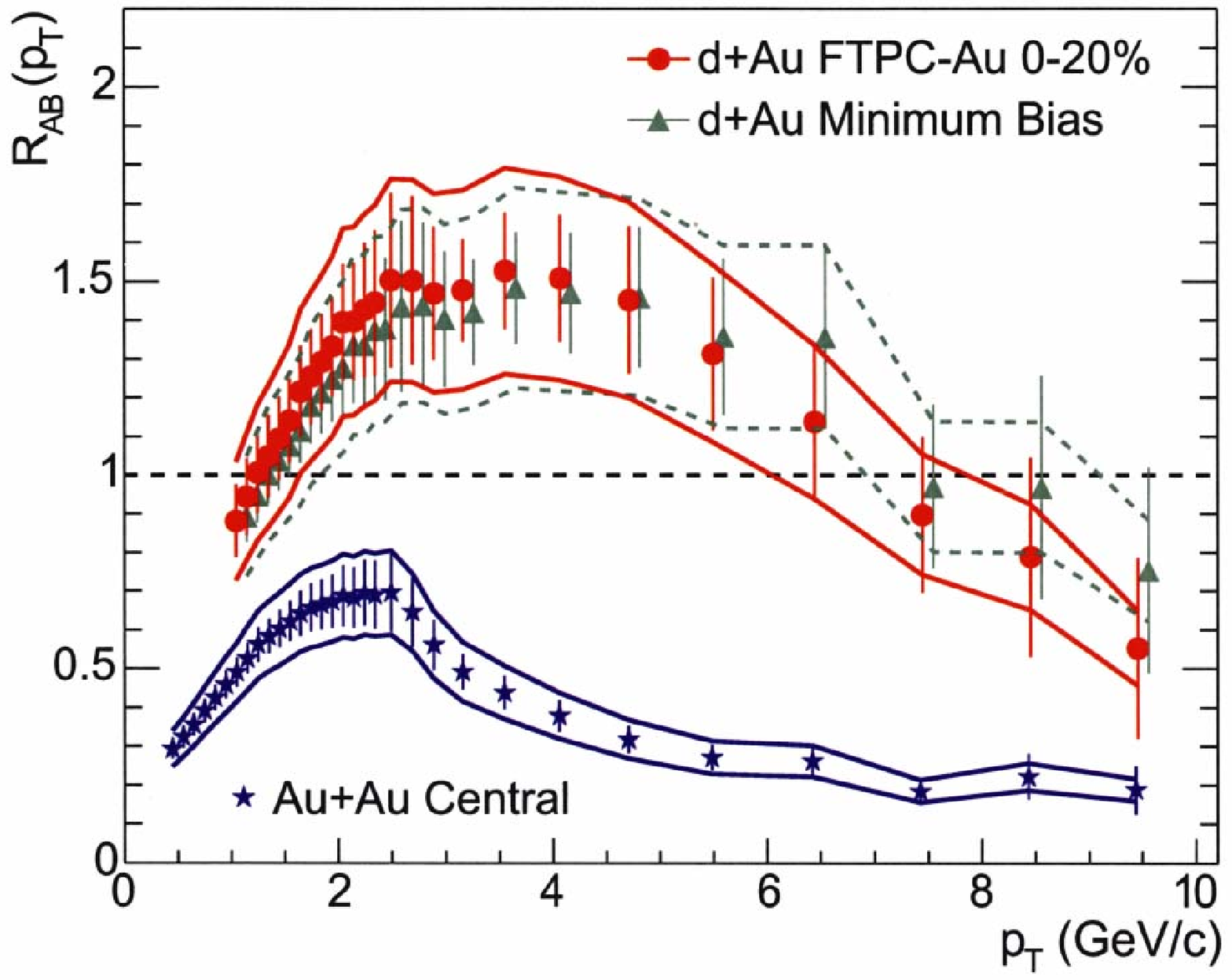}\vspace{-0.4cm}
\caption{$R_{AA}$ vs. $p_T$ for d+Au collisions at $\sqrt{s}=200 \: GeV$,
compared to central Au+Au results, for $\pi^0$ (left) and charged
hadrons (right). From \cite{190,191}.}
\label{fig:Figure47}
\end{center}
\end{figure}

This phenomenon occurs at intermediate values of the radiated
gluon energy, $\omega$, in between the limits known as the
Bethe-Heitler, and the factorization regimes \cite{186},
\begin{equation}
\omega_{BH} \approx \lambda \: q^2_T \ll \omega \ll \omega_{fact}
\approx L^2 \: q_T^2 /\lambda \le E
\label{eq:equation58}
\end{equation}
where $\lambda$ is the in-medium mean free path, $q_T^2$ the
(average) parton transverse momentum square, created per collision,
and $E$ the total $cm$ energy of the traveling charge. In the BDMPSZ
model \cite{186,187,188} the properties of the medium are encoded in
the transport coefficient, defined as the average induced transverse
momentum squared per unit mean free path,
\begin{equation}
\hat{q} = \left<q^2_T\right> / \lambda.
\label{eq:equation59}
\end{equation}
The scale of the radiated gluon energy distribution $\omega \: dN/d
\omega$ is set by the characteristic gluon energy \cite{186,187}
\begin{equation}
\omega_c=\frac{1}{2} \: \hat{q} \: L^2.
\label{eq:equation60}
\end{equation}
To see more explicitly how the various properties of the color
charged medium enter in $\hat{q}$ we rewrite it as
\begin{equation}
\hat{q}= \rho \int q^2_T \: dq^2_T \: \frac{d \sigma}{dq^2_T} \equiv
\rho \sigma \left<q^2_T\right> = \lambda^{-1} \left<q^2_T\right>
\label{eq:equation61}
\end{equation}
where $\rho$ is the color charge density of scattering centers,
$\sigma$ the effective binary cross section for interaction of the
considered leading parton at scale $q^2$ (which may depend on quark
flavor), and $\left<q^2_T\right>$ as above. Obviously, both $\sigma$ and
$\left<q^2_T\right>$ refer to detailed, intrinsic properties of the QCD medium,
globally described by density $\rho$. The leading parton cross
section with in-medium color charges should depend on the implied
resolution scale, $Q^2=\left<q^2_T\right>$, and can thus be obtained from
perturbative QCD \cite{186,187,188} only if $Q^2 > Q^2_{sat}$, the
saturation scale that we discussed in section~\ref{sec:Bulk_Hadron_Prod}. Likewise, $\left<q^2_T\right>$
itself reflects a medium property, the effective range of the color
force, which is different in confined and deconfined media. Hadron
size limits the force range in the former case, such that $\hat{q}$
is minimal in ground state hadronic matter also, of course, due to
the small energy density $\rho=\rho_0=0.15 \: GeV/fm^3$ \cite{189}.
This was, in fact, confirmed by a RHIC run with deuteron-gold
collisions, in which mid-rapidity hadrons traverse sections of cold
Au nucleus spectator matter. Fig.~\ref{fig:Figure47} shows results obtained for
$R_{dA}$ dependence on $p_T$, for $\pi^0$ from PHENIX \cite{190},
and for charged hadrons from STAR \cite{191}. For comparison, both panels also
include the corresponding $R_{AA}$ data for central Au+Au collisions
(all at $\sqrt{s}=200 \: GeV/c$), exhibiting the typical, drastic
high $p_T$ quenching of inclusive hadron production, clearly absent
in d+Au collisions. \\
\\
We have shown a first application of the BDMPSZ model, to RHIC
inclusive D meson production \cite{179}, in Fig.~\ref{fig:Figure43}. Before engaging
in further model application we note, first, that equations~\ref{eq:equation58}-\ref{eq:equation61} above
refer to the idealized conditions of an infinitely extended medium
of uniform composition. In reality, the fireball medium expands, at
time scale concurrent with the proper time incurred in the leading
partons propagation over distance $L$, such that all ingredients in
$\hat{q}$, exhibited in equation~\ref{eq:equation61}, vary with expansion time
\cite{192}. However, before turning to adaption to reality of the
infinite matter model, we wish to expose its prediction for the
final connection of the specific average partonic energy loss
$\left<\Delta E\right>$ and in-medium path length $L$ traversed:
\begin{equation}
\left<\Delta E\right> = \int \omega \frac{dN}{d \omega} \: d \omega \propto
\alpha_s \: C_R \: \hat{q} \: L^2.
\label{eq:equation62}
\end{equation}
This relation \cite{193} represents the eikonal approximation limit
of extended medium and high leading parton initial energy $E >
\omega_c$ (equation~\ref{eq:equation60}). The average energy loss is thus proportional
to the appropriate strong coupling constant $\alpha_s$, to the
Casimir factor corresponding to the leading parton (with value 4/3
for quarks, and 3 for gluons), as well as to $\hat{q}$ and $L^2$.

In order to verify the non-abelian nature of radiative parton energy
loss in a partonic QCD medium it would, of course, be most
convincing if a direct, explicit $L^2$ dependence of $\left<\Delta E\right>$
could be demonstrated. Such a demonstration is lacking thus far,
chiefly because of the obvious difficulty of simultaneously knowing
the primordial parton energy $E$, the transport coefficient
$\hat{q}$ and - in finite nuclear collision geometry - its variation
along the actual traversed path $L$ as the surrounding medium
expands with propagation time. Moreover, the partonic medium induced
energy loss $\Delta E$ of the primordial parton is not directly
observable. Even if we assume that high $p_T$ partons evolve into
the observed hadrons only {\it after} leaving the fireball medium
\cite{176}, their ensuing ''fragmentation'' to hadrons (which is
known from $p+p$ jet physics) results in several hadrons usually
comprising a ''leading'' hadron which transports a major fraction
$\left<z\right> \equiv \left<E^h/E^p\right>$ of the fragmenting parton energy $E^p$,
which, in turn, equals $E^p$ (primordial) - $\Delta E$, with $\Delta
E$ sampled from a probability distribution with mean $\left<\Delta E\right>$
according to equation~\ref{eq:equation62}. The observed leading hadron energy or
transverse momentum is thus subject to sampling, both, $z$ from the
fragmentation function, and $\Delta E$ from in-medium energy loss.
Finally, inclusive high $p_T$ leading hadron observation in A+A
collisions involves an average over all potential initial parton
production points, within the primordially produced density profile.
A specific distribution of in medium path lengths $f(L)$ arises, for
each such production point, which, moreover, depends on a model of
space-time fireball expansion. The final inclusive yield thus requires a
further, weighted volume average over $f(L)$ per production point.
Thus, typical of an inclusive mode of observation, the ''ideal''
relationship of equation~\ref{eq:equation62}, between radiative in-medium energy loss
$\Delta E$ and traversed path length $L$ gets shrouded  by double
averages, independently occuring at either side of the equation
\cite{176,179,189,194,195,196}. A detailed $L^2$ law verification
can not be expected from inclusive central collision data {\it
alone} (see next section).

However, the unmistakeably clear signal of a strong, in-medium high
$p_T$ parton quenching effect, gathered at RHIC by $R_{AA}$
measurement for a multitude of hadronic species (Figs.~\ref{fig:Figure41},~\ref{fig:Figure42},~\ref{fig:Figure20},~\ref{fig:Figure43},~\ref{fig:Figure45},~\ref{fig:Figure47}), in Au+Au collisions at $\sqrt{s}=200 \: GeV$, has resulted in first estimates of the transport coefficient $\hat{q}$,
the medium - specific quantity entering equation~\ref{eq:equation62}, in addition to
the geometry - specific path length $L$. In fact, the transport coefficient
can, to some extent, be analyzed independently, owing to
the fact that $\hat{q} \propto \varrho$ from equation~\ref{eq:equation61}. The density
$\varrho$ falls down rapidly during expansion, but it is initially
rather well constrained by the conditions of one-dimensional Bjorken
expansion that we have described in chapters~\ref{sec:Bulk_Hadron_Prod} and~\ref{chap:Elliptic_flow}. 
{\it The major contribution to partonic $\Delta E$ arises in the early
expansion phase (via a high $\hat{q}$), in close analogy to the
formation of the elliptic flow signal}. These two signals are, thus,
closely correlated: the primordial hydrodynamic expansion phase of
bulk matter evolution sets the stage for the attenuation, during
this stage of QCD matter, of primordially produced ''tracer''
partons, traversing the bulk matter medium as test particles.

\begin{figure}[h!]   
\begin{center}\vspace{-0.2cm}
\includegraphics[scale=0.4]{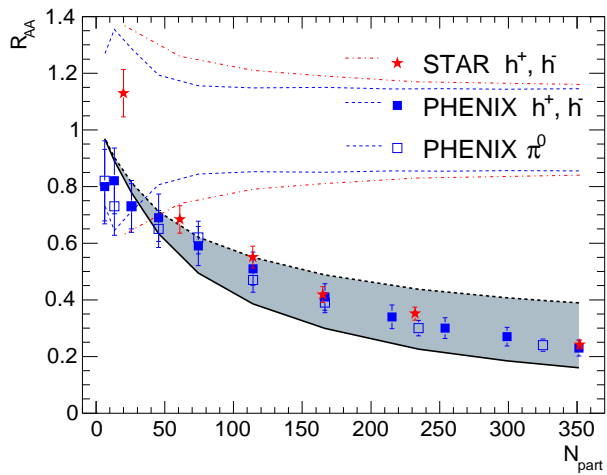}\vspace{-0.4cm}
\caption{The effective transport coefficient $\hat{q}=14 \: GeV^2/fm$ in the
parton quenching model (PQM) of \cite{195} determined from the
centrality dependence of $R_{AA}$ for $\pi^0$ and charged hadrons,
averaged over 4.5 $\le p_T \le 10 \: GeV/c$, in Au+Au at
$\sqrt{s}=200 \: GeV$.}
\label{fig:Figure48}
\end{center}
\end{figure}

The bias in partonic $\Delta E$ to the primordial expansion period
is borne out in an expression \cite{193,195} which replaces the
$\hat{q}$ coefficient, appropriate to an infinitely extended static
medium considered in the original BDMPSZ model, by an effective,
expansion time averaged
\begin{equation}
\hat{q}_{eff} = \frac{2}{L^2} \int^L_{t_0} \: dt \:(t-t_0)\: \hat{q}
\: (t)
\label{eq:equation63}
\end{equation}
to be employed in the realistic case of an expanding fireball
source. Due to the rapid fall-off of $\varrho$, in $\hat{q}=\varrho
\sigma \:\left<q_T^2\right>$ from equation~\ref{eq:equation61}, the integral depends, far more strongly, on $\hat{q} \: (t \approx t_0)$ than on total path
length $L$. Furthermore, inserting $\hat{q}_{eff}$ into the BDMPSZ
formula \cite{193,195} for the transverse momentum downward shift,
occuring in leading parton or hadron $p_T$ spectra (of power law
form $p_T^{-\nu}$, see Fig.~\ref{fig:Figure20})
\begin{equation}
\Delta \: p_T \approx - \alpha_s \sqrt{\pi \hat{q} L^2 p_T/\nu},
\label{eq:equation64}
\end{equation}
we see that the first order proportionality to $L^{2}$ is removed. The
downward $p_T$ shift is thus, primarily, a consequence of
$\hat{q}_{eff}$ which, in turn, is biased to reflect the ''ideal''
transport coefficient $\hat{q}$ at early evolution time. Within this
terminology,the $p_T$ shift (see equation~\ref{eq:equation64}) determines the experimentally
measured ratio $R_{AA}(p_T)$ which quantifies the effective
transport coefficient $\hat{q}_{eff}$ for the $p_T$ domain
considered. It can be related, as a cross check, to the initial
gluon rapidity density if the collision region expands according to
Bjorken scaling \cite{187,197}:
\begin{equation}
\hat{q} = \alpha_s \: \frac{2}{L} R^{-2}_A \: \frac{dN^g}{dy}.
\label{eq:equation65}
\end{equation}
A typical result of application of the model described above
\cite{195} is shown in Fig.~\ref{fig:Figure48}. Analogous to Fig.~\ref{fig:Figure41}, $R_{AA}$ for
neutral pions and charged hadrons is averaged over the range $4.5
\le p_T \le 10 \: GeV/c$, and shown as a function of centrality
(assessed by $N_{part}$) in minimum bias Au+Au collision at
$\sqrt{s}= 200 \: GeV$ \cite{190,191,198}. A path-averaged
$\hat{q}_{eff}$ of $14 \: GeV^2/fm$ is inferred from the fit, in
close agreement to the value found in \cite{195}.

\begin{figure}[h!]   
\begin{center}
\includegraphics[scale=0.4]{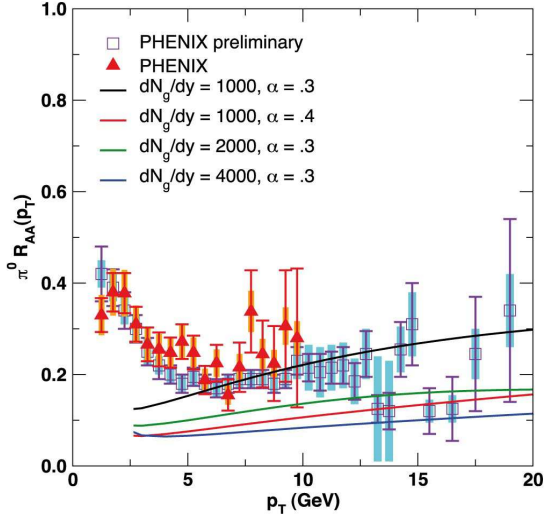}\vspace{-0.4cm}
\caption{Application of the WDHG transport model \cite{200} based on equation~\ref{eq:equation65} to PHENIX $R_{AA}$ data for $\pi^0$ \cite{175}, indicating
primordial $1000 \le \frac{dN^g}{dy} \le 2000$.}
\label{fig:Figure49}
\end{center}
\end{figure}

A more recent study \cite{199} of the PHENIX $R_{AA}$ data for
$\pi^0$ in central Au+Au collisions (Fig.~\ref{fig:Figure42}) is shown in Fig.~\ref{fig:Figure49}. The
analysis is carried out in the framework of the WHDG model
\cite{200}, which replaces the (effective) transport coefficient
$\hat{q}$ (employed in the BDMPSZ model \cite{186,187,188}, and
turned into the data analysis formalism of \cite{193,195}) by the
primordial gluon mid-rapidity density $dN^g/dy$, as the fundamental
parameter, via equation~\ref{eq:equation65}. The initial gluon density, in turn, being
related to the charged hadron mid-rapidity density \cite{61}. Fig.~\ref{fig:Figure49} 
shows that, within the still preliminary statistics at $p_T > 10
\: GeV/c$, the ''conservative'' estimate of $\alpha_s = 0.3$ and
$dN^g/dy= 1000$ does not appear to be the most appropriate choice,
the data rather requiring $1000 < dN^g/dy < 2000$. Overall, Fig.~\ref{fig:Figure49}
demonstrates a certain insensitivity of the data, to the basic
parameters of theoretical high $p_T$ quenching models of inclusive
hadron production at RHIC energy, that we have already inferred from
Fig.~\ref{fig:Figure43}, concerning choice of $\hat{q}$.

Radiative in-medium energy loss of
primordially produced partons, traversing the evolving bulk medium
as ''tracers'', must be extremely strong. This is indicated by the
inclusive attenuation ratios $R_{AA}$, which fall down to about 0.2
in central collisions thus almost reaching  the absolute lower limit
of about 0.15 that arises from the unavoidable fraction of
un-attenuated primordial surface ''corona'' nucleon-nucleon
interaction products \cite{201}. \\
\\
Not expected from early estimates of medium opacity based on the picture of a weakly coupled
perturbative QCD medium \cite{172, 202}, the interior sections of the
central Au+Au interaction volume must be almost ''black'' toward
high $p_T$ parton transport \cite{194,195} at $\sqrt{s}=200 \:GeV$,
also including charm quark propagation (Fig.~\ref{fig:Figure43}). The remaining
signal should thus stem, primarily, from the dilute surface
sections, and from the finite fraction of partons traversing the
interior with small, or zero radiative energy loss, as a consequence
of the finite width of the $\Delta E$ probability distribution
\cite{193,194}. Seen in this light, the smooth decrease of $R_{AA}$
with centrality (Fig.~\ref{fig:Figure48}) should reflect the combined effects, of a
decreasing surface to volume ratio, an increasing effective
$\hat{q}$ (due to interior density increase) that confronts the
increasing average geometrical potential path length $\left<L\right>$
(essentially enhancing its effect), and a thus diminishing fraction
of primordial high $p_T$ partons experiencing a small $\Delta E$.

Not surprisingly, the ideal non abelian  QCD relationship of $\Delta
E$ proportional to in-medium high $p_T$ parton path length $L^2$
can, thus, not be established from inclusive high $p_T$ quenching
data alone. We shall show in the next section that di-jet primordial
production can offer a  mechanism of higher resolution.
The inclusive $R_{AA}$ attenuation data, obtained at RHIC,
are seen to establish an unexpected, high opacity of the primordial
interaction volume, extending to high $p_T$ parton propagation.
The required, high transport coefficient $\hat{q} = 14 GeV^2/fm$ from Fig.~\ref{fig:Figure48}, confirms and extends the picture derived
from elliptic flow data \cite{167}: at top RHIC energy the plasma is
non-perturbatively, strongly coupled, a new challenge to lattice QCD
\cite{61}. The QGP may be largely composed of
colored, string-like partonic aggregates \cite{203}.

\section {Di-jet production and attenuation in A+A collisions}
\label{sec:Di_jet_production}

In order to analyze leading parton attenuation in a more constrained situation \cite{204}, one investigates parton tracer attenuation under {\it
further geometrical constraints} concerning the in-medium path length $L$, by
means of di-jet analysis, and/or by varying the primordial parton
density that enters $\hat{q}$ via equation~\ref{eq:equation61} in studies at different
$\sqrt{s}$ while maintaining the observational geometrical
constraints.

We shall concentrate here on di-jet attenuation data obtained in
Au+Au collisions at top RHIC energy, $\sqrt{s}=200 \: GeV$. At this
relatively modest energy the initial pQCD production cross section of leading
partons (as described in equation~\ref{eq:equation52}) reaches up to
$p_T = 25 \: GeV/c$. The ensuing DGLAP shower multiplication
initiates ''parton fragmentation'' to hadrons \cite{83,85,121}, each
carrying a momentum fraction $z_T = p_T/p_T$(primord.parton). The
created ensemble of hadrons $h$ belonging to the observed hadronic
jet can be wrapped up by the total {\it fragmentation function}
\begin{equation}
F_h \: (z, \sqrt{s}) = \sum_i \int \frac{dz}{z} C_i \: (z, \sqrt{s})
D_{part \rightarrow h} (z, \sqrt{s})
\label{eq:equation66}
\end{equation}
which summarizes the contributions arising from the different shower
partons $i$. Here, $C_i$ are the weight coefficients of the
particular process, and $D_{part(i) \rightarrow h}$ are the
individual fragmentation functions (FFs) for turning parton $i$ into
hadron $h$. Similar to the parton distribution functions (PDFs) in
equation~\ref{eq:equation52}, derived from deep inelastic electron-parton scattering
(DIS) and shown in Fig.~\ref{fig:Figure14}, the FFs are semi-empirical
non-perturbative QCD functions that have an intuitive probabilistic
interpretation. They quantify the probability that the primordial
parton produced at short distance $1/Q$ fragments into $i$ shower
partons, forming a jet that includes the hadron $h$ \cite{205,206}.

At Fermilab energy, $\sqrt{s}=1.8 \: TeV$, the jet spectrum reaches
up to $E_T \approx 400 \: GeV$, and a typical 100 $GeV$ jet
comprises about 10 hadrons which can be identified above background
by jet-cone reconstruction algorithms \cite{205}. This allows for a
complete determination of the corresponding fragmentation function,
and for a rather accurate reconstruction of the $p_T$ and $E_T$ of
the primordial parton that initiated the jet. Similar conditions
will prevail in jet spectroscopy of Pb+Pb collisions at LHC energy,
$\sqrt{s}=5.5 \: TeV$.

However, at RHIC energy a typical jet at $15 \le E_T \le 25 \: GeV$
features a fragmentation function comprised of a few hadrons with
$E_T$ in the 2-15 $GeV$ range. Considering the high background,
arising in the lower fraction of this energy domain from concurrent,
unrelated high $p_T$ hadron production processes, a complete
jet-cone analysis can not succeed. The RHIC experiments thus
confront back-to-back di-jet production with an analysis of the
azimuthal correlation between high $p_T$ hadrons. Defining the
observational geometry by selecting a high $p_T$ ''trigger'' hadron
observed at azimuthal emission angle $\varphi_{trig}$, the
associated production of high $p_T$ hadrons is inspected as a
function of $\Delta \varphi=\varphi_{ass} - \varphi_{trig}$. If the
trigger has caught a leading jet hadron one expects the hadrons of
the balancing back-to-back jet to occur at the side opposite to the
trigger, $ \Delta \varphi \approx \pi$. The trigger condition thus
imposes the definition of a ''near-side'' and an ''away side''
azimuthal domain. Furthermore, the relatively narrow rapidity
acceptance of the STAR and PHENIX experiments (centered at $y=0$)
selects di-jets with axis perpendicular to the beam direction.

Originating from a uniform distribution of primordial back-to-back
di-parton production vertices, throughout the primordial reaction
volume, the trigger selected di-jet partons thus experience an
(anti-)correlated average path length $\left<L\right>$ to arrive at the surface
while experiencing medium-specific attenuation, with $\left<L_{trig}\right>
\approx 2 R - \left<L_{away}\right>$, $R$ being the transverse medium radius.
No such geometric constraint exists in the study of inclusive high
$p_T$ hadron production. We thus expect information different from
the inclusive $R_{AA}(p_T)$ signal. The geometrical selectivity can
be even further constrained by fixing the direction of the impact
parameter (i.e. the reaction plane) in semi-central collisions
(recall chapter~\ref{chap:Elliptic_flow}), and observing the di-jet correlation signal in
dependence of the di-jet axis orientation relative to the reaction
plane.

\begin{figure}[h!]   
\begin{center}
\includegraphics[scale=0.32]{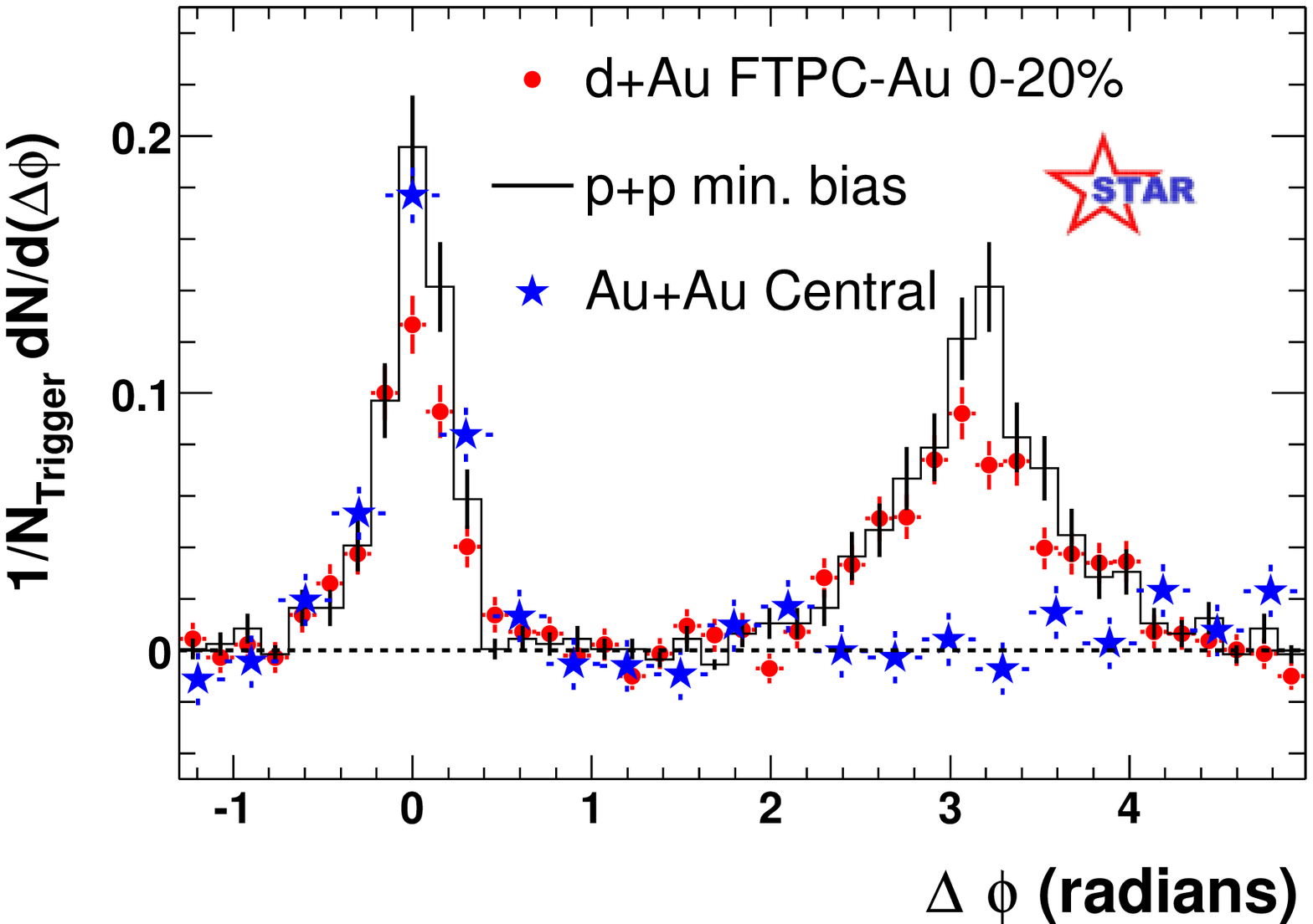}
\includegraphics[scale=0.32]{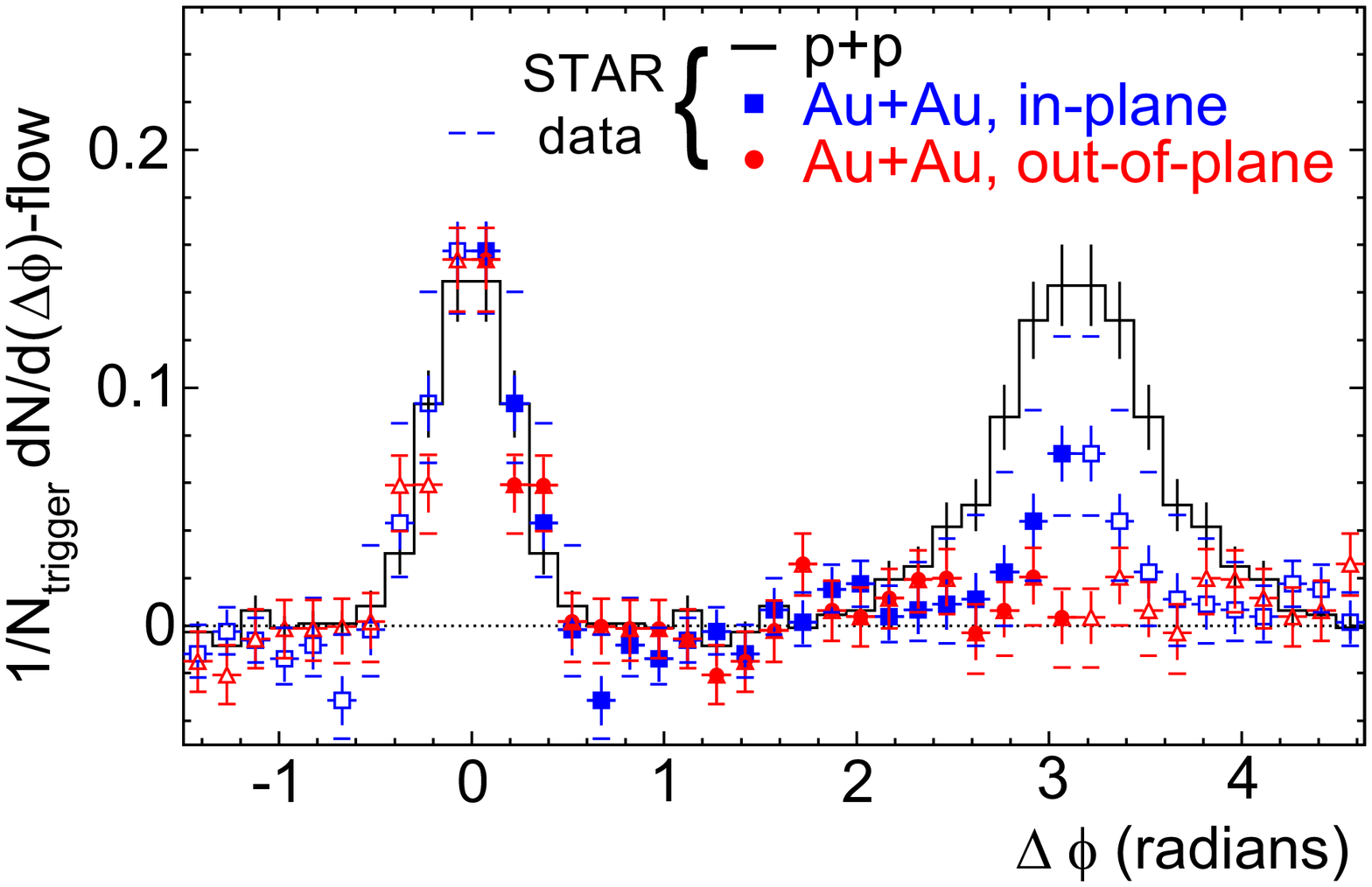}\vspace{-0.4cm}
\caption{Di-hadron correlation from back-to-back di-jet production in Au+Au
collisions at $\sqrt{s}=200 \: GeV$. The trigger particle is at
azimuth $\Phi=0$, with a $p_T > 4 \: GeV$ threshold. The away side
peak at $\Delta \Phi=\pi$ is observed (left panel) in p+p, d+A but
absent in central Au+Au. Right panel shows the correlation in Au+Au
for different orientations of the trigger  direction relative to the
reaction plane \cite{207}.}
\label{fig:Figure50}
\end{center}
\end{figure}

The very first di-hadron correlation measurements confirmed the
existence of strong in-medium attenuation. Fig.~\ref{fig:Figure50} shows the
azimuthal yield distributions, per trigger hadron, as observed by
STAR at $\sqrt{s}= 200 \: GeV$ \cite{207}. The left panel shows the
distribution of hadrons with $p_T \ge 2 \: GeV/c$ relative to a
trigger hadron with $p_T^{trig} \ge 4 \: GeV/c$. Data for p+p, d+Au
and central Au+Au are illustrated. At the ''near side'' (the trigger
position is $\Phi=0$) all three reactions exhibit a similar narrow
distribution of hadrons associated with the trigger hadron, typical
of a jet cone fragmentation mechanism. Note that the associated
near-side central Au+Au signal thus exhibits no signs of an
attenuation softened fragmentation function, indicating that the
trigger imposed high $p_T$ hadron should predominantly stem from
primordial jet production vertex points located near to the surface
of the reaction volume, in azimuthal trigger direction. Thus,
conversely, the balancing opposite jet has to develop while
traversing almost the entire transverse diameter of the interaction
volume. I.e. $\left<L_{oppos}\right> \approx 2 \: R$ thus emphasizing the
expectation that di-jet spectroscopy should allow for stricter
constraints on path length $L$ in comparison to single high $p_T$
hadron $R_{AA}$ analysis. In fact, no trigger related away side
signal of $p_T > 2 \: GeV/c$ hadrons is observed in Fig.~\ref{fig:Figure50} for
central Au+Au collisions, whereas p+p and central d+Au collisions
exhibit a clear away-side di-jet signal. \\
\\
We conclude that the trigger bias, selecting a single near side hadron of $p_T \ge 4 \:
GeV/c$ in central Au+Au collisions, responds to a primordial di-jet
of about $10 \: GeV$ per back-to-back parton. After traversal of in
medium average path length $L \rightarrow 2 \: R$ the fragmentation
function of the opposite side parton contains on average no hadron
at $p_T > 2 \: GeV/c$, indicating that it should have lost a
fraction $\left<\Delta E_T\right> \ge 5 \: GeV$. The medium is thus highly
opaque, but the total disappearance of the opposite side signal can
only provide for a lower limit estimate of $\left<\Delta E_T\right>$, within
the trigger conditions employed here. We shall show below that the
situation changes with more recent RHIC data \cite{208} that extend
the trigger hadron $p_T$ range toward $20 \: GeV/c$.

However, the right hand panel of Fig.~\ref{fig:Figure50} shows that an improved
constraint on partonic in-medium path length can already be obtained
by studying the di-jet back-to-back production geometry in
correlation with the orientation of the reaction plane that arises
from non-zero impact parameter in semi-central A+A collisions. We
have seen in Fig.~\ref{fig:Figure34} that such collisions exhibit an elliptical
primordial transverse density profile, with minor axis along the
impact vector $\vec{b}$, defining the reaction plane. Di-jets with
axis ''in-plane'' thus traverse a shorter in-medium path length as
orthogonal ''out-of-plane'' jets, the difference in average path
length being quantified by the spatial excentricity $\epsilon (b)$,
equation~\ref{eq:equation48}. Fig.~\ref{fig:Figure50} (right) shows the di-hadron correlation results in
semi-peripheral Au+Au collisions, as compared to the in-vacuum p+p
reference. At the 20-60\% centrality window employed here, out of
plane jet emission occurs along the major axis, the reaction volume
diameter still of magnitude $2R$, as in central collisions. The
trigger condition thus again selects opposite side path lengths
$L\rightarrow 2 \: R$, but the energy density should be lower than
in central collisions. Even so, the average opacity along the
away side parton path appears to be high enough to wipe out the
correlation signal. In-plane geometry, however, shows a partially
attenuated signal (as compared to the global p+p reference) at the
opposite side, corresponding to path lengths $L \approx R$. These
data thus provide for first information concerning the relation of
$\left<\Delta E\right>$ and average traversed path length \cite{176}. \\
\begin{figure}[h!]   
\begin{center}
\includegraphics[scale=0.69]{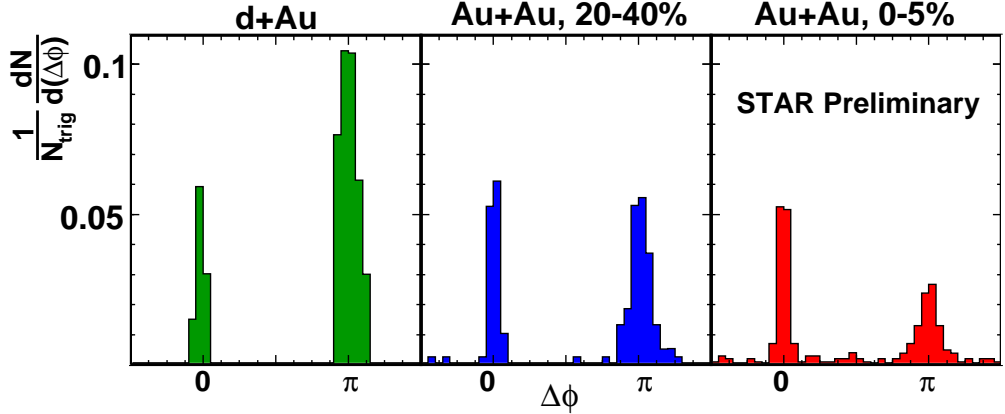}\vspace{-0.4cm}
\caption{Di-hadron correlation at high $p_T$ in central Au+Au collisions at
$\sqrt{s}=200 \: GeV$, compared to d+Au and peripheral Au+Au; for $8
\le p_T^{trig} \le 15 \: GeV$, and $p_T^{assoc} > 6 \: GeV$. From
\cite{208}.}
\label{fig:Figure51}
\end{center}
\end{figure}\\
Obviously, one wants to extend the above study to higher di-jet
energies, i.e. to measurement of di-hadron correlations at hadron
trigger $p_T \rightarrow 20 \: GeV/c$, conditions sampling the very
low primordial cross section, at $\sqrt{s}=200 \: GeV$, of jet
production at primordial $E_T \rightarrow 30 \: GeV$ \cite{208,209}.
Fig.~\ref{fig:Figure51} shows the corresponding jet correlations
selected with high $p_T$ trigger, $8 < p_T^{trig} < 15 \: GeV/c$,
and high $p_T$ associated hadrons, $p_T > 6 \: GeV/c$, in minimum
bias d+Au, semi-central Au+Au and central Au+Au at $\sqrt{s}=200 \:
GeV$. Very clear and almost background-free back-to-back jets are
observed {\it in all three cases}, in sharp contrast with the away
side jet total disappearance in central Au+Au at lower jet energy,
Fig.~\ref{fig:Figure50} (left panel). The near side trigger associated hadron yield
decreases only slightly from d+A to central Au+Au, while the away
side yield drops down by an attenuation factor of about 0.2, the
signal thus {\it not being completely extinguished}. We infer $R_{AA} (L \rightarrow 2R) \approx 0.2$.

In order to show how such data can be evaluated in a picture of
in-medium leading parton attenuation (as finally reflected in
leading hadron production observed in the above di-hadron
correlation data) we briefly consult the pQCD factorization
\cite{210} prediction for the inclusive production of a high $p_T$
hadron at central rapidity, in the nuclear collision $A+B
\rightarrow h+x$ \cite{196},
\begin{eqnarray}
\frac{d^3 \sigma_{AB \rightarrow hx}}{d^2 p_T dy} &=& K_{NLO}
\sum_{abc} \int d \vec{r} dx_a dx_b dz_c \: F_{a/A} (x_a, Q^2,
\vec{r}) \nonumber\\
& & F_{b/B} (x_b, Q^2, \vec{b} - \vec{r}) \frac{d^3 \sigma_
{ab-c}}{d^2 p_{T (c)} dy_c} (x_a,x_b, Q^2)\nonumber\\
& &  \frac{1}{z^2_c} \: D_{h/c} (z_c, Q^2)
\label{eq:equation67}
\end{eqnarray}
where the parton ($a,b$) distribution functions $F$ in nucleus $A,
B$ and the elementary pQCD cross section for $a+b \rightarrow c+x$
have been already implied in equation~\ref{eq:equation52}. Their spatial integral gets
convoluted with the fragmentation function $D$ that describes the
conversion of the leading parton $c$ to a hadron carrying a fraction
$0 < z_c < 1$ of its transverse momentum. $K$ is a factor introduced
as a phenomenological correction for ''next to leading order'' (NLO)
QCD effects. Within the (further) approximation that the leading
parton $c$ suffers medium induced gluon bremsstrahlung energy loss
but  hadronizes outside the interaction volume (in vacuum), the
in-medium quenching leads, merely, to a re-scaling of the
fragmentation function,
\begin{equation}
D_{h/c}^{med} \: =\: \int d \epsilon \: P (\epsilon)
\frac{1}{1-\epsilon} \: D^{vac}_{h/c} \: \left(\frac{z_c}{1-
\epsilon}, Q^2\right),
\label{eq:equation68}
\end{equation}
where the primary parton is implied to lose an energy fraction
$\epsilon= \Delta E / E_c$ with probability $P(\epsilon)$
\cite{196}. Therefore the leading hadron is a fragment of a parton
with reduced energy $(1- \epsilon)E_c$, and accordingly must carry a
larger fraction of the parton energy, $z_c/(1- \epsilon)$. If no
final state quenching is considered, $P(\epsilon)$ reduces to
$\delta (\epsilon)$. The entire effect of medium attenuation on the
leading parton is thus contained in the shift of the fragmentation
function.

\begin{figure}[h!]   
\begin{center}
\includegraphics[scale=0.5]{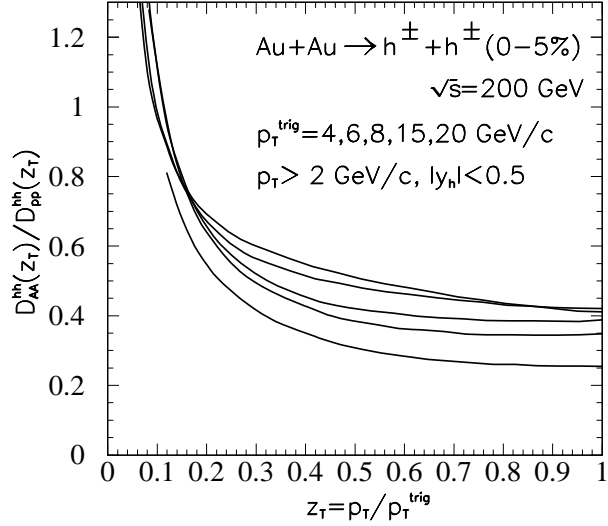}\vspace{-0.4cm}
\caption{The ratio of the hadron-triggered fragmentation function, equation~\ref{eq:equation68},
in central Au+Au and in p+p collisions, for different values of
$p_T^{trig}$ \cite{196}.}
\label{fig:Figure52}
\end{center}
\end{figure}

An application of this formalism \cite{196} is shown in Fig.~\ref{fig:Figure52}. The
in-medium modification of the hadron-triggered fragmentation
function (see equation~\ref{eq:equation68}) is evaluated for central Au+Au collisions at
$\sqrt{s}=200 \: GeV$. In adaptation to the modalities of RHIC
di-hadron correlation data, the opposite side fragmentation function
(for observation of trigger-related hadrons with $p_T > 2 \: GeV/c$)
is studied in dependence of the trigger selected $p_T$ window. Its
attenuation is quantified by the ratio $D$ (with quenching) to $D$
(without quenching), as a function of the fraction $z_T$, of
opposite side hadron $p_T$ to trigger hadron $p_T$. Referring to the
observational conditions implied in Fig.~\ref{fig:Figure50} (left) $p_T^{trig}
\approx 4-6 \: GeV/c$ and opposite side $p_T > 2 \: GeV$, the
predicted suppression amounts to a factor of about 0.35, whereas the data 
imply a factor smaller than 0.2. Likewise, comparing to the data of Fig.~\ref{fig:Figure51}, the (harder) trigger conditions 
should lead to a  suppression of about 0.45 from Fig.~\ref{fig:Figure52} but the observed value is close to 0.2. 
The predictions appear to fall short of the actually observed
suppression. This calculation employs an ansatz for the transport
coefficient $\hat{q}$ similar to equation~\ref{eq:equation65}, recurring to the
primordial gluon density at $\tau_0 \le 0.6 \: fm/c$ which, in turn,
is estimated by the charged hadron mid-rapidity density \cite{197}.
However, this pQCD based argument can also not reproduce the magnitude of
the effective transport coefficient (equation~\ref{eq:equation63}), $\hat{q}_{eff}
\approx 10-15 \: GeV^2/fm$, shown in refs. \cite{194,195,199} to be
{\it required} by the large observed suppression (see Fig.~\ref{fig:Figure43}). \\
\\
It has been argued \cite{199,211} that these models need refinement
by introducing more realistic dynamics, and/or by completely
abandoning the pQCD ansatz for hard parton in-medium transport
\cite{212,213}. We shall return to these novel suggestions, of how
to treat dynamical, non equilibrium quantities of the ''strongly
coupled'' parton plasma of non perturbative QCD (toward which
equilibrium lattice theory can only give hints), in our final
conclusion (chapter~\ref{chap:Conclusion}). In the meanwhile, we note that the expected
non abelian behaviour, $\Delta E \propto L^2$, could not be verified
quantitatively, as of yet \cite{211}, because of the unexpectedly
high opacity of the fireball interior sections at $\sqrt{s}=200 \:
GeV$, in combination with the limited jet energy range that is
available at RHIC energy. It appears possible, however, to extend
the analysis of the ''back side jet re-appearance'' data
\cite{208,209} (Fig.~\ref{fig:Figure51})  toward this goal. The situation should
improve at LHC energy, $\sqrt{s}=5.5 \: TeV$, where primordial
100-200 $GeV$ jets are abundant, such that the opposite side jet can
be reconstructed with explicit use of the complete Fermilab jet cone
recognition algorithms \cite{205} even if their in-medium $E_T$ loss
ranges up to 50 $GeV$.

\begin{figure}[h!]   
\begin{center}
\includegraphics[scale=0.33]{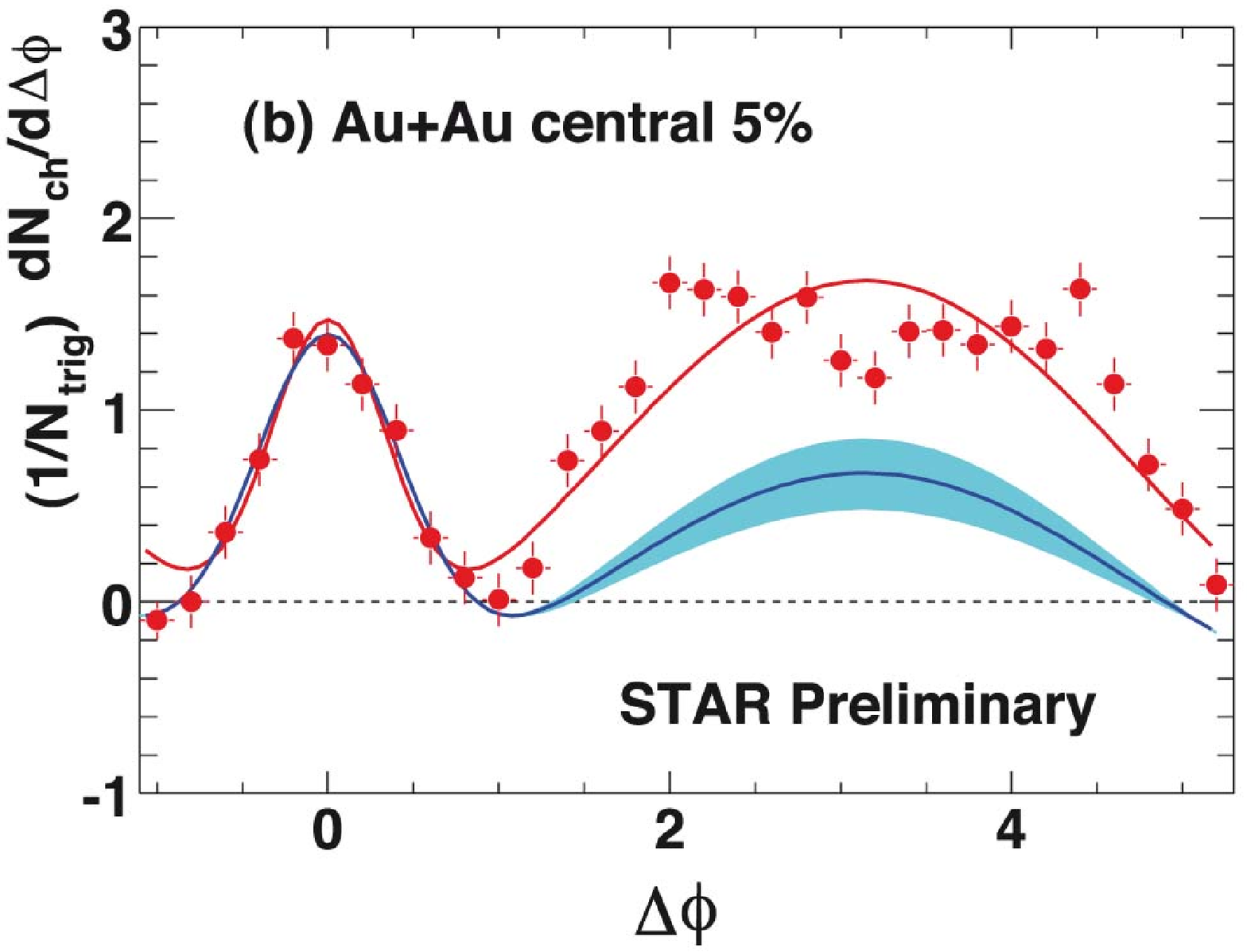}
\includegraphics[scale=0.33]{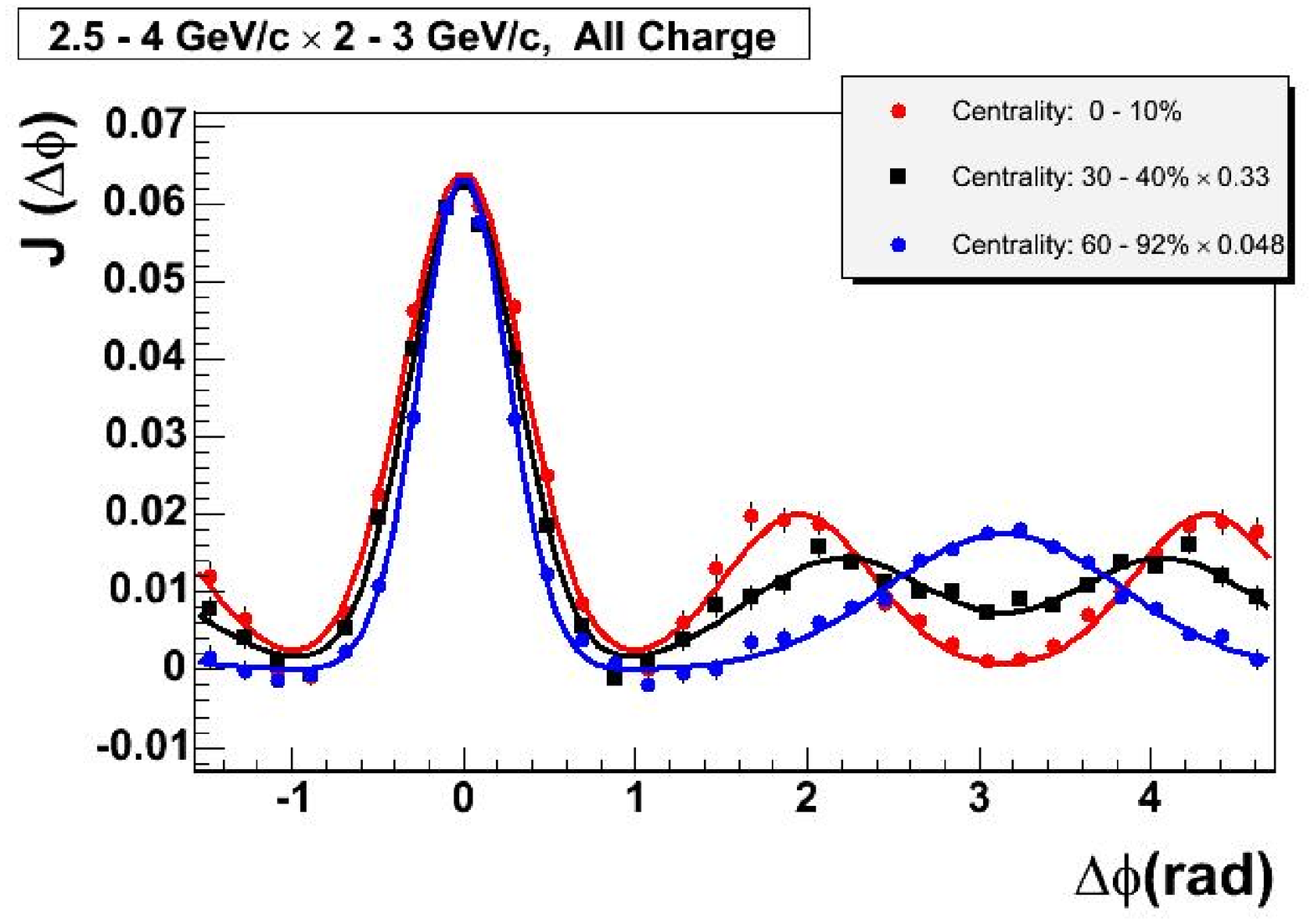}\vspace{-0.4cm}
\caption{Di-hadron correlation: away side emission pattern in central Au+Au
collisions, compared to pp data by STAR \cite{214} (left panel) and
to peripheral Au+Au (right panel) by PHENIX \cite{215}.}
\label{fig:Figure53}
\end{center}
\end{figure}

A further prediction of the model employed in Fig.~\ref{fig:Figure52} has been confirmed by the RHIC experiments. The high medium opacity at top RHIC energy leads to an
intriguing emission pattern of low $p_T$ opposite side hadrons.
Clearly, the trigger-selected $E_T$ flux, of up to 20 $GeV$, toward
the away-side, can not remain unnoticeable. Inspection of the
attenuated fragmentation functions \cite{196} in Fig.~\ref{fig:Figure52} reveals an
{\it enhanced} emission of bremsstrahlung gluon hadronization
products at $z_T \le 0.1$ This fraction of in-medium jet-degradation
products has in fact been observed, as is shown in Fig.~\ref{fig:Figure53}. The left
panel shows STAR results \cite{214} for the di-hadron correlation in
central Au+Au at $\sqrt{s}=200 \: GeV$, with near-side hadron
trigger $4 < p_T < 6 \: GeV/c$, and opposite side observation
extended to soft hadrons, $0.15 < p_T < 4 \: GeV/c$. A prominent
double-peak structure is indicated, symmetric about $\Delta \Phi=\pi$.
The right panel shows high resolution PHENIX results \cite{215} for
Au+Au at three centralities, from peripheral to central collisions.
For the former, the typical p+p-like away side peak (c.f. Fig.~\ref{fig:Figure47}) is
recovered, while a double peak appears in the latter case, shifted
from $\Delta \Phi = \pi$ by $\pm \delta \Phi \approx 70^0$. A
hypothetical mechanism comes to mind \cite{213}, of sideward matter
acceleration in a ''Mach-cone'' directed mechanism of compressional
shock waves initiated by the in-medium energy loss of the opposite
side leading jet parton, which traverses the medium at
''super-sonic'' velocity, i.e. at $v > v_s$, the appropriate speed
of sound in a parton plasma.\\
\\
If confirmed by pending studies of the away-side multi-hadron
correlation, that might ascertain the implied conical shape of the
soft hadron emission pattern (about the direction $\Delta \Phi
\approx \pi$ of the leading parton), this mechanism might lead to
the {\it determination of the sound (or shock wave) velocity} of a
strongly coupled parton plasma: a third characteristic QGP matter
property, in addition to viscosity $\eta$ (from elliptic flow) and
$\hat{q}$ (from high $p_T$ parton attenuation). We note that the
implied concept, of measuring the shock wave transport velocity of a
strongly interacting medium, dates back to the 1959 idea of
Glassgold, Heckrotte and Watson \cite{216}, to study shock wave
emission in central p+A collisions at AGS and PS proton energy, of
about 30 $GeV$. In hindsight we can understand the inconclusiveness
of such searches: the ''low energy'' incident proton does not
preserve its identity, breaking up into relatively soft partons in
the course of traversing the target nucleus. Their coupling to the
surrounding cold baryonic target matter is weak, and possible
collective transverse compressional waves are swiftly dissipated by
the cold, highly viscous \cite{61} hadronic medium. The shock
compression and Mach-cone emission mechanism was then revived by
Greiner and collaborators \cite{217} for central $^{20}Ne + U$
collisions at Bevalac energy \cite{218}. The recent RHIC
observations in Fig.~\ref{fig:Figure53} have been called {\it ''conical flow''}
\cite{219}; they demonstrate, again, the strong coupling to the
medium, of even a $20 \: GeV/c$ leading parton \cite{220}.

%% file: Chapter_5.tex
\chapter{Vector meson and direct photon production: penetrating probes}
\label{chap:Vector_Mesons}

This chapter is devoted to three observables that have been profoundly studied already at the SPS, 
as tracers of the created fireball medium:
\begin{enumerate}
\item $J/\Psi$ ''charmonium'' production, as suppressed in a high
$T$ QGP medium,
\item ''direct'' photons as black body $T$ sensors of the QGP state;
\item $\rho$ meson in medium production, studied by di-lepton
decay.
\end{enumerate}

These three observables represent an internally connected set of
ideas of high density QCD matter diagnostics: all serve as medium
tracers but in a complementary way. Matsui and Satz \cite{41} realized
that at the modest top SPS energy, $17.3 < \sqrt{s} < 20 \: GeV$,
the production rate of $c \overline{c}$ pairs (that would in part
develop toward charmonium production, $J/\Psi, \Psi^`$, etc.) was so
low that only the primordial, {\it first generation} nucleon-nucleon
collisions in an A+A event would create a measurable yield (nothing
after thermalization). I.e. the primordial $c \overline{c}$ yield
would be well-estimated by $A^{4/3} \ \sigma_{pp} (c
\overline{c})$. Initially produced in a superposition of states
including color triplets and octets \cite{221} the emerging $c
\overline{c}$ pairs thus co-travel with the developing high energy
density fireball medium, as tracers, on their way toward $J/\Psi$ or
$\Psi^`$ and D, $\bar{D}$ formation. Attenuation, characteristic of medium
properties, will break up leading $c \overline{c}$ pairs resulting
in a suppression of the eventually observed $J/\Psi$ yield: another
''quenching'' observable.

As the suppression of $J/\psi$ is related to medium temperature and
density, the extent of charmonium quenching should also be related to the
rate of black body thermal fireball radiation, by photon emission
via elementary $q \overline{q} \rightarrow g \gamma$ and $qg
\rightarrow q \gamma$ processes in the plasma \cite{222,223}. Photons leave the
interaction volume essentially un-rescattered, and their radiative
intensity, proportional to $T^4$, makes them an ideal probe of the
initial fireball temperature. {\it This thus could be an ideal diagnostics of the early deconfined matter}, but it is
difficult to disentangle from a multitude of concurrent low $p_T$ photon
sources \cite{224}, most prominently $\pi^0 \rightarrow 2
\gamma$ decay, with cross sections higher by several orders of
magnitude. The thermal photon signal thus becomes more promising the
higher the initial temperature $T_i$ which might reach up to 500-600
$MeV$ at LHC energy.

Similar to photons, in medium created lepton pairs \cite{225} escape
essentially un-attenuated as was shown for the Drell-Yan process, $q
\overline{q} \rightarrow L \overline{L}$ by CERN experiment
NA38/NA50 \cite{226}. Thermal di-lepton production in the mass
region $\le 1 \: GeV$ is largely mediated by light vector mesons.
Among these, the $\rho$ meson is of particular interest due to its
short lifetime (1.3 $fm/c$), making it an ideal tracer (via its
in-medium di-lepton decay) for the modification of hadrons composed
of {\it light quarks}, in the vicinity of $T=T_c$. This modification
signal is thus complementary to the {\it heavy quark} charmonium
$J/\Psi$ suppression (break up) effect that sets in at $T \ge 1.5-2
T_c$ (see below). Moreover, in addition to the deconfinement breakup 
mechanism by QCD plasma Debye screening of the color force potential acting 
on the $c \overline{c}$ pair \cite{41}, the QCD chiral symmetry restoration mechanism \cite{227} can be
studied via in-medium modification of the $\rho$ spectral
function as $T \rightarrow T_c$. Note that the in vacuum $\rho$ mass
and width properties are owed to non-perturbative QCD condensate
structures \cite{1,228} which spontaneously break the chiral
symmetry of the QCD Lagrangian, at $T \rightarrow 0$. These
properties should change, in the vicinity of $T_c$, and be 
reflected in modifications of the di-electron or di-myon decay spectra -
unlike the suppression effect on the $J/\Psi$ which simply dissolves
primordial $c \overline{c}$ pairs before they can hadronize, a
yes-no-effect whose onset with $\sqrt{s}$ or centrality serves as a
plasma thermometer, by observing $R_{AA}$ or $R_{CP} < 1$.

\section{Charmonium Suppression}
\label{sec:charmonium_supression}

Due to the high charm and bottom quark masses, the ''quarkonium''
states of $c \overline{c}$ and $b \overline{b}$ can be described in
non-relativistic potential theory \cite{229,230}, using
\begin{equation}
V(r) = \sigma r - \frac{\alpha}{r}
\label{eq:equation69}
\end{equation}
as the confining potential \cite{231}, with string tension
$\sigma=0.2 \: GeV^2$ and gauge coupling $\alpha = \pi/12$. We are
interested in the states $J/\Psi$ (3.097), $\chi_c$ (3.53) and
$\Psi^`$ (3.685) which are the 1S, 1P and 2S levels. The decay of
the latter two feeds into the $J/\Psi$, accounting for about 40\% of
its yield. The radii derived from equation~\ref{eq:equation69} are $0.25 \: fm$, $0.36 \:
fm$ and $0.45 \: fm$, respectively, well below hadron size at least
for the $J/\Psi$ and $\chi_c$ states.

With increasing temperature, $\sigma (T)$ decreases, and at
deconfinement $\sigma (T_c) = 0$. For $T \ge T_c$ we thus expect
\begin{equation}
V(r) = - \frac{\alpha}{r} \: exp \: [-r/r_D \: (T)]
\label{eq:equation70}
\end{equation}
where $r_D (T)$ is the QCD Debye screening radius. It was initially
estimated from a SU (2) gauge theory of thermal gluons \cite{41}, to
amount to about 0.2-0.3 $fm$ at $T/T_c = 1.5$. In this picture, the
screened potential (equation~\ref{eq:equation70}) can still give rise to bound $c \overline{c}$
states provided their radius is smaller than $r_D$. The pioneering
study of Matsui and Satz \cite{41} concluded that screening
in a QGP system would dissolve the $J/\Psi$, or its c$\bar{c}$ precursor, at $T \ge 1.3 \: T_c$
whereas the $\chi_c$ and $\Psi^`$ states would be suppressed already
directly above $T_c$.

The corresponding energy density for $J/\Psi$ suppression, employing
\begin{equation}
\epsilon/\epsilon_c \approx (T/T_c)^4 \approx 2.9
\label{eq:equation71}
\end{equation}
(obtained with $\epsilon_c \approx 1 \: GeV/fm^3$ from lattice QCD),
would thus amount to about $2.9 \: GeV/fm^3$. This motivated an
extensive experimental effort at the CERN SPS Pb beam at
$\sqrt{s}=17.3 \: GeV$. We have seen in chapter~\ref{sec:Bulk_Hadron_Prod} that the Bjorken
estimate \cite{45} of average transverse energy density reached in
central Pb+Pb collisions \cite{43,44} amounts to $\epsilon = (3.0
\pm 0.6) \: GeV/fm^3$, with higher $\epsilon$ to be expected in the
interior fireball sections: encouraging conditions.

However, the above approach assumes the validity of a two-body
potential treatment at {\it finite} $T$, near a conjectured critical
point of QCD. More recently the quarkonium spectrum was calculated
{\it directly} in finite temperature lattice QCD \cite{232}, with
the striking result that the $J/\Psi$ dissociation temperature in a
realistic non-perturbative treatment of the QCD plasma state moves
up to about $T = 2 \: T_c$, whereas $\chi_c$ and $\Psi^`$
dissociation is expected \cite{230} to occur at $T = (1.1-1.2)
T_c$.

In addition to high $T$ breakup of $c\bar{c}$ or $J/\Psi$, we have to take account of the
so-called ''normal suppression'' of charmonium yields, observed in
proton-nucleus collisions \cite{233}. This effect is due to a
re-scattering dissociation of the primordially produced,
pre-hadronic $c \overline{c}$ system upon traversal of (cold)
hadronic matter \cite{234}. It can be studied in p+A collisions where the data on $J/\Psi$ production relative to pp collisions can be described by the survival probability
\begin{equation}
S_{pA} \equiv \frac{\sigma_{pA}}{A \sigma_{pp}} = \int d^2 b \int dz
\rho_A (b,z)  exp \left\{-(A-1) \int^{\infty}_z
 dz^` \rho_A (b,z^`)
\sigma_{abs}\right\}
\label{eq:equation72}
\end{equation}

where $\sigma_{abs}$ is the effective cross section for the
''absorption'' (break-up) of the $c \overline{c}$ in cold nuclear matter, and
$\rho_A$ is the transverse nuclear density profile. The data
\cite{233} suggest $\sigma_{abs}=4.2 \: mb$. The generalization of
equation~\ref{eq:equation72} to the nucleus-nucleus case \cite{235} gives a good description
of the $J/\Psi$ suppression (relative to binary pp scaling) in S+U
and peripheral Pb+Pb collisions at top SPS energy \cite{226}. It has
thus become customary to quantify the $J/\Psi$
suppression in central A+A collisions by relating the observed yield, not directly to the scaled pp yield (thus obtaining $R_{AA}$), but to a hypothetical ``normal absorption" yield baseline, established by equation~\ref{eq:equation72}. All \emph{further} absorption is called ``anomalous suppression".

\begin{figure}[h!]   
\begin{center}
\includegraphics[scale=0.5]{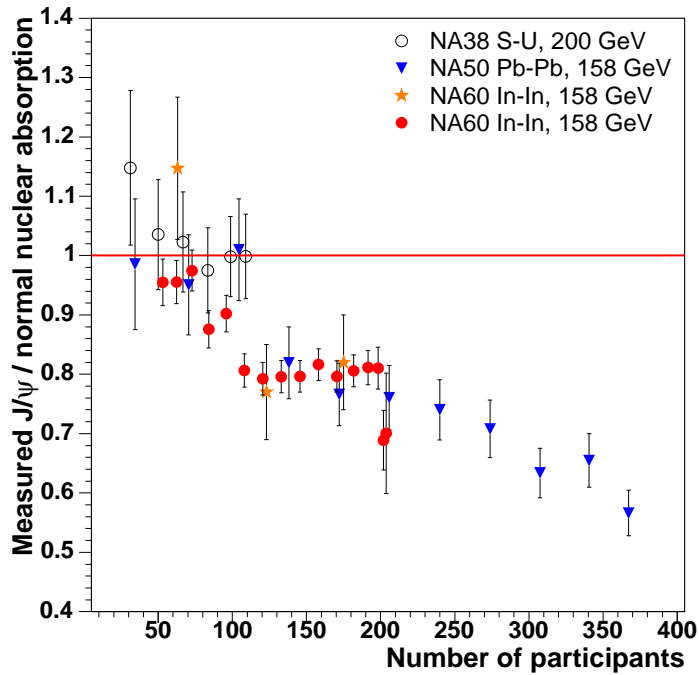}\vspace{-0.4cm}
\caption{$J/\Psi$ production measured in minimum bias collisions of S+U at
$\sqrt{s}=20 \: GeV$ and Pb+Pb and In+In at $\sqrt{s}=17.3 \: GeV$.
The yield is scaled by ''normal nuclear absorption'', equation~\ref{eq:equation72} \cite{236}.}
\label{fig:Figure54}
\end{center}
\end{figure}

\begin{figure}[h!]   
\begin{center}
\includegraphics[scale=0.5]{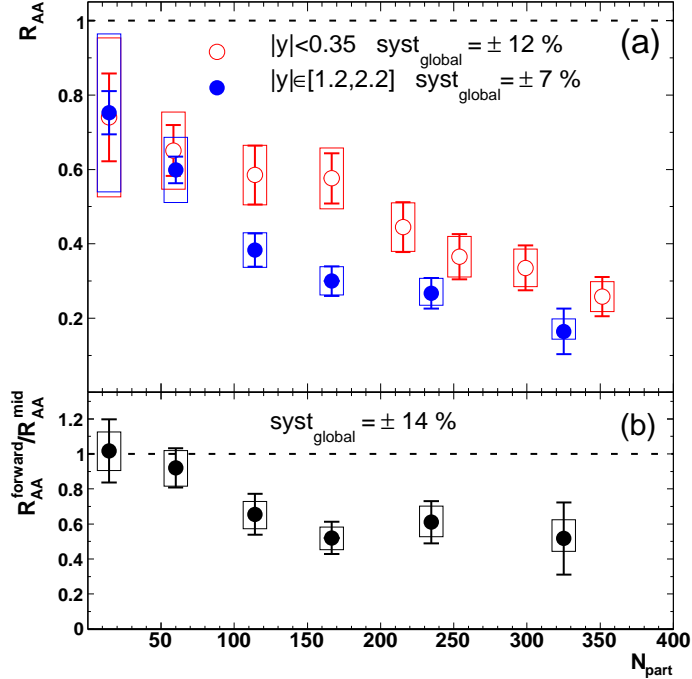}\vspace{-0.4cm}
\caption{$R_{AA}$ for $J/\Psi$ production in minimum bias Au+Au collisions at
$\sqrt{s}=200 \: GeV$, by PHENIX \cite{237} at mid-rapidity and at
1.2 $<y < 2.2$.}
\label{fig:Figure55}
\end{center}
\end{figure}

Fig.~\ref{fig:Figure54} shows the results gathered at $\sqrt{s} = 17.3 \: GeV$ by the
NA38 - NA50 - NA60 di-muon spectrometer \cite{236}, for minimum bias
S+U ($\sqrt{s}= 20 \: GeV$), In+In and Pb+Pb. Up to $N_{part}
\approx 100$ all yields gather at the ''normal absorption''
expectation from p+A scaling. A plateau at 0.8 follows for
intermediate $N_{part}$ values up to about 200 (which corresponds to
central In+In collisions, so the NA60 data end here), and a final
falloff toward 0.6 for central Pb+Pb collisions. It appears natural
to interpret the former step as an indication of $\Psi^`$
suppression, the final step as $\chi_c$ dissociation. No genuine $J/\Psi$ suppression is indicated. The
expectation from lattice calculations \cite{232} that $J/\Psi$
dissociation does not occur until $T \approx 2 \: T_c$ (and, thus,
$\epsilon \approx 16 \: \epsilon_c$) is thus compatible with Fig.~\ref{fig:Figure54}. We know from equation~\ref{eq:equation71} that
$T \le 1.3 \: T_c$ at top SPS energy. The data are thus compatible with no break-up of the $J/\Psi$ at the SPS, 
unlike at top RHIC energy where one expects $T
\approx 2 T_c$ \cite{61,96}.\\
\\
The RHIC data obtained by PHENIX \cite{237} are shown in Fig.~\ref{fig:Figure55}.
Minimum bias Au+Au collisions are covered at mid-rapidity, as well
as at $1.2 < y < 2.2$, and plotted vs. $N_{part}$. Due to a parallel
measurement of $J/\Psi$ production in p+p collisions \cite{238} the
PHENIX experiment is in the position to show $R_{AA}$, but this is done without
re-normalization to p-A absorption. $J/\Psi$ is suppressed in central
collisions, $R_{AA} \le 0.2$. Note that $R_{AA}$ {\it can not drop down} below about
0.15, due to unsuppressed surface contributions. The suppression is
thus stronger than at top SPS energy\footnote{Dropping the unfortunate distinction between normal and anomalous absorption one gets $R_{AA}=0.35$ for the central Pb+Pb collisions in Fig.~\ref{fig:Figure54}, almost a factor 2 above the RHIC value.} - in fact it is almost maximal.
We conclude that in central Au+Au at $\sqrt{s} = 200 \: GeV$ the
charmonium signal gets significantly quenched, in accord with the
inferences about the primordial temperature that we presented in
section~\ref{subsec:Transvers_phase_space} and~\ref{chap:Elliptic_flow}, to amount to about 300 $MeV$, i.e. $T/T_c \approx 2$ as implied for $J/\Psi$ dissociation by the lattice calculations
\cite{232}.

The above interpretation is still a matter of controversy
\cite{239}. 
It remains unclear whether successive, distinct stages of normal and anomalous $J/\Psi$ suppression are compatible with the dynamical evolution in central A+A collisions.
A further open question refers to the difference in
Fig.~\ref{fig:Figure55} of the $\left<y\right> = 0$ and $\left<y\right> = 1.7$ data at intermediate
centrality \cite{239} interpretation of which should be suspended
until the advent of higher statistics data.\\
\\
A different aspect of charmonium production in A+A collisions
requires attention: at the stage of universal hadronization
a certain fraction of bound $c \overline{c}$ mesons results, as the
outcome of the density of uncorrelated $c$ and $\overline{c}$ quarks
in the QGP medium as it enters hadronization \cite{240,241}. The
stage of statistical hadronization (recall chapter~\ref{chap:hadronization}) is omitted in
the charmonium suppression models \cite{41,230,234,235}, which
proceed in two steps (only): primordial nucleon-nucleon collisions
produce an initial input of $c \overline{c}$ pairs, proportional to
$\sigma_{c \overline{c}}^{NN} \times N^{AA}_{coll} \: (b)$. In
vacuum, the low relative momentum fraction of these $c \overline{c}$
pairs (about 1\%) would evolve into charmonium hadrons, after a
formation time of several $fm/c$. In the concurrently developing QGP
medium, however, the initial $c \overline{c}$ correlated pairs may
dissociate owing to the absence (due to color
screening \cite{41}) of vacuum-like bound states \cite{232}, at high
$T$. At $T \ge 2 \: T_c$ {\it all fireball charmonium production
would thus cease}, all $c \overline{c}$ pairs ending up in singly
charmed hadrons. {\it This picture \cite{230} is incomplete.}

Even if all $c \bar{c}$ pairs break up at RHIC during the early phase of
central collisions, the single charm
and anti-charm quarks flow along in the expanding medium of
equilibrated light and strange quarks. This picture is supported
\cite{94} by the observed elliptic flow of charm merging into the
universal trend. Charm can not, however, be chemically (flavor) equilibrated
at the low plasma temperature, where it is essentially neither newly
created nor annihilated \cite{241}. The initially produced $c$ and
$\overline{c}$ quarks (a few per central Au+Au collision at RHIC, a
few tens at LHC) finally undergo statistical hadronization at
$T=T_c$ along with all other light and strange quarks. To deal with
the non-equilibrium overall charm abundance an extra charm fugacity factor
$\gamma_c$ is introduced into the statistical model \cite{108}
calculation (for details see \cite{241}). $J/\Psi$ and $\Psi^`$ are
thus created in non-perturbative hadronization, with multiplicities proportional to
$\gamma^2_c$ and phase space weights, along with all other charmed
hadrons. This ''regeneration'' model also agrees with the RHIC data of
Fig.~\ref{fig:Figure55}, albeit within a large systematic uncertainty \cite{241}.

We note that the term regeneration is, in fact, {\it
misleading}. The statistical hadronization process does {\it not}
recover the initial, small fraction of correlated $c \overline{c}$
pairs that would end up in $J/\Psi$ in vacuum. It arises from the
{\it total density} of primordially produced $c$ and $\overline{c}$,
uncorrelated in the hadronizing fireball volume.

The statistical hadronization $J/\Psi$ production process, sketched
above, thus has the unfortunate property of providing a trivial
background charmonium yield, unrelated to the deconfinement signal
\cite{41} referring to the primordial $J/\Psi$ yield. 
Only about 1\% of the primordial $c \overline{c}$ yield
results in charmonia, in vacuum. The in-medium deconfinement process
breaking up the $c \overline{c}$ correlation on its way to
charmonia, thus constitutes a mere 1\% fraction of the total charmed
quark and anti-quark number. The regeneration process is insensitive
to this 1\% fraction, deconfined or not. At $T_c$, charm
hadronization reacts only to the total abundance of $c$ and
$\overline{c}$, as imprinted into the dynamical evolution by the
perturbative QCD $c \overline{c}$ production rate of initial
nucleon-nucleon collisions. At RHIC, it turns out \cite{241} that
the $c$ and $\overline{c}$ density is low, giving rise to
substantial canonical suppression (recalling equations~\ref{eq:equation39}-\ref{eq:equation43} in
chapter~\ref{chap:hadronization}) of the two charm quark charmonia, relative to $D$ mesons,
during hadronization. With a tenfold $c, \: \overline{c}$ density at
LHC, grand canonical charmonium production will set in, thus
probably overshooting the primordial yield reference,
$\sigma_{NN}^{J/\Psi} \times N_{coll}$. Thus we expect $R_{AA} > 1$
at the LHC. The role of a critical deconfinement ''thermometer'' is
lost for $J/\Psi$ at LHC, but the bottonium $Y$ states can take
over, being deconfined well above $T=300 \: MeV$ \cite{242}.

The RHIC result \cite{237} for $J/\Psi$ in central
Au+Au collisions (Fig.~\ref{fig:Figure55}), namely that $R_{AA} \rightarrow 0.2$,
represents the lucky coincidence that the initial temperature, $T
\approx 300 \: MeV$, is high enough to dissolve the correlated $c
\overline{c}$ charmonium precursor states, while the
$J/\Psi$ suppression is not yet overshadowed by the trivial
hadronization yield of $J/\Psi$.

\section{Direct Photons}
\label{sec:Direct_Photons}

\begin{figure}[h!]   
\begin{center}
\includegraphics[scale=0.4]{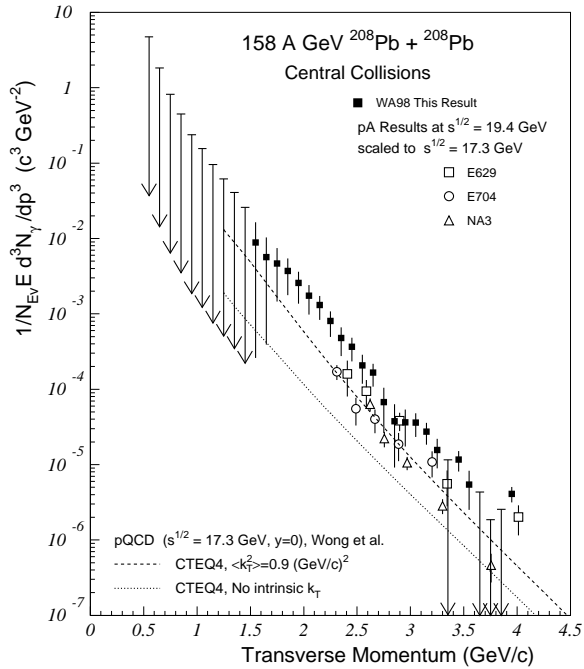}\vspace{-0.4cm}
\caption{The WA98 direct photon transverse momentum spectrum for central
Pb+Pb collisions at $\sqrt{s}=17.3 \: GeV$. Also indicated are
scaled $pA$ results above $2.0 \: GeV/c$ and pQCD estimates
\cite{243}. From \cite{224}.}
\label{fig:Figure56}
\end{center}
\end{figure}

Photons are produced during all stages of the dynamical evolution in
A+A collisions. About 98\% stem from final electromagnetic hadron
decays, not of interest in the present context, other then by noting
that their rate has to be painstakingly measured experimentally, in
order to obtain ''direct'' photon spectra at low $p_T$ by
subtracting the decay fraction from the total. This was done
\cite{224} by WA98 in central Pb+Pb collisions at top SPS energy; we
show their result in Fig.~\ref{fig:Figure56}. \\
\\
Only upper limits could be obtained at $p_T \le 1.5 \: GeV/c$ due to overwhelming 
background from $\pi^0$ and $\eta$ decay, thus also obscuring the major spectral domain in
which to look for a direct photon QCD plasma black body radiation
source, i.e. for thermal photons of a fireball at $T$ between $T_c$
and about 250 $MeV$ \cite{223}. Several data from p+p and p+A
collisions at nearby $\sqrt{s}$ and scaled up to central Pb+Pb are
included in Fig.~\ref{fig:Figure56}, at $p_T \ge 2 \: GeV/c$, and one sees the Pb+Pb
data in clear excess of such contributions from primordial
bremsstrahlung and hard, pQCD initial partonic collisions
\cite{243}. This excess ranges up to about $3.5 \: \: GeV/c$, in
Fig.~\ref{fig:Figure56}. Above, the hard pQCD initial collision yield dominates
\cite{244} over thermal production.

In contrast to all other primordial high $p_T$ pQCD yields (e.g.
$J/\Psi$, charm, jet leading partons) this photon yield is {\it not}
attenuated in the medium of A+A collisions. We have shown in Fig.~\ref{fig:Figure44} the
$R_{AA}$ for the PHENIX central Au+Au direct photon results
\cite{181} at RHIC $\sqrt{s}=200 \: GeV$, obtained in a background
substraction procedure \cite{245} similar to the one undertaken by
WA98. This procedure gives reliable data at $p_T > 4.0 \: GeV/c$, at
RHIC, and we observe $R_{AA}=1$. Hard initial photons are not
attenuated, and there is no sign of any other direct photon
contribution besides the primordial pQCD yield which, in fact, is
shown (by $R_{AA}=1$) to obey binary scaling. However, there is no
hint to plasma thermal radiation (except for a trend at the very
limit of statistical significance, at $p_T < 4.5 \: GeV/c$) in this
high $p_T$ window.

\begin{figure}[h!]   
\begin{center}
\includegraphics[scale=0.85]{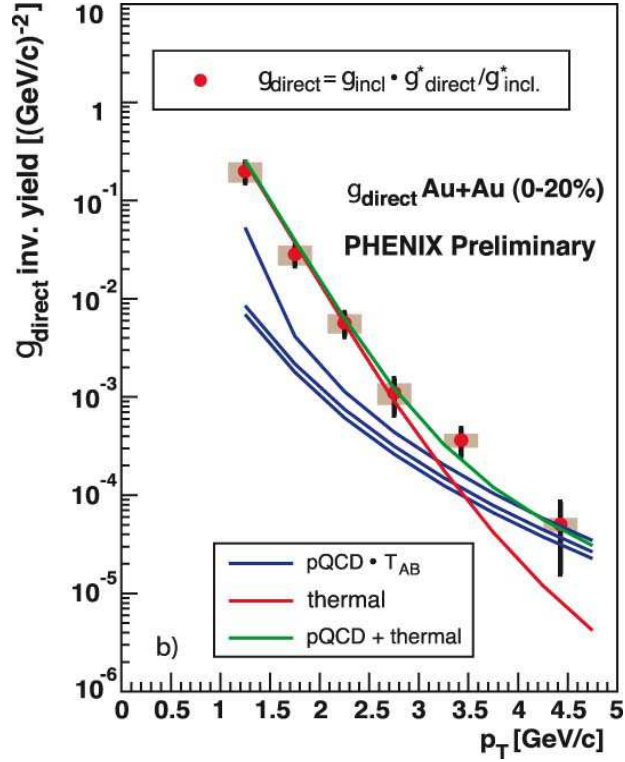}\vspace{-0.4cm}
\caption{Internal conversion measurement of direct photons in central Au+Au
collisions at $200 \: GeV$ \cite{246}. Predictions by pQCD
\cite{248} and thermal hydrodynamic \cite{249} models are included.}
\label{fig:Figure57}
\end{center}
\end{figure}

The WA98 SPS data, with thermal radiation enhancement indicated in
the interval $1.5 < p_T < 3.5 \: GeV/c$, thus remained as the sole
evidence until, more recently, the PHENIX experiment gained low
$p_T$ data \cite{246} exploiting the fact that any source of real
photons emits also virtual photons $\gamma^*$ leading to internal
conversion to an $e^+e^-$ pair (the Dalitz effect). To identify this
yield the invariant mass distribution of $e^+e^-$ pairs is analyzed
outside the phase space limits of $\pi^0$ Dalitz decay; the decay
pairs of all remaining hadron sources ($\eta, \Delta$) is subtracted
as a ''cocktail''. The remaining pair yield is then converted
assuming $\gamma^*_{dir}/\gamma^*_{inclusive}=
\gamma_{dir}/\gamma_{inclusive}$ (see ref. \cite{246} for detail),
thus finally obtaining data representative of $\gamma_{dir}$ in this
approach. Fig.~\ref{fig:Figure57} shows the corresponding $p_T$ distribution which
covers the interval $1.3 \le p_T \le 4.5 \: GeV/c$, within which the
conventional direct photon extraction method did not give
significant results \cite{181}.

The PHENIX experiment has also obtained direct photon spectra in p+p
and d+Au at $\sqrt{s}=200 \: GeV$ \cite{247} which are both well
accounted for \cite{246} by a next to leading order (NLO) pQCD
photon production model \cite{248}. These data were already employed
in deriving $R_{AA}=1$ for central Au+Au collisions, as shown in
Fig.~\ref{fig:Figure44} and referred to, above. The pQCD fits derived from p+p and
d+A are shown in Fig.~\ref{fig:Figure57} after binary scaling to Au+Au (pQCD $\times
T_{AA}$). They merge with the yield at $p_T \ge 4 \: GeV/c$ but
demonstrate a large excess yield below $3 \: GeV/c$. That excess is
well described by adding a thermal photon component resulting from
the hydrodynamic model of d'Enterria and Peressounko \cite{249}. It
traces the dynamical evolution during the early stages of
equilibrium attainment, in which the photon luminosity of the
emerging QGP matter phase is maximal. The hydrodynamic model
provides for the space-time evolution of the local photon emission
rate \cite{223} which includes hard thermal loop diagrams to all
orders, and Landau-Migdal- Pomeranchuk (LPM) in-medium interference
effects. Similar, in outline, to previous models that combined
algorithms of plasma photon radiation luminosity with hydrodynamic
expansion \cite{250}, the model \cite{249} fits the data in Fig.~\ref{fig:Figure57}.
It implies a picture in which the early stage of approach toward
thermal equilibrium at RHIC is governed by a symbolic, initial,
effective ''temperature'' of about 550 $MeV$ which, after
equilibration at $t \approx 0.6 \: fm/c$, {\it corresponds to $T
\approx 360 \: MeV$ in the primordial plasma} \cite{249}: close to
the consensus about initial $T$ as derived from $J/\Psi$
suppression, jet attenuation, elliptic flow and transverse energy
coupled with the 1-dimensional Bjorken expansion model.\\
\\
However, we note that the employed theoretical framework, hard
thermal loop (HTL) QCD perturbation theory of a weakly coupled
plasma state, as combined with hydrodynamics, has a tendency to call
for rather high initial $T$ values. This applies both to the above
analysis \cite{249} of RHIC data which is based on the thermal field
theory model of Arnold, Moore and Yaffe \cite{223}, and to previous
analysis \cite{250} of the WA98 SPS data of Fig.~\ref{fig:Figure56}. Direct photon
production is even more strongly biased toward the primordial, high
$T$ evolution than jet attenuation (that may be proportional to
$T^3$ \cite{212}). Thus, by implication of the model \cite{249} that
fits the low $p_T$ RHIC data of Fig.~\ref{fig:Figure57}, the yield is highly
sensitive to the (pre-equilibrium) formation period, $0.15 < t < 0.6
\: fm/c$, where the HTL model might not be fully applicable. This illustrates the present state of the art.
The model(s) based on perturbative QCD require extreme initial ''temperatures'' to produce the high photon 
yield, indicated by the RHIC experiment. The strongly coupled nature \cite{213} of the non perturbative
local equilibrium  QGP state at RHIC, $T \approx 300 \: MeV$, mayprovide for an alternative approach to
plasma photon production.

\section {Low mass dilepton spectra: vector mesons in-medium}
\label{sec:Low_mass_dilepton_spectra}

We have dealt with dilepton spectra throughout the above discussion
of $J/\Psi$ and direct photon production, as messenger processes
sensitive to the energy density prevailing (during and) at the end
of the primordial equilibration phase.
The third tracer observable, low mass vector meson dilepton decay
in-medium, samples - on the contrary - the conditions and modalities
of hadron deconfinement in the vicinity of $T=T_c$. SPS energy is
ideally suited for such studies as the QGP fireball is prepared near
$T_c$ whereas, at RHIC, it races through the $T_c$ domain with
developed expansion flow. The major relevant data thus stem from the
CERN SPS experiments NA45 \cite{251} and NA60 \cite{136}, which have
analyzed $e^+e^-$ and $\mu^+ \mu^-$ production at invariant mass
from 0.2 to 1.4 $GeV$.

\begin{figure}[h!]   
\begin{center}
\includegraphics[scale=0.34]{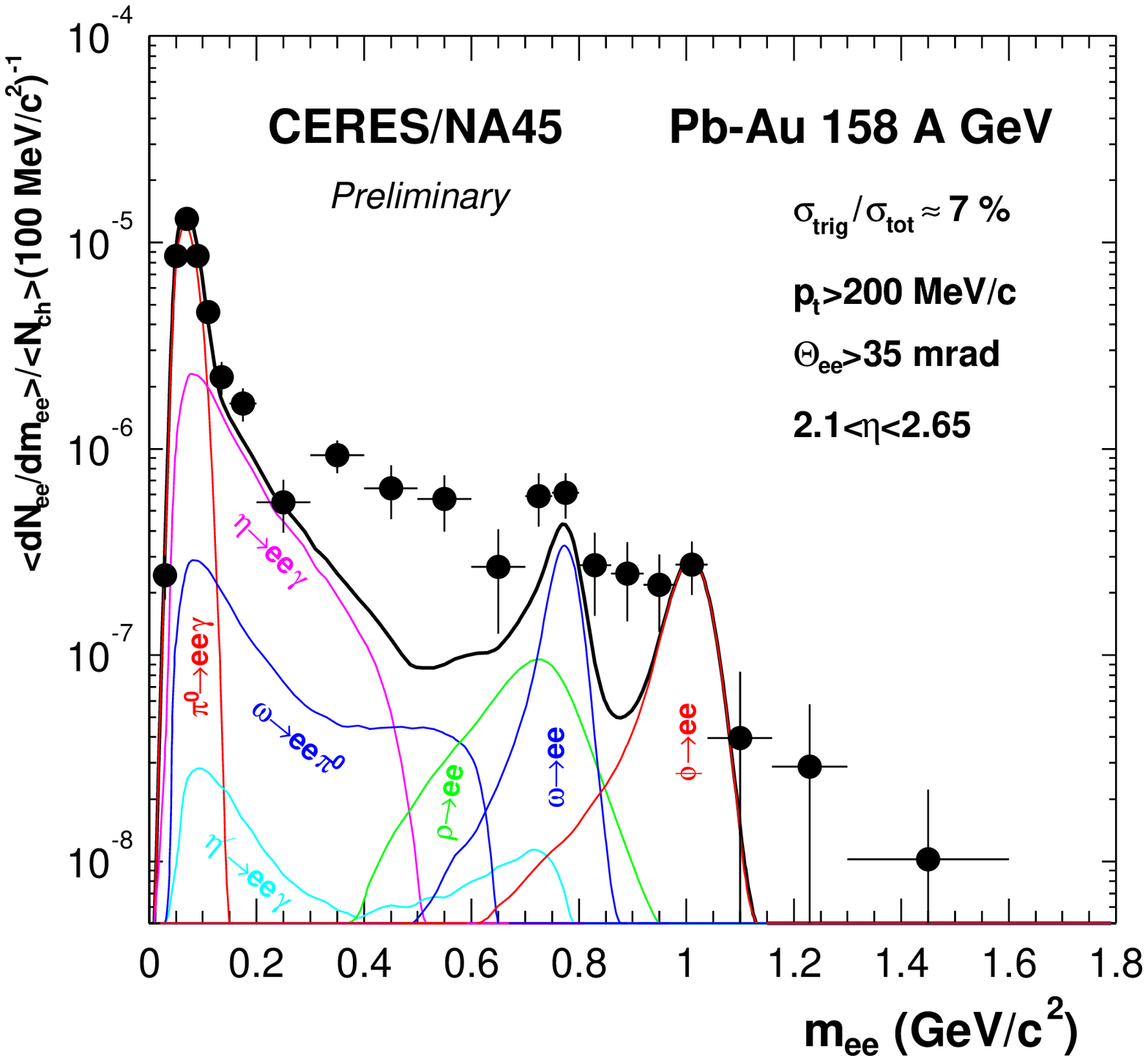}
\includegraphics[scale=0.34]{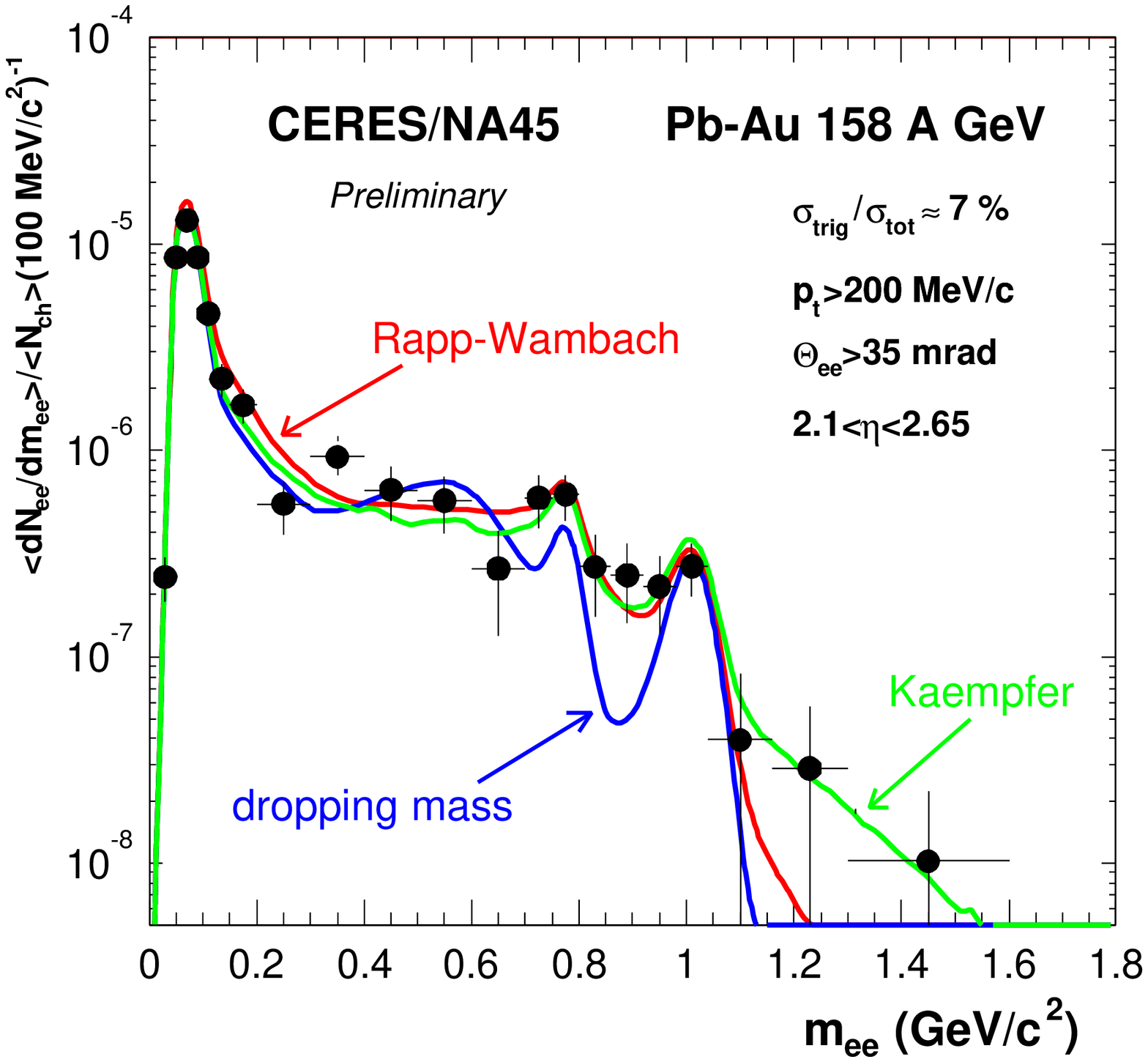}
\caption{Di-electron mass spectrum for central Pb+Au collisions at
$\sqrt{s}=17.3 \: GeV$ with the hadron decay cocktail (left) and
further in medium contributions (right) \cite{251}; see text for
detail.}
\label{fig:Figure58}
\end{center}
\end{figure}

Fig.~\ref{fig:Figure58} shows NA45 data \cite{251} for $e^+e^-$ production in central
Pb+Au collisions at $\sqrt{s}=17.3 \: GeV$. In searching for
modifications of $\varrho$ properties and of $\pi^+ \pi^-$ annihilation
via virtual intermediate $\varrho$ decay to $e^+e^-$, in the high
density environment near the hadron-parton coexistence line at
$T=T_c$, the various background sources have to be under firm
control. The Dalitz decays of $\pi^0, \eta$ and $\eta^`$ and the in
vacuo $e^+e^-$ decays of $\varrho, \omega$ and $\Phi$, which occur
after hadronic freeze-out to on-shell particles, form a hadronic
''cocktail'' (left panel in Fig.~\ref{fig:Figure58}) that is generated from yields
provided by the grand canonical statistical model \cite{108}. Within
detector resolution, $\pi, \omega$ and $\Phi$ leave distinct peaks
but the observed invariant mass distribution is not accounted for.

One needs also to account for background from Drell-Yan
lepton pair production and open charm decay, both scaling with
''number of collisions'' $A^{4/3}$. The latter contribution arises
from primordial $c \overline{c}$ charm production,
$\left<c\right>=\left<\overline{c}\right>$ which leads to synchronous formation of
$\left<D\right>=\left<\overline{D}\right>$ at hadronization; subsequent decays $D
\rightarrow$ lepton + X, $\overline{D} \rightarrow$ antilepton + Y
create $L \overline{L}$ pairs. This procedure is straight forward as
no significant medium attenuation occurs besides the statistical
charm redistribution conditions at hadronization (section~\ref{sec:charmonium_supression}),
governing $D, \overline{D}$ production \cite{241}.\\
\\
Onward to non-trivial backgrounds in the invariant mass plot of
Fig.~\ref{fig:Figure58}, we recall the presence of thermal lepton pairs from virtual
photon production in the early plasma phase, in parallel to real
''direct'' photon emission \cite{252}. The spectrum of such pairs
contains an average decay factor exp$\{- \frac{M_{ll}}{T}\}$, with
$T$ the (initial) plasma temperature. With $T \ge 220 \: MeV$
assumed for top SPS energy \cite{43,44}, this contribution is a
background candidate over the entire invariant mass interval covered
by the data. In general Drell-Yan, open charm decay and plasma
radiation contributions are smooth, partially closing the ''holes''
in the hadronic cocktail undershoot of the data. This smoothing
effect is helped, finally, by consideration of modifications
concerning the $\varrho$ meson spectral function near $T=T_c$, which
\cite{133,134,135} both affects the immediate $\varrho \rightarrow
e^+e^-$ decay invariant mass region (through the fraction of
in-medium $\varrho$ decays vs. the in vacuum decay fraction after
hadronic freeze-out) and, even more importantly, the contribution of
in-medium $\pi^+ \pi^-$ annihilation to dileptons. The latter
contribution accounts for the most obvious deviation between cocktail and data at
$0.3 \le m_{ll} \le 0.7 \: GeV$ in Fig.~\ref{fig:Figure58} (left panel).

The right hand panel of Fig.~\ref{fig:Figure58} shows the results of various
theoretical models which address the sources of the significant
dilepton excess over hadronic (in vacuum) and Drell-Yan cocktails,
labeled ''Rapp-Wambach'' \cite{133,135}, ''dropping mass''
(Brown-Rho \cite{134}) and ''Kaempfer'' \cite{252}. We shall return
to these models below but note, for now, that the extra yield in
central A+A collisions chiefly derives from $\pi^+ \pi^-$
annihilation via the (in medium modified) $\varrho$ resonance,
and from modification of the $\varrho$ peak itself.

\begin{figure}[h!]   
\begin{center}
\includegraphics[scale=0.24]{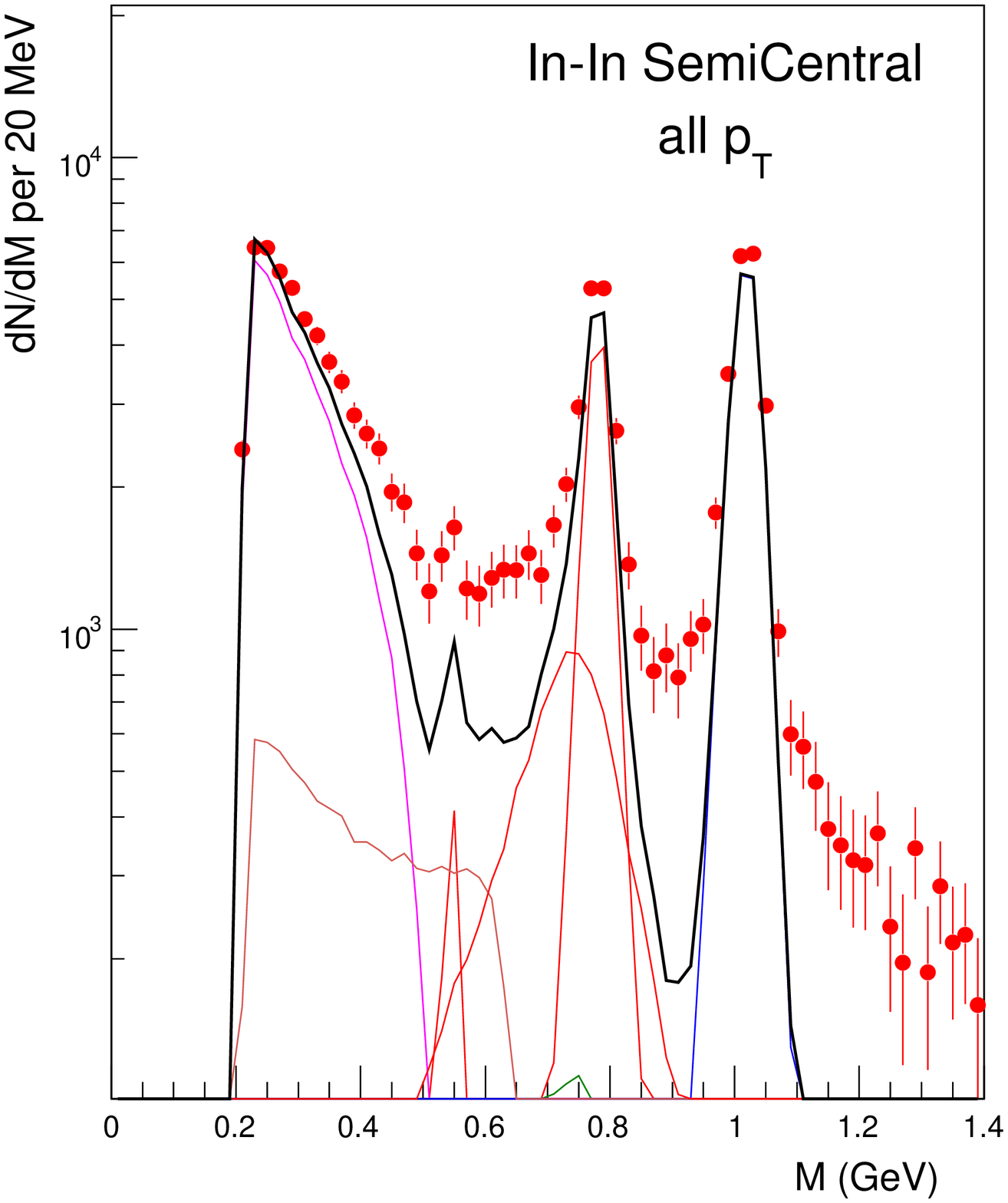}
\includegraphics[scale=0.36]{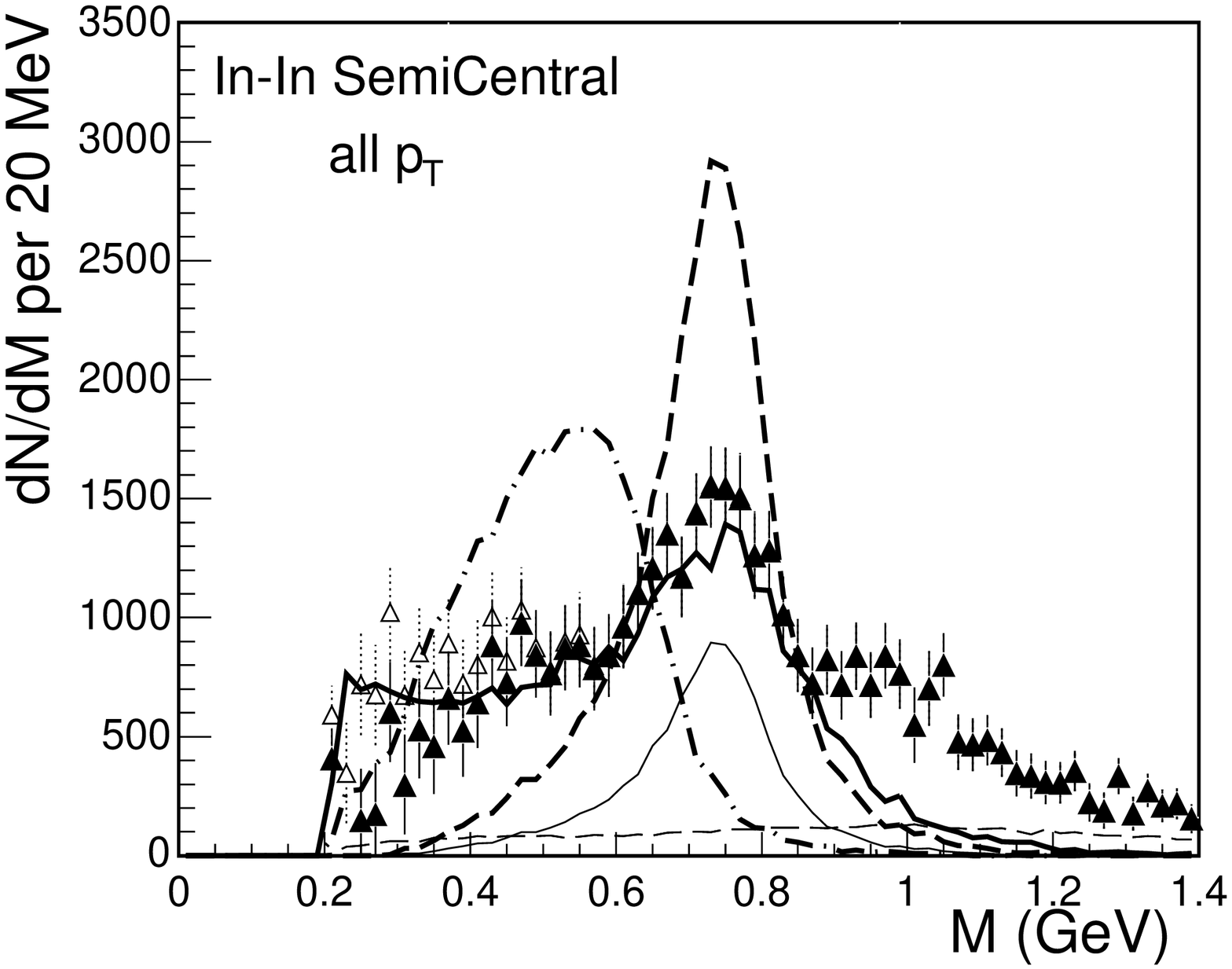}
\caption{(left) Di-muon invariant mass spectrum in semi-central In+In
collisions at $\sqrt{s}=17.3 \: GeV$, with hadron final state decay
cocktail; (right) excess mass spectrum after cocktail subtraction,
confronted with fits from the broadening \cite{133,135,255} and the
dropping mass \cite{228} models. From \cite{136}.}
\label{fig:Figure59}
\end{center}
\end{figure}

With improved statistics and background control, the A+A specific
extra dilepton yield below $M \approx 1.4 \: GeV/c^2$ can be
represented by itself, after cocktail subtraction. This has been
first accomplished by NA60 \cite{136,253} and, more recently, also
by NA45 \cite{254}. We show the former results in Fig.~\ref{fig:Figure59}. The left
panel shows the di-muon invariant mass spectrum in semi-central
Indium-Indium collisions at top SPS energy $\sqrt{s}=17.3 \: GeV$,
compared to the hadronic cocktail, qualitatively in agreement with
the left hand panel of Fig.~\ref{fig:Figure58} but with superior resolution and
statistics. The cocktail subtraction procedure (see \cite{136} for
details) leads to an invariant mass spectrum of di-muon excess in
In+In, shown in the right side panel of Fig.~\ref{fig:Figure59}: an experimental landmark
accomplishment. The $\varrho$ vacuum decay contribution to the
hadronic cocktail has been retained and is shown (thin solid line)
to be a small fraction of the excess mass spectrum, which exhibits a
broad distribution (that is almost structureless if the cocktail
$\varrho$ is also subtracted out), widening with collision
centrality \cite{136}. The best theoretical representation of the
excess yield again results (like in Fig.~\ref{fig:Figure58}, right panel) from the
broadening model \cite{133,135,255} where the $\varrho$ spectral
function is smeared due to various coupling mechanisms within the
medium via the vector dominance model, prior to hadronic freeze-out.
In the VDM $\varrho$ couples, both, to pion pair annihilation, and
to excited baryon states like the $N^*$ (1520), via their $N \pi
\pi$ decay branches.\\
\\
The above data still imply serious conceptual questions.
It is unclear to exactly which stage of the dynamical evolution (in
the vicinity of $T=T_c$) the excess dilepton yield should
correspond. As the observations appear to be coupled to in-medium
$\varrho$ meson ''metabolism'' we need to identify a period of
temporal extension above the (in vacuo) $\varrho$ half life (of 1.3
$fm/c$), safely 2 $fm/c$. This period should be located {\it in the
vicinity of hadro-chemical freeze-out}. From top SPS to RHIC energy, hadro-chemical freeze-out should
closely coincide with hadron formation, i.e. it should occur near
the parton-hadron coexistence line, at $T=T_c$. The microscopic
parton cascade model of reference \cite{85} implements the Webber \cite{121}
phenomenological, non-perturbative QCD hadronization model (recall
chapter~\ref{sec:Origin_of_hadro}) which proposes pre-hadronization clusters of color
neutralization (Fig.~\ref{fig:Figure31}) as the central hadronization step. In it, so
one might speculate, the transition from pQCD to non perturbative
QCD creates the chiral condensates $\left<q \overline{q}\right>$, spontaneously
breaking chiral symmetry \cite{256} (see below), and creating
hadronic mass. The overall process, from pQCD color neutralization
to on-shell hadrons, takes about 2.5 $fm/c$ \cite{85,86} at top SPS
energy. This could, thus, be the period of excess dilepton yield
creation. However, the relation of the models employed above
\cite{133,134,135,255,256} to this primordial spontaneous creation
of chiral condensates is still essentially unknown \cite{256}.

Thus, at present, the 1990's paradigm of a direct observation of the
chiral phase transition in QCD has been lost. The Brown-Rho model
\cite{134} predicted the $\varrho$ mass to drop to zero at $T=T_c$,
occuring as a certain power of the ratio $\left<q \overline{q}\right>^{med}/\left<q
\overline{q}\right>^{vac}$ of the chiral condensate in medium and in
vacuum which approaches zero at the chiral phase transition
temperature, then expected to coincidence with the deconfinement
temperature. This ''dropping mass'' model is ruled out by the data
in Fig.~\ref{fig:Figure58} and~\ref{fig:Figure59}. This is, perhaps, a further manifestation of the
fact that the deconfined QGP state at $T \ge T_c$ is not a simple
pQCD gas of quarks and gluons \cite{213}. In fact, lattice
calculations \cite{232,257} find indications of surviving light $q
\overline{q}$ pair correlations in the vector channel at $T \ge
T_c$. Thus the two most prominent symmetries of the QCD Lagrangian,
non abelian gauge invariance (related to confinement) and chiral
invariance (related to mass) might exhibit {\it different} critical
patterns at $T=T_c$ and low baryo-chemical potential. This
conjecture is best illustrated by the observation that the broad,
{\it structureless} NA60 excess dilepton spectrum of Fig.~\ref{fig:Figure59} (after
cocktail $\varrho$ subtraction) is equally well reproduced by a $T
\approx 160-170 \: MeV$ calculation in hadronic (equilibrium) matter
\cite{133,253,254}, and by a thermal QGP fireball of $q
\overline{q}$ annihilation at this average temperature \cite{252},
as illustrated here by the model curve labeled ''Kaempfer'' in
Fig.~\ref{fig:Figure58} (right panel). This observation has invited the concept of
{\it parton-hadron duality} near $T_c$ \cite{258}, which might be
provocatively translated as ''the QCD chiral transition properties
can not be unambiguously disentangled from the deconfinement transition effects
at $T_{c}$'' \cite{256}.

We may be looking at the wrong domain of
$[T, \mu_B]$ space: too high $\sqrt{s}$ and, thus, too high $T$, too
low $\mu_B$. Already at top SPS energy the medium is dominated by
the deconfinement, not by the chiral QCD transition. After
hadronization the medium is still at $T$ close to $T_c$ but the
density drops off within a few $fm/c$, not allowing for an equilibrium mean field state of
chirally restored hadrons. It is thus perhaps not
surprising that the data are seen to be dominated by simple
broadening and lack of structure: perhaps that is all that happens
to hadrons at $T_c$.

The chiral restoration transition should thus be studied at
higher $\mu_B$ and lower $T$ such that the dynamics achieves high
baryon densities but still merely touches the critical
(deconfinement) temperature. In fact, at $\mu_B \rightarrow 1 \:
GeV$ and $T<100 \: MeV$ the {\it chiral transition} should be of
first order \cite{259}. Here, in fact, the chiral condensate mass
plays the role of the order parameter (in analogy to the
magnetization in a spin system), which approaches zero as $T
\rightarrow T_c$. We might thus speculate that, unlike at top SPS to
LHC energy (where deconfinement dominates), the chiral QCD first
order phase transition will dominate the phenomena occuring near the
hadron-parton borderline, in the vicinity of $\sqrt{s}=4-6 \: GeV$
\cite{142}. This requires a new experimental program, with low
energy running at the SPS \cite{260}, at RHIC \cite{261} and at the
GSI FAIR project \cite{262}.

%% file: Chapter_6.tex
\chapter{Fluctuation and correlation signals}
\label{chap:Fluctuations}

Fluctuation and correlation signals in A+A collisions can be
evaluated in single events, due to the high multiplicity of produced
particles.  Depending on the physics context we may be interested to
see either a small, or a large nonstatistical fluctuation effect.
For example in view of the universality of hadronization (chapter~\ref{chap:hadronization})
we would have difficulty with an event by event pion to baryon ratio
(essentially $\mu_B^{-1}$) fluctuating by, say, 50\%. Conversely,
searching for critical fluctuations in the vicinity of a predicted
critical point of QCD \cite{146,147} we would be frustrated if
event-wise $\left<p_T\right>$, $\frac{dN}{dy}$ (low $p_T$ pion) \cite{263} or
strange to non-strange ratios like $K/\pi$ \cite{148,264} would not
exhibit any significant fluctuation beyond statistics. Overall,
event by event fluctuation observables also carry a message
concerning the robustness of our assumptions about equilibrium
attainment. It turns out that equilibrium properties are not,
merely, central limit consequences of ensemble averaging. Rather to the
contrary, each $A \approx 200$ central collision event at $\sqrt{s}
\ge 10 \: GeV$ appears to offer, to the dynamical evolution of bulk
properties, a sufficiently complete macroscopic limit. Such that we
can view the event-wise bulk dynamics as ''self analyzing''.\\
\begin{figure}[h!]   
\begin{center}
\includegraphics[scale=1.1]{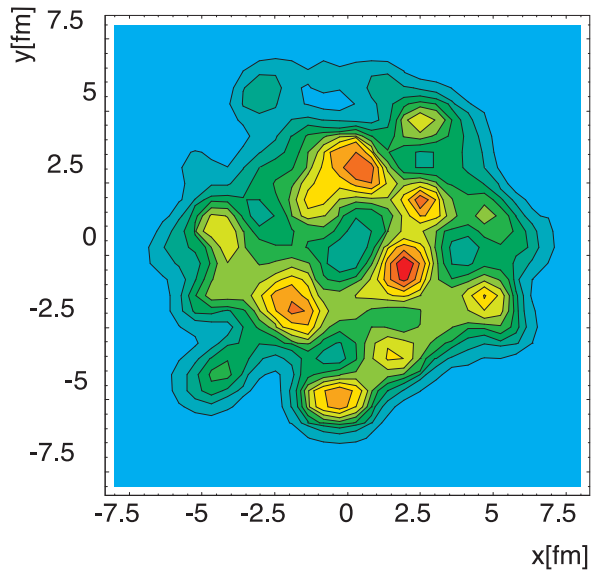}\vspace{-0.4cm}
\caption{Initial transverse energy density distribution of a single central
Au+Au collision event at $\sqrt{s}=200 \: GeV$ \cite{265}.}
\label{fig:Figure60}
\end{center}
\end{figure}

\section{Elliptic flow fluctuation}
\label{sec:elliptic_flow_fluctuations}

That this ascertation is far from trivial  is illustrated in Fig.~\ref{fig:Figure60}.
It shows \cite{265} the primordial transverse energy density
projection (the stage basic to all primordial abservables) at $t
\approx 0.2 \: fm/c$, in a central Au+Au collision at $\sqrt{s}=200
\: GeV$, exhibiting an extremely clumpy and nonhomogeneous
structure, apparently far away from equilibrium.

\begin{figure}[h!]   
\begin{center}
\includegraphics[scale=0.32]{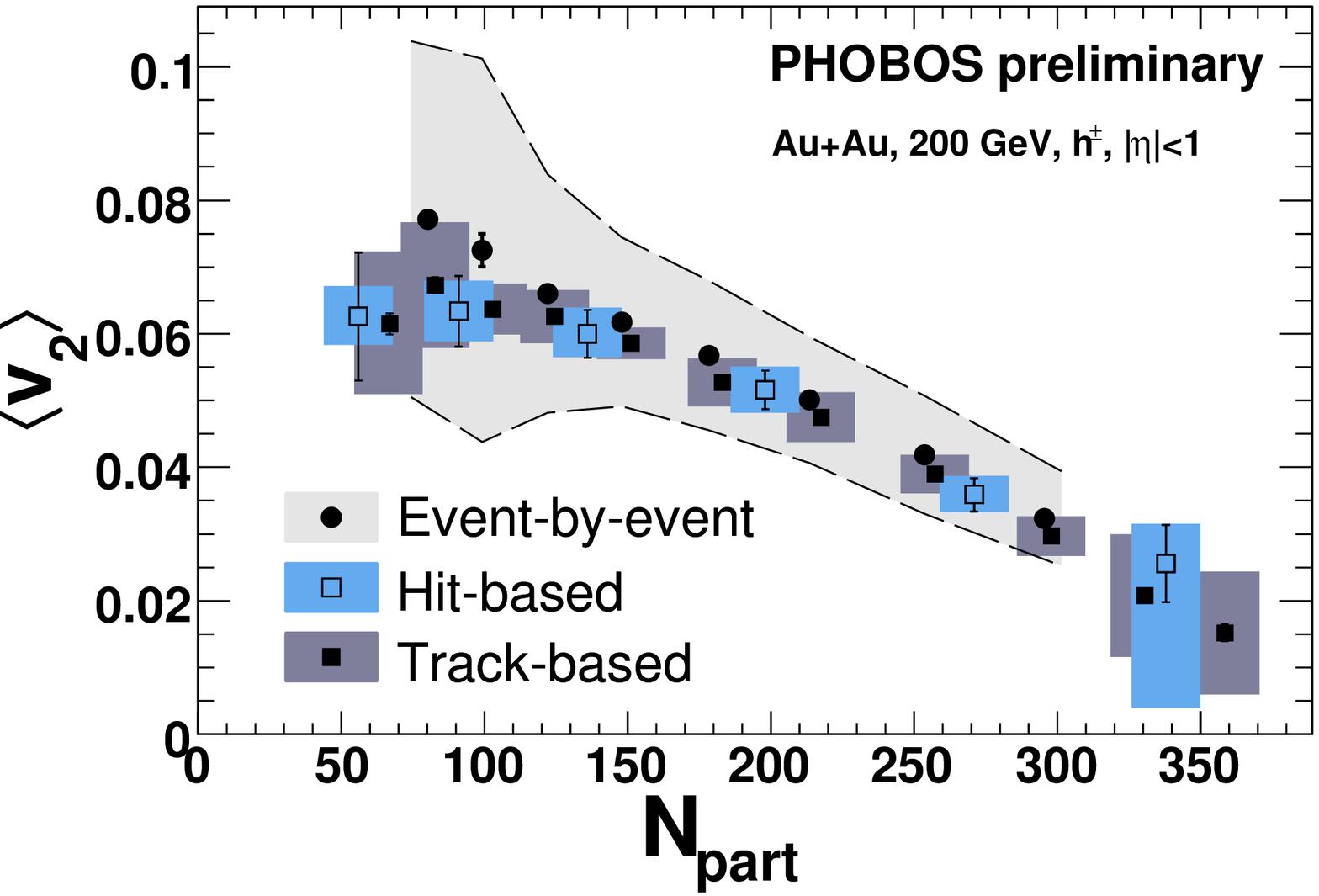}\vspace{-0.4cm}
\includegraphics[scale=0.32]{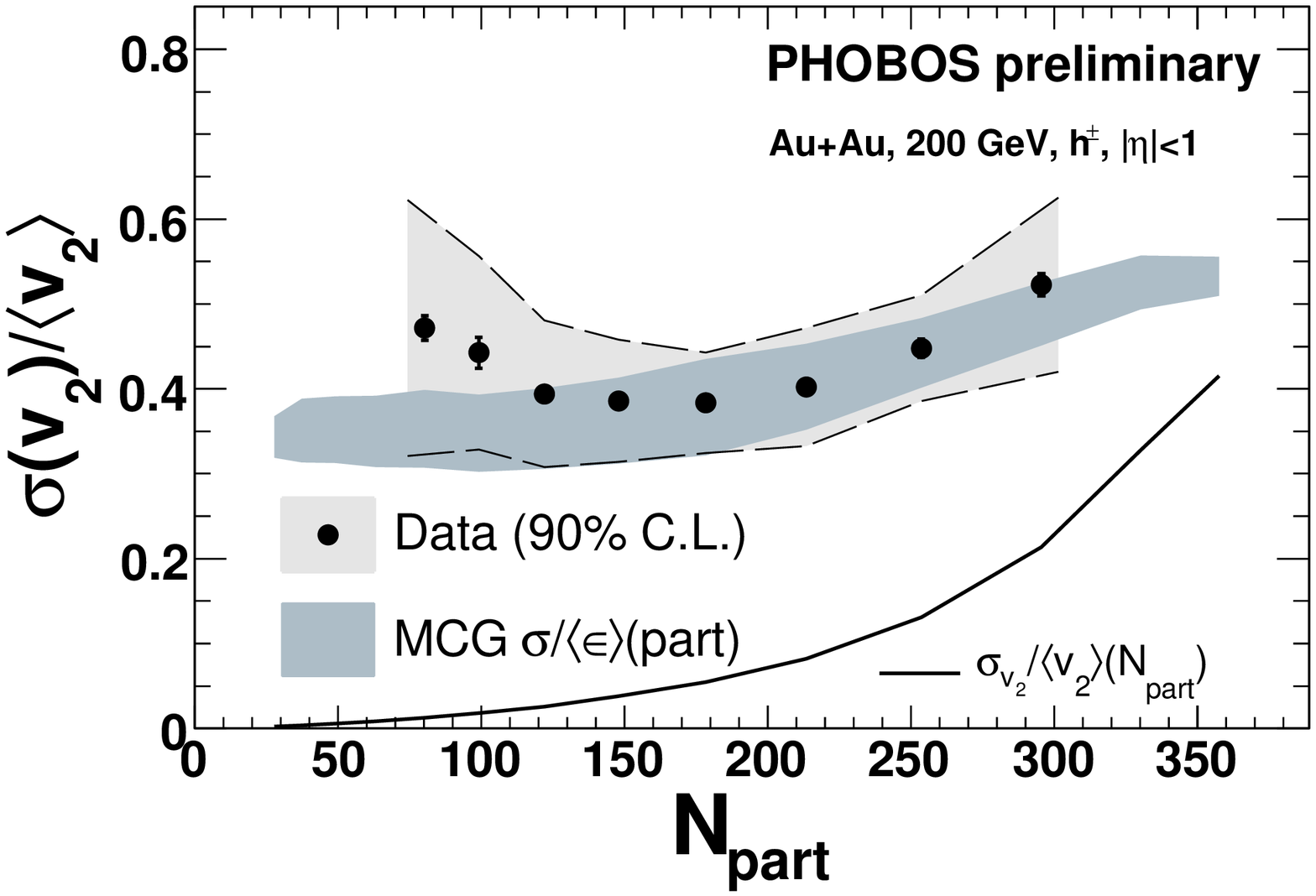}
\caption{(left) Event by event elliptic flow analysis by PHOBOS gives an
average $\left<v_2\right>$  that agrees with the result of ensemble analysis,
in Au+Au at $200 \: GeV$, for charged hadrons. (right) The
event-wise relative $v_2$ fluctuation vs. $N_{part}$, compared to
the event-wise relative fluctuation of the participant excentricity,
$\sigma (\epsilon_{part})/\left<\epsilon_{part}\right>$. The closed line gives $v_2$ variation due to $N_{part}$ number fluctuation. From \cite{268}.}
\label{fig:Figure61}
\end{center}
\end{figure}

The imaging of this initial geometrical pattern of the
collision volume by a hydrodynamic evolution even applies at the
level of individual events, such as illustrated in Fig.~\ref{fig:Figure60}. The PHOBOS
Collaboration has shown \cite{266,267,268} that an event by event
analysis of the elliptic flow coefficient $v_2$ is possible (see
ref. \cite{266} for detail), by means of a maximum likelyhood
method. For initialization they sample the seemingly random initial
spatial density distribution of single Au+Au collision events
by the "participant excentricity" of individual Monte Carlo events,
\begin{equation}
\epsilon_{part} = \frac{\sqrt{(\sigma_y^2 - \sigma_x^2)^2 + 4
\sigma_{xy}^2}}{\sigma_y^2+ \sigma_x^2}
\label{eq:equation73}
\end{equation}
where $\sigma_{xy}= \left<xy\right>-\left<x\right>\left<y\right>$. The {\it average} values of
$\epsilon_{part}$ turn out to be similar to $\epsilon_x$ from equation~\ref{eq:equation48}, 
as expected, but the relative fluctuation width
$\sigma(\epsilon)/\left<\epsilon\right>_{part}$ turns out to be considerable.
It is the point of this investigation \cite{267,268} to show that
the observed relative event by event flow fluctuation equals the
magnitude of the relative excentricity fluctuation. This is shown
\cite{268} in Fig.~\ref{fig:Figure61}. The left panel demonstrates that the {\it
average} $\left<v_2\right>$ obtained vs. $N_{part}$ from the event-wise
analysis agrees with the previously published \cite{156}
event-averaged PHOBOS data. The right panel shows that the
event-wise relative fluctuation of $v_2$ is large: it amounts to
about 0.45 and is equal to the relative fluctuation of
$\epsilon_{part}$, i.e.
\begin{equation}
\sigma(v_2)/\left<v_2\right> \approx \sigma
(\epsilon_{part})/\left<\epsilon_{part}\right>.
\label{eq:equation74}
\end{equation}
The initial geometry appears to drive the hydrodynamic evolution of
the system, not only on average but event-by-event \cite{268}, thus providing for an example
of the self-analyzing property mentioned above. The
$v_2$ signal thus emerges as the most sensitive and specific
diagnostic instrument for the primordial conditions and their
subsequent evolution: it reveals even the random (Fig.~\ref{fig:Figure60}) initial
fluctuations. In comparison the analysis with thermal photons is
only sensitive to the primordial temperature \cite{249}, and
restricted by very small cross sections and significant background
from other sources and evolution times. It also does not give
viscosity information.\\

\section{Critical point: fluctuations from diverging susceptibilities}
\label{sec:Critical_Point}

Recalling the goal defined in the introduction we seek observables
that help to elaborate points or regions of the QCD phase diagram,
Fig.~\ref{fig:Figure1}. We have seen several observables that refer to the QCD plasma
environment at $T \ge 300 \: MeV$, $\mu \approx 0$ (elliptic flow,
jet attenuation, $J/\Psi$ suppression, direct photon production),
which could be first resolved within the primordial time interval
$\tau \le 1 \: fm/c$ accessible at RHIC energy. The LHC will extend
the reach of such observables toward $T \approx 600 \: MeV$, $x_F
\le 10^{-3}$, at $\mu_B=0$. On the other hand, relativistic A+A
collisions at lower energy permit a focus on the hypothetical QCD
parton-hadron coexistence line, $T=T_c (\mu_B)$, with the domain $\mu_B
\rightarrow 500 \: MeV$ being accessible at the SPS. Characteristic
observables are radial flow, hadro-chemical freeze-out, and chiral
symmetry restoration effects in dilepton vector meson spectra.
Focusing on this domain, we discuss fluctuations potentially
associated with the existence of a critical point
\cite{8,9,10,11,146,147}.

At the end of chapter~\ref{sec:Low_mass_dilepton_spectra} we mentioned the conclusion from chiral
symmetry restoration models \cite{15,259} that at high $\mu_B$ the
phase transformation occuring at $T_c \: (\mu_B)$ should be a {\it
chiral first order phase transition}. On the other hand, lattice QCD
has characterized \cite{16} the phase transformation at $\mu_B
\rightarrow 0$, to be merely a rapid cross-over. Thus, the first
order nature of the phase coexistence line in Fig.~\ref{fig:Figure1} has to end, with
decreasing $\mu_B$, in a QCD critical point, tentatively located by
recent lattice QCD calculations \cite{9,10,11} in the interval
$\mu_B=300-500 \: MeV$. The existence of such a point in the [$T,
\mu_B$] plane would imply fluctuations analogous to critical
opalescence in QED \cite{146,147,263}. Beyond this second order
phase transition point the coexistence line would be the site of a
rapid cross-over \cite{16}. This overall theoretical proposal places
potential observations related to the critical point itself, and/or
to the onset of first order phase transition conditions at higher
$\mu_B$, within the domain of the lower SPS energies, $\sqrt{s} \le
10 \: GeV$. Note that, at such low energies, the initialization of
thermal equilibrium conditions should occur in the vicinity of
$T_c$, unlike at RHIC and LHC, and that the central fireball spends
considerable time near the coexistence line, at $300 \le \mu_B \le
500 \: MeV$.\\
\begin{figure}[h!]   
\begin{center}
\includegraphics[scale=1.0]{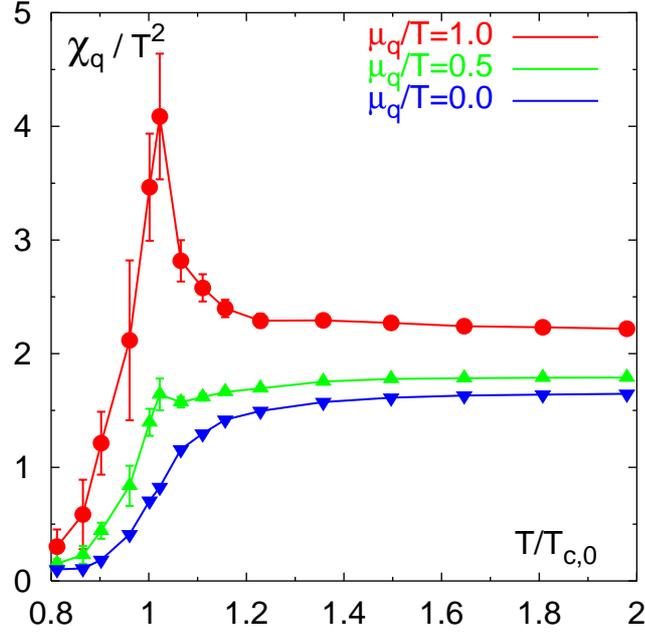}\vspace{-0.4cm}
\caption{Quark number density susceptibility vs.\ temperature for light quarks
in 2 flavor lattice QCD at finite $\mu_B$. The calculation refers to
$T_c=150 \: MeV$ and quark chemical potential $\mu_q/T_c=0, \: 0.5
\: \ \mbox{and} \ \: 1.0$, respectively \cite{270}. Smooth lines interpolate the calculated points; error bars indicate lattice statistics.}
\label{fig:Figure62}
\end{center}
\end{figure}\\
To analyze potential observable effects of a critical point, we
recall briefly the procedure in finite $\mu_B$ lattice theory that
led to its discovery. One method to compute thermodynamic functions
at $\mu_B > 0$ from the grand canonical partition function $Z(V,T,
\mu_q)$ at $\mu_q=0$ is to employ a Taylor expansion with respect to
the chemical quark potential \cite{10,11,270}, defined by the
derivatives of $Z$ at $\mu=0$. Of particular interest is the quark
number density susceptibility,
\begin{equation}
\chi_{u,d} = T^2 \left(\frac{\delta^2}{\delta(\mu/T)^2} \:
\frac{p}{T^4} \right)
\label{eq:equation75}
\end{equation}
which can also be written as
\begin{equation}
\chi_q = T^2 \left(\frac{\delta}{\delta(\mu_u/T)} +
\frac{\delta}{\delta(\mu_d/T)} \right) \: \frac{n_u+n_d}{T^3}
\label{eq:equation76}
\end{equation}
with $\chi_q=(\chi_u + \chi_d)/2$ and quark number densities
$n_u, \: n_d$. We see that the susceptibility refers to the quark
number density fluctuation. The lattice result \cite{10,270} is
shown in Fig.~\ref{fig:Figure62}, a calculation with two dynamical flavors assuming
$T_c=150 \: MeV$ and three choices of chemical quark potential,
$\mu_q=0,75$ and 150 $MeV$, respectively, corresponding to $\mu_B =
3 \: \mu_q = 0,225$ and 450 $MeV$. These choices correspond to
LHC/RHIC energy, top SPS energy and $\sqrt{s} \approx 6.5 \: GeV$,
respectively. At $\mu_B=0$ one sees a typical smooth cross-over
transition at $T=T_c$ whereas a steep maximum of susceptibility 
occurs with $\mu_B=450 \: MeV$. This suggests the presence of a
critical point in the ($T, \: \mu_B$) plane \cite{270} in the
vicinity of (150 $MeV$, 450 $MeV$). For final confirmation one would
like to see this maximum disappear again, toward $\mu_q >T_c$, but
this is beyond the convergence domain of the employed Taylor
expansion (see ref. \cite{9} for alternative approaches).\\
\begin{figure}[h!]   
\begin{center}
\includegraphics[scale=0.4]{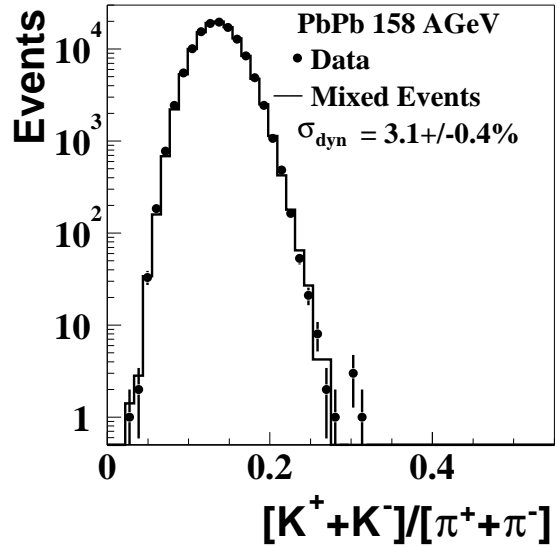}\vspace{-0.4cm}
\caption{Event by event fluctuation of the $K^-/\pi^{+-}$ ratio in central
collisions of Pb+Pb at $\sqrt{s}=17.3 \: GeV$, relative to mixed
event background (histogram) \cite{272}.}
\label{fig:Figure63}
\end{center}
\end{figure}\\
From Fig.~\ref{fig:Figure62} we also expect a divergence of the strangeness
susceptibility, for which no results from a 3 dynamical flavors
calculation  at finite $\mu_B$ exist to date. A lattice calculation
at $\mu_B=0$, $T\approx 1.5 \: T_c$ suggests \cite{271} that the
$u,d,s$ quark flavors densities fluctuate {\it uncorrelated} (but we
do not know whether that is also true at $\mu_B=\mu_B^{crit})$. This
could thus be observed in event by event analysis, in particular as
a fluctuation of the Wroblewski ratio $\lambda_s = 2(s +
\overline{s})/(u + \overline{u} + d + \overline{d})$ which is
approximated by the event-wise ratio $(K^+ + K^-)/(\pi^+ + \pi^-)$.
This was first measured by NA49 in central collisions of Pb+Pb at
top SPS energy; the result \cite{272} is shown in Fig.~\ref{fig:Figure63}. The data
result from a maximum likelyhood analysis of track-wise specific
ionization in the domain $3.5 \le y \le 5$ slightly forward of
mid-rapidity. The width $\sigma_{data}$ is almost perfectly
reproduced by the mixed event reference, such that the difference,
\begin{equation}
\sigma_{dyn} = \sqrt{(\sigma^2_{data} \: - \: \sigma^2_{mix}})
\label{eq:equation77}
\end{equation}
amounts to about 3\% of $\sigma_{data}$ only, at $\sqrt{s}=17.3 \:
GeV$. This analysis has more recently been extended to all energies
available thus far, at the SPS \cite{273} and at RHIC \cite{274}.
Fig.~\ref{fig:Figure64} shows that $\sigma_{dyn}$ stays constant from top SPS to top
RHIC energy but exhibits a steep rise toward lower energies that
persists down to the lowest SPS energy, $\sqrt{s}=6.2 \: GeV$.
Fig.~\ref{fig:Figure33} shows \cite{107} that at this energy $\mu_B=450 \: MeV$, thus
corresponding to the susceptibility peak in Fig.~\ref{fig:Figure62}. Again, as we
noted about the peak in Fig.~\ref{fig:Figure62}: if these data indicate a critical
point effect in the vicinity of $\mu_B=450 \: MeV$ the relative
fluctuation should {\it decrease} again, toward yet higher $\mu_B$
and lower $\sqrt{s}$. These data will, hence, be re-measured and
extended to lower $\sqrt{s}$ by experiments in preparation
\cite{260,261,262} at CERN, RHIC and GSI-FAIR. This will also help
to evaluate alternative tentative explanations invoking fluctuating
canonical suppression \cite{120}, or strangeness trapping
\cite{275}. Finally, the position of the critical point
needs to be ascertained by lattice theory.\\
\begin{figure}[h!]   
\begin{center}
\includegraphics[scale=1.3]{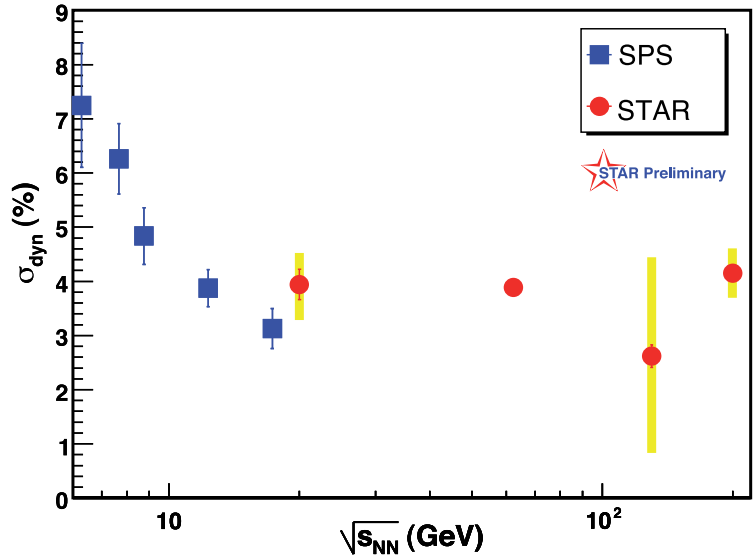}\vspace{-0.4cm}
\caption{The relative deviation of the event by event $K/\pi$ fluctuation
width from the mixed event background width, $\sigma
_{dyn}$ (see equation~\ref{eq:equation77}); at SPS \cite{273} and RHIC
\cite{274} energies.}
\label{fig:Figure64}
\end{center}
\end{figure}

\section{Critical fluctuation of the sigma-field, and related
pionic observables}
\label{sec:Critical_Fluctuations}

Earlier investigations of critical QCD phenomena that might occur in
high energy nuclear collisions were based on QCD chiral field theory
\cite{276}. The QCD critical point is associated with the chiral
phase transition in so far as it appears as a remnant of a
tri-critical point \cite{147} corresponding to the ''ideal'' chiral
limit that would occur if $m_u=m_d=0$. Therefore the existence of a
second-order critical point, at $\mu_B>0$, is a fundamental property
of QCD with small but non-zero quark masses \cite{277}. The
magnitude of the quark condensate, which plays the role of an order
parameter of the spontaneously broken symmetry (generating hadronic
mass), has the thermal expectation value
\begin{equation}
\left<\overline{q} q\right> = \frac{1}{Z} \sum_n \left<n \mid \overline{q} q \mid n
\right> exp (-E_n/T)
\label{eq:equation78}
\end{equation}
with the partition function of hadronic states $E_n$
\begin{equation}
Z=\sum_n exp(-E_n/T).
\label{eq:equation79}
\end{equation}
The low energy behaviour of the matrix elements $\left<n \mid
\overline{q}  q \mid n\right>$ can be worked out in chiral perturbation
theory \cite{277}. At the QCD critical point the order parameter
fluctuates strongly. Its magnitude $\left<\overline{q} q\right>$ is identified
with an isoscalar quantity, the so-called $\sigma$-field. The
critical point communicates to the hadronic population via the
$\sigma \leftrightarrow \pi \pi$ reaction, generating fluctuating
fractions of the direct pion yield present near $T=T_c$, which thus
gets imprinted with a fluctuation of transverse momentum (in the low
$p_T$ domain) stemming from $\sigma$ mass fluctuation, downward
toward the critical point. At it the isoscalar field ideally
approaches zero mass, in order to provide for the long wavelength mode
required by the divergence of the correlation length \cite{147}.

\begin{figure}[h!]   
\begin{center}
\includegraphics[scale=0.4]{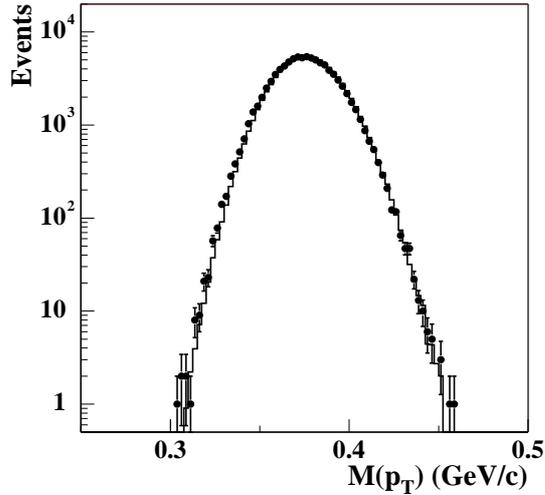}\vspace{-0.4cm}
\caption{Event by event fluctuation of average charged hadron $p_T$ in the
interval $4.0<y<5.5$, in central Pb+Pb collisions at $\sqrt{s}=17.3
\: GeV$. Mixed event background given by histogram \cite{278}.}
\label{fig:Figure65}
\end{center}
\end{figure}

Note the relatively fragile structure of the argument. In an
ideal, stationary infinite volume situation the sigma field would
really become massless, or at least fall below the $\pi^+ \pi^-$
threshold; thus its coupling to $\pi^+ \pi^-$ becomes weak, and
restricted to very small $p_T$. Furthermore, such primary soft
pions, already small in number, are subject to intense subsequent
re-absorption and re-scattering in the final hadronic cascade
evolution \cite{114}. In fact, experimental investigations of event
by event $p_T$ fluctuations in central A+A collisions, covering the
entire $\sqrt{s}$ domain from low SPS to top RHIC energy have not
found significant dynamic effects \cite{278,279,280,281}. Fig.~\ref{fig:Figure65}
illustrates the first such measurement by NA49 \cite{278} in central
Pb+Pb collisions at $\sqrt{s}=17.3 \: GeV$, at forward rapidity
$4<y<5.5$, showing the distribution of event-wise charged particle
average transverse momentum, a perfect Gaussian. It is very closely
approximated by the mixed event distribution, ruling out a
significant value of $\sigma_{dyn}$ from equation~\ref{eq:equation77}. More sensitive
measures of changes, event by event, in the parent distribution in
transverse momentum space, have been developed \cite{282,283}. NA49
has employed \cite{278,279} the measure $\Phi (p_T)$, defined as
\cite{282}
\begin{equation}
\Phi (p_T) = \sqrt{\frac{\left<Z^2\right>}{\left<N\right>}} - \sqrt{\overline{z^2}}
\label{eq:equation80}
\end{equation}
where $z_i=p_{Ti}- \overline{p}_T$ for each particle, with
$\overline{p}_T$ the overall inclusive average, and for each event
$Z= \sum_{N} z_i$ is calculated. With the second term the trivial
independent particle emission fluctuation is subtracted out, i.e.
$\Phi$ vanishes if this is all. Indeed, the data of Fig.~\ref{fig:Figure65} lead to
$\Phi$ compatible with zero. Furthermore, a recent NA49 study
\cite{279} at mid-rapidity, covers the range from $\sqrt{s}=17.3$ to
6.3 $GeV$ (where the $K/\pi$ ratio fluctuation in Fig.~\ref{fig:Figure64} exhibits
the much-discussed rise, and even the ensemble average in Fig.~\ref{fig:Figure28}
shows the unexplained sharp peak) but finds no significant $\Phi$
signal.\\
\\
Alternatively, one can base a signal of dynamical $p_T$ fluctuation
on the binary {\it correlation} of particle transverse momenta in a
given event, i.e. on the co-variance $\left<p_{Ti} \: p_{Tj}\right>$
\cite{281,283} of particles $i,j$ in one event. Of course, the
co-variance receives contributions from sources beyond our present
concern, i.e. Bose-Einstein correlation, flow and jets (the jet
activity becomes prominent at high $\sqrt{s}$, and will dominate the
$p_T$ fluctuation signal at the LHC). In co-variance analysis, the
dynamical $p_T$ fluctuation (of whatever origin) is recovered via
its effect on correlations among the transverse momentum of
particles. Such correlations can be quantified employing the
two-particle $p_T$ correlator \cite{281,284}
\begin{equation}
\left<\Delta p_{Ti} \: \Delta p_{Tj}\right> = \frac{1}{M_{pairs}} \:
\sum^n_{k=1} \: \sum^{N(k)}_{i=1} \: \sum^{N(k)}_{j=i+1} \: \Delta
p_{Ti} \Delta p_{Tj}
\label{eq:equation81}
\end{equation}
where $M_{pairs}$ is the total number of track pairs of the events
$k$ contained in the entire ensemble of $n$ events, $N(k)$ is the
number of tracks in event $k$, and $\Delta p_{Ti}=p_{Ti}-
\overline{p}_T$ where $\overline{p}_T$ is the global ensemble mean
$p_T$. The normalized dynamical fluctuation is then expressed
\cite{281} as
\begin{equation}
\sigma (p_T)_{dyn} = \sqrt{\left<\Delta p_{Ti} \: \Delta p_{Tj}\right>} \: / \:
\overline{p_T}.
\label{eq:equation82}
\end{equation}
It is zero for uncorrelated particle emission.

\begin{figure}[h!]   
\begin{center}
\includegraphics[scale=1.3]{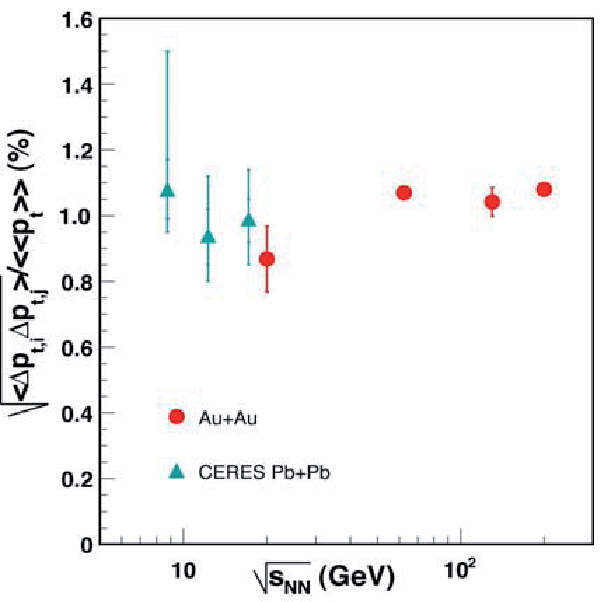}\vspace{-0.4cm}
\caption{Dynamical $p_T$ event by event fluctuation analysis by
$\sigma(p_T)_{dyn}$ of equation~\ref{eq:equation82}, vs. $\sqrt{s}$, showing SPS
\cite{280} and RHIC \cite{281} data.}
\label{fig:Figure66}
\end{center}
\end{figure}

Fig.~\ref{fig:Figure66} shows the analysis of $p_{T}$ fluctuations based on 
the $p_T$ correlator, for central Pb+Au SPS collisions by CERES \cite{280} 
and for central Au+Au at four RHIC energies by STAR \cite{281}. The signal is at the 1\%
level at all $\sqrt{s}$, with no hint at critical point phenomena.
Its small but finite size could arise from a multitude of sources,
e.g. Bose-Einstein statistics, Coulomb or flow effects,
mini-jet-formation, but also from experimental conditions such as
two-track resolution limits \cite{284}. We note that even if a
critical opalescence effect, related to a fluctuating chiral
condensate at $T=T_{crit}$, couples to the primordial, low $p_T$
pion pair population \cite{147,285}, this signal might be dissipated
away, and thus ''thermalized'' to the thermal freeze-out scale of
about 90-110 $MeV$, as a pion experiences about 6 re-scatterings
during the hadronic cascade \cite{114}. On the other hand the
hadro-chemical $K/\pi$ ratio fluctuation (Fig.~\ref{fig:Figure64}) would be preserved 
throughout the cascade (chapter~\ref{chap:hadronization}).\\

\section{Bose-Einstein-correlation}
\label{sec:Bose_Einstein}

Identical boson pairs exhibit a positive correlation enhancement
when $\Delta {\vec{p}_{ij}} \rightarrow 0$, an intensity correlation
analogous to the historical Hanbury-Brown and Twiss effect (HBT) of
two photon interferometry \cite{286} employed in astrophysics. In
nucleus-nucleus collisions this is an aspect of the symmetry of the
N-pion wave function that describes pion pairs at the instant of
their decoupling from strong interaction. The momentum space
correlation function $C(q,K), \: q=p_1 - p_2, \: K = \frac{1}{2}
(p_1+p_2)$ is the Fourier transform of the spatial emission function
$S(x, K)$ which can be viewed as the probability that a meson pair
with momentum $K$ is emitted from the space-time point $x$ in the
freezing-out fireball density distribution \cite{96,287}.

The aim of HBT two particle interferometry is to extract from the
measured correlator $C(q, K)$ as much information about $S(x,K)$ as
possible. In addition to the traditional HBT focus on geometrical
properties of the source distribution (''HBT radii'') there occurs
information on time and duration of emission, as we have already
employed in Fig.~\ref{fig:Figure19}, illustrating $\tau_f$ and $\Delta \tau$ in
central Pb+Pb collisions at top SPS energy \cite{90}. Moreover, even
more of dynamical information is accessible via the effects of
collective flow on the emission function. In this respect, the
observations make close contact to the hydrodynamic model
freeze-out hyper-surface (chapter~\ref{sec:Hadronic_freeze_out}) and the contained flow
profiles. HBT thus helps to visualize the {\it end} of the
collective radial and elliptic flow evolution. This implies that we
expect to gather evidence for the, perhaps, most consequential
property of A+A collisions at high $\sqrt{s}$, namely the
primordially imprinted Hubble expansion.

We do not engage here in a detailed exposition of HBT formalism and
results as comprehensive recent reviews are available \cite{96,287}.
Briefly, the measured quantity in $\pi^+ \pi^+, \pi^- \pi^-$ or
$K^+K^+$ interferometry is the correlator
\begin{equation}
C(q,K) = \frac{d^6N/dp^3_1dp_2^3}{d^3N/dp_1^3 \: d^3N/dp_2^3}
\label{eq:equation83}
\end{equation}
of pair yield normalized to the inclusive yield product. The
correlator is related to the emission function which is interpreted
as the Wigner phase space density of the emitting source \cite{288}:
\begin{equation}
C(q,K) \approx 1 + \frac{\mid \int d^4x S(x,K) e^{iqx}\mid^2}{\mid
\int d^4 x S(x,K) \mid^2}.
\label{eq:equation84}
\end{equation}
Due to the experimental on-shell requirement $K_0 = \sqrt{K^2+m^2}$
the 4-vector components of $K$ are constrained on the left hand
side. Thus, relation~\ref{eq:equation84} can not simply be inverted.

To proceed one employs a Gaussian ansatz on either side of equation~\ref{eq:equation84}. The
experimental data are parametrized by source ''radius'' parameters
$R_{ij}(K)$,
\begin{equation}
C(q,K) = 1 + \lambda (K) exp \: [ - \sum_{ij} R^2_{ij} (K) q_i q_j
\: ]
\label{eq:equation85}
\end{equation}
employing $\lambda(K)$ {\it essentially as a fudge factor} (related
nominally to possible coherence of emission effects in the source,
which would dilute or remove the Bose-Einstein statistics effect of
$C \rightarrow 2$ for $q \rightarrow 0$ but have never been seen in nuclear collisions).
In equation~\ref{eq:equation85} the sum runs over three of the four components of $q$, due
again to the on-shell requirements \cite{287}. For the emission
function $S(x,K)$ a Gaussian profile is assumed about an ''effective
source center'' $\overline{x} (K)$, thus
\begin{equation}
S(x,K) \rightarrow S(\overline{x} (K), K) \: \mbox{x} \: G
\label{eq:equation86}
\end{equation}
where $G$ is a Gaussian in coordinates $\tilde{x}^{\mu} (K)$
relative to the center coordinates $\overline{x}^{\mu} (K)$.
Inserting equation~\ref{eq:equation86} into \ref{eq:equation84} one finally obtains
\begin{equation}
C(q,K) = 1+ \lambda(K) \: exp \: [ -q_{\mu} q_v \left<\tilde{x} ^{\mu} \:
\tilde{x}^{\nu}\right> ]
\label{eq:equation87}
\end{equation}
where $\left<\tilde{x} ^{\mu} \: \tilde{x}^{\nu}\right>$ are the elements of
the space-time variance of the correlation function, which
re-interpret the ''radii'' $R_{ij}^2$ in equation~\ref{eq:equation85}. Assuming azimuthal
symmetry (central collisions), cartesian parametrizations of the
pair relative momentum $q$ coordinates  (corresponding to fixation
of the space-time variance in equation~\ref{eq:equation85}) have been introduced by Yano,
Koonin and Podgoretskii \cite{289}, and, alternatively, by Pratt
\cite{290}. The latter, {\it out-side-longitudinal} coordinate
system has the ''long'' direction along the beam axis. In the
transverse plane, the ''out'' direction is chosen parallel to
$K_T=(p_{1T} + p_{2T})/2$, the transverse component of the pair
momentum $K$. The ''side'' direction is then orthogonal to the out-
and long-direction but, moreover, it has the simplest geometrical
interpretation (see ref. \cite{287} for detail), to essentially
reflect the transverse system size \cite{288}. The parameters of
equation~\ref{eq:equation87} are thus defined; as an example we quote, from identification
of equation~\ref{eq:equation85} with \ref{eq:equation87}, the resulting geometrical definition of the
''side'' radius,
\begin{equation}
R_{side}^2 (K) =\left<\tilde{y}(K)^2\right>.
\label{eq:equation88}
\end{equation}
Overall, this model of Fourier related correlators, $C(q,K)$ the
experimentally accessible quantity (see equation~\ref{eq:equation83}), and $S(x,K)$ the
to-be-inferred spatial freeze-out fireball configuration, leads to
the Gaussian ansatz \cite{287}
\begin{eqnarray}
C(q,K) =1+ \lambda (K) \: exp [-R^2_{out} (K) q_{out}^2 - R_{side}^2
(K) q^2_{side} - \nonumber\\
R^2_{long} (K) q^2_{long} + \mbox{cross terms}]
\end{eqnarray}
which is fitted to the experimental correlation function (equation~\ref{eq:equation83}). The
experiment thus determines the variances $R_{out}, \: R_{side} \:
\mbox{and} \: R_{long}$. In the so-called ''local co-moving system''
(LCMS), defined as the frame in which $p_{z,1}=-p_{z,2}$, i.e.
$\beta_{long}=0$, we obtain in addition to equation~\ref{eq:equation88}
\begin{eqnarray}
R_{out}(K)^2 & = & \left<(\tilde{x}(K) - \beta_T \tilde{t}(K))^2\right>
 \nonumber\\
R^2_{long}(K)& = & \left<\tilde{z}(K)^2\right>
\end{eqnarray}
and finally, for azimuthal symmetry in central collisions with
$\left<\tilde{x}^2\right> \approx \left<\tilde{y}^2\right>$ we find the ''duration of
emission'' parameter illustrated in Fig.~\ref{fig:Figure19} \cite{90}:
\begin{equation}
\left<\tilde{t} \: ^2\right> \approx \frac{1}{\beta_T} \: (R^2_{out} -
R^2_{side}).
\label{eq:equation89}
\end{equation}
The resulting reduction of the initial 8 coordinates of the meson
pair stems, in summary, from the on-shell requirement, from
azimuthal symmetry and from approximate Bjorken invariance in the
LCMS system \cite{287,288}. One has to be aware of the latter
idealizations. Note that in an expanding source all HBT parameters
depend on $K$, the pair mean momentum (see below).

\begin{figure}[h!]   
\begin{center}
\includegraphics[scale=0.6]{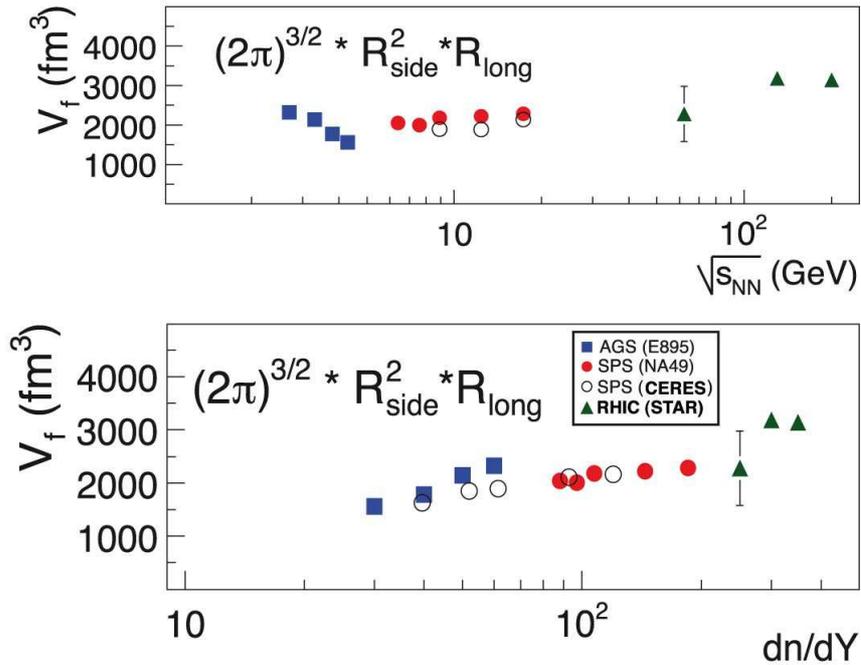}\vspace{-0.4cm}
\caption{Coherence freeze-out volume $V_f$ from $\pi^-$pair Bose-Einstein
correlation analysis in central Au+Au and Pb+Pb collisions, (upper
panel) plotted vs. $\sqrt{s}$, (lower panel) vs. mid-rapidity
charged particle density $dN^{ch}/dy$ \cite{291}.}
\label{fig:Figure67}
\end{center}
\end{figure}

We make first use of the above parametrization in Fig.~\ref{fig:Figure67}. From the
purely spatial radii $R_{side} \: \mbox{and} \:  R_{long}$ one can
define a mid-rapidity volume at pionic decoupling, $V_f=(2 \pi) ^{2/3} \:
R^2_{side} \: R_{long}$ which is shown \cite{291} at $K=0.2 \: GeV$
for central Pb+Pb and Au+Au collisions from AGS \cite{292} via SPS
\cite{90,293} to RHIC \cite{294} energy. The upper panel shows the
$\sqrt{s}$ dependence, the lower illustrates the dependence on the
charged particle rapidity density $dN_{ch}/dy$ which one might
intuitively expect to be related to the freeze-out volume
\cite{295}. We see, firstly, that the plot vs. $\sqrt{s}$ exhibits a
non-monotonous pattern at the transition from AGS to SPS energies
\cite{287,295}, whereas the plot vs. $dN/dy$ rather features a rise
toward a plateau that ends in a steep increase at RHIC energies.
Second, the tenfold increase in charged particle rapidity density is
reflected in only a doubling of the ''volume'' $V_f$.\\
\\
The latter observation reminds us of the fact that the geometrical
parameters obtained via the above analysis do not refer to the
global source volume if that volume undergoes a collective Hubble
expansion \cite{96,287}. A pion pair emitted with small relative
momentum into the azimuthal direction $\vec{K}_T$ is likely to stem
(only) from the fraction of the expanding source that also moves
into this direction. This coupling between position and momentum in
the source becomes more pronounced, both, with increasing $K_T$
and increasing sources transverse velocity $\beta_T$ from radial
expansion. We have seen in Fig.~\ref{fig:Figure24} that the latter increases
dramatically with $\sqrt{s}$, such that the coherence volume $V_f$
comprises a decreasing fraction of the total fireball. It should
thus rise much more slowly than proportional to the global $dN/dy$
\cite{96,287}. 

\begin{figure}[h!]   
\begin{center}
\includegraphics[scale=0.4]{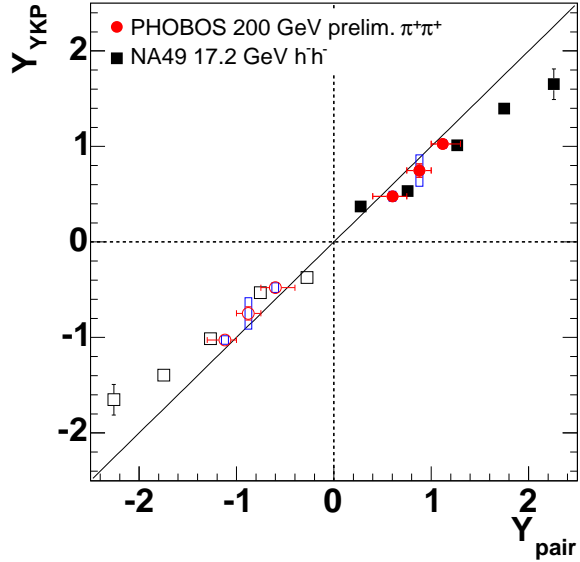}\vspace{-0.4cm}
\caption{Emitting source rapidity ($Y_{YKP}$) as a function of pion pair
rapidity ($Y_{\pi \pi}$). From $\pi^-$pair correlation analysis in
central Pb+Pb collisions at $\sqrt{s}=17.3 \: GeV$ \cite{90}, and in
central Au+Au collisions at $\sqrt{s}=200 \: GeV$ \cite{294}.}
\label{fig:Figure68}
\end{center}
\end{figure}

\begin{figure}[h!]   
\begin{center}
\includegraphics[scale=0.28]{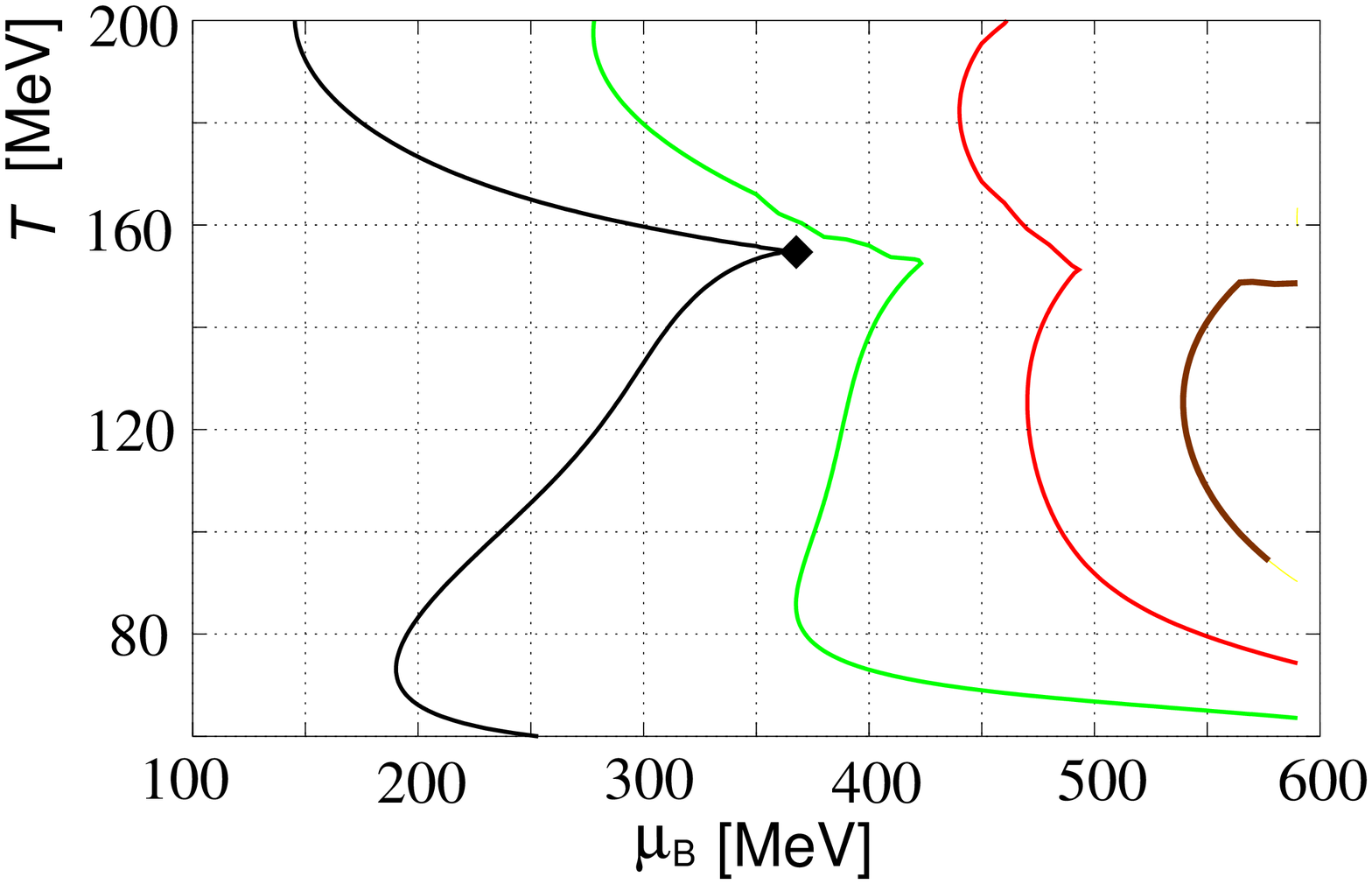}\vspace{-0.4cm}
\includegraphics[scale=0.28]{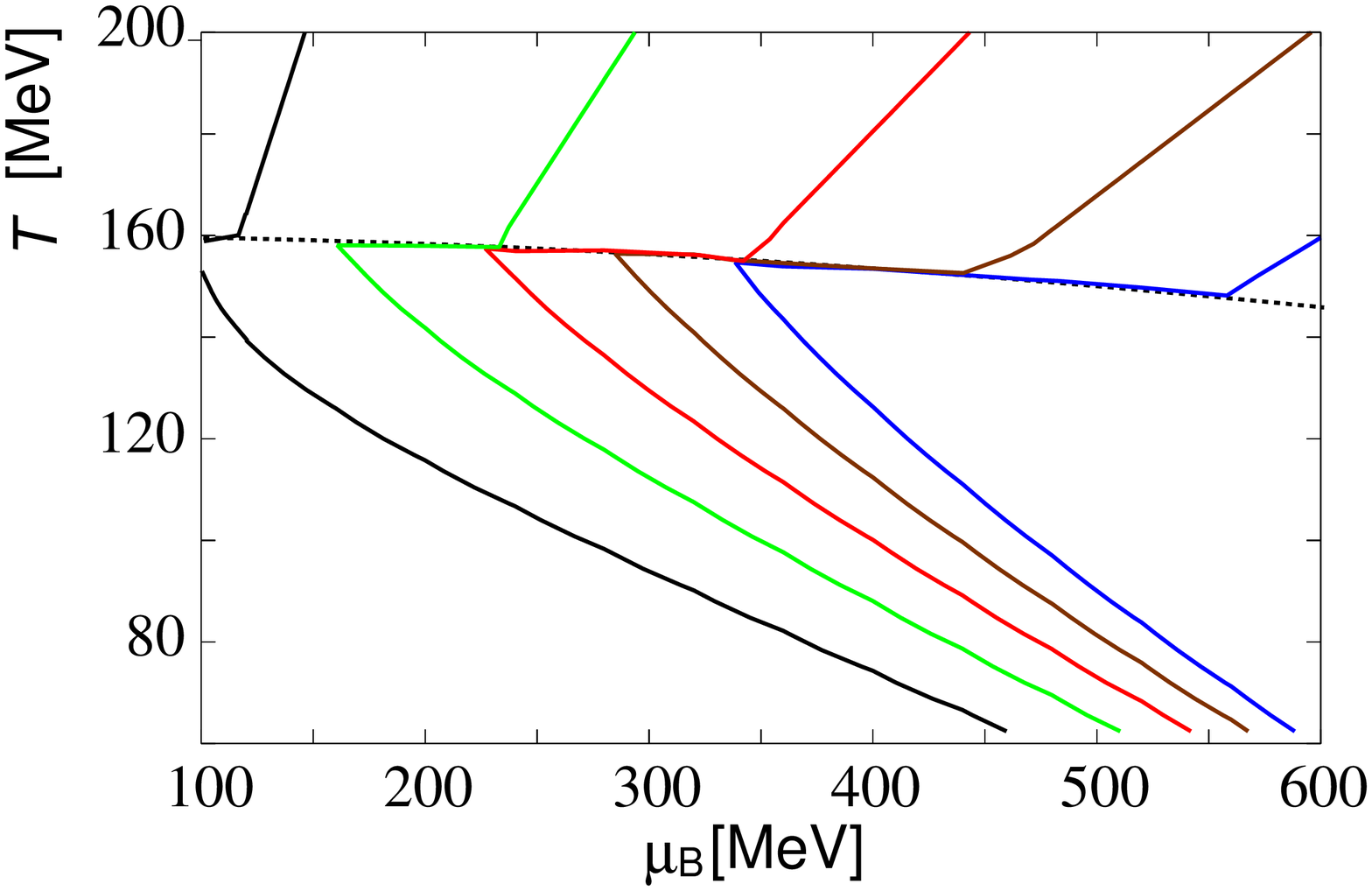}
\caption{(left) Influence of a critical point on hydrodynamic model isentropic
expansion trajectories characterized by various values of entropy to
net baryon number $s/n_B$. (right) The same trajectories without a critical point but a first
order transition all along the phase boundary \cite{296}.}
\label{fig:Figure69}
\end{center}
\end{figure}

A striking experimental confirmation of the Hubble expansion pattern
in central A+A collisions is shown in Fig.~\ref{fig:Figure68}. The illustrated HBT
analysis of NA49 \cite{90} at $\sqrt{s}=17.3 \: GeV$, and of PHOBOS
\cite{294} at $\sqrt{s}=200 \: GeV$, employs the alternative
parametrization of the correlation function $C(q,K)$ introduced
\cite{289} by Yano, Koonin and Podgoretskii (YKP). Without
describing the detail we note that the YKP correlation function
contains the ''YK velocity'' $\beta_{YK}$ describing the source's
longitudinal collective motion in each interval of pion pair
rapidity,
\begin{equation}
Y_{\pi \pi} = \frac{1}{2} \: ln (\frac{E_1+E_2 + p_{z1} +
p_{z2}}{E_1 + E_2 -p_{z1} - p_{z2}}).
\label{eq:equation89b}
\end{equation}
Defining the ''YKP rapidity'' corresponding to $\beta_{YK}$ by
\begin{equation}
Y_{YKP} = \frac{1}{2} \: ln \frac{1+ \beta_{YK}}{1- \beta_{YK}} \: +
y_{cm}
\label{eq:equation90}
\end{equation}
leads to the results in Fig.~\ref{fig:Figure68}, indicating a strong correlation of
the collective longitudinal source rapidity $Y_{YKP}$ with the
position of the pion pair in rapidity space.\\
\\
We are interested in a possible non-monotonous
$\sqrt{s}$ dependence of freeze-out parameters such as $V_f$ because
of recent predictions of the relativistic hydrodynamic model that
a QCD critical point should act as an attractor of the isentropic
expansion trajectories in the $[T, \: \mu_B]$ plane \cite{101,296}.
Fig.~\ref{fig:Figure69} shows such trajectories characterized by the ratio of entropy
to net baryon number, $S/n_B$, which is conserved in each
hydro-fluid cell throughout the expansion. Note that the $S/n_B$
ratio is determined from the EOS during the primordial
initialization stage in the partonic phase; the relationship between
$S/n_B$ and $\sqrt{s}$ is thus not affected by the presence or
absence of a critical end point (CEP) at $T=T_c$ which, however, drastically
influences the trajectory at later times as is obvious from
comparing the left and right panels, the latter obtained with a
first order phase transition at all $\mu_B$ but no CEP. In this
case, the $S/n_B$ = const. trajectories in the partonic phase all
point to the origin in the $[T, \: \mu_q]$ plane because $\mu_q/T
\propto \mbox{ln} (\frac{S}{n_B})$  in an ideal parton gas; whereas
they point oppositely below $T=T_c$ because $T^{3/2}/\varrho \propto
\mbox{ln} (\frac{S}{n_B})$ in a hadron gas. This organization is
dramatically upset for the cases $S/n$=100, 50, and 33 by the
assumption of a CEP, tentatively placed here at $T=155 \: MeV, \:
\mu_B = 368 \: MeV$.\\
\\
A first conclusion from this model is that it should not be
essential to fine-tune $\sqrt{s}$ to make the system pass near the
CEP because the attractor character would dominate over a
substantial domain, e.g. at hadro-chemical freeze-out conditions
$250 \le \mu_B \le 500 \: MeV$ in the example of Fig.~\ref{fig:Figure69}. Please note
that hadro-chemical freeze-out is not treated correctly in this
model, to occur at $T_H$ which is $160 \le T_H \le 130 \: MeV$ from
Figs.~\ref{fig:Figure1} and~\ref{fig:Figure33}. The trajectories shown here {\it below} $T_H$ are
thus not very realistic. The expected pattern could cause the
''plateau'' behaviour of several observables at hadronic freeze-out
over the range of SPS energies {\it that corresponds to the above
interval of $\mu_B$} and $T_H$, e.g. $\left<m_T\right>$ and $T$ in Figs.~\ref{fig:Figure21} and~\ref{fig:Figure22}, elliptic flow $v_2$ at mid-rapidity in Fig.~\ref{fig:Figure36}, and coherent
hadronic freeze-out volume $V_f$ from HBT in Fig.~\ref{fig:Figure67}.

A consistent search for critical point effects implies, at first, a correct
treatment of the hadronic expansion phase in hydro-models as above
\cite{101,296}, properly distinguishing chemical composition freeze-out,
and eventual ''thermal'' freeze-out at hadronic decoupling.
From such a model
predictions for the $\sqrt{s}$ or $S/n_B$ systematics could be
provided for the HBT source parametrization implied by equations~\ref{eq:equation85}-\ref{eq:equation87}. 
I.e. the hydrodynamic model provides the correlator
$S(x,K)$ in cases with, and without a critical point which, as
Fig.~\ref{fig:Figure69} shows, leads to considerable changes in the system trajectory
in the domain near hadronization and onward to hadronic thermal
freeze-out, at each given $S/n$ or $\sqrt{s}$. On the experimental
side, this trajectory is documented by $[T, \mu_B]$ at hadronic
freeze-out from grand canonical analysis \cite{140} (chapter~\ref{chap:hadronization})
which also yields $S/n_B$. Furthermore, as we shall show below, HBT
in combination with the analysis of $p_T$ or $m_T$ spectra will
describe the situation at thermal hadron freeze-out yielding $T_f,
\: \beta_T$ at the surface and the ''true'' transverse radius
$R_{geom}$ of the source, in addition to the coherence volume $V_f$
illustrated in Fig.~\ref{fig:Figure67} (which also documents the present,
insufficient data situation). Of course, we have to note that the lattice
results concerning the critical point are not yet final
\cite{270}, and neither is the hydrodynamic treatment \cite{296}
concerning the EOS in the vicinity of the CEP.\\
\begin{figure}[h!]   
\begin{center}
\includegraphics[scale=0.55]{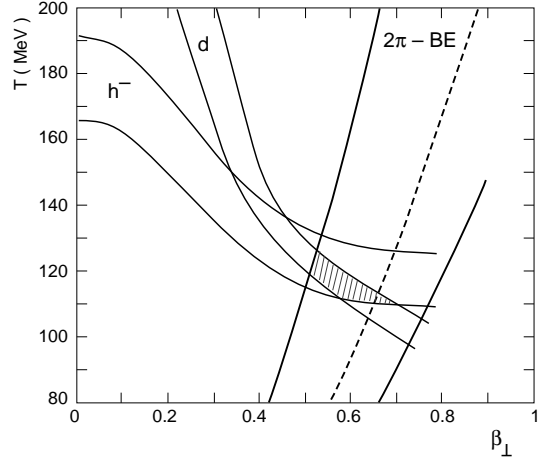}\vspace{-0.4cm}
\caption{Allowed regions of freeze-out temperature vs.\ radial
expansion velocity for central Pb+Pb collisions at $\sqrt{s}=17.3 \:
GeV$ and mid-rapidity, combining negative hadron and deuterium
spectral data analysis with BE $\pi^-$ correlation results on $K_T$
dependence of $R_{\perp} \approx R_s$ \cite{90}.}
\label{fig:Figure70}
\end{center}
\end{figure}\\
We turn to combined analysis of the two final processes of expansive
evolution, formation of $p_T$ spectra and decoupling to
Bose-Einstein pair correlation, in order to show how the
experimental information mentioned above can be gathered. At the
level of ''hydrodynamically inspired'' analytic models for the
emission function $S(x,K)$ several approaches \cite{106,297} have
established the connection, arising from radial flow, between
hadronic $m_T$ spectra and the $K_T$ dependence of the HBT radii
$R_{side}, \: R_{out} \:  \mbox{and} \: R_{long}$, equation~\ref{eq:equation88},~\ref{eq:equation90}, which
fall down steeply with increasing $K_T$ \cite{291,292,293,294}. A combined
analysis lifts continuous ambiguities between the thermal freeze-out
temperature $T_f$ and the radial expansion velocity $\beta_T$ (for
which a radial profile $\beta_T=\frac{r}{R_{side}} \: \beta_0$ is
assumed), that exist in the blast wave model derivation of both the
$p_T$ spectra, equation~\ref{eq:equation26}, and the $K_T$ dependence of $R_s$.

\begin{figure}[h!]   
\begin{center}
\includegraphics[scale=0.9]{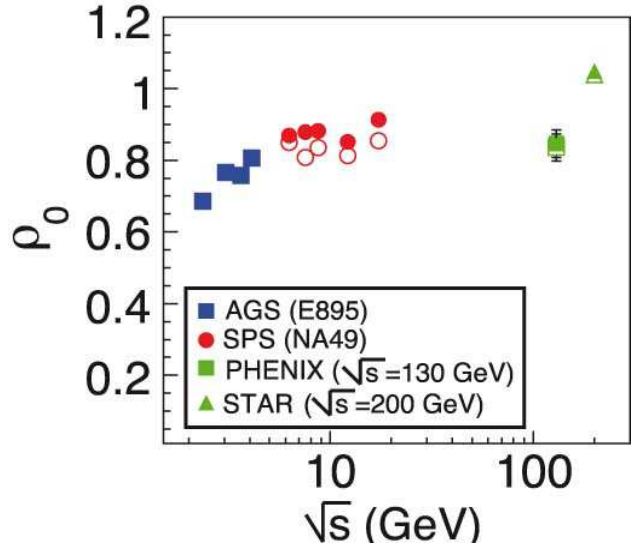}\vspace{-0.4cm}
\caption{The surface velocity (here denoted as $\rho_0$) of radial expansion at decoupling in central
Au+Au and Pb+Pb collisions vs. $\sqrt{s}$. From combined analysis of
hadron $p_T$ spectra and $\pi^-$pair correlation \cite{291,297}.}
\label{fig:Figure71}
\end{center}
\end{figure}

This was first demonstrated in a NA49 study of pion correlation in
central Pb+Pb collisions at $\sqrt{s}=17.3 \: GeV$ \cite{90} that is
shown in Fig.~\ref{fig:Figure70}. The ambiguous anti-correlation
of fit parameters $T_f$ and $\beta^2_T$ can be somewhat constrained if
several hadronic species are fit concurrently. This is obvious in Fig.~\ref{fig:Figure70} from the overlap of
parametrization regions for negative hadron, and for deuterium $m_T$
spectra. An {\it
orthogonal constraint} on $\beta_T^2/T_f$ results from a blast wave
model fit of the HBT $K_T$ dependence of $R_T \approx R_{side}$
\cite{298}, employed here. We see that the three independent $1 \:
\sigma$ fit domains pin down $T_f$ and $\beta_T$ rather sharply, to
$[115 \: MeV, 0.55]$. A relativistic correction \cite{297}
leads to $T_f = 105 \: MeV$, $\beta_T = 0.60$. The $\sqrt{s}$ dependence of
$\beta_T$ at the freeze-out surface, from such an analysis
\cite{291,297} is shown in Fig.~\ref{fig:Figure71}.  The data again exhibit a plateau
at SPS energies, which remains to be understood \cite{299}.
In the light of the considerations above, the plateau might turn out to reflect a critical point focussing effect.\\

%% file: Chapter_7.tex
\chapter{Summary}
\label{chap:Conclusion}

We have seen that many physics observables in relativistic nucleus-nucleus collisions can, at RHIC energy $\sqrt{s}=200GeV$, be related to the primordial dynamical phase, from initial QCD parton shower formation to local momentum space equilibration. The time interval is 0.35 to 0.65 fm/c.
This domain can be investigated at RHIC due to the short interpenetration time, of 0.15fm/c. From among the bulk hadron production signals, total and midrapidity charged particle rapidity densities, $dN_{ch}/dy$, reflect the primordial parton saturation phenomenon~\cite{72}. It leads to an unexpectedly slow increase of multiplicity with fireball participant number $N_{part}$, and with $\sqrt{s}$. This observation signals the onset of
nonperturbative QCD, a coherent shower multiplication by multigluon coherence~\cite{62,65,70,71,75}. It is expected to be even more dominant in LHC
Pb+Pb collisions at 5.5TeV. Furthermore, elliptic flow, a collective bulk hadron emission anisotropy, also originates from the primordial, nonisotropic spatial density profile of shower-produced partons \cite{93,96}. A hydrodynamic evolution sets in at $t < 1fm/c$, implying the existence of a primordial equation of state (EOS) for partonic matter in local equilibrium. Moreover, the experimental elliptic flow data at RHIC are well described by "ideal fluid" hydrodynamics, i.e. by a very small shear viscosity $\eta$ and a small mean free path $\lambda$~\cite{94,95}.

These observations indicate the existence of a (local) equilibrium state at very early times in the bulk parton matter stage of the dynamical evolution. This quark-gluon plasma (QGP) state of nonperturbative QCD was predicted by QCD lattice theory~\cite{11}. In Au+Au collisions at RHIC energy this state appears to be realized under the following conditions~\cite{61}.

The energy density $\epsilon$ amounts to $6-15 GeV/fm^3$, far above the parton-hadron QCD confinement region at $\epsilon= 1GeV/fm^3$, and at about 55 times the density of nuclear ground state matter, $\rho_0 = 0.14GeV/fm^3$.
Translating to macroscopic units we are dealing with a matter density of about $1.3 \cdot 10^{19} kg/m^3$, the density prevailing in the picosecond era of
the cosmological evolution.
The corresponding temperature amounts to T=300-330 MeV (about $3.6 \cdot 10^{12}$
degrees Kelvin), far above the "Hagedorn limit"~\cite{38} for the hadronic phase ($T_H=165MeV$).
From an analysis of the production ratios of the various hadronic species (from pions to Omega hyperons) arising from this primordial partonic state at hadronization, the statistical hadronization model (SHM) determines its
baryo-chemical potential~\cite{108}, $\mu_B = 20 MeV$ at RHIC energy. This value indicates that one is already close to the near-zero net baryon number conditions, $\mu_B \approx 0$, prevailing in the corresponding big bang evolution, where the density of particles exceeds that of antiparticles by a fraction of about $10^{-9}$ only.

Overall we thus obtain entries in the QCD phase diagram of Fig.~\ref{fig:Figure1}. RHIC creates a parton plasma at about $T=300MeV$ and $\mu_B = 20MeV$. It turns out to behave like an ideal fluid and features an extremely short mean free path
$\lambda$, thus inviting a description by relativistic hydrodynamics. The small shear viscosity $\eta$ (or the small viscosity to entropy ratio $\eta / s$) are highlighted by the striking observation that even the fluctuations of primordial parton density in individual, single events,
are preserved in the single event variation of elliptic flow~\cite{267,268}.
Moreover, the observed scaling of elliptic flow with hadron valence quark
number~\cite{94} - and thus not with hadronic mass as in radial flow~\cite{103} -
confirms the implied origin of the elliptic flow signal from the partonic phase.

At the LHC the phase of early QCD plasma formation is expected to shift to yet higher energy density, in the vicinity of $T=600MeV$ and $\mu_b = 5 MeV$.One is thus getting nearer to the domain of QCD asymptotic freedom, and might expect a falloff of the partonic cross section which is extremely high at RHIC~\cite{61}, as reflected by the small $\eta/s$ and $\lambda$ values.

The observed features of the QCD plasma produced at RHIC energy have invited the terminology of a "strongly coupled" quark-gluon plasma (sQGP~\cite{165}). Further evidence for the strongly coupled, non perturbative nature of the primordial partonic state stems from the various observed, strong in-medium attenuation effects on initially produced high $p_T$ partons. In Au+Au collisions at $\sqrt{s}=200GeV$ this high medium opacity leads to a universal quenching of the high $p_T$ hadron yield~\cite{175}
including, most remarkably, heavy charm quark propagation to D mesons~\cite{175,178}. We have shown in chapter~\ref{chap:In_medium_high_pt} that the interior of the collisional fireball at midrapidity is almost "black" at $t <1fm/c$.This is also reflected in a strong suppression of the back-to-back correlation of
hadrons from primordially produced di-jets~\cite{207,208}, and in a similarly strong suppression of the $J/\Psi$ yield~\cite{237} which we have shown in chapter~\ref{chap:Vector_Mesons}
to be ascribed to an in-medium dissolution of the primordially produced
$c \bar{c}$ pairs~\cite{41} at $T$ about 300MeV.

The underlying high medium opacity can be formally expressed~\cite{188,193,195}
by an effective parton transport coefficient $\hat{q}$ (equations~\ref{eq:equation59} and~\ref{eq:equation63})
which quantifies the medium induced transverse momentum squared
per unit mean free path $\lambda$. The value of $\hat{q}$ derived from analysis of the various attenuation phenomena turns out to be almost an order of magnitude higher than what was expected from former, perturbative QCD based models~\cite{196}. Analogously, $\eta/s$ has turned out to be much smaller than the previous perturbative QCD expectation~\cite{92}. The two quantities may be related~\cite{300} via the heuristic expression
\begin{equation}
\frac{\eta}{s} \approx 3.75 \ C \ \frac{T^{3}}{\hat{q}}
\label{eq:equation91}
\end{equation}
with C a to be determined constant; C = 1/3 from~\cite{300}. This relation shows that a larger value of $\hat{q}$ implies a small value for the ratio $\eta/s$. The latter has a lower bound by the general quantum gauge field theory limit $\eta/s$ $\ge$ $(4\pi)^{-1}$ \cite{171}, a value not too far from the estimate $\eta/s$ = 0.09 $\pm$ 0.02 derived from relativistic hydrodynamics applied to elliptic flow $v_{2}$~\cite{94, 170}. As a consequence, $\hat{q}$ can not grow beyond a certain upper bound that should be established at LHC energy.

These considerations are an example of the recent intense theoretical search for alternative methods of real-time strong coupling calculations, complementary to lattice QCD.
In this regime, lattice QCD has to date been the prime non-perturbative calculational tool. However, understanding collective flow, jet quenching and primordial photon radiation requires real time dynamics, on which lattice QCD information is at present both scarce and indirect. Complementary methods for real-time strong coupling calculations at high temperature are therefore being explored.
For a class of non-abelian thermal gauge field theories, the conjecture of a correspondence between anti-de Sitter space-time theory and conformal field theory (the so-called AdS/CFT conjecture) has been shown~\cite{301} to present such an alternative. It maps nonperturbative problems at strong coupling onto calculable problems of classical gravity in a five-dimensional anti-de Sitter (ADS$_{5}$) black hole space-time theory~\cite{302}. In fact, this formalism has been recently applied~\cite{212} in a calculation of the transport coefficient $\hat{q}$ that governs in-medium jet attenuation, resulting in an effective, expansion time averaged $\hat{q}_{eff}$ = 5 GeV$^{2}$/fm at T = 300 MeV corresponding to top RHIC energy, rather close to the experimental estimates (c.f.\ Figs.~\ref{fig:Figure41} and~\ref{fig:Figure48}).

Thus, it does not seem to be too far-fetched to imagine~\cite{301} that the quark-gluon plasma of QCD, as explored at RHIC, and soon at the LHC (and theoretically in lattice QCD), and the thermal plasma of certain supersymmetric conformal gauge field theories (for example N = 4 "Super-Yang-Mills" (SYM) theory as employed in~\cite{212, 301}) share certain fundamental common properties. 

The focus on early time in the dynamical evolution of matter in nucleus-nucleus collisions is specific to RHIC energy as the initial interpenetration period of two Lorentz contracted mass 200 nuclei amounts to $0.15 fm/c$ only. The subsequent evolution is thus reflected in a multitude of observables. It is, at first, described as a one-dimensional Hubble expansion~\cite{61}, setting the stage for the emergence of the medium specific quantities addressed above ( gluon saturation,direct photon production, hydrodynamic elliptic flow, jet quenching and $J/\Psi$ suppression).
These observables tend to settle toward their eventually observed patterns at
$t \le 1.0-1.5 fm/c$, owing to the fact that they are most prominently determined
under primordial conditions of high temperature and density. For example,
photon production reflects $T^4$, and the transport coefficient $\hat{q}$ falls
with $T^3$~\cite{212}.

On the contrary, at the energy of former studies at the CERN SPS, $6 \le \sqrt{s} \le 20GeV$, such early times stay essentially unresolved as the initial interpenetration takes upward of 1.5fm/c. A fireball system in local (or global) equilibrium thus develops toward $t = 3fm/c$, at $T$ about
220MeV, closer to the onset of hadronization~\cite{85,86}. Also the baryo-chemical potential is much higher than in RHIC collisions, $250 \le \mu_B \le 450MeV$. However, we thus gain insight into the QCD physics
of the hadronization, and high $\mu_B$ domain of the phase diagram sketched in Fig.~\ref{fig:Figure1}, in the vicinity of the conjectured parton-hadron coexistence line of QCD.

For reference of such data, e.g.\ statistical species equilibrium (chapter~\ref{chap:hadronization}), dilepton studies of light vector meson "melting" at the phase boundary (chapter~\ref{sec:Low_mass_dilepton_spectra}), and hadronic event-by-event fluctuations (chapters~\ref{sec:Critical_Point} and~\ref{sec:Critical_Fluctuations}), to theoretical QCD predictions, a recent progress of lattice QCD~\cite{8,9,10} is of foremost importance.
The technical limitation of lattice QCD to the singular case of vanishing chemical potential, $\mu_{B}$ = 0 (which arises from the Fermion determinant in the lattice formulation of the grand canonical partition function), has been overcome recently. Three different approaches have been outlined, the respective inherent approximation schemes touching upon the limits of both the mathematical and conceptual framework of lattice theory, and of present day computation power even with multi-teraflop machines. First results comprise the prediction of the parton-hadron phase boundary line, which interpolates between the well studied limits of the crossover-region at $\mu_{B}$ $\rightarrow$ 0, $T$ $\ge$ $T_{c}$ and the high net baryon density, low T region for which a first order character of the phase transition has been predicted by chiral QCD models~\cite{15}. We have illustrated this line in Fig.~\ref{fig:Figure1}, and we have shown in chapter~\ref{chap:hadronization} that hadronic freeze-out occurs at, or near this line at $\sqrt{s} \ge 17.3 GeV$ (top SPS energy).
The coexistence line includes an intermediate (T, $\mu_{B}$) domain featuring a critical point of QCD at which the first order line at higher $\mu_{B}$ terminates, in a critical domain of (T, $\mu_{B}$) in which the transition is of second order. One thus expects the nature of the confining hadronization transition $-$ an open QCD question $-$ to change from a crossover to a second order, and onward to a first order characteristics in a relatively small intervall of $\mu_{B}$ that is accessible to nuclear collision dynamics at the relatively modest $\sqrt{s}$ of about 5 to 15 GeV. This domain has as of yet only received a first experimental glance, but the top AGS and low SPS energy experiments exhibit phenomena that can be connected to the occurrence of a critical point and/or a first order phase transition, notably the ``SPS plateau" in $\left< m_T \right>$, the non-monotoneous $K^{+}/\pi^{+}$ excitation function, and the eventwise fluctuations of this ratio (chapters~\ref{subsec:Bulk_hadron_transverse_spectra}, \ref{chap:hadronization}, \ref{sec:Critical_Point} and~\ref{sec:Critical_Fluctuations}). A renewed effort is underway at RHIC, at the CERN SPS and at the future GSI FAIR facility to study hadronization, in-medium meson modification induced by the onset of QCD chiral restoration, as well as critical fluctuations and flow, in the low $\sqrt{s}$ domain.

%% file: Bibliography.tex
{}